\documentclass[final,3p,times]{elsarticle}

\usepackage{hyperref,multirow,graphicx,booktabs,amsmath,amssymb,mathrsfs,bookmark}

\journal{arXiv}

\begin{document}

\begin{frontmatter}

\title{Computational Socioeconomics}

\author[1,2,3]{Jian Gao}
\ead{gaojian08@hotmail.com}

\author[4,5]{Yi-Cheng Zhang\corref{cor1}}
\ead{yi-cheng.zhang@unifr.ch}
\cortext[cor1]{Corresponding author at: Department of Physics, University of Fribourg, CH-1700 Fribourg, Switzerland.}

\author[1,3,6]{Tao Zhou\corref{cor2}}
\ead{zhutou@ustc.edu}
\cortext[cor2]{Corresponding author at: CompleX Lab, University of Electronic Science and Technology of China, Chengdu 611731, PR China.}

\address[1]{CompleX Lab, University of Electronic Science and Technology of China, Chengdu 611731, PR China.}
\address[2]{MIT Media Lab, Massachusetts Institute of Technology, Cambridge, MA 02139, USA.}
\address[3]{Big Data Research Center, University of Electronic Science and Technology of China, Chengdu 611731, PR China.}
\address[4]{Institute of Fundamental and Frontier Science, University of Electronic Science and Technology of China, Chengdu 610054, PR China.}
\address[5]{Department of Physics, University of Fribourg, CH-1700 Fribourg, Switzerland.}
\address[6]{Institution of New Economic Development, Chengdu 610110, PR China.}

\begin{abstract}
Uncovering the structure of socioeconomic systems and timely estimation of socioeconomic status are significant for economic development. The understanding of socioeconomic processes provides foundations to quantify global economic development, to map regional industrial structure, and to infer individual socioeconomic status. In this review, we will make a brief manifesto about a new interdisciplinary research field named \emph{Computational Socioeconomics}, followed by detailed introduction about data resources, computational tools, data-driven methods, theoretical models and novel applications at multiple resolutions, including the quantification of global economic inequality and complexity, the map of regional industrial structure and urban perception, the estimation of individual socioeconomic status and demographic, and the real-time monitoring of emergent events. This review, together with pioneering works we have highlighted, will draw increasing interdisciplinary attentions and induce a methodological shift in future socioeconomic studies.
\end{abstract}


\end{frontmatter}

\tableofcontents

\clearpage

\section{Introduction}
\label{Intro}

Many branches of science have experienced the paradigm shift from qualitative to quantitative studies. Even the most representative one for quantitative sciences, physical science, has undergone a long period for qualitative explorations in its early stages. For example, more than two thousand years ago, Aristotle raised the famous \emph{four elements theory}, which claims that the four classical elements, namely earth, water, air and fire, are the material basis of the physical world. At almost the same time, some Chinese ancient philosophers proposed the \emph{Wu Xing theory} (i.e., the Chinese five elements theory), which is a fivefold conceptual scheme that uses the proportion of ingredients and movements of the five elements (i.e., metal, wood, water, fire and earth) to explain a wide array of phenomena, from the cosmic cycles to the validity of a dynasty. For about two thousand years, the ancient Greek system, contributed by Aristotle and some others, represents the most advanced understanding of the world, which is indeed one of the most influential theory in human history. Up to the end of the middle ages, thanks to the quantitative analyses and experimental verifications, these ancient theories, such as Aristotle's four elements theory and kinetic theory, were progressively replaced by modern scientific theories like the Atomic theory and the Newton's laws.

In contrast to physical science that concentrates on the study of matter and its motion through space and time, social science investigates the social structure based on the activities of and relations between human beings, including sociology, economics, politics, linguistics, jurisprudence, and many other branches. In comparison with physical science, the way from qualitative to quantitative studies is more difficult for social science. On the one hand, the objects under social science study are much more complex than those under physical science study. An individual person is one of the most important units for social science study, playing an analogous role to an atom in physical science \cite{Ball2004}. However, human behaviors exhibit heterogeneity and burstiness: different people have much different behavioral patterns and even the same person shows far different behaviors in different spaces and times \cite{Barabasi2010}. Therefore, except a certain success in analyzing the flow of human crowds \cite{Hughes2003,Helbing2010}, to treat human beings as atoms will kill many interesting social phenomena. Some other objects under study are naturally not easy to be characterized numerically, such as policies and legal provisions. On the other hand, social science study inevitably suffers from uncertainty and incompleteness. The factors affecting social development are countless, and thus any seemingly coverall theory cannot include all relevant factors and be self-contained. In addition, every single factor is unstable and not independent, being affected by other factors and the external environment. The above intrinsic complexity makes it infeasible to quantitatively test and verify any social theory through controllable repeated experiments in a closed environment, while such experimental verification is indeed the methodological cornerstone that pushes forward physical science and other branches of natural science \cite{Popper2005}. At the same time, social science is not good at quantitative predictions to the future, with many predictions from experts and complicated theories being no better than wild guesses \cite{Silver2012}. What a pity is that such incorrect predictions cannot subvert the corresponding social theories (much different from physical science) since the mistakes are attributed to the unknow/undetected factors or emergent events \cite{Taleb2007}, instead of the flaws of the theories themselves.

Up to now, along with the development of quantitative methods, social science has successfully learned how to be wise after the event. That is to say, we can always find some theoretical models (possibly together with some cosmetic changes) to provide qualitatively correct or even quantitatively accurate explanations after the event. However, these theories are usually powerless in predicting the future. Confronting such straits, social scientists should not turn back to the qualitative description, but insist on quantitative explanation and prediction, and evaluate the validity of a theory based on its explanatory power and prediction accuracy before the event. In fact, social science study recently shows higher and higher level of quantification and becomes increasingly dependent on real data \cite{Lazer2009,Shah2015}. However, the traditional way to obtain real data has many limitations. For example, survey data from questionnaires and self-reports usually contains a small number of samples and suffers from social desirability bias (i.e., subjects tend to give socially acceptable answers, instead of the real facts) \cite{Fisher1993}. Larger-scale and more precise data, such as data from economic census, usually consumes huge resources and lacks timeliness. In many poor countries and regions, population-scale economic census is not feasible. Fortunately, thanks to the digital wave that sweeps across the whole world \cite{Mayer2013}, social scientists have an unprecedented opportunity to develop a quantitative methodology. Indeed, it is for the first time in history, data in the processes of social and economic development, as well as the data of human activities, are recorded by more and more sensing devices, online platforms and other data acquisition terminals. However, these data are not well-structured and are different from the normally handled data in social science. Typical examples include satellite remote sensing data, mobile phone data, social media data, and so on. On the one hand, to understand and analyze these data asks for advanced techniques in data mining and machine learning, which is a considerable challenge to traditional social scientists. On the other hand, these data are of larger size, almost in real time and with higher resolution, which can reduce the sparsity and bias in small-size data, and reduce the invisible parts in the developing processes (e.g., data points in two consecutive censuses are usually across a few years, and the changes in between are not visible). Therefore, based on these large-scale novel data, we can in principle make great progress in perceiving socioeconomic situations, evaluating and amending known theories, enlightening and creating new theories, detecting abnormal events, predicting future trends, and so on.

The above-mentioned challenges and corresponding attempts have led to the emergence of a new scientific branch, which studies various phenomena in socioeconomic development by using quantitative methods that based on large-scale real data, with particular attention to the economic development problems related to social processes and the social problems related to economic development. We name it as \emph{Computational Socioeconomics}, which is immature, but future-pointing and burgeoning. The computational socioeconomics can be considered as a new branch of socioeconomics resulted from the transformation of methodology, or as a new branch of computational social science by emphasizing on socioeconomic problems.

In the above definition, three keywords are worth paying close attention to. The first one is ``quantitative methods'', which emphasizes the usage of numerical values, rather than qualitative description, in characterizing problems and presenting results. In the 5th century BC, the ancient Greek doctor Hippocrates (who is often referred to as the ``Father of Medicine'') proposed the \emph{four temperaments theory}, which suggests that there are four fundamental personality types: sanguine, choleric, melancholic, and phlegmatic, and the personality type of an individual is determined by the excess or lack of four body fluids: blood, yellow bile, black bile, and phlegm \cite{Merenda1987}. Such a qualitative theory, analogous to the impacts of the four elements theory on physical science, has ruled social psychology (in particular the studies on personality) for more than two thousand years. In despite of some reasonable ingredients, the four temperaments theory has stayed on the level of qualitative description, and thus failed to accumulate scientifically solid achievements in its long-time development. Only after modern psychologists obtained quantitative evaluations of the Big Five personality traits via standard scales, personality analysis became an important research domain that plays central roles in many issues of social psychology \cite{Gosling2003}. Such example show the importance and necessity of the development of quantitative methods. The second one is ``real data'', which emphasizes that any theoretical model should respect real data and use the explanatory power and prediction accuracy for real data as the evaluation criteria for its validity. Economics shows a high level of quantification, with most theoretical models being precisely described by a group of elegant equations. Accordingly, given the values of necessary parameters, many targeted economic variables are calculable. However, the majority of economic theories have cocooned themselves in a quantitative fantasyland consisted of ideal assumptions while largely ignored real data. It eventually makes the classical economic theories beautiful rather than practical. For the short term, it cannot predict the upcoming economic crisis \cite{Battiston2016} (but it can always find out graceful and reasonable theoretical explanations after the crisis \cite{Reinhart2009}). For a long time, it failed to provide effective strategies on economic development for more than a hundred of developing countries over the world \cite{Lin2011}. The third one is ``large-scale'', which emphasizes the importance of population-scale data (i.e., the data that can directly reflect the entire population under study, instead of a small sample). A very small data set may not only bring statistical bias, but result in completely wrong conclusions. For example, a widely accepted theory by academic community, which has also been validated by various experiments on small-scale social networks, is that the interacting strength between two connected individuals (which can be measured by the frequency and duration of mobile communication or the number of comments, replies and mentions on a social platform, and so on) decays as the increase of the range of their link (the range of a link is defined as the shortest distance between its two endpoints after the removal of this link, and a large link range indicates that the two corresponding endpoints locate in two distant communities with few overlapping nodes) \cite{Granovetter1973,Onnela2007}. A very recent experiment on 11 population-scale social networks, however, shows that the interacting strengths through very long-range links are not weaker than those through short-range links \cite{Park2018}, which fundamentally challenges our traditional understanding of social network organization.

\begin{figure}[t]
  \centering
  \includegraphics[width=0.4\textwidth]{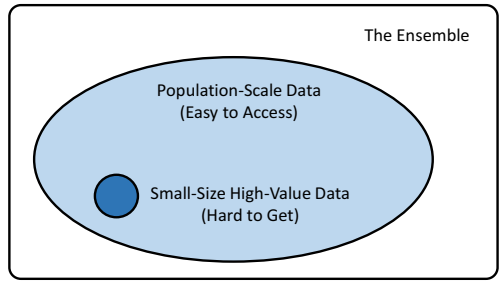}
  \caption{Illustration of the relationship between the entire population, the easily accessible population-scale data and the small-size high-value data. The small-size data set contains some high-value data dimensions that do not appear in the original population-scale data.}
\label{Fig_1}
\end{figure}

In comparison with routine methods in social science, the increasing diversity and volume of data lead to methodological changes in two aspects. Firstly, simple statistical tools are not suitable for analyzing unstructured data, such as remote sensing images, street views, social networks, textual content, and so on. Therefore, researchers are badly in need of artificial intelligence, in particular advanced techniques of data mining and machine learning, such as deep learning \cite{Lecun2015}. Secondly, with population-scale data, sampling is not a necessary method to estimate the statistical properties of the whole population. Instead, one can concentrate on a small-size subset sampled from the original data in hand, and add new dimensions of data. These new data dimensions are usually of high values, which can be obtained from traditional ways like manual labelling and questionnaire survey. Using such a small sample as the training data, one can learn a model to infer new dimensions of data from the original dimensions. Applying such model to the whole data set, new dimensions of data for all individuals appeared in the original dataset can be obtained in principle. Such method integrates some routine methods like sampling, labelling and surveying, while it is much more powerful. For example, it is relatively easy to obtain the population-scale data on mobile communication and mobility (all can be obtained from mobile phones), in contrast, it is very hard to know the household income of every family since a poor country cannot support a population-scale economic census and such data is usually treated as official secrets that are not open for public or research institutions. Under the circumstances, we can obtain household incomes of a certain number of families (what we need is just a tiny fraction of all families) via routine questionnaires. These much smaller data set can be used as training data, based on which we can apply machine learning techniques to build a model that can predict household income of a family from the mobile phone data of the family members \cite{Blumenstock2016}. Although the inferred data is not perfect, it can be very close to the real data under a certain well-designed algorithm. Notice that, a significant advantage is that the high-value data for almost every individual can be obtained at a very low cost. As shown in Figure~\ref{Fig_1}, combining the accessible population-scale data, a small sample of high-value but hard-to-get data, and a properly selected or well-designed algorithm to infer the high-value data for individuals other than the sample is a novel and representative method in the computational socioeconomics study, showing the deep integration of social science and computer science methods.

Long-term speaking, no matter computational socioeconomics will become a mature branch of science with distinct borderlines or it will completely integrate into the framework of traditional social science, the above-mentioned novel perspective and methodology, driven by big data and artificial intelligence, will definitely become the mainstream in the future and change the landscape of science research in a profound and irrevocable way. Inspired by this positive judgment, we decide to present this review article. In addition, there are three technical reasons for us to write this review. Firstly, computational socioeconomics is an emerging research domain with research findings published in disparate journals and conference proceedings across many disciplines. Therefore, it is necessary to collect these results together. Secondly, we would like to sort and classify representative results according to the objects and data sets under study, so that it is easy for readers to see the landscapes of both methods and achievements. A proper taxonomy can largely reduce the difficulty to master the related knowledge and methods. Although the presented one is built just according to the current progresses of this field, it will evolve to be a more systematic and reasonable one along with further studies. Thirdly, in the nascent stage of computational socioeconomics, different research articles used different expressions to describe essentially the same problems and methods, and thus it is valuable to unify the problem description and the symbolic system. In a word, we hope this review will become a handbook for researchers who are willing to contribute to the development of computational socioeconomics. Furthermore, the paradigm shift in methodology, as presented in this review, is not only relevant to socioeconomics, but also to most branches of social science and to many other qualitative disciplines beyond social science.

The remainder of this review article is organized as follows. The second section will discuss some important problems at the macroscopic scale, such as the world economic development, the competitive powers of countries, the inequality problem, and so on. The third section will mainly concentrate on the urban scale and introduce some novel ways to solve problems related to the regional economic development, such as how to precisely perceive regional socioeconomic status and how to choose the suitable development paths and strategies for a city. The fourth section will focus on individual level, discussing how to make use of some unobtrusive data to estimate the individual socioeconomic status, including income, employment situation, and even health condition. The fifth section will go a little beyond the scope of computational socioeconomics, and to discuss how the frequently-used data in the previous sections can be utilized to benefit the emergency management and disaster assistance. We cover such issue because the emergency management is an increasingly important social problem and the reported methods are consistent to the methods introduced in the previous sections. In this review, many different data resources have played important roles, among which the following three are the most important: remote sensing satellites, mobile phones and social media platforms. For each of the three data resources, there are some certain representative analytics tools and methods, so it is very effective to sort the results according to the data resources and corresponding methods. Finally, in the last section, we will summarize representative progresses, explore the tendency of the development of computational socioeconomics, discuss the challenges and opportunities in this emerging field, and outline some potentially interesting and significant open issues.

\section{Global development, inequality and complexity}
\label{Sec2}

\subsection{World development and poverty mapping}

Revealing the status of economic development is one of the long-standing problems in socioeconomics \cite{Kuznets1955,Gao2016}. Recently, data with improved quantity and quality have been used to map nations' economic characteristics such as poverty, which comes with economic development and is a major cause of societal instability. Based on an international poverty line at USD 1.25 a day in 2008 \cite{Ravallion2009}, 1.2 billion people (21\%) lived in poverty in 2012 \cite{Ravallion2013}. Reducing poverty is thus a key target of the Millennium Development Goals (MDGs). To approach this goal, the first step is to accurately map the spatial distribution of poverty \cite{Hulme2007}. New data and tools have been utilized to better reveal, explain and predict global poverty and economic inequality. In this section, we will briefly introduce literature that map poverty from satellite imagery, infer socioeconomic status from mobile phone (MP) data and fight against poverty with combined data.

\subsubsection{Remote sensing observes poverty}

Remote sensing (RS) is the acquisition of information by using sensor technologies to detect objects on earth, which is originally used in earth science disciplines \cite{Paul1981}. In recent years, high resolution data from RS, for example, nighttime lights (NTLs) satellite imagery, has been used to supply information about economic activity, especially in developing countries where traditional economic census data are insufficient \cite{Ghosh2013}. With a great potential for recording the presence of humanity on the surface of earth, NTLs data can provide an unambiguous indication of the spatial distribution of economic development. Indeed, NTLs have been found to be a powerful predictor of ambient population density and economic activity. Nightsat \cite{Elvidge2007} is a concept for a satellite system, which is capable of global observation. The Nightsat can capture the location and density of lighted infrastructures within human settlement.

One of the pioneering works by Elvidge et al. \cite{Elvidge1997} suggested that NTLs data can be used as a proxy for socioeconomic development in developing countries. Lighted area has a high correlation with the gross domestic product (GDP) and electric power consumption. Moreover, lighted area is strongly correlated with GDP for 21 countries. Latter, by combing lighted area with ancillary statistical information of a city, Doll et al. \cite{Doll2000} investigated the potential of NTLs data for quantitative estimation of global socioeconomic parameters. They found that the country-level total lighted area exhibits significantly high correlations with GDP and $\text{CO}_2$ emission. Sutton and Costanza \cite{Sutton2002} estimated the amount of light energy (LE) from satellite images with global coverage at a high spatial resolution. They found that LE is correlated with GDP at the national level and can serve as a more accurate indicator of economic activity. That is because LE is more spatially explicit and can be directly observed and easily updated almost in real time.

Together with census and survey data, NTLs data have been applied in mapping poverty. Ebener et al. \cite{Ebener2005} applied regression methods using NTLs imagery to model the distribution of wealth within 171 countries at the national level and 26 countries at the subnational level. They showed that NTLs data is correlated with GDP per capita and other socioeconomic indicators. Noor et al. \cite{Noor2008} computed asset-based poverty by applying the principal component analysis (PCA) to NTLs data and household survey data of 37 countries in Africa. They found that the mean brightness of NTLs data can offer a robust and inexpensive alternative to asset-based poverty indices derived from survey data, suggesting that it is possible to explore and track economic inequity at subnational levels by leveraging NTLs data. For Uganda, Rogers et al. \cite{Rogers2006} presented a discriminant analysis model to predict poverty after combining satellite imagery and household survey data. They estimated the poverty index by the likelihood of each pixel falling within a specified poverty class. They found that external and independent data have descriptive power for poverty mapping. These novel data sources are likely to outperform socioeconomic datasets that are internally correlated and exploited by small area methods.

\begin{figure}[t]
  \centering
  \includegraphics[width=0.6\textwidth]{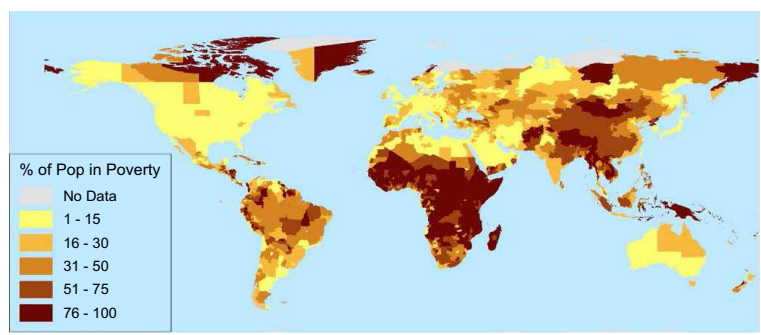}
  \caption{Percentage of population in poverty for subnational administrative units. The world poverty map was estimated based on satellite data-derived poverty index. Figure from \cite{Elvidge2009}.}
\label{Fig_3_1_1}
\end{figure}

A spatially disaggregated global poverty map of 233 countries was produced by Elvidge et al. \cite{Elvidge2009}. The poverty levels were estimated by dividing the LandScan population count data \cite{Dobson2000} by the brightness from NTLs data (see Figure~\ref{Fig_3_1_1}). The produced poverty indices correlate very strongly with other widely accepted measures, suggesting that satellite imagery can enhance the knowledge of socioeconomic conditions around the world at a fine spatial resolution. Later, Ghosh et al. \cite{Ghosh2010} proposed a model to estimate the world-wide economic activity. In their model, a grid of nonagricultural economic activity was created according to the NTLs, while a grid of agricultural activity was created according to the LandScan population grid. Then, by integrating the two grids, a disaggregated map of total economic activity was produced, which can provide an alternative means for measuring global economic activity and predicting future socioeconomic trends.

To better estimate true income growth from NTLs, Henderson et al. \cite{Henderson2012} developed a statistical framework to estimate two parameters. One is a coefficient that maps NTLs growth into a proxy for GDP growth, and the other is an optimal weight to combine this proxy with national account data. After applying the method to countries with very low-quality national account data, Henderson et al. \cite{Henderson2012} demonstrated the key role of NTLs data in analyzing growth at the subnational and supranational levels. With the NTLs data, income data and papulation data of 748 regions across 54 countries in Africa, Mveyange \cite{Mveyange2015} estimated the regional income inequality by calculate two standard measures of inequality, the Gini index and the mean log deviation (MLD) measure \cite{Khandker2009}. After presenting the empirical model, they showed that the estimated inequality index has significant and positive correlations with income-based regional inequality indicators, suggesting that NTLs are good proxies to estimate regional inequality. These results are especially meaningful in the lack of reliable and consistent subnational income data.

Cauwels et al. \cite{Cauwels2013} explored the dynamics and spatial distribution of global NTLs for 160 different countries. They found that the center of light moves eastwards about 60 km per year, and there is a tendency of global centralization of light. After introducing spatial light Gini coefficients, they found a universal pattern of human settlements across different countries. Ghosh et al. \cite{Ghosh2013} summarized literature that leveraged NTLs to develop a variety of alternative measures of human well-being. They introduced the application of NTLs to estimate various human well-being indicators (e.g., GDP, poverty, informal economic activity and remittances), develop the night light development index (NLDI), map the human ecological footprint, measure the electrification rates and estimate the ICT Development Index (IDI). Recently, Bennett et al. \cite{Bennett2017} summarized the methods to correlate NTLs with socioeconomic parameters including urbanization, economic activity and population. They highlighted the value of NTLs for detecting, estimating and monitoring socioeconomic dynamics.

\begin{figure}[t]
  \centering
  \includegraphics[width=0.6\textwidth]{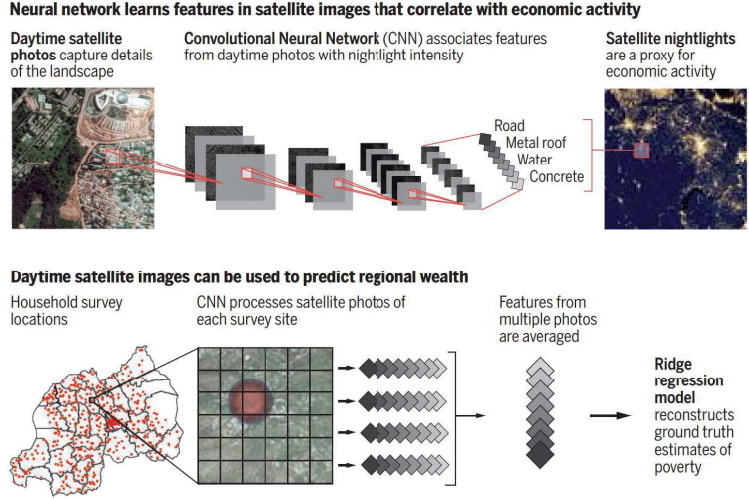}
  \caption{Predicting poverty from satellite images using Convolutional Neural Network (CNN). Figure from \cite{Blumenstock2016}.}
  \label{Fig_3_1_2}
\end{figure}

NTLs data are successful in revealing economic activity, however, it is not effective for less developed areas due to the uniformly dark of satellite imagery in these areas. To this point, Jean et al. \cite{Jean2016} applied deep learning algorithms to learn the relationship between NTLs and daytime satellite imagery. The former can predict the wealth distribution while the latter contains rich information about landscape features. They employed a multi-step transfer learning approach \cite{Pan2010} to train a convolutional neural network (CNN) \cite{Xie2016}. In particular, a linear chain transfer learning graph was constructed. First, they transferred knowledge from the object recognition on the ImageNet (Problem 1) \cite{Krizhevsky2017}, an object classification image dataset of over 14 million images from 1000 different categories, to the prediction of NTL intensity from daytime satellite imagery (Problem 2). They chose the model trained on ImageNet as the starting CNN model \cite{Chatfield2014}, and then constructed the fully convolutional model. Formally, given an unrolled $(h \times w \times d)$-dimensional input $x\in \mathbb{R}^{hwd}$, the fully connected layers perform a matrix-vector product,
\begin{equation}
    \hat{x} = f(Wx+b),
\end{equation}
where $W\in \mathbb{R}^{hwd}$ is a weight matrix, $b$ is a bias term, $f$ is a nonlinear function, and $\hat{x} \in \mathbb{R}^{k}$ is the output. Then, they transferred knowledge from Problem 2 to the prediction of poverty from daytime satellite imagery (Problem 3), for which the amount of training data is limited. The illustration of the method is summarized by Blumenstock \cite{Blumenstock2016} (see Figure~\ref{Fig_3_1_2}), and technical details are presented in the early work by Xie et al. \cite{Xie2016}. The image features extracted from the daytime imagery can explain up to 75\% of the variation in the average household asset across five African countries. Moreover, the method is able to reconstruct survey-based indicators of regional poverty with high accuracy. Using only publicly available data, the method has broad potential applications in tracking and targeting poverty in developing countries.

Other RS data and machine learning approaches can also be used in quantifying poverty-environment relationships. By applying the principal component analysis (PCA) and spatial models in the field of geostatistics, Sedda et al. \cite{Sedda2015} demonstrated the correlations between the normalized difference vegetation index (NDVI, a measure of vegetation greenness in RS \cite{Imran2014}), intensity of poverty, and health for a large area of West Africa. They found that high NDVI is associated with low poverty and child mortality. Their results highlight the utility of satellite-based metrics for poverty analysis. With high-resolution daytime satellite imagery, the UN Global Pulse Lab Kampala built a proxy indicator for poverty based on the household's roof counting. The research project entitled ``Measuring Poverty with Machine Roof Counting'' \cite{UNGP2019} developed image processing software to count the roofs and identify the type of roof that a house has. Watmough et al. \cite{Watmough2016} applied a random forests approach to study the relationships between welfare and geographic metrics for over 14,000 villages in India. They found that geographic metrics account for 61\% and 57\% of the variation in the lowest and highest welfare quintile, respectively. These methods help estimate socioeconomic status in less developed countries where household surveys remain lacking.

\subsubsection{Mobile phones reveal socioeconomic status}

Mobile phones (MPs), serving as ubiquitous sensors, are increasingly common in developing economies. Compared to coarse-grained remote sensing, MPs are able to capture an enormous information and provide cost-effective data at the individual level, such as the frequency and timing of communication events \cite{Onnela2007,Hong2009,Zhao2011b}, the traveling patterns \cite{Gonzalez2008}, the histories of consumption and expenditure \cite{Blumenstock2010}, and so on. With MP logs that are related to housing, education, health, etc., socioeconomic status can be inferred by employing regression models and machine learning approaches at the aggregated subnational and national levels.

To explore the relationship between MP usages and wealth in developing countries, Blumenstock et al. \cite{Blumenstock2010} presented a novel method that contains three steps: (1) modeling the relationship between assets and expenditures using Demographic and Health Survey (DHS) data; (2) conducting a phone-survey with a small subset of MP users to collect information on asset ownership; (3) obtaining call detail records (CDRs) for the individuals in the phone survey and creating a single dataset that use call histories to predict annual expenditures. By analyzing the data from Rwanda, they found that household expenditures are positively correlated with MP usages, mainly with the numbers of international calls, the number of different districts contacted, and the average airtime credit purchase. Airtime credit is money in MP number account, ready to spend on texts, calls and data. These results suggest that the annual expenditures of MP users can be predicted only using their anonymous phone usage data. Blumenstock and Eagle \cite{Blumenstock2012} later found that MP usages in Rwanda are not uniform. They provided a quantitative description about the demographic and socioeconomic structure of MP usages, for example, phone owners are considerably richer and predominantly male. Moreover, Blumenstock et al. \cite{Blumenstock2016b} showed that Rwandans use MP network to transfer their airtime credit to those affected by disasters. In particular, transfers tend to be sent to rich individuals and between pairs of individuals with a strong history of reciprocal.

Individual MP data can be aggregated to estimate socioeconomic status at the national level. By analyzing CDRs and airtime credit purchase histories, Gutierrez et al. \cite{Gutierrez2013} mapped the relative income of individuals, the diversity and inequality of income, and the socioeconomic segregation for fine-grained regions in C{\^o}te d'Ivoire. In particular, they quantified the variation in purchase amounts of each user by using the Coefficient of Variation (CV),
\begin{equation}
    CV = \sigma / \mu,
\end{equation}
where $\sigma$ and $\mu$ are the standard deviation and the mean of the purchase amounts. They found that urban areas clearly stand out in diversity, showing the opportunity to obtain real-time and low-cost socioeconomic statistics. Also for C{\^o}te d'Ivoire, Smith et al. \cite{Smith2013} demonstrated how aggregated CDRs can be mined to derive proxies of socioeconomic indicators. They found strongly negative correlations between the communication activity within a region and the multidimensional poverty index (MPI) \cite{UN2010}, a survey-based indicator that measures a region's actual poverty. Further, they derived a linear model to estimate the poverty level using the diversity of communication. Their work suggests CDRs as an invaluable source for poverty estimation, even without the knowledge of individual behavior.

MP data from C{\^o}te d'Ivoire has also been used to explore the relations between national communication network and socioeconomic dynamics. Mao et al. \cite{Mao2015} introduced the CallRank indicator--the PageRank centrality \cite{Brin1998} calculated over the MP communication network--to quantify the relative importance of an area and tested the correlation between network features and socioeconomic indicators. They found that the outgoing call ratio consistently correlates with local socioeconomic statistics such as low poverty rate and high annual income. Moreover, the Gini index exhibits significant correlations with CallRank and other CDRs-based indicators. Further, to quantify the strength of the \emph{rich-club effect} \cite{Zhou2004,Flammini2006}, they measured the weighted rich-club coefficient of the MP communication network,
\begin{equation}
\rho^{w}(r) = \frac{\phi^{w}(r)}{\phi^{w}_{\text{null}}(r)},
\end{equation}
where $\phi^{w}(r) = W_{>r} / \sum_{l=1}^{E_{>r}} w_{l}^{\text{rank}}$, and $\phi^{w}_{\text{null}}(r)$ corresponds to the null model generated by randomizing the original MP network while preserving its degree distribution. Here, each node has a richness parameter $r$ as the average annual income of the region, $E$ is the total number of links, $E_{>r}$ is the number of links to the region, $W_{>r}$ is the sum of the weights attached to these links, and $w_{l}^{\text{rank}} \ge w_{l+1}^{\text{rank}}$ with $l=\{ 1,2,\ldots,E-1 \}$ are the ranked weights of links on the network. If $\rho^{w}(r)>1$, network shows the \emph{rich-club effect} in comparison with the null model. The extent to which $\rho^{w}(r)$ is larger than 1 indicates the strongness of the \emph{rich-club effect}. After analyzing the CDRs, Mao et al. \cite{Mao2015} found that rich areas form rich club in MP communication, where rich areas communicate more frequently with each other.

\begin{figure}[t]
  \centering
  \includegraphics[width=0.5\textwidth]{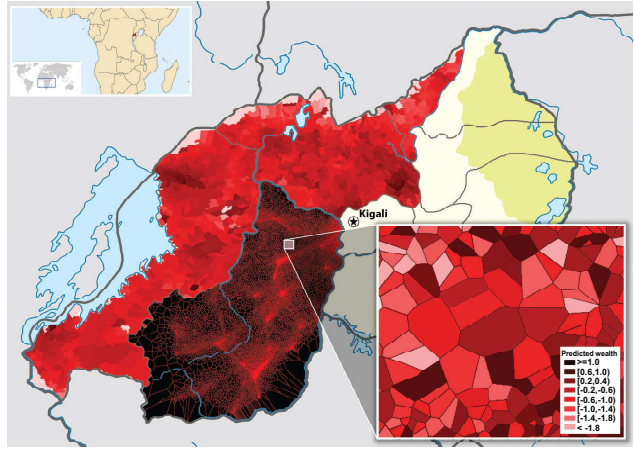}
  \caption{High-resolution map of poverty and wealth predicted from mobile phone call records of 1.5 million users in Rwanda. Figure from \cite{Blumenstock2015}.}
  \label{Fig_3_1_3}
\end{figure}

By analyzing anonymized records of interactions on Rwanda's MP network and the follow-up phone surveys of some individual subscribers, Blumenstock et al. \cite{Blumenstock2015} predicted the wealth of MP users. They demonstrated that the predicted attributes of individuals can accurately reconstruct the distribution of the entire nation's wealth. Specifically, they used a two-step approach in feature engineering and model selection, where the first step generates a thousand metrics from the MP data, and the second step eliminates irrelevant metrics and selects a parsimonious model using the elastic net regularization \cite{Zou2005}. After applying this machine learning approach to analyze the survey data, they found that individual wealth can be well predicted and individuals in relative poverty can be accurately identified. Then, they generated out-of-sample predictions for 1.5 million MP users and produced the wealth map of Rwanda at a very high resolution (see Figure~\ref{Fig_3_1_3}). Further, they found a strong correlation between the government ``ground truth'' data and the predicted wealth data after aggregating them to the district level. Their method is promising to map the distribution of wealth and other socioeconomic indicators for the full national population. Other works that leveraged MP data to infer socioeconomic status at the regional or urban levels will be introduced in the following sections.

\subsubsection{Combined data for better inference}

Novel sources of data with a high spatial resolution have been used to provide an up-to-date indication of living conditions. For example, remote sensing (RS) data capture information about physical properties of the land, which are cost-effective but relatively coarse in urban areas. By contrast, call detail records (CDRs) from mobile phones (MPs) have high spatial resolution in urban areas but the resolution is usually insufficient in rural areas due to the sparsity of towers. Therefore, some recent works estimate socioeconomic status by combining data from different domains such as LandScan population \cite{Dobson2000}, RS and MPs.

\begin{figure}[t]
  \centering
  \includegraphics[width=0.7\textwidth]{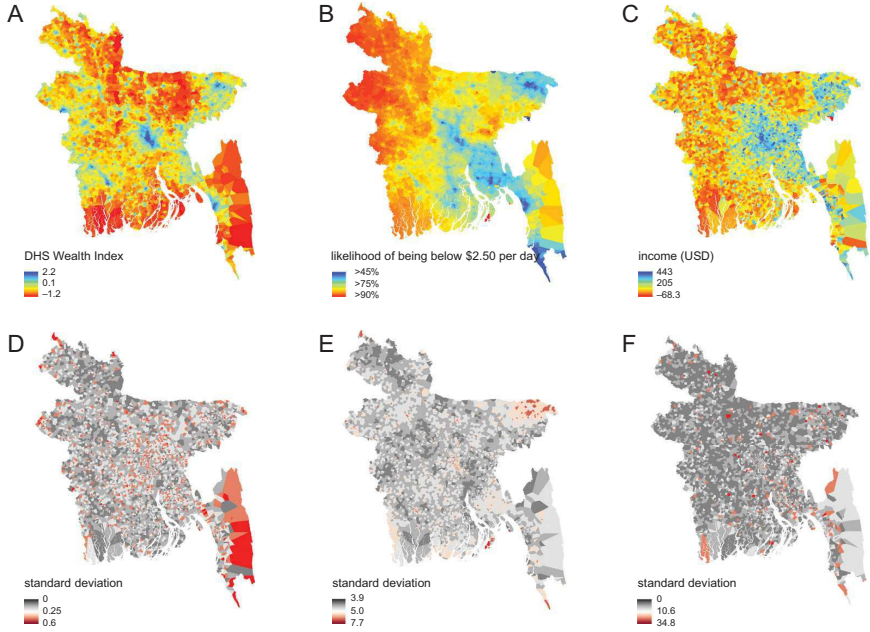}
  \caption{Maps of predicted living standards based on call detail records (CDRs) and remote sensing (RS) data in Bangladesh. Mean wealth index (A) with uncertainty (D); mean likelihood being below \$2.50/day (B) with uncertainty (E); and mean income (C) with uncertainty (F). The maps show the posterior mean and standard deviation from CDR-RS models for the WI and income data (A,C), and the RS model for the PPI (B). Red color indicates poorer areas in prediction maps, and higher error in uncertainty maps. Figure from \cite{Steele2017}.}
  \label{Fig_3_1_4}
\end{figure}

While RS-only and CDRs-only models perform comparably in mapping poverty, Steele et al. \cite{Steele2017} demonstrated that their combination can produce better predictive maps of socioeconomic status in Bangladesh. Specifically, they employed hierarchical Bayesian geostatistical models (BGMs) \cite{Blangiardo2015} that combine RS data, CDRs and traditional survey-based data to map three commonly used indicators of living standards, namely, Wealth Index (WI), Progress out of Poverty Index (PPI) and reported household income (Income). The BGMs are built on the scale of the Voronoi polygons, which approximate the mobile tower coverage areas using Voronoi tessellation \cite{Okabe1992}. They applied BGMs to predict the poverty metrics (WI, PPI and Income) for each Voronoi polygon as a posterior distribution with completely modeled uncertainty around estimates. Then, they generated prediction maps with associated uncertainty using the posterior mean and standard deviation (see Figure~\ref{Fig_3_1_4}). Their method using combined CDRs¨CRS data exhibits a better predictive power (highest $R^{2}=0.78$) for the observed data than RS-only method ($R^{2}=0.71$) and CDRs-only method ($R^{2}=0.70$). Similarly, Njuguna and McSharry \cite{Njuguna2017} built a linear model to predict MPI based on the combination of CDRs, RS and LandScan datasets in Rwanda. They extracted four meaningful features that proxy socioeconomic status from the combine dataset, specifically, nighttime lights (NTLs) per capita from RS data, mobile ownership per capita from CDRs, average daily call volume per phone from CDRs, and population density from LandScan data. They proposed a simple linear regression model using the four features to predict MPI, as
\begin{equation}
\log(\text{MPI}) = \beta_{0} + \beta_{1}x_1 + \beta_{2}x_2 + \beta_{3}x_3 + \beta_{4}x_4,
\end{equation}
where $x_i$ stands for the value of the corresponding feature. This model can explain 76\% of the variance in MPI across 295 sectors in Rwanda. These results suggest that combination of multiple data sources can yield socioeconomic estimates at a high spatial resolution.

Exhaust from digital and physical commodities can provide rich information about socioeconomic status, and thus proxy indicators can be built by leveraging these novel data sources. For example, United Nation Global Pulse launched the project entitled ``Building Proxy Indicators of National Wellbeing with Postal Data'' \cite{UNGlobalPulse2016}, which investigates the potential of using the international postal flow network to approximate indicators of countries' socioeconomic profiles. The project collected 14 million electronic postal records of 187 countries from 2010 to 2014. The dataset covers 680,000 post offices and forms the world's largest postal network. Results show that indicators gathered from the postal network correlate well with fourteen widely used socioeconomic indicators such as GDP and Human Development Index (HDI). This work demonstrates that structural features of world flow networks can be used to produce proxy indicators of socioeconomic status.

Meanwhile, Hristova et al. \cite{Hristova2016} examined how digital traces and the network structure can reveal the socioeconomic profiles of different countries. They measured the position of each country in six different global networks (trade, postal, migration, international flights, IP and digital communications) and built proxies for a number of socioeconomic indicators including GDP per capita and HDI ranking and other twelve indicators. In particular, they applied the multilayer network model \cite{Kivel2014} to characterize the strength of these international ties, where six networks representing six types of international ties are considered as six layers of the multiplex network with each pair of nodes possibly having one relationship in each layer. Formally, the multiplex network \cite{Battiston2014} is denoted as
\begin{equation}
{\cal{M}}= \left\{ G^1(V^1,E^1), \ldots ,G^\alpha(V^\alpha,E^\alpha), \ldots ,G^m(V^m,E^m) \right\},
\end{equation}
where each layer contains a set of edges $E$ and a set of nodes $V$, and $m=6$ is the total number of networks. The multiplex neighborhood of a node $i$ is defined as the union of its neighborhoods on each layer:
\begin{equation}
N_{\cal{M}}(i) = \left\{ N_\alpha(i) \cup N_\beta(i) \ldots \cup N_m(i) \right\},
\end{equation}
where $N_\alpha(i)$ is the neighbourhood of node $i$ in layer $\alpha$. The global multiplex degree of node $i$ is defined as $k^{\text{glob}}(i) = |N_{\cal{M}}(i)|$, and the weighted global multiplex degree is defined as
\begin{equation}
k^{\text{glob}}_w(i) = \displaystyle \sum_{j \in N_{\cal{M}}(i)}\displaystyle\sum_{G \in \cal{M}} \frac{e_{ji}}{n \times m},
\end{equation}
where $n$ is the total number of nodes. The network metrics have predictability to several socioeconomic indicators. The global multiplex degree is the best-performing degree in terms of consistently high performance across all fourteen indicators. In particular, the global degree exhibits the most highly negative correlation with the HDI ranking (Spearman's rank correlation $\rho \approx 0.8$). These results show that a nation's socioeconomic proxy indicators can be constructed based on different global networks after combining the data from multiple sources.

\subsection{Economic complexity and fitness of nations}

Understanding how economies develop to prosperity is a long-standing challenge in economics. In traditional literature, as an aggregated monetary indicator, GDP has been widely used to identify the stages of economic development of countries. Recently, a novel index named economic complexity has been proposed as the root in the gaps of economic development. In particular, the new steam of literature introduce a variety of non-monetary metrics based on international trade networks to quantitatively assess a country's potential for future economic growth. In this section, we will briefly introduce recent works on economic complexity index, fitness index, and some variant indices, as well as their applications to predict world economic development.

\subsubsection{Product space and economic complexity}

Economic development has been traditionally measured by aggregated variables like GDP, however, such averages can not capture the increasing diversity that is associated with economic development. An insight raised recently is that the mix and diversity of products and industries are highly suggestive to economic growth. Hausmann et al. \cite{Hausmann2007} introduced the level of sophistication--the income level of a country's exports--to the characterization of products and demonstrated that it can predict subsequent economic growth. Specifically, they first construct an index called PRODY, which represents the income level associated with a product. The PRODY index for product $p$ is given by
\begin{equation}
    PRODY_{p} = \sum_{c} \frac{(x_{cp}/X_c)}{\sum_{c'}(x_{{c'}p}/X_{c'})} Y_{c},
    \label{Eq:PROD}
\end{equation}
where $x_{cp}$ is the total export of product $p$ by county $c$, $X_c = \sum_p x_{cp}$ is the total export of country $c$, and $Y_c$ is the GDP per capita of country $c$. Indeed, the PRODY index is a weighted average of the per capita GDPs of countries exporting a given product. Then, they construct the PRODY index, which represents the income level associated with a country's export basket. The PRODY index for country $c$ is given by
\begin{equation}
    EXPY{c} = \sum_{p} \left( \frac{x_{cp}}{X_c} \right) PRODY_{p}.
    \label{Eq:EXP}
\end{equation}
Indeed, the PRODY index is a weighted average of the PRODY for the country, where the weights are the shares of the products in the total exports of the country. After analyzing the international trade data covering over 5,000 products and 124 countries, Hausmann et al. \cite{Hausmann2007} found that countries with high initial sophistication of export baskets (EXPY) tend to perform better in subsequent economic growth. These results suggest that countries have economically meaningful differences in the specialization patterns of exporting baskets, and countries export more sophistication products are likely to grow more rapidly.

\begin{figure}[t]
  \centering
  \includegraphics[width=0.45\textwidth]{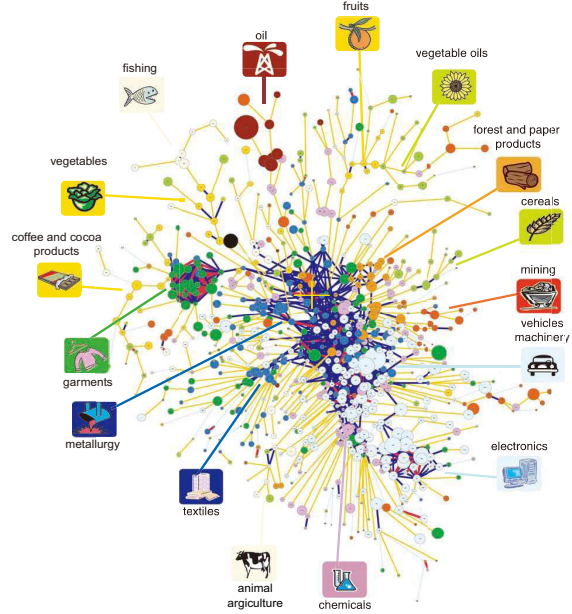}
  \caption{The network representation of the product space built on the international trade data. Links are color coded with their proximity value. The sizes of the nodes are proportional to world trade, and their colors are chosen according to the product classification. Figure after \cite{Hidalgo2007}.}
  \label{Fig_3_2_1}
\end{figure}

Later, Hidalgo et al. \cite{Hidalgo2007} illuminated this viewpoint through analyzing the network of relatedness between products, named product space, which is built based on the international trade data. Products are considered to have high relatedness if they have a high probability to be co-exported by many countries in the international trade. Formally, the proximity between products $i$ and $j$ is defined as
\begin{equation}
    \phi_{ij} = \min \left\{ P(\text{RCA}x_i|\text{RCA}x_j), P(\text{RCA}x_j|\text{RCA}x_i) \right\},
\end{equation}
where $P(\text{RCA}x_i|\text{RCA}x_j)$ is the conditional probability that country $x$ is a significant exporter of product $i$ given that it has been a significant exporter of product $j$. The significant exporter of a product is identified by the revealed comparative advantage (RCA) \cite{Balassa1965}. The RCA value is defined as the share of product $p$ in the export basket of country $c$ to the share of product $p$ in the world trade. Specifically, the $\text{RCA}_{cp}$ of country $c$ in product $p$ is defined by
\begin{equation}
    \text{RCA}_{cp} = \left.{\frac{x_{cp}}{\sum_{p'}x_{cp'}}}\middle/ \frac{\sum_{c'}x_{c'p}}{\sum_{p'}\sum_{c'}x_{c'p'}}\right.
    ,
\label{Eq:RCA}
\end{equation}
where $x_{cp}$ is the total export of product $p$ by country $c$. If $\text{RCA}_{cp} \ge 1$, country $c$ is a significant exporter of product $p$. Larger proximity $\phi_{ij}$ means higher relatedness between products $i$ and $j$. Based on the proximity measure, the product space is generated and visualized (see Figure~\ref{Fig_3_2_1}). It can be seen that the product space has a core-periphery structure with more-sophisticated products locating in the core and less-sophisticated products occupying the periphery (see Ref. \cite{Borgatti2000,Holme2005} for the definition of core-periphery structure in networks). Richer and poorer countries tend to export products that are located in the core and periphery, respectively. More significantly, countries move through the ``product space'' by developing products that are related to what they currently have. These results provide explanations to the fact that economic development a path-dependent process \cite{Neffke2011} and not all countries face the same opportunities in development.

In particular, it is hard for poor countries to move toward new products with high sophistication since these countries tend to occupy the peripheries of the product space with current exports of less-sophisticated products. Using the concept of product space to explore the international trade data, Abdon and Felipe \cite{Abdon2011} studied the opportunity for economic growth and structural transformation of Sub-Saharan Africa (SSA) countries. They found that the majority of SSA countries are trapped in the export of products that are unsophisticated, standard and poorly connected in the product space. This makes the structural transformation of a region being particularly difficult, because the nearby products are in the periphery and the current capabilities are not enough to jump into more sophisticated products. To solve this problem, governments must implement policies and provide public inputs that can give incentives for the private sector to invest in the more sophisticated activities.

Further, Hidalgo and Hausmann \cite{Hidalgo2009} quantified the economic complexity of nations based on international trade data and demonstrated its central role in a country's economic development. In particular, they proposed the Method of Reflections (MR) to characterize the structure of ``country-product'' bipartite network and showed that the variables produced by the MR method can be interpreted as indicators of economic complexity. Formally, the bipartite network can be represented by an adjacency matrix $M_{cp}$, where $M_{cp}=1$ if country $c$ is a significant exporter ($\text{RCA}_{cp} \ge 1$) of product $p$, and $M_{cp}=0$ if otherwise. The economic complexity index (ECI) of country $c$ is then defined as
\begin{equation}
    \text{ECI}_c = \frac{K_c-\langle \vec{K}\rangle}{std(\vec{K})}=\frac{N^{2}K_{c}-N\sum_{c}{K_c}}{\sqrt{N\sum_{c}{(N K_c-\sum_{c}{K_c}})^2}},
\end{equation}
where $N$ is the number of countries, $\langle \cdot \rangle$ and $std(\cdot)$ are functions of mean and stand deviation that operate on the elements of vector $\vec{K}$, and $\vec{K}$ is the eigenvector associated with the 2nd largest eigenvalue of the matrix
\begin{equation}
    \tilde{M}_{cc'} = \sum_p {\frac{M_{cp}M_{c'p}}{k_{c,0} k_{p,0}}}.
\end{equation}

Indeed, the matrix $\tilde{M}_{cc'}$ is defined through a set of linear iterative equations by connecting countries who have similar products, weighted by the inverse of the ubiquity of product ($k_{p,0} = \sum_{c}M_{cp}$) and normalized by the diversity of country ($k_{c,0} = \sum_{p}M_{cp}$). Formally, putting the equation $\text{k}_{p,N} = \sum_{c} M_{cp} k_{c,N-1} / k_{p,0}$ (the average ubiquity of product) into the equation $\text{k}_{c,N} = \sum_{p} M_{cp} k_{p,N-1} / k_{c,0}$ (the average diversity of country) can generate the equation
\begin{equation}
k_{c,N} = \sum_{c'} k_{c',N-2} \sum_{p} \frac{M_{cp}M_{c'p}}{k_{c,0}k_{p,0}},
\end{equation}
where $N \ge 2$ is the number of iteration. The economic complexity of country $c$ is given by $\text{ECI}_{c} = \sum_{c'} \tilde{M}_{cc'} \text{ECI}_{c'}$, where $\text{ECI}_{c'}$ is country $c$'s complexity in the previous iteration step. For more mathematical details, readers are encouraged to read the book on economic complexity wrote by Hausmann et al. \cite{Hausmann2014}. Empirical results showed that countries' ECIs are highly correlated with their income levels are predictive of their future growth. Indeed, economic development is a process that requires acquiring more complex sets of capabilities to move towards new activities associated with higher levels of productivity. Therefore, efforts should focus on generating the conditions that allow complexity to emerge, so that sustained growth and prosperity in economic development will appear.

From a network perspective, uncovering the characteristics of the ``country-product'' bipartite network is very important for understanding economic development. Hausmann and Hidalgo \cite{Hausmann2011} proposed an analytic framework to account for the nature of the bipartite network structure. They found that countries differ in their product diversification and in the ubiquity of their exported products. Countries with more capabilities are able to produce less ubiquitous products. This logic explains the negative relationship between the diversification of countries and the average ubiquity of the products that they produce. Later, Bustos et al. \cite{Bustos2012} studied the presence and absence of industries in international and domestic economies. They found that ``country-product'' bipartite networks are significantly nested \cite{Patterson1986}, and the dynamics of nestedness can predict the evolution of industrial ecosystems (see Refs. \cite{Bascompte2003,Lin2018} for details on nestedness in networks). Moreover, the nestedness tends to be constant over time, making the pattern of industrial appearances predictable. Felipe et al. \cite{Felipe2012} applied MR to rank 5107 products and 124 countries in the international trade. They found that countries' export shares of products of different complexity vary with the level of their income per capita. Specifically, export shares of the most complex products increase with income, while the export share of the less complex products decrease with income. Moreover, MR can distinguish products that require more complex or simpler capabilities, and the complexity rankings of countries exhibit a high correlation with their technological capabilities.

\subsubsection{Fitness index and economic dynamics}

The Fitness index employs a statistical approach to define a new set of metrics to quantify the fitness of countries and the complexity of products through coupled nonlinear maps. Based on the analysis of the ``country-product'' bipartite networks of international trade, Caldarelli et al. \cite{Caldarelli2012} proposed a new method based on biased Markov chain process to rank countries in a more conceptually consistent way, where a two-parameter bias is used to account for the bipartite network structure. Formally, the Markov process is given by
\begin{equation}
\left\{
\begin{array}{l}
w_c^{(N+1)}(\alpha,\beta)=\sum_{p}G_{cp}(\beta)w_p^{(N)}(\alpha,\beta)\\
w_p^{(N+1)}(\alpha,\beta)=\sum_{c}G_{pc}(\alpha)w_c^{(N)}(\alpha,\beta)
\end{array}
\right.
,
\end{equation}
where $w_c$ is the fitness of country $c$, $w_p$ is the complexity of product $p$, $N$ is the interaction step, and $G$ is the Markov transition matrix given by
\begin{equation}
\left\{
\begin{array}{l}
G_{cp}(\beta)=\frac{M_{cp}k_c^{-\beta}}{\sum_{c'}M_{c'p}k_{c'}^{-\beta}}\\
G_{pc}(\alpha)=\frac{M_{cp}k_p^{-\alpha}}{\sum_{p'}M_{cp'}k_{p'}^{-\alpha}}
\end{array}
\right.
,
\end{equation}
where $\alpha$ and $\beta$ are free parameters. In a vectorial formalism, country $c$'s fitness is ${\bf w}_{c}^{(N+1)}(\alpha,\beta)=T(\alpha,\beta) {\bf w}_{c}^{(N)}(\alpha,\beta)$, where the ergodic stochastic matrix $T$ is defined as $T_{cc'}(\alpha,\beta)=\sum_{p}G_{cp}(\beta)G_{pc'}(\alpha)$. The complexity of product $p$ is ${\bf w}_{p}^{(N+1)}(\alpha,\beta)=S(\alpha,\beta) {\bf w}_{p}^{(N)}(\alpha,\beta)$, where the ergodic stochastic matrix $S$ is $S_{pp'}(\alpha,\beta)=\sum_{c}G_{pc}(\alpha)G_{cp'}(\beta)$ (see Ref. \cite{Caldarelli2012} for mathematical details). After analyzing these equations, Caldarelli et al. \cite{Caldarelli2012} revealed a strongly nonlinear entanglement between the diversification of a country and the ubiquity of its products in determining the competitiveness of countries and the complexity of products. In particular, having more-sophisticated products in the portfolio contributes more to the competitiveness of a country than having many less-sophisticated products.

Moving forward, Tacchella et al. \cite{Tacchella2012} developed a so-called Fitness-Complexity Method (FCM) using coupled nonlinear maps, whose fixed point can define new metrics for the fitness of countries and the complexity of products. In their iterative algorithm, fitness of countries and complexity of products interact in a nonlinear and self-consistent mathematical way. Specifically, the fitness of a country is proportional to the number of its products weighted by their complexity. In turn, the complexity of a product is inversely proportional to the number of countries exporting it weighted by the inverse of their fitness (similar methods have also been proposed for search engine \cite{Kleinberg1999} and online reputation systems \cite{Zhou2011}). Formally, the coupling between the fitness $F_c$ of country $c$ and the complexity $Q_p$ of product $p$ is given by the nonlinear iterative scheme:
\begin{equation}
    \left\{
    \begin{aligned}
        \tilde{F}_{c}^{(N)} & = & \sum_{i}{M_{cp}Q_{p}^{(N-1)}} \\
        \tilde{Q}_{p}^{(N)} & = & \frac{1}{\sum_{c}{M_{cp}\frac{1}{F_{c}^{(N-1)}}}}
    \end{aligned}
    \right.
    ,
\end{equation}
where $\tilde{F}_{c}^{(N)}$ and $\tilde{Q}_{p}^{(N)}$ are respectively normalized in each step by $F_{c}^{(N)}=\tilde{F}_{c}^{(N)} / \langle \tilde{F}_{c}^{(N)}\rangle$ and $Q_{p}^{(N)}=\tilde{Q}_{p}^{(N)} / \langle \tilde{Q}_{p}^{(N)}\rangle$, given the initial condition $F_{c}^{(0)}=1$ and $Q_{p}^{(0)}=1$. The nonlinear iteration goes until the stationary state is reached (see Ref. \cite{Pugliese2016} for the convergence property), in which $F$ reflects the fitness of countries and $Q$ reflects the complexity of products. Indeed, FCM is based on the idea that (i) a diversified country gives limited information on the complexity of products, and (ii) a poorly diversified country tends to have a specific product of a low level sophistication. Therefore, a nonlinear iteration is needed to bound the complexity of industries by the fitness of the less competitive provinces having them. After applied to the international trade data, FCM performs better than MR in capturing the bipartite network structure, in defining an effective non-monetary matric for economic complexity, and in quantifying a country's potential for growth.

Meanwhile, Cristelli et al. \cite{Cristelli2013} argued that nonlinear dependence is the fundamental element and the nonlinear approach is consistent with the structure of the unweighted ``country-product'' bipartite network. Moreover, they analyzed the case of including weights in the matrix $M_{cp}$ through $M_{cp}=x_{cp}/\sum_{c'}x_{c'p}$, where $x_{cp}$ is the total export of product $p$ by country $c$. After comparing MR and FCM in both economic and mathematical aspects, they found that FCM is more conceptually consistent and well-grounded from an economic point of view. Taking into account the triangular structure of the bipartite network, Tacchella et al. \cite{Tacchella2013} discussed how to define suitable non-monetary metrics for both the complexity of products and the diversification of countries. In particular, they argued the conceptual flaws of MR by using three toy models and demonstrated that FCM is able to grasp the level of competitiveness of a country by defining the simplest metrics that seem to be consistent with the triangular-like pattern.

\begin{figure}[t]
  \centering
  \includegraphics[width=0.75\textwidth]{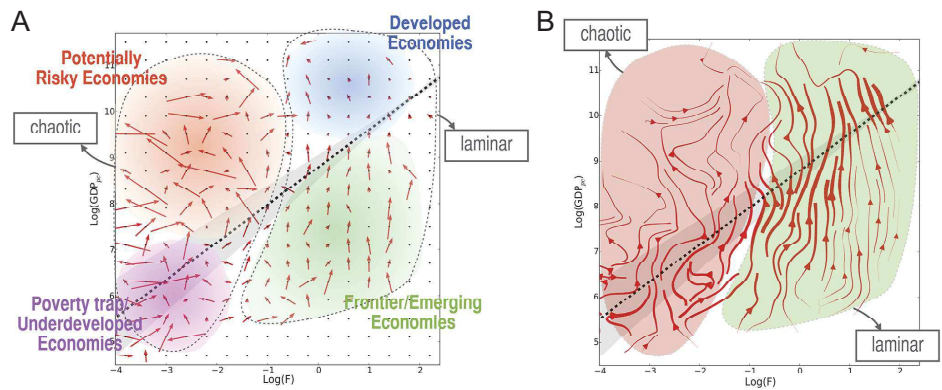}
  \caption{The heterogeneous dynamics of countries in the Fitness-GDPpc plane. (a) A finer coarse graining of the dynamics highlights two regimes. One regime is the laminar region (right), where fitness is the driving force of the growth, and the evolution of countries in this region is highly predictable. The other regime is the chaotic region (left), where the issues are very close to the problems of predictability for dynamical systems and to develop a predictive scheme using tools like regressions is no more appropriate. (b) The continuous interpolation of the coarse grained dynamics. The predictability in the two regimes, laminar and chaotic, is better illustrated. Figure from \cite{Cristelli2015}.}
  \label{Fig_3_2_2}
\end{figure}

This branch of studies has provided new perspectives to cast economic prediction into the conceptual scheme of forecasting the evolution of a dynamical system, for example, weather dynamics. Cristelli et al. \cite{Cristelli2015} compared the non-monetary metrics, in particular the fitness of countries, with their monetary figures, say GDPpc. They showed that FCM is able to quantify the hidden growth potential of countries. More interestingly, they demonstrated that the pattern of countries' evolution in the Fitness-GDPpc plane is strongly heterogeneous with two regimes of very different predictability features (see Figure~\ref{Fig_3_2_2}). Specifically, there is a strongly predictable area of economic development, named the laminar regime, while the predictability is low in the so-called chaotic regime. Two kinds of evolution patterns can be observed in the laminar regime, where emerging economies develop rapidly and developed economies enjoy stable growth. In the chaotic regime, the dynamics of countries are highly diverse and unstable, leading to the difficulty in predicting the economic development. In this case, tools like regressions are no more appropriate in developing a predictive scheme.

To address this issue, Cristelli et al. \cite{Cristelli2015} defined a selective predictability scheme to assess future evolution of countries by resembling the method of analogues \cite{Lorenz1969}, which was developed to predict the evolution of a dynamical system given the knowledge of the past but without the laws of motion. The framework provides insights to the regime-dependent economic predictability and opens new paths to economic forecasting. Recently, Tacchella et al. \cite{Tacchella2018} applied this scheme to predict the five-year GDP growth. In the Fitness-GDPpc plane, they repeatedly sampled analogues with a Gaussian kernel (centred on the present state of a country) and performed a bootstrap of previously observed evolution (weighted by the distance of the analogues starting points), resulting in the global distribution of possible outcomes. They further refined the forecast by taking into account the strong self-correlation of GDP growth. Specifically, the forecast based on the global distribution is combined with the forecast that assumes a past five-year growth by a certain weighted averaging. This scheme outperforms the International Monetary Fund (IMF) five-year GDPpc forecast \cite{IMF2016} by more than 25\% in accuracy. Moreover, the method's forecasting errors are predictable and not correlated with IMF errors, showing its complementarity to traditional approaches.

\subsubsection{Variant indices and development analysis}

Many recent studies have highlighted the importance of complexity and capabilities in economic development. The pioneering work by Hidalgo and Hausmann \cite{Hidalgo2009} introduced MR to extract the competitiveness of countries and the complexity of product from the ``country-product'' bipartite networks with the assumption that there are linear interactions between the two metrics. Tacchella et al. \cite{Tacchella2012,Tacchella2013} proposed FCM and emphasized the necessary of nonlinear coupling between the fitness of countries and the complexity of products. Mariani et al. \cite{Mariani2015} quantitatively compared the ability of MR and FCM in ranking countries and products by their importance in networks. Based on the international trade data of 132 countries and 723 products, they found that FCM outperforms MR in ranking both products and countries. In particular, FCM captures the nestedness of the bipartite network and ranks nodes better by their importance.

Mariani et al. \cite{Mariani2015} proposed a modified FCM (MFCM for short), in which the nonlinear coupling is governed by a tunable parameter. By adjusting the parameter, we can find a better tradeoff between the favor on countries with diversified exports and the penalization on products with a large number of exporting countries. Formally, MFCM is defined by the equations
\begin{equation}
    \left\{
    \begin{aligned}
        \tilde{F}_{c}^{(N)} (\gamma) & = & \sum_{i}{M_{cp}Q_{p}^{(N-1)}} \\
        \tilde{Q}_{p}^{(N)} (\gamma)  & = & \left[ \sum_{c}{M_{cp}(F_{c}^{(N-1)})^{-\gamma}} \right] ^{-1/\gamma} \\
    \end{aligned}
    \right.
    ,
\label{Eq:GFit}
\end{equation}
where $\gamma$ is the tunable parameter. When $\gamma = 1$, MFCM degenerates to FCM. The correlation between the product complexity $\tilde{Q}(\gamma)$ and the product ubiquity $k_{p,0}$ decreases with the increase of $\gamma$. When $\gamma \gtrsim 2$, the ranking of product complexity $\tilde{Q}(\gamma)$ by MFCM is perfectly correlated with that by FCM, however the ranking is volatile (very sensitive to noise). For this reason, MFCM with larger $\gamma$ can only be applied to high-quality data instead of noisy data. When input data is reliable, MFCM is able to produce better rankings of countries and products.

Wu et al. \cite{Wu2016} showed some rigorous mathematical properties of the fitness-complexity metric for nested networks. They introduced a simpler variant of FCM, named Minimal Extremal Metric (MEM), where the complexity of a product $p$ is equal to the fitness of the least-fit country that exports it. Formally, MEM defines the fitness of country $c$ and the complexity of product $p$ by
\begin{equation}
    \left\{
    \begin{aligned}
        \tilde{F}_{c}^{(N)} & = & \sum_{i}{M_{cp}Q_{p}^{(N-1)}} \\
        F_{c}^{(N)} & = & \tilde{F}_{c}^{(N)} / \langle \tilde{F}_{c}^{(N)}\rangle \\
        Q_{p}^{(N)} & = & \min \limits_{i:M_{cp}=1} {F_{c}^{(N)}} \\
    \end{aligned}
    \right.
    .
\end{equation}
Obviously, in MEM, only the fitness of the least-fit country $\min \limits_{i:M_{cp}=1} {F_{c}^{(N)}}$ contributes to the product complexity $Q_{p}^{(N)}$. In the limit $\gamma \to \infty$, MEM is a special case of MFCM. Results based on the analysis of the international trade data show that MEM can reproduce the nested structure of the ``country-product'' bipartite network but it is highly sensitive to noise in data.

Morrison et al. \cite{Morrison2017} provided both theoretical and numerical evidence for the intrinsic instability in the nonlinear map employed by FCM. Using the preferential attachment model (see Refs. \cite{Barabasi1999,Foschi2014}) and two real-world datasets (trade and patent), they showed that FCM is unstable to even small perturbations in the network, while MR does not suffer from this problem. That is because the nonlinear iterative approach in FCM amplifies the effects of countries with low fitness on the complexity of a product and highlights economies producing exclusive niche products, which are produced by a very few countries but not necessarily the most sophisticated. Adding a product exported by only a single country may lead to a global reorganization of the fitness landscape. Therefore, FCM has a serious problem when applied to dynamic economical systems with new products entering markets.

With new methodologies, attentions have been paid to better understand economic development, innovation and industrialization. Based on the international trade data, Zaccaria et al. \cite{Zaccaria2014} built a hierarchically directed network by measuring the taxonomy of products through computing the excess frequency of co-occurrence of two products comparing to the random binomial case. Formally, the taxonomy between products $p$ and $p'$ is defined by projecting the ``country-product'' matrix $M_{cp}$ to a unipartite space as (similar to \cite{Zhou2007})
\begin{equation}
    B_{pp'}=\frac{1}{ \max \left\{\sum_{c}M_{cp}, \sum_{c}M_{cp'}\right\} } \sum_{c} \frac{M_{cp}M_{cp'}}{\sum_p M_{cp}}.
\end{equation}
The taxonomy network presents the temporal connections between products and suggests the most relevant products for the development of countries. Indeed, the structure of the taxonomy network is suggestive to the potential growth of countries. Later, Saracco et al. \cite{Saracco2015} proposed a dynamical network approach to model the process of country's innovation and competition on the evolution of the export baskets. Their dynamical model can accurately reproduce the main features observed in the evolution of the ``country-product'' bipartite network. Moreover, their model suggests that countries can follow different paths in the ``product space'' \cite{Hidalgo2007,Zaccaria2014} to gradually diversify their export baskets.

Focusing on the time evolution of trade volume, average complexity and competitiveness, Zaccaria et al. \cite{Zaccaria2016} compared the exports of different sectors in Netherlands. They found that high-tech related sectors have high average complexity but low competitiveness, while sectors heavily relying on raw materials have a low complexity but high competitiveness such as Energy and Horticulture sectors. Indeed, not only products but also services are important in explaining economic stability and predicting future growth. Stojkoski et al. \cite{Stojkoski2016} found that services have in general higher economic complexity than products. The sophistication and diversification of service exports can provide an additional route for economic growth in both developing and developed countries. Countries that are not able to diversify service portfolio may face diminishing growth prospects.

\begin{figure}[t]
  \centering
  \includegraphics[width=0.98\textwidth]{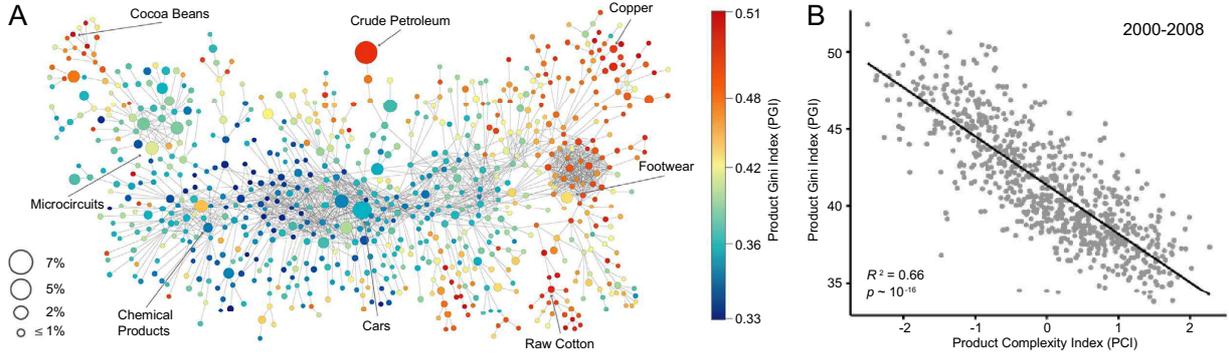}
  \caption{The product space and income inequality. (A) In the product space, nodes are colored according to the Product Gini Index (PGI) as measured during 1995-2008. The sizes of nodes are proportional to the volume of the international trade during 2000-2008. The networks are based on a proximity matrix representing 775 SITC-4 product classes exported during 1963-2008. The link strength (proximity) is based on the conditional probability that the products are co-exported. (B) The relationship between the Product Complexity Index (PCI) and the Product Gini Index (PGI) in the 2000-2008. Figure from \cite{Hartmann2017}.}
  \label{Fig_3_2_3}
\end{figure}

Hartmann et al. \cite{Hartmann2017} found that countries exporting more complex products have lower levels of income inequality. In particular, economic complexity index (ECI) outperforms GDP in explaining income inequality. Based on the international trade data, they calculated the Product Complexity Index (PCI) using the method proposed by Hidalgo and Hausmann \cite{Hidalgo2009}. Further, they estimated the level of income inequality associated with products by introducing the Product Gini Index (PGI), which is a weighted average of the Gini coefficients of the countries that export a product (see Figure~\ref{Fig_3_2_3}A for the PGIs of products in the product space). There is a strong and negative correlation between PCI and PGI, showing that sophisticated products tend to have low levels of inequality (see Figure~\ref{Fig_3_2_3}B). Moreover, countries with high (low) level of ECI are more likely to specialize in high-PCI (low-PCI) products, suggesting that the productive structure of a country may condition its range of income inequality. Recently, Mealy et al. \cite{Mealy2019} interpreted economic complexity metrics by showing that ECI and PCI are equivalent to a spectral clustering algorithm, which divides a similarity network into two parts. Moreover, these measures are closely related to many dimensionality reduction methods such as correspondence analysis and diffusion maps. Their findings shed some new light on the empirical success of ECI and PCI in explaining specialization patterns of countries in economic growth.

Pugliese et al. \cite{Pugliese2017} analyzed the role of complexity in economic development and found that economies with differentiated products face a lower barrier in the transition towards industrialization. They extended the concept of poverty trap to include the two factors of economic complexity and GPDpc (see also Ref. \cite{Cristelli2015}). They defined an index of development and industrialization, named Complex Index of Relative Development (CIRD), by the equation:
\begin{equation}
    \text{CIRD}_{c,t}=\beta (\log F_{c,t}) + (1-\beta) \log (GDPpc_{c,t}),
\end{equation}
where $F_{c,t}$ is the fitness of country $c$ at time $t$, $GDPpc_{c,t}$ is the GDPpc of country $c$ at time $t$, and $\beta$ is a tunable parameter. The use of the CIRD index allows to study development as a monodimensional process. In particular, $\text{CIRD}_{c,t} \approx -2$ is a threshold for countries to exit the poverty trap, and the increase of the input growth reaches its maximum at this critical point. The CIRD index facilitates our understanding of industrialization dynamics and is helpful for development analysis. Sbardella et al. \cite{Sbardella2017} analyzed the relationship between wage inequality and industrialization using fitness and GDPpc. They found that movement of wage inequality along with the industrialization follows a longitudinally persistent pattern. This finding is comparable to theories proposed by Kuznets \cite{Kuznets1955}, who hypothesized that countries with an average level of development suffer the highest levels of wage inequality.

Along with the literature, some online platforms have been developed and launched to help understand the evolution of countries' productive structures and economic development. For example, Simoes and Hidalgo \cite{Simoes2011} launched a data visualization site, named Observatory of Economic Complexity (OEC) (https://atlas.media.mit.edu). The OEC combines a number of international trade datasets and serves more than millions of interactive visualizations including imports and exports, origins and destinations, product space, economic complexity rankings based on MR, income inequality, and so on. Meanwhile, the GROWTHCOM Project launched a data platform (http://www.growthcom.eu), which provides visualization tools of the product network \cite{Zaccaria2014} and the countries' trajectories in the fitness-GDPpc plane \cite{Cristelli2015}.

\subsection{Spatial demography and culture evolution}

High resolution and near real-time data from new sources like remote sensing (RS), mobile phone (MP) and social media (SM) are complementary to traditional costly data with a long-time delay in inferring population distributions and demographics. Moreover, these so-called socioeconomic big data, together with methods from interdisciplinary fields including statistical physics and computer sciences, have been used to predict international migration and quantify world culture evolution. In this subsection, we will briefly introduce some methods using new data sources to map world population, estimate international migration and study culture evolution.

\subsubsection{World population distribution}

Knowing the spatial distribution of population on earth is critical for many socioeconomic applications such as accurate environmental impact assessments, human health adaptive strategies and disease burden estimation \cite{Linard2012}. Developed countries have substantial resources to create accurate and contemporary population datasets with high spatial resolution \cite{Patel2017}, however, relevant data are often scarce, outdated and unreliable in low-income countries due to economic constrains. In addition, acquiring census data in a timely and accurate manner is very difficult due to the rapid change of population and some administrative challenges. As a results, our knowledge of population distribution in many areas of the world remains poor thus far. Fortunately, technologies developed during the past decades have opened new ways for us to estimate and map world population distribution in a more timely manner and with a relatively lower cost.

Some large-area gridded world population distribution datasets have been built based on multiple data resources. Tobler et al. \cite{Tobler1997} developed the first version of Gridded Population of the World (GPW) database by transforming population counts from census units to a grid. The Global Rural Urban Mapping Project (GRUMP) utilizes higher resolution inputs and renders outputs at a 30 arc-second resolution (approximately 1km). In addition to census data, spatial covariate datasets are also used to estimate populations. For example, the LandScan Global Population Project \cite{Dobson2000} produced the world-wide 1998 LandScan population database at a 30 arc-second resolution based on the land cover database derived from satellite imagery and urban area vector data \cite{Bhaduri2007}. Tatem et al. \cite{Tatem2007} produced the 100m gridded population map by combining land cover information and census data under the Malaria Atlas Project. The semi-automated population distribution mapping at unprecedented spatial resolution produces more accurate results at a spatial resolution of about 100m in East Africa.

Cheriyadat et al. \cite{Cheriyadat2007} generated human settlement maps based on high-resolution satellite imagery. Their algorithm employed gray level co-occurrence matrices \cite{Martinao2003} to generate texture and edge patterns from satellite imagery that are useful in urban land cover classification. Liao et al. \cite{Liao2010} presented a high-accuracy population mapping method that integrates genetic programming (GP) \cite{Kishore2001} and genetic algorithms (GA) \cite{Holland1992} with geographic information systems (GIS). Specifically, they applied GIS to identify relevant factors (e.g., land-cover types and transport infrastructure) and use GP and GA to transform census data to population grids. Deng et al. \cite{Deng2010} estimated small-area population by incorporating GIS, remote sensing (RS) and demographic data into a popular demographic model. They demonstrated that the derived spatial factors can significantly improve the accuracy of small-area population estimation.

Gaughan et al. \cite{Gaughan2013} constructed an accurate and high-resolution population distribution dataset for Southeast Asia. They modeled population distributions for 2010 and 2015 by combining satellite-derived settlement maps, land cover information, and ancillary datasets on infrastructure. Stevens et al. \cite{Stevens2015} presented a new semi-automated dasymetric modeling approach, where RS and geospatial data are combined to model the dasymetric weights and the random forest model is used to generate a gridded prediction of population density at about 100m resolution. Patel et al. \cite{Patel2015} presented a novel method to map multitemporal settlement and population from Landsat imagery using Google Earth Engine, which is an online environmental data monitoring platform that provides analysis capabilities on Landsat data by leveraging cloud computing services. They demonstrated that the integration of GEE-derived urban extents improves the quality of population mapping.

Spatial covariates derived from satellite imagery and land cover are typically static in nature and are not direct measures of people's presence on earth \cite{Patel2017}. Thanks to the rapid adoption of Internet and mobile devices in developing countries, there is a great potential of using digital records to do population mapping. For example, call detail records (CDRs) can overcome many limitations of census-based data since MPs have a high penetration rate across the world. For urban areas, Pulselli et al. \cite{Pulselli2008} developed a technique to monitor population density in real time based on MP chatting, given that the intensity of activity in the area covered by an antenna is proportional to the number of MP users. Based on MP location data, Dan and He \cite{Dan2010} proposed a dynamic distribution model to estimate urban population density using an improved K-means clustering algorithm \cite{Lloyd1982}. Kang et al. \cite{Kang2012} discussed several fundamental issues on using CDRs to estimate population distributions. After analyzing the CDRs of nearly two million MP subscribers, they found that the number of calls other than the total daily call volume serves as a good estimator of population distribution.

\begin{figure}[t]
  \centering
  \includegraphics[width=0.52\textwidth]{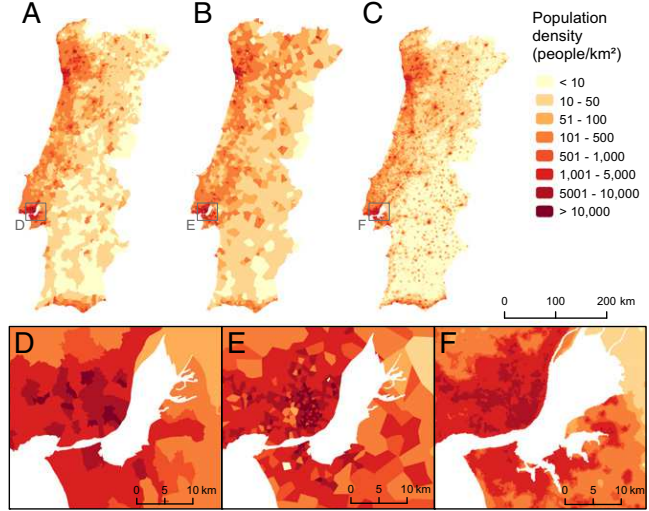}
  \caption{Comparison of predicted population density datasets with baseline data for mainland Portugal. (A) Population density derived from the national census. (B) Population estimated by the mobile phone method. (C) Population density estimated by the remotely sensing method. (D-F) Close-ups around the capital city Lisbon. Figure from \cite{Deville2014}.}
  \label{Fig_3_3_1}
\end{figure}

Recently, using both RS and MP data, Deville et al. \cite{Deville2014} produced spatially and temporarily explicit estimations of population densities at national scales (see Figure~\ref{Fig_3_3_1}). Based on over one billion CDRs from Portugal and France, they estimated the population density of an administrative unit $c_i$ using a two-step method that relies on the density $\sigma_{v_j}$ of MP users, where $v_j$ is the Voronoi polygon \cite{Okabe1992} associated with tower $j$. The nighttime density $\sigma_{c_i}$ for unit $c_i$ is calculated by
\begin{equation}
    \sigma_{c_i}=\frac{1}{A_{c_i}} \sum_{v_j} \sigma_{v_i}A_{(c_i \cap v_j)},
\label{Eq:sigma}
\end{equation}
where $A_{c_i}$ is the area of unit $c_i$, and $A_{(c_i \cap v_j)}$ is the intersection area of unit $c_i$ and the Voronoi polygon $v_j$. The density $\sigma_{c_i}$ is compared with the census-derived population densities $\rho_{c_i}$ through
\begin{equation}
    \rho_{c}= \alpha \sigma_{c}^{\beta},
\label{Eq:rho}
\end{equation}
where $\rho_{c}=[\rho_{c_1}, \rho_{c_2}, \ldots, \rho_{c_n}]$ and $\sigma_{c}=[\sigma_{c_1}, \sigma_{c_2}, \ldots, \sigma_{c_n}]$. By transforming Eq.~(\ref{Eq:rho}) to $\log (\rho_{c}) = \log (\alpha) + \beta \log (\sigma_{c})$, the two parameters $\alpha$ and $\beta$ can be fitted by a linear regression on training data. Further, they combined the MP method with the RS method proposed by Stevens et al. \cite{Stevens2015}, who used the random forest model to generate gridded predictions of population density. Formally, the population density in pixel $i$ is estimated by
\begin{equation}
    \rho_{i}^{RS}= \frac{w_i}{\sum_j w_j}P,
\end{equation}
where $w_i$ is the weight assigned to pixel $i$ and $P$ is the total population. Combining MP and RS data can produce population datasets with a high spatial and temporal resolution.

Douglass et al. \cite{Douglass2015} created high-resolution maps of population distribution by combing telecommunications data, satellite imagery and census data in Milan, Italy. They fitted population and call data by applying an elementary model that is similar to Eq.~(\ref{Eq:rho}). They found that the total out-call volume has the strongest correlation (about 0.68) with the grid-level population. Further, they employed a random forest regression to predict population using features of land cover measures, call activity measures and their combinations. They found that building land cover and calls made out at 10am are the top-two predictors that are sufficient to provide accurate predictions. Lulli et al. \cite{Lulli2016} proposed a function to capture similarities between individual call profiles (ICPs). The similarity of ICPs is captured by combining the Euclidean similarity and the Jaccard similarity. Then, they built a clustering algorithm to provide clusters of individuals based on the similarity between ICPs. Using an automatic classifier to label the clusters, their method can estimate the number of residents, commuters and visitors in a given region. At the urban scale, Khodabandelou et al. \cite{Khodabandelou2016} estimated population density by applying Eq.~(\ref{Eq:sigma}) and Eq.~(\ref{Eq:rho}) based on the mobile network traffic metadata. Their method can estimate both static and dynamic populations across different cities.

Calling activities are powerful in mapping populations, however, it is usually not easy to obtain due to privacy concerns \cite{Kosinski2013}. For example, some highly sensitive traits and attributes can be inferred from digital records of human behavior \cite{Tsavli2015}. The increasingly available social media data presents alternative opportunities in estimating population distribution. Twitter has gained worldwide popularity, making the geotagged tweets show detailed depictions of human activity. Leetaru et al. \cite{Leetaru2013} explored over 1.5 billion tweets posted by over 70 million users. They found a high correlation (0.79) between geotagged tweets and the NASA City Lights imagery. The most accurate feature is the self-reported user location field, exhibiting a correlation 0.72 with the geotagged baseline. Their work demonstrates the potential of geotagged tweets in world population mapping.

Very recently, volunteered geographic information (VGI) collected from the Internet (e.g., check-in data \cite{Yang2017}) has been used to estimate population at a fine scale. Yao et al. \cite{Yao2017} presented a framework to map population distribution at the building level by integrating national census data with two geospatial data sources. One is the points-of-interest (POIs) provided by Baidu Map Services, and the other is the real-time Tencent user densities (RTUD). They employed the random forest algorithm \cite{Breiman2001} to analyze the two geospatial datasets and downscale the street-level population distribution to the grid level. Then, they proposed an iterative gravity model that can efficiently estimates the population density in each building and study area. Their method achieves a high correlation to the official census data.

The WorldPop collection recently brings together publications describing detailed and open-access spatial demographic datasets built using transparent approaches \cite{Tatem2017}. For the Latin America and the Caribbean region, Sorichetta et al. \cite{Sorichetta2015} opened an archive of high-resolution gridded population datasets for 2010, 2015 and 2020 based on the most recent official population count data for 28 countries. Gaughan et al. \cite{Gaughan2016} opened mainland China population maps for 1990, 2000 and 2010 after analyzing temporally-explicit census data using an ensemble prediction model. Lloyd et al. \cite{Lloyd2017} described the datasets and production methodology for the 3 and 30 arc-second resolution global gridded population data. The basis of the archive contains four tiled raster datasets and other layers.

\subsubsection{International migration}

International migration is one of the major reasons of demographic, economic and political changes. Literature suggested some determinants of migration such as family and personal networks \cite{Boyd1989} and revealed the impact of immigrants on the host country's economy \cite{Friedberg1995}. There are some bottlenecks in studying migration such as data availability, data quality, data collection rules, and inconsistencies in measurement. For example, a person may involve multiple migrations during a given year, but most systems considered the number of migrations instead of migrants, resulting in the overestimate of the amount of immigrants. Moreover, ``migration'' defined by different countries may differ substantially, which results in the inconsistencies among international data. In addition, which country is reporting the data will lead to significant different patterns of migration \cite{Beer2010}. Census and registered migration data are helpful for the estimation of international migrations. By combining census migration data and patient registration data, Raymer et al. \cite{Raymer2007} developed a log-linear model to estimate elderly migration flows in England and Wales. Their model extends the spatial interaction model (see Ref. \cite{Willekens1999} for details) by adding a third variable of interest, such as health status in migration data. Formally, the log-linear model with an offset is given by
\begin{equation}
    \log (\lambda_{ijk})= \log (\alpha_i) + \log (\beta_j) + \log (m_{ijk}),
\end{equation}
where $\lambda_{ijk}$ is the expected migration flow from origin $i$ to destination $j$ for level $k$ of the third variable, $\alpha_i$ and $\beta_j$ are respectively related to the origin and destination's characteristics, and $m_{ijk}$ is the auxiliary information on migration flows.

\begin{figure}[t]
  \centering
  \includegraphics[width=0.4\textwidth]{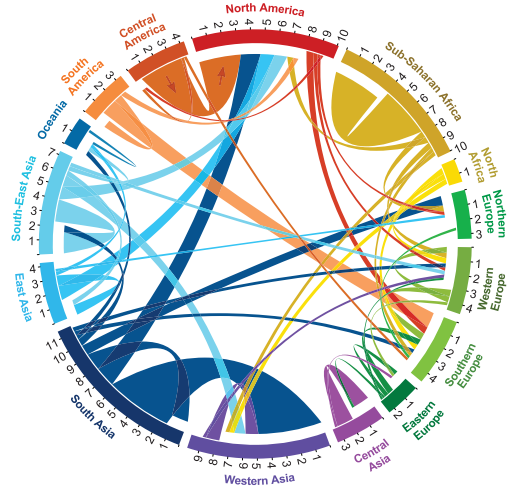}
  \caption{Circular plot of migration flows within and between world regions during 2005 to 2010. Tick marks show the number of migrants (inflows and outflows) in millions. Only flows containing at least 170,000 migrants are shown. Figure from \cite{Abel2014}.}
  \label{Fig_3_3_2}
\end{figure}

Cohen et al. \cite{Cohen2008} developed a generalized linear model (GLM) \cite{Mccullagh1989} to predict international migrants using only geographic and demographic variables. They found that the number of migrants per year depends on population of origin and its population density. De Beer et al. \cite{Beer2010} presented a methodology to estimate total immigration and emigration numbers for 19 European countries. Abel and Sander \cite{Abel2014} provided the spatial structure of international migration flows between 196 countries from 1990 to 2010 (see Figure~\ref{Fig_3_3_2}). The bilateral migration flows are based on refugee statistics, population registers, and place-of-birth responses to census questions. They employed an iterative proportional fitting algorithm \cite{Deming1940} to estimate the global migration flows. They found that the percentage of 5-year flows has been relatively stable at about 0.6\% of world population since 1995. Moreover, African migrants move predominantly within the African continent, Asian and Latin American migration flows are spatially focused, and long-distance flows usually go to higher income level countries with negligible return flows.

The increasingly available geolocated digital records from intelligent devices and online platforms offer the opportunity to better quantify migration flows. Using a large sample of Yahoo! e-mail data, Zagheni and Weber \cite{Zagheni2012} estimated the age and gender-specific migration rates. The locations of users are estimated by the country where their most messages were sent. The self-reported age and gender of users are then linked to their locations. They found that the estimated age profiles of migrants are consistent with the official data, and the mobility of females grows at a faster pace. Using the similar Yahoo! data of over 100 million users, State et al. \cite{State2013} developed a statistical model to identify migrants and tourists. After generating a global mobility map, they found that the European Economic Area has high levels of pendularity, and pendular migrations are in closely located countries. State et al. \cite{State2014} investigated international migration of professional workers by analyzing millions of geotagged career histories on LinkedIn. They found that the percentage of professional migrants to the US decreases from 2000 to 2012, while Asia has been a major professional migration destination during the past twelve years. Kikas et al. \cite{Kikas2015} extracted international migration from the Skype login events, showing that international migration can be estimated based on some social network features such as the percentages of international calls. Barchiesi et al. \cite{Barchiesi2015} extracted the location of users from geotagged photographs on Flickr and inferred their trajectories. The estimated number of visitors to the UK correlates with the official estimates for 28 countries.

Twitter provides a rich source of geotagged data to estimate international migration. Based on about one billion tweets, Hawelka et al. \cite{Hawelka2014} estimated the volume of international travelers according to the country of residence. They revealed spatially cohesive regions after analyzing the community structure of the Twitter-based international mobility network. By analyzing geotagged tweets produced by about 500,000 users, Zagheni et al. \cite{Zagheni2014} evaluated recent trends of migrations in OECD countries. They applied a difference-in-differences approach \cite{Bertrand2004} to reduce selection bias when inferring trends in out-migration rates. Their method can predict turning points in migration trends. Fagiolo and Mastrorillo \cite{Fagiolo2013} analyzed the topological structure of a international migration network (IMN) and its evolution from 1960 to 2000, where nodes are countries and links are the stock of migrants. They found that link weights follow a power-law distribution with a stable exponent at about 1.3. Moreover, IMN is highly clustered, disassortative, with a modular structure and of small-world property. In addition, most topological features of IMN an be explained by GLM, suggesting that socioeconomic, geographical and political factors are important in shaping the structure of migration networks.

International migration issues are prominent on economy and policy agendas. Fagiolo and Mastrorillo \cite{Fagiolo2014} studied how international migrations affect bilateral trade. They found that IMN and trade are strongly correlated with each other, and high centrality in IMN can increase the bilateral trade of countries. These results also indicate that the number of international immigrants can boost bilateral trade. Lee et al. \cite{Lee2014} suggested the research themes to focus on the growth of migration flows driven by humanitarian crises and the connections between migration and inequality. The Global Migration Group \cite{GMG2016} has provided guidance to support the collection, tabulation, analysis, dissemination and use of international migration data to monitor the implementation of the Sustainable Development Goals.

\subsubsection{Culture evolution}

Culture is the essential character of human society, and it serves as a driving force for human development. Quantifying cultural evolution is a challenging task due to the lack of suitable data. Recently, the development of information technologies has made large-scale data available for culture evolution studies \cite{Bail2014} such as digitized books \cite{Michel2011}, baby names\cite{Berger2012}, languages \cite{Ronen2014}, recipes \cite{Zhu2013}, and biographies \cite{Yu2016b}. Moreover, human languages, as an important part of culture, have also been studied using novel data resources (e.g., social media data \cite{Eisenstein2014}) besides evolution models \cite{Larson2010}. Here, we will introduce applications of new data sources and methods in quantifying culture evolution.

Part of the evolution of human society is recorded by books. By analyzing a corpus of 5 million Google digitized books, Michel et al. \cite{Michel2011} observed cultural trends with over two billion culturomic trajectories. Focusing on linguistic and cultural phenomena reflected in the English language between 1800 and 2000, they provided insights about the size of English lexicon, collective memory, and evolution of grammar. In particular, the polarization of the states before the Civil War was revealed by the trajectories of using ``the North'', ``the South'', and finally ``the enemy''. Zeng and Greenfield \cite{Zeng2015} analyzed massive culture-wide content using the Google Ngram Viewer. They found that cultural values shift along with specific ecological changes (urbanisation, wealth and formal education) in Chinese society. In particular, the frequencies of words related to adaptive individualistic values (indexed by words such as ``choose'', ``compete'' and ``get'') increases from 1970 to 2008. Bail \cite{Bail2014} summarized text extraction methods to classify different types of culture and map cultural environments from text-based data. These new tools were further combined with conventional qualitative methods to track cultural element evolution. Figuring out sales trajectories of books will help understand the cultural evolution. Yucesoy et al. \cite{Yucesoy2018} revealed a universal sales pattern of bestsellers, and further proposed a model that can explain the time evolution of book sales.

\begin{figure}[t]
  \centering
  \includegraphics[width=0.6\textwidth]{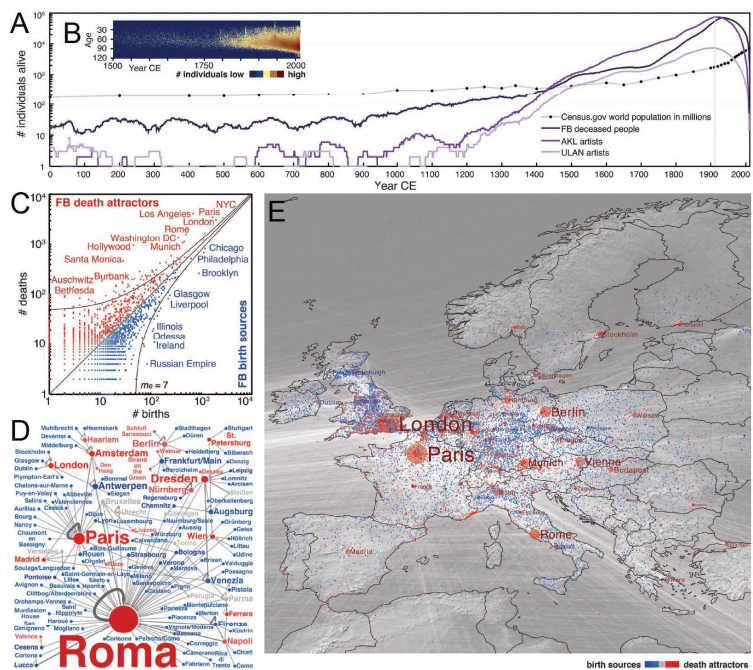}
  \caption{Interactions between culturally relevant locations over two millennia. (A) Notable individuals with birth and death locations. (B) Demographic life table for the Freebase.com (FB) dataset indicating death age frequency. (C) Birth-death scatter plot for locations in FB. (D) Illustration of birth-death flows of antiquarians in the 18th century. (E) Migration in Europe based on FB. Figure from \cite{Schich2014}.}
  \label{Fig_3_3_3}
\end{figure}

Human behavior massively reflects cultural information. Schich et al. \cite{Schich2014} reconstructed the aggregated mobility of over 150,000 intellectual individuals (see Figure~\ref{Fig_3_3_3}) and then measured cultural interactions on a historical time scale. They developed quantitative methods to identify statistical regularities of individuals based on spatiotemporal birth and death information of notable individuals collected from Freebase.com (FB) and other sources (see Ref. \cite{Yu2016b} for a similar dataset of globally famous biographies). They found that the distribution of distances between the birth and death locations of notable individuals remains unchanged over eight centuries. By employing network tools and complexity theory, they further identified the characteristic statistical patterns. In particular, Europe can be characterized by two different cultural regimes. One is winner-takes-all regime, where massive centralization is toward centers, and the other is fit-gets-richer regime, where many sub-centers compete in federal clusters. This work provides a macroscopic perspective of cultural history. Recently, Yang et al. \cite{Yang2016} explored cultural mapping based on user behavioral data collected from location-based social networks. From check-ins messages and check-ins at a city's POIs, they extracted three key cultural features, namely, language usage, daily activity pattern, and intercity crowd mobility patterns. Then, they proposed a cultural clustering method to capture cultural features and generate cultural maps that match traditional survey-based ones.

Names can be used to study the underlying mechanism for cultural evolution. Hahn and Bentley \cite{Hahn2003} analyzed 1000 most commonly used baby names in the US in each decade of the twentieth century. They found that the frequency distribution of baby names obeys a power law for over 100 years, and the distribution can be explained by a simple process where names are randomly copied. Bentley et al. \cite{Bentley2007} explained the steady turnover of modern baby names using a random-copying model. Female names in each decade have a higher turnover rate than male names, implying more innovation in naming girls. The random-copying model can characterize collective copying behavior in culture evolution. Berger et al. \cite{Berger2012} analyzed names given to babies born from 1882 to 2006 in the US. They found that the popularity of names is affected nonlinearly by the similar names that became popular recently. Xi et al. \cite{Xi2014} found the sustained decline of inequality level among baby names with time. The reason behind this observation may be that people have more chances to know others' names, and new names need to be more distinctive and novel. Further, they proposed a stochastic model in which social influence and individual preference determine individual choice of names. Recently, Barucca et al. \cite{Barucca2015} analyzed the correlations of newborns' names in different states of the US from 1910 to 2012. They found a clear division of states into two homogeneous groups, where either group has similarity in their distributions of names. However, a transformation occurred at the end of the 20th century, where new clusters emerged in naming babies. Kim and Park \cite{Kim2005} investigated the distribution of family names in Korea, finding that the growth rates of smaller family names are higher. Lee et al. \cite{Lee2016} analyzed statistics of given names in Korea, Quebec, and the US. They found that the average popularities of given names show similar patterns of rise and fall at about one generation.

Language evolution is an important aspect of culture evolution. From a modeling perspective, Nowak et al. \cite{Nowak2002} showed that some certain evolutionary dynamics can describe both the cultural evolution of language and the biological evolution of universal grammar. Abrams and Strogatz \cite{Abrams2003} developed a simple model of language competition that explains historical data on the decline of some endangered languages. They derived a linguistic parameter that quantifies the threat of language extinction for the model. From an empirical perspective, Lieberman et al. \cite{Lieberman2007} quantified the evolving dynamics of language by analyzing the regularization of English verbs over the past 1,200 years. They explored how the rate of regularization depends on the frequency of word usage and found that the half-life of irregular verbs is proportional to the square root of their frequency. Based on the dataset of 107 million tweets, Eisenstein et al. \cite{Eisenstein2014} investigated the fundamental changes in the nature of written language. After employing a latent vector auto-regressive model to identify high-level patterns in the diffusion of linguistic change over the US, they found that language evolution in computer-mediated communication reproduces existing fault lines in spoken American English. Recently, Newberry et al. \cite{Newberry2017} quantified the strength of selection relative to stochastic drift in language evolution. After inferring selection towards the irregular forms of some past-tense verbs, they found that stochastic drift is stronger for rare words, suggesting that stochasticity plays an under-appreciated role in language evolution.

Vocabulary growth in natural languages follows scaling laws. For example, the character frequency distribution follows Zipf's law \cite{Zipf1949} in the relation $Z(r) \approx r^{\alpha}$, where $r$ denotes the rank of a word by its frequency $Z(r)$, and $\alpha$ is the Zipf's exponent. The number of distinct characters follows Heaps' law \cite{Heaps1978} as $N(t) \approx t^{\lambda}$, where $N(t)$ denotes the number of distinct words when the text length is $t$, and $\lambda \leq 1$ is the Heaps' exponent. Zipf's law and Heaps' law have been widely observed in Indo-European language family and keywords in journals \cite{Zhang2008}. Indeed, these two laws are mathematically related, say Zipf's law leads to Heaps' law \cite{Lu2010}. After analyzing over 15 million words in books, Petersen et al. \cite{Petersen2012} found that only the more common words obey the classic Zipf's law, and the annual growth fluctuation of word usages decreases with the corpus size. Based on the Google Ngram database of books, Gerlach and Altmann \cite{Gerlach2013} proposed a stochastic model for vocabulary growth that can generalize Zipf's and Heaps' law to two-scaling regimes. They found that the main historical change is the composition of specific words, where the list of core words is finite and decays exponentially in time with about 30 words per year for English. Pechenick et al. \cite{Pechenick2017} analyzed the English fiction corpus and found that the Zipf's distribution has changed little from 1820 to 2000.

Some languages like Chinese, Japanese and Korean do not obey Zipf's law or Heaps' law. For these languages with very limited dictionary sizes (the number of characters is much smaller than the number of words), L\"u et al. \cite{Lu2013b} found that $Z(r)$ follows a power law with $\alpha \approx 1$, and $N(t)$ grows with the text length in three stages: $N(t)$ grows linearly at the beginning, then turns to a logarithmical form, and saturates in the end. After analyzing four Chinese texts, Deng et al. \cite{Deng2014} found that Zipf's law perfectly holds for sufficiently short texts of Chinese characters, However, rank-frequency relations display a two-layer structure for long texts, with a Zipfian power-law regime for high-frequent characters in the first layer and an exponential-like regime for less-frequent characters in the second layer. Yan and Minnhagen \cite{Yan2015} proposed a neutral model to predict character frequency distributions in \emph{Chinese characters}, where the maximum entropy prediction is used to describe a text written in Chinese. They demonstrated that the same Chinese texts written in \emph{words} and \emph{Chinese characters} are both well predicted by their three characteristic values (the total number of words, the number of distinct words, and the number of repetitions of the most common word). Yan et al. \cite{Yan2013} further built a node-weighted network of Chinese characters, in which the weights of nodes are the frequencies of character usages, and the directed links correspond to the relations of direct components of characters. They developed a distributed node weight (DNW) strategy for learning Chinese characters and analyzed learning strategies using the dynamical processes. Results showed that the DNW strategy can significantly improve the efficiency of learning major Chinese textbooks.

\begin{figure}[t]
  \centering
  \includegraphics[width=0.85\textwidth]{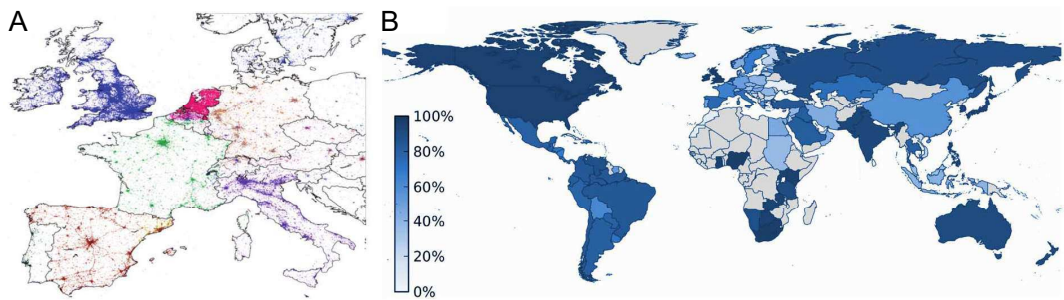}
  \caption{Geographic distribution of languages based on Twitter data. (A) Raw Twitter signal. Each color represents a language. Densely populated areas are easily identified, showing that languages are separated among European countries. (B) Dominant language usage. The color indicates the fraction of users adopting the official language in tweets. Figure from \cite{Mocanu2013}.}
  \label{Fig_3_3_4}
\end{figure}

Language can be used to reveal linguistic and cultural borders. Bryden et al. \cite{Bryden2013} explored the interlink between language and social network structure based on Twitter data. They found that the hierarchy of communities on social networks can be characterized by their most significantly used words. The community of a user can be predicted by the used words in tweets. Based on co-editing activities of Wikipedia, Samoilenko et al. \cite{Samoilenko2016} studied the linguistic neighbourhoods between language communities. They found that similar interests of Wikipedia editors between cultural communities can be explained by bilingualism, linguistic similarity of languages, and shared religion. Further, they proposed a method that can extract cultural borders from the co-editing activities. Mocanu et al. \cite{Mocanu2013} studied worldwide linguistic indicators and trends (see Figure~\ref{Fig_3_3_4}) by analyzing a large-scale dataset of geotagged tweets. They found that Twitter penetration is highly heterogeneous and it is strongly correlated with GDP. Moreover, tweets can be used to study linguistic homogeneity at the country level, map language distributions in regions, and identify linguistically specific communities in urban areas.

Language can also reflect the ability of cultural influence. Ronen et al. \cite{Ronen2014} proposed a quantitative measure for a language's global influence based on the structure of three global language networks (GLNs). The GLNs are constructed by identifying significant links between languages with respect to the population of speakers expressed in three datasets. Formally, the correlation $\phi_{ij}$ between languages $i$ and $j$ is given by
\begin{equation}
\phi_{ij} = \frac{M_{ij}N-M_{i}M_{j}}{\sqrt{M_{i}M_{j}(N-M_i)(N-M_j)}},
\end{equation}
where $M_{ij}$ is the number of multilingual users (or book translations) between languages $i$ and $j$, $M_{i}=\sum_j M_{ij}$, and $N$ is the total number of users (or book translations). The statistical significance of the correlation is given by the $t$ statistic,
\begin{equation}
t_{ij} = \frac{\phi_{ij}\sqrt{D-2}}{\sqrt{1-\phi_{ij}^2}},
\end{equation}
where $D-2$ is the degree of freedom and $D=\max(M_i, M_j)$. Empirical results show that the position of a language in the GLNs contributes to the visibility of its speakers and the global popularity of the cultural content they produce. Gon{\c{c}}alves et al. \cite{Gonccalves2018} explored a large corpus of geotagged tweets and the Google Books datasets corresponding to books published in the US and the UK. After studying how the world-wide varieties of written English are evolving, they found that the past two centuries have clearly resulted in a clear shift in vocabulary and spelling conventions from British to American. The result suggests the capacity to culturally influence the rest of the world gradually shifts from the UK to the US.

Food is an integral part of cultures. Counihan and van Esterik \cite{Counihan2013} analyzed food-related activities and presented a crosscultural study of personal identities and social groups. They introduced empirical and theoretical tools to understand food systems at multiple levels. Data of cuisines have been used to study food culture. Ahn et al. \cite{Ahn2011} explored cultural diversity by analyzing the variety of regional cuisines. After introducing a flavor network to capture the ingredient combinations in recipes, they found that Western cuisines tend to use compound sharing ingredients, supporting the food pairing hypothesis that ingredients having similar flavor compounds may taste well together \cite{Blumenthal2008}. By contrast, East Asian cuisines show a tendency to avoid food pairing. Zhu et al. \cite{Zhu2013} explored the similarity of regional cuisines in China based on online recipes. They found that geographical proximity plays a more crucial role than climate proximity in determining regional cuisine similarity. Further, they proposed an evolution model of Chinese cuisines that achieves the similar tendency as the real dataset. Their work extends our understanding of the evolution of Chinese regional cuisines and cultures.

Food preference can reflect cultural diversity and cross-cultural relations. Based on a server log data from a large recipe platform, Wagner et al. \cite{Wagner2014} explored the evolution of food preferences. They found that ingredients partly drive recipe preferences, and ingredient preference distributions have less regional differences than recipe preference distributions. Moreover, weekday preferences differ from weekend preferences. Abbar et al. \cite{Abbar2015} studied US-wide dietary choices by analyzing dining experiences tweeted by 0.21 million users. They found that the caloric values of tweeted foods have a high correlation (0.77) with the state-wide obesity rates. Moreover, users in higher-educated areas tweeted about food with less caloric. Based on twitted food names and demographic variables, they built a model that can well predict county-wide obesity and diabetes statistics. Laufer et al. \cite{Laufer2015} explored the cross-cultural relations based on 31 European food cultures recorded by Wikipedia. They mined cultural relations through the collective description and popularity of culinary practices within and across different Wikipedia language communities. They found that shared internal states (e.g., beliefs and values) are positively correlated with shared culinary practices, and neighbouring countries tend to have similar cultural practices.

\section{Regional socioeconomic status and urban perception}
\label{Sec3}

\subsection{Economic activity and socioeconomic status}

High-resolution data and improved methods allow us to reveal economic activity and socioeconomic status in subnational, regional and urban scales. In this section, we will introduce recent works on predicting regional economic activity from nighttime lights (NTLs), mapping slums from very high resolution (VHR) imagery, inferring regional socioeconomic status from mobile phone (MP) data, and quantifying regional economic development based on social media (SM) data.

\subsubsection{Nighttime lights reflect economic activity}

Nighttime lights (NTLs) data have been widely used to infer socioeconomic status and predict income per capita at the regional and urban scales. Sutton et al. \cite{Sutton2007} proposed a regression model to estimate subnational GDP of India, China, Turkey, and the US. They estimated urban population using a log-log relationship between population and the areal extent of lit areas in NTLs imagery derived from the Defense Meteorological Satellite Program-Operational Line Scan System (DMSP-OLS) \cite{Elvidge2001}. They predicted GDP for the subnational administrative units by adding the estimated urban population into the regression model, with an assumption that urban population is the most critical factor for economic activity. They found that spatial disaggregation of estimates dramatically improved the aggregated national estimates of GDP based on NTLs.

Regional and urban socioeconomic status can be inferred from the changes in electric power consumption patterns reflected by NTLs that are derived from the DMSP-OLS data. Chand et al. \cite{Chand2009} studied the socioeconomic development of states and cities in India by looking at the spatial and temporal changes in electric power consumption. They found that the number of NTLs overall increases up to 26\% in all states from 1993 to 2002, but there is a decline in some states. The increase in population correlates with both the increase in NTLs ($R^2=0.59$) and the electric power consumption ($R^2=0.56$). For the Republic of Kazakhstan, Propastin and Kappas \cite{Propastin2012} leveraged NTLs to monitor socioeconomic indicators (e.g., population, electricity consumption and GDP) at different spatial resolutions. Linear regression models were used to estimate population and electricity consumption at the settlement level. They revealed a strong correlation between NTLs and GDP. In particular, the regression model can explain 76\% of the spatial variability in GDP among 17 provinces and 94\% of the inter-annual variation in total GDP of Kazakhstan during 1994-1999.

Luminosity from NTLs satellite imagery has been used as a proxy for economic statistics. Chen and Nordhaus \cite{Chen2011} studied how much luminosity can contribute to the construction of GDP measures. They proposed an analytic method to quantify the relationship between luminosity from the DMSP-OLS data and GDP of North America. They found that luminosity is likely to add values as a proxy to estimate economic output for countries and regions. Due to a high measurement error of luminosity, however, the added values are limited for countries and regions with poor quality data. For more than 200 cities in China, Ma et al. \cite{Ma2012} comparatively used three regression models to study the responses of stable NTLs from the DMSP-OLS data to changes in urbanization variables (e.g., population, GDP, electric power consumption, and built-up area). They found that NTLs can help estimate urbanization dynamics.

Mellander, et al. \cite{Mellander2015} studied the relationship between NTLs and economic activity at a fine level. They used a geo-coded socioeconomic dataset consisting of spatially matched population and establishment counts in Sweden. After matching the dataset with light emissions, they used correlation analysis and geographically weighted regressions (GWR) to examine the relationship. The GWR model is given by
\begin{equation}
    y_i = \beta_0(i) + \beta_1(i)x_{1i} + \beta_2(i)x_{2i}+ \ldots + \beta_n(i)x_{ni} + \varepsilon_i,
\end{equation}
where $i$ denotes location, $y$ is the dependent variable of NTLs, $x$ is a population-based or industry-based socioeconomic variable, and $\varepsilon$ is the error term. They found that NTLs is a good proxy for population density and a weaker proxy for economic activity at the micro-level. The link between NTLs and economic activity is slightly overestimated for large urban areas but underestimated for rural areas. Moreover, economic activity has a stronger correlation with radiance light than saturated light.

\begin{figure}[t]
  \centering
  \includegraphics[width=0.45\textwidth]{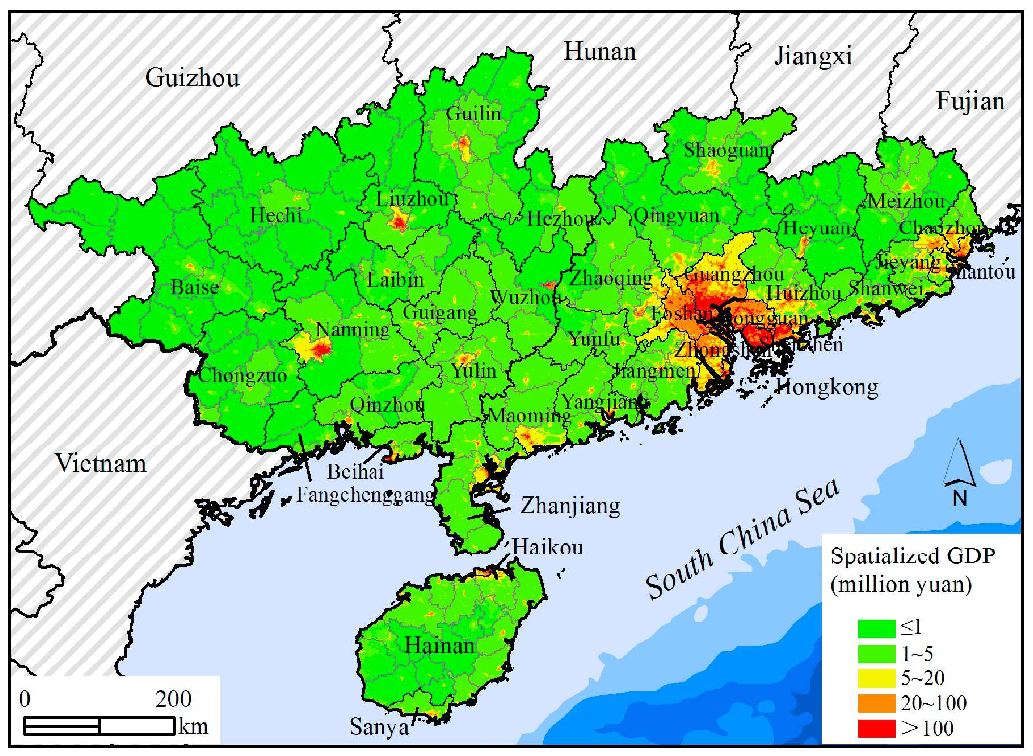}
  \caption{The pixel-level (500m $\times$ 500m) GDP map for South China in 2014. The GDP map was produced by using the corrected NPP-VIIRS data and the regression model. Figure from \cite{Zhao2017}.}
  \label{Fig_4_1_1}
\end{figure}

NTLs satellite imagery has been used to study the economic differences and GDP spatialization at the regional levels in China. For example, Zhao et al. \cite{Zhao2011} examined the relations between China's environmental change and economic development from 1996 to 2000. A proxy evaluator of ecosystems is net primary production (NPP), which is the amount of carbon and energy that enters ecosystems. A proxy evaluator of economic development is GDP, which is estimated based on NTLs. They found that changing patterns of NPP and GDP vary by regions, and the relations are greatly affected by spatial locations. Later, Zhao et al. \cite{Zhao2017} studied economic differences among diverse geomorphological types in South China based on NTLs from the Visible Infrared Imaging Radiometer Suite (VIIRS) \cite{Baugh2013}. They found that the total NTLs exhibit a high correlation with both prefecture GDP ($R^2=0.8935$) and county GDP ($R^2=0.9243$), suggesting the capability of NTLs in estimating economic development. Further, they proposed a GDP spatialization model and produced a pixel-level GDP map for South China (see Figure~\ref{Fig_4_1_1}). Meanwhile, Dai et al. \cite{Dai2017} explored the suitability of using NTLs and regression methods to estimate GDP at the provincial and city levels in China. Empirical analysis suggested that the VIIRS data outperforms the DMSP-OLS data, and the polynomial model outperforms the linear regression model.

\subsubsection{Very high resolution imagery maps poverty}

Indicators derived from both nighttime lights (NTLs) and very high resolution (VHR) imagery have been used to map poverty at fine scales. By analyzing stable NTLs, Weng et al. \cite{Wang2012} estimated poverty for 31 provinces in China. Specifically, they used the principal component analysis (PCA) to develop a so-called Integrated Poverty Index (IPI) based on 17 socioeconomic indexes and computed the average light index (ALI) for each region based on the total number of lit pixels. They found a high correlation ($r=0.85$) between IPI and ALI for the 31 provinces, suggesting the capacity of NTLs imagery to analyze poverty at the regional level. Using a supervised learning approach, Engstrom et al. \cite{Engstrom2017} linked features derived from VHR satellite images to survey-based poverty rates at the local level in Sri Lanka. They found that satellite-based features are highly predictive to poverty. Moreover, measures of built-up area and building density are strongly correlated with welfare in both urban and rural areas.

Slums are common in low and middle income countries with poor quality of basic services (e.g., water supply, electricity and sanitation). Detecting and monitoring slum areas is valuable for implementing policies to improve living conditions. However, mapping the spatial distribution of urban slums is a challenging problem due to the reasons that there is no universal model of slums and slum conditions can take various forms. The meeting in 2008 \cite{Sliuzas2008} brought together the methodological expertise on slum monitoring and reviewed methods for slum identification based on VHR imagery. Using object-oriented classification of VHR images, Rhinane et al. \cite{Rhinane2011} developed a novel approach to detect slums in Casablanca, which integrates spectral, spatial and contextual information to map the urban land, achieving a high accuracy 0.85 in extracting slum areas.

VHR images have been increasingly used to inventory the location and physical composition of slums. Shekhar \cite{Shekhar2012} applied the object oriented analysis \cite{Cheng2003} to VHR images to detect slums in Pune, India. First, they generated segments by automatically dividing images into coherent objects. Then, they used the feature extraction method to identify the characteristic features for object-classes. Finally, they used contextual information to separate slum objects from non-slum built-up objects. Their approach exhibits the overall accuracy 87\% in the classification of slums. Kohli et al. \cite{Kohli2012} developed an ontological framework to conceptualize slums based on input from 50 domain-experts covering 16 different countries. They identified the morphology of built environment at the environs, settlement and object levels. By including all potentially relevant indicators, their ontological framework provides a comprehensive basis for image-based classification of slums.

\begin{figure}[t]
  \centering
  \includegraphics[width=0.4\textwidth]{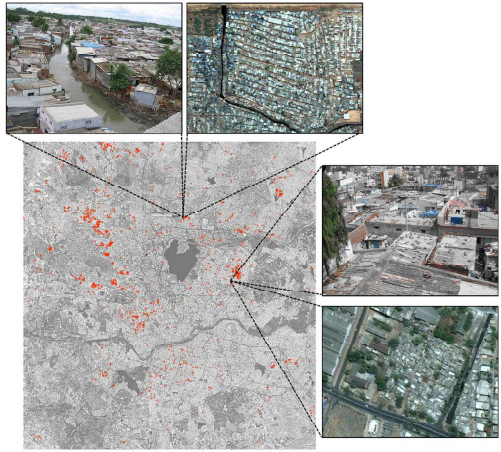}
  \caption{Slum map in Hyderabad, India. The slum locations (red areas) are identified by the lacunarity-based slum detection algorithm. Two different subsets of the original satellite image together with georeferenced photographs are shown as ground truth: the Rasolpoora slum in the northern and the Nagamiah Kunta slum in the eastern. Figure from \cite{Kit2012}.}
  \label{Fig_4_1_2}
\end{figure}

Recent literature have applied advanced image processing techniques to map slums from VHR images with minimal operator intervention. Kit et al. \cite{Kit2012} developed the concept of lacunarity to identify slums in Hyderabad, India. First, they produced high resolution binary image using two binarization methods, the principal component analysis (PCA)-based method and the line detection-based method \cite{Martinez2005}. Then, they calculated lacunarity based on the binary image following Malhi and Rom\'an-Cuesta \cite{Malhi2008}. Formally, the lacunarity $\Lambda$ of a subset $P$ of the original binary image is given by
\begin{equation}
    \Lambda = \frac{\sigma_r}{\bar{x}_{r}^{2}} + 1 ,
\end{equation}
where $\sigma_r$ is the variance and $\bar{x}_{r}$ is the arithmetic mean of the number of filled pixels within all $r$-sized unique square subsets of the larger subset $P$ (see Ref. \cite{Kit2012} for details). The line detection algorithm performs better than the PCA-based method in providing suitable binary datasets for lacunarity analysis. The best method can reach an accuracy 0.8333 in slum identification when $\Lambda \in [1.10, 1.15]$. Figure~\ref{Fig_4_1_2} shows the slum map of Hyderabad generated by the lacunarity-based slum detection algorithm.

Kit et al. \cite{Kit2013} soon improved the lacunarity-based slum detection algorithm by combing two advanced image analysis methods (the Canny edge detection \cite{Canny1986} and the line-segment-detection (LSD) straight line detection \cite{Gioi2012}) to reduce errors in slum identification. Their method identifies the plausibly and spatially explicit slum locations, which can be verified by a series of ground truthing visits. In particular, such method can capture the changing patters of slum areas from 2003 to 2010 in Hyderabad, India. Gruebner et al. \cite{Gruebner2014} mapped urban slums in Dhaka, Bangladesh, from the visual interpretation of Quickbird data from 2006 to 2010. To avoid small and isolated slums, they filtered the 2006 slums in GIS and defined the changes of 2010 slums over the 2006's polygons to retain border consistency. Accordingly, they produced a slum distribution dataset for the Dhaka metropolitan area.

Engstrom et al. \cite{Engstrom2015} mapped slum areas in Accra, Ghana, by utilizing features extracted from the VHR Quickbird images acquired in 2002. They demonstrated that the satellite image-derived slum areas exhibits an overall accuracy of 94.3\% when comparing to the field-based slum map from the UN Habitat/Accra Metropolitan Assembly (UNAMA). However, the accuracy drops when comparing to two census derived slum maps. Moreover, they found a moderate correlation ($r = 0.67$) between satellite image-derived classification of slums and the census derived slum index, and the correlation increases ($r = 0.88$) after taking into account population density. Kohli et al. \cite{Kohli2016} studied the spatial uncertainties related to slum delineations, which are observed from VHR images in Ahmedabad (India), Nairobi (Kenya) and Cape Town (South Africa). They found that the slum identification and delineation for the three contexts are significantly different, suggesting the existential and extensional uncertainty of slums.

VHR imagery allows the monitoring of slums and the analysis of deprived areas. Kuffer et al. \cite{Kuffer2016b} utilized the gray-level co-occurrence matrix (GLCM) variance to distinguish slums areas in VHR imagery. They showed that the GLCM variance combined with the normalized difference vegetation index (NDVI) can separate slum areas with an overall accuracy 87\%, 88\% and 84\% for Mumbai (India), Ahmedabad (India) and Kigali (Rwanda), respectively. The overall accuracy can be increased to 90\% by adding spectral information to the GLCM within a random forest classifier \cite{Breiman2001}. Wurm et al. \cite{Wurm2017} explored the capabilities of X-band Synthetic Aperture Radar (SAR) data to estimate the extent of poverty in slum areas using the Kennaugh element framework in image preprocessing \cite{Schmitt2015}. Employing a random forest classifier, they tested different spatial image features at various window sizes to map slums. Results show that GLCM performs very well on slum mapping as it addresses a large spatial neighborhood of the pixels.

Recently, Kuffer et al. \cite{Kuffer2016} provided a literature review of slum mapping regarding four dimensions: contextual factors, physical slum characteristics, data and requirements, and slum extraction methods. They argued that the diversity and dynamics of slums have not been well captured due to the complex and diverse morphology of slums. Thereby, a more systematic exploration of physical slum characteristics is required (see Ref. \cite{Kuffer2016} for details). They demonstrated that texture-based methods show good robustness, while machine-learning algorithms exhibit the highest reported accuracy. Mahabir et al. \cite{Mahabir2018} suggested to develop a more comprehensive framework by considering emerging sources of geospatial data (e.g., social media) and combining multiple emerging approaches in technology (e.g., geosensor networks).

\subsubsection{Mobile phones track socioeconomic levels}

Scientists have explored the relations between social structure and economic development. Woolcock \cite{Woolcock1998} provided a brief intellectual history of social capital and economic development. Adler and Kwon \cite{Adler2002} synthesized studies on social capital undertaken in various disciplines and developed a common conceptual framework. Granovetter \cite{Granovetter2005} suggested several underlying mechanisms on how social structure affects economic outcomes, for example, social networks influence the flow and the quality of information, and social networks are an important source of reward and punishment. Recently, empirical works have demonstrated that social network analysis of large-scale mobile phone (MP) data can be applied to monitor socioeconomic development.

\begin{figure}[t]
  \centering
  \includegraphics[width=0.7\textwidth]{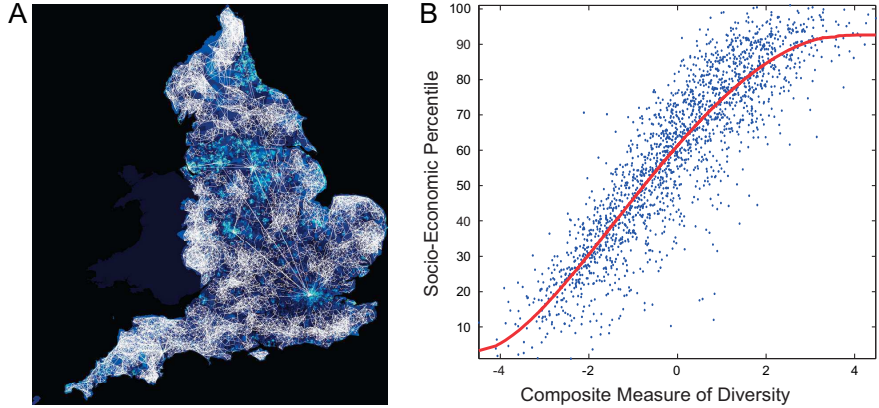}
  \caption{Regional communication diversity and socioeconomic ranking for the UK. (A) Communication networks of regions and regional socioeconomic rank based on IMD. Saturation and width of links correspond to the volume of communications. High rank to low rank of the IMD is represented by light blue to dark blue. (B) The relation between social network diversity and socioeconomic rank. The network diversity was a composite of Shannon entropy and Burt's measure of structural holes. The fractional polynomial fit to the data is shown in red. Figure from \cite{Eagle2010}.}
  \label{Fig_4_1_3}
\end{figure}

Based on MP data and socioeconomic metric from national census, Eagle et al. \cite{Eagle2010} investigated the relation between the structure of communication network and economic development at the population level in the UK (see Figure~\ref{Fig_4_1_3}). The socioeconomic metric is the 2004 UK government's index of multiple deprivation (IMD), which is a composite measure of relative prosperity of communities. They calculated the socioeconomic profile of a region by aggregating the population-weighted average of the IMD for each telephone exchange area. The communication network data covers over 90\% of MP users during August 2005, based on which they calculated two diversity metrics of communication ties. The social diversity $D_{\text{social}}(i)$ is defined as the Shannon entropy $H(i)$ associated with individual $i$'s communication behavior normalized by its number of contacts $k$. Formally,
\begin{equation}
D_{\text{social}}(i) = \frac{H(i)}{\log (k)} = \frac{- \sum_{j=1}^{k} p_{ij} \log (p_{ij})}{\log (k)},
\end{equation}
where $p_{ij}$ is the proportion of individual $i$'s call volume that involves individual $j$. A regions's social diversity is then calculated by averaging the social diversities of individuals in that region. The spatial diversity $D_{\text{spatial}}(i)$ is defined by replacing call volume with geographic distance. Formally,
\begin{equation}
D_{\text{spatial}}(i) = \frac{- \sum_{a=1}^{A} p_{ia} \log (p_{ia})}{\log (A)},
\end{equation}
where $A$ is the number of exchange areas, and $p_{ia}$ is the proportion of time that individual $i$ spent on communicating with area $a$. They found that the IMD socioeconomic rank is strongly correlated with both the social diversity $D_{\text{social}}$ ($r=0.73$) and the spatial diversity $D_{\text{spatial}}$ ($r=0.58$). The strong correlation ($r=0.72$) persists if using Burt's measure of ``structural holes'' (see Ref. \cite{Burt1995} for details). Moreover, a composite diversity measure can exhibit even stronger correlation ($r=0.78$) with socioeconomic status (see Figure~\ref{Fig_4_1_3}B). This work takes a significant step towards inferring regional socioeconomic status from MP data.

The ubiquitous adoption of MPs in emerging economies provides a new way to track socioeconomic status. Based on CDRs of 0.22 million MP users in an advanced economy and 0.19 million MP users in a developing economy, Rubio et al. \cite{Rubio2010} studied human mobility patterns in regions of different socioeconomic levels. They found that individuals in the developing economy have smaller average traveled distance, their social networks have smaller geographical sparsity, and these patterns have no significant changes from workweeks to weekends. Later, Frias-Martinez and Virseda \cite{Frias2012} explored the relations between behavioral features extracted from large-scale CDRs and socioeconomic indices from country-wide census data in a Latin American country. They found that socioeconomic levels are strongly correlated with expenses, reciprocity of communications, physical distance with contacts, mobility patterns, and some others. Moreover, a multivariate linear regression including MP usage variables can accurately predict census-based variables such as the socioeconomic level ($R^2=0.83$). These results suggest MP-derived human mobility patterns can be used to predict socioeconomic indices at fine scales.

A body of literature have leveraged CDRs to estimate regional socioeconomic status. A widely used CDRs data contain 2.5 billion calls and SMS exchanges from anonymous customers in C{\^o}te d'Ivoire. Smith-Clarke et al. \cite{Smith2014} estimated socioeconomic levels of regions in C{\^o}te d'Ivoire. They derived some socioeconomic-related features from the communication flows, including activity, gravity residual, network advantage, and introversion. They found that regions with higher call volumes from other regions are more likely to have a higher socioeconomic level. Further, they proposed a simple linear model that estimates socioeconomic status for 255 sub-prefectures in C{\^o}te d'Ivoire. {\v{S}}{\'c}epanovi{\'c} et al. \cite{Scepanovic2015} extracted different spatial-temporal mobility patterns from the same CDRs in C{\^o}te d'Ivoire and used them to predict socioeconomic indices. They showed that the spatial-variance of calling frequency can identify electricity lacking rural and regions, the spatial-variance of the probability density functions of the radius of gyration (see Ref. \cite{Gonzalez2008} for the definition) can identify a region's wealth, and the number of a region's migration workers is negatively correlated ($r=-0.7681$) with the multidimensional poverty index (MPI).

Recently, MP data have been combined with other data sources to study socioeconomic stratification. Leo et al. \cite{Leo2016} analyzed a coupled datasets of MP communications and bank transactions for over one million people in Mexico. They constructed a social network based on call/SMS interactions and estimated economic indicators based on bank transactions. After calculating the cumulative distributions of individual average monthly purchase (AMP) and debt (AMD), they found that both wealth and debt are unevenly distributed among people. Further, they studied the social stratification by categorizing users into nine socioeconomic classes using the cumulative AMP function. They observed that people are more densely connected to others of their own class. To quantify this observation, they calculated the ``rich-club'' coefficient \cite{Zhou2004},
\begin{equation}
\rho(P_{>}) = \frac{\phi(P_{>})}{\langle \phi_{rn} \rangle (P_{>})},
\end{equation}
where $\phi(P_{>}) = 2L_{P_>} / N_{P_>}(N_{P_>}-1)$ and $\langle \phi_{rn} \rangle (P_{>})$ is the average density. Here, $L_{P_>}$ and $N_{P_>}$ are respectively the number of links and nodes remaining in the communication network after removing nodes with their AMP value $P_u$ smaller than a given threshold $P_>$ (see Ref. \cite{Leo2016} for details). They found that the rich-club coefficient grows rapidly with $P_>$, suggesting an assortative socioeconomic correlation.

Moreover, Leo et al. \cite{Leo2016} studied the spatio-socioeconomic correlations by calculating the average geodesic distance between any pairs of socioeconomic classes,
\begin{equation}
\langle d_{\text{geo}}(s_i, s_j) \rangle = \frac{1}{|E(s_i, s_j)|} \sum_{\substack{ (u,v) \in E \\ u \in s_i, v \in s_j}} d_{\text{geo}}^{\text{zip}}(u,v),
\end{equation}
where $|E(s_i, s_j)|$ is the number of links between nodes in classes $s_i$ and $s_j$, and $d_{\text{geo}}^{\text{zip}}(u,v)$ is the geodesic distance between zip locations of individuals $u$ and $v$. They found that the distance $\langle d_{\text{geo}}(s_i, s_j) \rangle$ is always minimal between individuals of the same class, suggesting that individuals from the same socioeconomic class live relatively the closest. In addition, there is a positive correlation between individuals' socioeconomic levels and their typical commuting distances. After further exploring the same coupled dataset, Leo et al. \cite{Leo2016b} found a strong correlation between identified socioeconomic classes and typical consumption patterns.

\subsubsection{Social media reveals socioeconomic status}

Social media (SM) data have many appealing advantages including low acquisition cost, wide geographical coverage and real-time update, which enable the feasibility to estimate socioeconomic status at regional and urban scales. For example, Twitter provides a huge number of tweets with user locations being directly tagged or can be mined out from content information. Cheng et al. \cite{Cheng2010} proposed a probabilistic framework to automatically identify words related to locations in tweets and then infer a Twitter user's location at the city level from the content. They showed that about one hundred tweets are enough for their method to infer a user's location. This method can place on average 51\% users within 100 miles of their actual locations.

The contents of SM posts have been used to track socioeconomic well-beings. Quercia et al. \cite{Quercia2012} studied the relations between sentiment expressed in tweets and census-based socioeconomic well-being of communities in London. Specifically, they calculated the word count sentiment score \cite{Kramer2010} by counting the number of positive and negative words. Formally,
\begin{equation}
S_{i}^{\text{WC}} = \frac{p_i - \mu_p}{\sigma_p} - \frac{n_i - \mu_n}{\sigma_n},
\end{equation}
where $p_i$ ($n_i$) is the fraction of positive (negative) words for user $i$, $\mu_p$ ($\mu_n$) is the mean of $p$ ($n$) across all users, and $\sigma_p$ ($\sigma_n$) is the corresponding standard deviation. Then, the gross community happiness (GCH) of a community is calculated by averaging the sentiment scores of users in that community. The GCH is highly correlated with the community's socioeconomic well-being, suggesting the effectiveness of using tweets to track community well-being.

Mahmud et al. \cite{Mahmud2012} inferred home locations of Twitter users at different granularities using an algorithm that ensembles statistical and heuristic classifiers \cite{Jimenez1998}. The algorithm achieves a higher performance in predicting Twitter users' locations compared with the state-of-the-art algorithms. Hasan et al. \cite{Hasan2013} analyzed human activity patterns based on tweets with location information. By finding the distributions of different activity categories over a city geography, they characterized aggregate activity patterns and determined the purpose-specific activity distribution maps. Moreover, the timing distribution of visiting different places depends on activity category. Hasan and Ukkusuri \cite{Hasan2014} further proposed a data-driven modeling approach based on topic models \cite{Blei2003} to infer urban activity pattern from geotagged tweets. Results demonstrated that their model can extract user-specific activity patterns and predict missing activities.

Using Twitter data generated during weekdays in Inner London, Lansley and Longley \cite{Lansley2016} applied an unsupervised learning algorithm to classify geo-tagged tweets into 20 distinctive and interpretive topic groupings. They found that users' socioeconomic characteristics can be inferred from their behaviours on Twitter. In particular, users whose neighbourhoods are of higher socioeconomic levels tend to tweet optimistically and discuss business, networking and leisure. Huang and Wong \cite{Huang2016} explored to what extent Twitter data can be used to support the activity pattern analysis of users with different socioeconomic status. Activity patterns of Twitter users in Washington, D.C. were analyzed, and their socioeconomic levels were inferred by incorporating census data. Results showed that socioeconomic status remarkably affects users' activity patterns. Moreover, the urban spatial structure is a key factor that affects the variation in activity patterns among users from different communities. In particular, the mid-income group other than the most affluent group may have the shortest travel. Moreover, affluent residents are more internationally oriented than mid-income and poor residents.

\begin{figure}[t]
  \centering
  \includegraphics[width=0.72\textwidth]{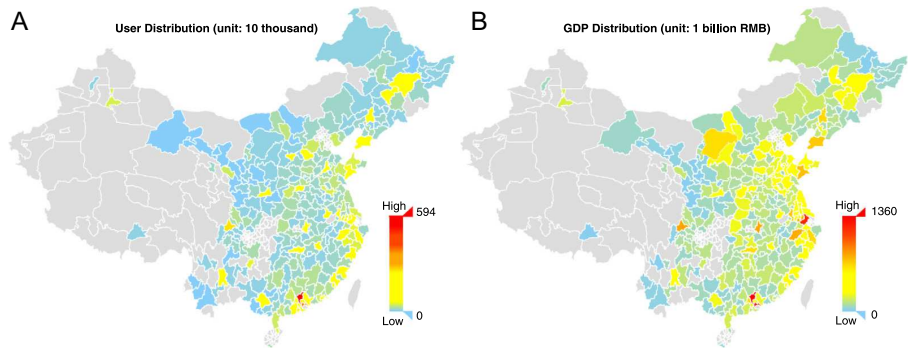}
  \caption{The spatial distributions of (A) the number of registered users in Weibo and (B) the values of GDP in the 282 prefecture-level cities of China in 2012. Figure from \cite{Liu2016}.}
  \label{Fig_4_1_4}
\end{figure}

Liu et al. \cite{Liu2016} collected the registered location information of nearly 200 million Weibo users from 2009 to 2012 and explored the relationship between online activities and socioeconomic indices. Specifically, the online activity is estimated by the number of registered users (UN), and the socioeconomic indices are resident population (RP), GDP and GDP per captia. Figure~\ref{Fig_4_1_4} presents the spatial distributions of registered Weibo users (left) and the values of GDP (right). After calculating two correlation coefficients (Pearson coefficient $r$ \cite{Stigler1989} and Spearman's rank coefficient $\rho$ \cite{Myers2010}), they showed that UN is strongly correlated with socioeconomic indices. For example, the strengths of correlation between UN and GDP are $r=0.88$ and $\rho=0.90$. These results demonstrate that socioeconomic status can be inferred from online social activity at the city-level. Of particular significance, they further proposed a method to detect a few abnormal cities, whose GDP is much higher than others with the same number of registered users. These GDP winners have less-diverse economic structure and highly dependent on some specific resources. In fact, these cities' economics experienced a huge loss after 2013 due to the market price fluctuation of non-renewable energy resources and rare earths.

\begin{figure}[t]
  \centering
  \includegraphics[width=0.68\textwidth]{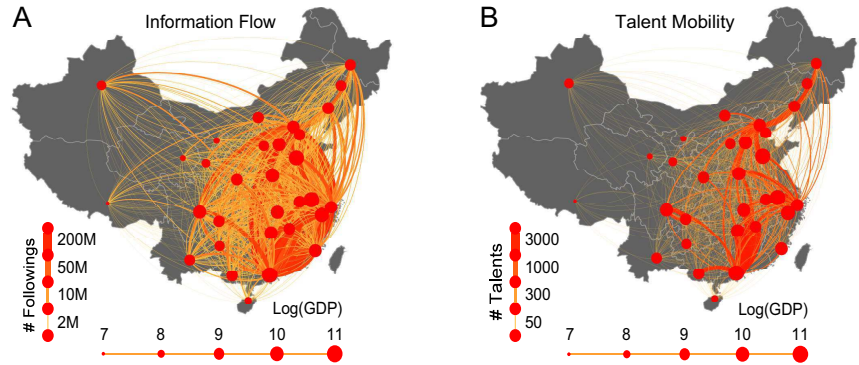}
  \caption{Networks of online information flow and offine talent mobility. Nodes represent provinces, where the size of node corresponds to the province's GDP in natural logarithmic form in 2016, and the layout of node corresponds to the geographical location of the province's capital city. (A) The online information flow network, where the link weight corresponds to the number of followings on Weibo. (B) The offline talent mobility network, where the link weight corresponds to the number of moved talents as recorded by their resumes. Figure from \cite{Wang2019b}.}
  \label{Fig_4_1_5}
\end{figure}

The structure of location-based social networks (LBSN) has been linked to socioeconomic development. Wang et al. \cite{Wang2019b} estimated regional economic status based on the structures of information flow and talent mobility networks (see Figure~\ref{Fig_4_1_5}). Specifically, the online information flow network is built on the following relations among about 433 million Weibo users (see also Ref. \cite{Liu2016}), and the offline talent mobility network is built on the resumes of about 142 thousand anonymized Chinese job seekers with higher education (see Ref. \cite{Yang2018b} for details). They calculated ten network structural features such as spatial and topological diversities and then linked them to regional economic indices. They found that structural features of both networks are relevant to economic status, while the talent mobility network exhibits a stronger predictive power for regional GDP. Further, they constructed a composite index of structural features, which can explain up to about 84\% of the variance in regional GDP.

Based on data from the SM platform Gowalla \cite{Nguyen2012} with friendship information and geo-locations, Holzbauer et al. \cite{Holzbauer2016} studied the relations between regional economic status and quantitative measures of social ties in the US during 2009-2012. They found that cross-state long ties are strongly correlated with three economic measurements, namely, GDP ($r=0.921$), the number of patents ($r=0.788$), and the number of startups ($r=0.892$), while short ties are much less predictive. This finding highlights the role of long ties in supporting regional innovation and economic development. Recently, Norbutas and Corten \cite{Norbutas2018} explored the relations between network structure and economic prosperity of 438 municipalities in Netherlands by analyzing data of over 10 million users on the Dutch online social network Hyves. They found that network diversity in terms of geographical distance \cite{Scellato2010} other than contacts' topological diversity \cite{Eagle2010} exhibits a positive correlation with economic prosperity, while network density at the community level and network modularity \cite{Guimera2004,Newman2006} are negative predictors of economic status.

SM data have also been used to measure socioeconomic deprivation of regions (e.g., low level of economic status and lack of education) and quantify landscape values (e.g., the values shaped by the recreational and cultural services and benefits provided by landscapes). Venerandi et al. \cite{Venerandi2015} proposed a method to automatically mine deprivation from two datasets of urban elements in physical environment at a fine level in UK. The two datasets are respectively collected from Foursquare, a mobile social-networking application with check-ins, and OpenStreetMap, an openly global accessible map with geographical positions, names and categories. They defined the offering advantage to identify distinctive urban elements of each neighborhood (see Ref. \cite{Venerandi2015} for details) and built accurate classifiers of urban deprivation that can be verified by the census-based IMD. Later, van Zanten et al. \cite{Zanten2016} analyzed data from three online SM platforms (Panoramio, Flickr and Instagram). They found that data from these three platforms reveal similar patterns of landscape values. In particular, a significant portion of observed variation across different platforms can be explained by variables describing accessibility, population density, income, mountainous terrain, proximity to water, and so on.

\subsection{Industrial structure and development path}

Data from many new sources have been used to quantify economic structure and analyze industrial diversification, including large-scale social media data, labor market data, trade data, publicly listed firm data, and so on. In this subsection, we will briefly introduce the quantification of regional industrial structure, the role of relatedness on economic diversification, the collective learning effects, and the strategies for regional economic development.

\subsubsection{Economic structure and relatedness}

Economic development is not only a process of continuously improving the production of the same goods and the occupation of the same industries \cite{Lin2011}, but also one that requires structural transformation toward new economic activities associated with higher levels of productivity \cite{Hausmann2007,Hidalgo2007}. This implies that economic development and industrial diversification is a path-dependent process where structural transformation plays an important role. Revealing industrial structure of regions and quantifying relatedness of industries are critical for understanding development paths of regions and evolution patterns of regional economic diversification.

Data from LBSNs have been used to reveal regional economic structure. Based on location information of Weibo users, Liu et al. \cite{Liu2016} proposed an effective method to uncover the city-level macro economic structure in China. They employed the linear least square method to model the relations between user number (UN) and GDP for 282 prefecture-level cities. They found that cities below the fitting line are likely to be driven by the tertiary industries, while cities above the fitted line tend to focus on the secondary industry. They quantified the deviation of cities from the fitted line by calculating the measure $Doo = l_i - y_i$, where $l_i$ is the value of fitted line for city $i$, and $y_i$ is the corresponding GDP in the logarithmic form. Further, they used the $Doo$ measure to predict the macro-economic structure, in particular GDP, by employing the support vector regression (SVR) \cite{Cortes1995,Smola2004}. They found that the user number (UN) performs better in predicting GDP than some macroeconomic indices such as population and average GDP. This work shows the capacity of online social activity in revealing industrial structure and estimating economic status at the regional level.

\begin{figure}[t]
  \centering
  \includegraphics[width=0.6\textwidth]{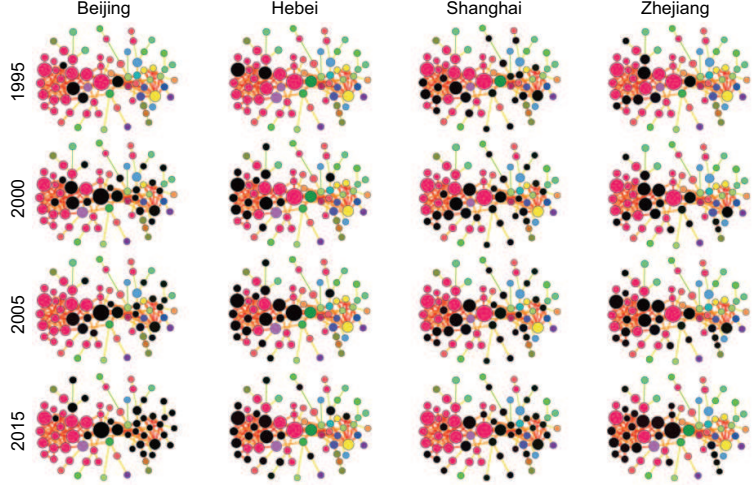}
  \caption{Evolution of China's regional industrial structure from 1995 to 2015. The industry spaces of four provinces are illustrated, including Beijing, Hebei, Shanghai and Zhejiang. Black circles highlight industries that are present in the industry space of the corresponding province. Figure from \cite{Gao2017}.}
  \label{Fig_4_2_1}
\end{figure}

Based on the China's publicly listed firm data from 1990 to 2015, Gao et al. \cite{Gao2017} quantified the regional industrial structure of China by constructing a network of related industries, named industry space. First, they estimated the proximity $\phi_{\alpha,\beta}$ between industries $\alpha$ and $\beta$ by calculating the cosine similarity. Formally, let $x_{i,\alpha,t}$ and $x_{i,\beta,t}$ be the number of firms in province $i$ operating respectively in industries $\alpha$ and $\beta$ at year $t$, the proximity $\phi_{\alpha,\beta,t}$ is given by:
\begin{equation}
\phi_{\alpha,\beta,t} =\frac{ \sum_{i}{ x_{i,\alpha,t} x_{i,\beta,t} }}{\sqrt{\sum_{i}{(x_{i,\alpha,t})^2}} \sqrt{\sum_{i}{(x_{i,\beta,t})^2}}}.
\end{equation}
Then, based on the proximity $\phi$, they built the industry space that highlights the relatedness between 70 industries at the sub-sectoral level. The China's industry space exhibits both a core-periphery structure and a dumbbell structure with a big tightly knit core of manufacturing industries and some small tightly knit cores of service- and information-related activities. Figure~\ref{Fig_4_2_1} presents the evolution of industrial structure of four provinces in China (Beijing, Hebei, Shanghai and Zhejiang) from 1995 to 2015, with black circles showing the industries of presence in the industry space. Specifically, the presence of industry $\alpha$ in province $i$ at year $t$ is identified by the revealed comparative advantage being over 1 (i.e., $\text{RCA}_{i,\alpha,t} \geq 1$) \cite{Balassa1965}. Beijing and Shanghai gradually occupied Internet and financial services, while Hebei and Zhejiang gradually occupied manufacturing industries. By analyzing the same data, Gao and Zhou \cite{Gao2018} found that provinces located along the coast tend to be industrial sophisticated with a high level of economic complexity. Moreover, the provinces' ranks by their economic complexity are relatively stable during the considered period.

National bureau of industrial enterprises can also be used to portray the production space as the representation of industrial structure. Based on data of four-digit manufacturing sectors from the China's State Statistical Bureau covering the period of 1999-2007, Guo and He \cite{Guo2017} calculated the inter-sector relatedness and produced the production space consisting of 424 manufacturing sectors. The production space in 1999 has a core-periphery structure with a major core of electric apparatus, electronic and telecommunications equipment, and a small sub-core cluster consisted of food products, chemical and non-metallic mineral products. The small sub-core cluster developed into an important and dense core of the production space in 2007. They further found that China's regions undergo substantial structural change from 1999 to 2007 with different magnitudes, and industrial evolution has a strong tendency of path dependencies, where regional development is rooted in and subject to the preexisting economic structure.

Economic relatedness contributes significantly to regional industrial diversification. By analyzing Italian trade data during 1995-2003, Boschma et al. \cite{Boschma2009} presented strong evidence in support of the fact that related variety contributes to regional economic growth. They grouped products into related variety sets $S_r$ based on the industrial classification and then calculated the related variety index $RV_i$ of product $i$ by
\begin{equation}
RV_i = \sum_{r=1}^{R} P_r H_r ,
\end{equation}
where $P_r = \sum_{i} p_i$ is the total exports of region $r$, and $H_r$ is the entropy within the related variety set $S_r$. Formally, the entropy $H_r$ of region $r$ is given by
\begin{equation}
H_r = - \sum_{i \in S_r} \frac{p_i}{P_r} \log_2 \frac{p_i}{P_r}.
\end{equation}
They found that a region benefits from extra-regional knowledge originated from related sectors that are already present in that region. Later, Boschma and Frenken \cite{Boschma2012b} demonstrated that technological relatedness affects the process of knowledge spillovers, which benefits regions with different but technologically related activities. As a result, new industries are likely to emerge from related industries. However, the process occurs primarily at the regional level as knowledge spillovers are geographically bounded. Using trade data of 50 Spanish provinces during 1995-2007, Boschma et al. \cite{Boschma2012} further investigated whether related variety affects regional growth. They calculated two measures of relatedness between industries: the related variety index \cite{Boschma2009} and the proximity index \cite{Hidalgo2007}. They found that Spanish provinces with a variety of related industries exhibit higher rates of economic growth.

By analyzing the US patent data during 1977-1999, Castaldi et al. \cite{Castaldi2015} showed that related technologies can enhance the innovation of a new technology and unrelated variety can enhances technological breakthroughs. Boschma et al. \cite{Boschma2015} investigated technological relatedness at the city level and technological change in 366 US cities by analyzing the US patent data during 1981-2010. They found that the level of relatedness with existing technologies increases the entry probability of a new technology in a city. Balland et al. \cite{Balland2015} discussed the co-evolutionary dynamics between proximity and knowledge ties. They found that proximities might gradually increase due to the past knowledge ties. In particular, the co-evolutionary dynamics can be captured by the processes of learning (cognitive proximity), decoupling (social proximity), agglomeration (geographical proximity), integration (organizational proximity) and institutionalization (institutional proximity). Acemoglu et al. \cite{Acemoglu2016} measured the strength of technological flows between technology subcategories using data of 1.8 million US patents and their citation properties during 1975-2004. They found that related pre-existing technological developments have a strong predictive power for future innovations.

A body of literature have demonstrated that relatedness plays an important role in economic development. Indeed, recent empirical evidences have generalized the principle of relatedness \cite{Hidalgo2018}, which describes the probability that an economy develops or loses an economic activity as a function of the density of its related activities in that economy. Jun et al. \cite{Jun2017} studied the role of relatedness in the evolution of bilateral trade. They found that produce relatedness, importer relatedness and exporter relatedness can increase a country's exports of a product. Boschma \cite{Boschma2017b} provided a valuable future research agenda regarding the relatedness as a driver of regional diversification. They suggested to focus on the role of economic and institutional agents. Davids and Frenken \cite{Davids2018} recently showed that the type of knowledge being mobilized and produced determines the relative importance of proximity dimensions. They proposed a framework that combines the proximity dimensions with different types of knowledge in the innovation process.

\subsubsection{Collective learning in economic development}

In addition to related varieties, geographic knowledge also plays a crucial role in regional economic development. Boschma et al. \cite{Boschma2013} demonstrated that capabilities that enable the development of new industries are regional specialized, supporting the hypothesis that knowledge decays strongly with distance in its diffusion process \cite{Keller2002}. Due to the localized nature of knowledge diffusion, neighboring regions should share more similar knowledge and exhibit a geographically correlated pattern in producing structure and economic growth \cite{Bahar2014}. Scientists have revealed the role of geographic neighbors and highlighted it as an alternative channel for development. Indeed, recent literature have focused on the effects of collective learning--the learning that takes place at the scale of teams, organizations, regions, and nations--by highlighting two learning channels, namely, the inter-industry learning (from related industries), and the inter-regional learning (from neighboring regions)\cite{Lawson1999,Gao2017}.

The effects of geographic knowledge spillovers on firm survival and industry development have been studied based on multiple data at regional, firm and plant levels. Acs et al. \cite{Acs2007} analyzed annual data of 11 million establishments in the US private sectors during 1989-1998. By incorporating knowledge spillovers through a geographical variation model, they investigated the relations between regional human capital stocks and new-firm survival. They found that knowledge spillovers lead to higher rates of new-firm survival. Holmes \cite{Holmes2011} studied the geographic expansion of Wal-Mart stores in the US by analyzing store-level data on sales. They found that locations of new Wal-Mart stores tend to be close to regions where Wal-Mart already had a high density of stores. Broekel and Boschma \cite{Broekel2012} analyzed data from 59 organizations in the Dutch aviation industry. They uncovered that geographical proximity serves as a driver of network formation and it is a stimulus for firm innovative performance after controlling for the effects of other proximities.

After analyzing a dataset summarizing individual work history, Jara-Figueroa et al. \cite{Jara2018} found that the growth and survival of new firms in a location increase when they hire workers with location-specific and industry-specific knowledge instead of occupation-specific knowledge. Moreover, industry-specific knowledge plays a more important role for pioneer than for non-pioneer firms. Using network clustering techniques, Alabdulkareem et al. \cite{Alabdulkareem2018} analyzed the dataset detailing the importance of 161 workplace skills for 672 occupations in the US. They found that skills exhibit a polarization into two clusters: the social-cognitive skills of high-wage occupations and the sensory-physical skills of low-wage occupations. Moreover, workers in occupations relying heavily on one skill cluster are likely to move to other occupations within the same skill cluster, says polarized skill network constrains career mobility of workers.

Based on the international trade data during 1962-2000, Bahar et al. \cite{Bahar2014} studied the effects of neighboring countries on the evolution of a country's exporting basket. They measured the similarity in countries' export structure by defining an export similarity index (ESI) through the Pearson correlation coefficient. Formally, the ESI between countries $c$ and $c'$ is given by
\begin{equation}
S_{c,c'} = \frac{\sum_p (r_{c,p}-\bar{r}_c) \sum_p (r_{c',p}-\bar{r}_c')}{\sqrt{ \sum_p (r_{c,p}-\bar{r}_c)^2 \sum_p (r_{c',p}-\bar{r}_c')^2 }},
\label{Eq:SCC}
\end{equation}
where $r_{c,p} = \ln (\text{RCA}_{c,p}+ \varepsilon )$, and $\bar{r}_c$ is the average value over all products for country $c$. Here, $\text{RCA}_{c,p}$ is the revealed comparative advantage (RCA) \cite{Balassa1965} of country $c$ and product $p$, which is calculated by Eq.~(\ref{Eq:RCA}). They found that neighboring countries have significantly larger ESI value than non-neighbors, and ECI is negatively correlated with geographical distance. After using regressions to discount the effects of product relatedness, they further found that a country's probability to export a new product increases significantly (on average, 65\% larger) if it has neighboring countries that are already successful exporters of that product.

Based on the survey data of 295 firms in 8 European regions, Broekel and Boschma \cite{Broekel2016} studied the geographical and cognitive structure of knowledge links. They found that firms' knowledge exchange have differences in their cognitive and geographical dimensions. In particular, connecting with technologically related and similar organizations as well as organizations at various geographical levels (regional and non-regional) can enhance the innovations of firms. By analyzing the data of US state-level exports during 2000-2012, Boschma et al. \cite{Boschma2017} found that a state in the US has a higher probability (about 58\%) of developing a new industry if it has a neighbouring state specialized in that industry. Further, they tested if neighboring regions have more similar export patterns by including ESI given by Eq.~(\ref{Eq:SCC}) into the regression model. They found that the ESI between a pair of states raises by 0.43 standard deviations if the two states share a border in the US.

\begin{figure}[t]
  \centering
  \includegraphics[width=0.65\textwidth]{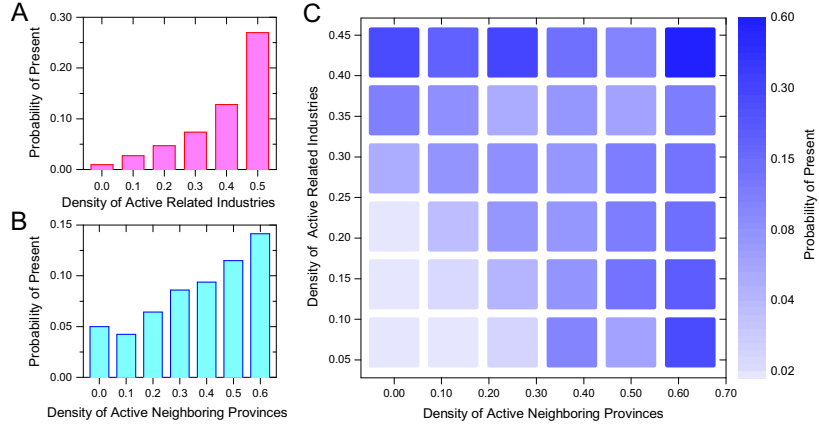}
  \caption{Quantifying the collective learning effects. (A) and (B) are the corresponding marginal probability distributions of new industries present in the next five years, given the density ($\omega$) of active related industries and the density ($\Omega$) of active neighboring provinces, respectively. (C) Joint probability of a new industry developing revealed comparative advantage in a province in the next five years, given $\omega$ in horizontal-axis and $\Omega$ in vertical-axis. The color marks the joint probability of new industries present after dividing the two densities into bins. Figure after \cite{Gao2017}.}
  \label{Fig_4_2_2}
\end{figure}

In a word, previous literature have demonstrated two collective learning channels in regional economic development: the inter-industry learning that involves learning from related industries and the inter-regional learning that involves learning from neighboring regions. Using publicly listed firm data describing the evolution of China's economy between 1990 and 2015, Gao et al. \cite{Gao2017} formalized these two collective learning effects. For inter-industry learning, they calculated the density of active related industries ($\omega$) by counting the number of related industries that are already present (i.e., $\text{RCA}\geq 1$) in that province (see also Ref. \cite{Boschma2013}). The density $\omega_{i,\alpha,t}$ for industry $\alpha$ in province $i$ at year $t$ is given by
\begin{equation}
    \omega_{i,\alpha,t} =\frac{\sum_{\beta}{\phi_{\alpha,\beta,t} U_{i,\beta,t}}}{\sum_{\beta}{\phi_{\alpha,\beta,t}}},
\end{equation}
where the binary variable $U_{i,\beta,t}=1$ if province $i$ has advantage in industry $\alpha$ at year $t$ (i.e., $\text{RCA}_{i,\beta,t}\geq 1$), and $U_{i,\beta,t}=0$ otherwise. They found that the probability for a province to develop a new industry in the next five years increases with $\omega$ (see Figure~\ref{Fig_4_2_2}A), supporting the inter-industry learning effect. For inter-regional learning, they calculated the density of active neighboring provinces ($\Omega$) by counting the number of neighboring provinces that have developed advantage in an industry. The density $\Omega_{i,\alpha,t}$ for province $i$ in industry $\alpha$ at year $t$ is given by
\begin{equation}
    \Omega_{i,\alpha,t} = \left.{ \sum_{j} \frac{U_{j,\alpha,t}}{D_{i,j}} }\middle/
    \sum_{j} \frac{1}{D_{i,j}} \right.,
\end{equation}
where $D_{i,j}$ is the geographic distance between two provinces $i$ and $j$. They found that the probability that a province will develop a new industry in the next five years increases with $\Omega$ (see Figure~\ref{Fig_4_2_2}B), supporting the inter-regional learning effect.

Furthermore, Gao et al. \cite{Gao2017} explored the interaction between inter-regional and inter-industry learning effects. They calculated the joint probability that a new industry will emerge in a province as a function of both densities $\omega$ and $\Omega$. They found that the probability for a province to develop a new industry in a five-year period increases with both $\omega$ and $\Omega$ (see Figure~\ref{Fig_4_2_2}C). After using a probit model to check the robustness of the results, they demonstrated that the inter-regional and inter-regional learning effects are jointly significant. Interestingly, the regression coefficient of the two densities' interaction term is negative and significant, suggesting the presence of diminishing returns (see Ref. \cite{Gao2017} for details). The observation means that, when one learning channel is sufficiently active (inter-industry or inter-regional), the marginal contribution of the other one is reduced. In other words, the two collective learning channels are substitutes for economic development. These empirical findings have been tested and generalized for countries at various stages of development based on different types of data. For example, Gao et al. \cite{Gao2017c} analyzed over 300 million Brazilian labor records and found evidences in support of the collective learning effects in Brazilian regional economic development.

As geographic knowledge diffusion requires direct forms of human interaction \cite{Arrow1969}, the construction of high-speed rails (HSRs) is likely to facilitate market integration and knowledge spillovers. Using data of China's HSRs, Zheng and Kahn \cite{Zheng2013} demonstrated that bullet trains help improve the life quality of urban population as HSRs entry allows individuals to access the megacity without living within its boundaries. To explore the impact of HSRs on regional economic activities, Li et al. \cite{Li2016b} developed the geographically network weighted regression that incorporates the changes in network-based travel time from HSRs. They found that HSRs have significantly changed the spatial redistribution of economic activities in regions of China. Later, based on data of prefectural-level cities in China during 1990-2013, Ke et al. \cite{Ke2017} explored how HSRs affect the economic growth of cities. They found that the local economic gains are greater for cities connected by HSRs. Meanwhile, Qin \cite{Qin2017} found a mild impact of HSRs upgrades on economic growth in China's prefecture-level cities, while the peripheral regions along the upgraded HSRs (e.g., counties close to high-speed rail stations) experienced an investment-driven reduction (3-5\%) in GDP and GDP per capita after 2007.

The effects of modern transportation (e.g., HSRs and flights) on economic development and knowledge spillovers have also been studied in developed countries. For the European Union, Kim et al. \cite{Cheng2015} explored the contribution of HSRs in promoting economic integration. They found that local economic development is necessarily leaded by transport improvements alone, especially when this involves cross-border links. By analyzing the northwest European HSRs and the UK's first HSR, Vickerman \cite{Vickerman2018} found that transport infrastructure by itself does not likely have a transformative effect on economy, but it can contribute to such effect after being coupled with policy interventions such as policies related to complementary planning and policies towards labour markets. Ahlfeldt and Feddersen \cite{Ahlfeldt2018} analyzed the economic impact of the German HSR  and found that HSR has a causal effect (on average about 8.5\%) on GDP growth in the regions of intermediate stops. Moreover, the strength of spillovers halves every 30 minutes of travelling time and diminishes to zero after about 200 minutes. Besides HSRs, a reduction in travel cost brought by cheaper flights can also facilitate knowledge spillovers reflected by scientific collaborations. Catalini et al. \cite{Catalini2016} analyzed a scientist-level dataset covering all US chemistry faculty members during 1991-2013. They found that scientific collaborations increase by 50\% after the Southwest Airlines opens a new route, showing that face-to-face interactions can enhance scientific collaborations.

To address endogenous concerns of inter-regional learning, Gao et al. \cite{Gao2017} applied the differences-in-differences (DID) analysis \cite{Bertrand2004} and used the introduction of HSRs as an adequate instrument. The underlying intuition is that HSRs entry reduces the barriers to the inter-regional learning but should not affect the inter-industry learning. Specifically, they used the DID analysis to test whether provinces connected by HSRs increased their industrial similarity and experienced a boost in the productivity of shared industries. The industrial similarity $\varphi_{i,j,t}$ between a pair provinces $i$ and $j$ at year $t$ is measured by
\begin{equation}
  \varphi_{i,j,t}=\frac{ \sum_{\alpha}{ y_{i,\alpha,t} y_{j,\alpha,t} }}{\sqrt{\sum_{\alpha}{(y_{i,\alpha,t})^2}} \sqrt{\sum_{\alpha}{(y_{j,\alpha,t})^2}} },
\end{equation}
where $y_{i,\alpha,t}=\ln(\text{RCA}_{i,\alpha,t}+1)$ and $y_{j,\alpha,t}=\ln(\text{RCA}_{j,\alpha,t}+1)$. They found that the industrial similarity ($\varphi_{i,j}$) decays strongly with the geographic distance ($D_{i,j}$), and HSRs entry significantly increases the industrial similarity between provinces connected by HSRs. Moreover, the labor productivity (measured by the revenue per worker) increases in the provinces connected by HSRs, supporting the hypothesis that HSRs entry promotes inter-regional learning.

\subsubsection{Development paths and strategies}

Regional industrial diversification has been suggested as a strong path-dependent process, where economic relatedness plays a significant role. Based on plant-level data of 70 Swedish regions during 1969-2002, Neffke et al. \cite{Neffke2011} identified related industries using the revealed relatedness (RR) measure \cite{Neffke2008}. They found that the probability that an industry will enter (exit) a region increases (decreases) with the number of related industries already present in that region. Neffke et al. \cite{Neffke2012} further studied the effects of technological relatedness on plant survival in Sweden during 1970-2004. They found that the plant survival rates are increased by the presence of technologically related local industries. Further, Neffke and Henning \cite{Neffke2013} investigated how industry's skill relatedness affects the diversification of firms by calculating the RR measure based on the labor flow data covering about 4.5 million workers in 400 industries in Sweden during 2004-2007. They found that firms tend to diversify into industries that require skills strongly related to the firms' existing industries. These works suggest the predictive power of skill relatedness for firm diversification.

Some literature have also explored the role of relatedness in regional development, industrial structural change and firm survival in China \cite{He2016b}. Howell et al. \cite{Howell2018} analyzed the data of over 13 million entrepreneurial firms in China during 1998-2007. They found that local related variety has a stronger positive effect than other types of agglomeration on new firm survival. Moreover, the intensity and location of governmental support affect post-entry performance and survival of firms. He et al. \cite{He2017} analyzed the annual survey of industrial firms in China during 1998-2005. They found that private enterprises rely more on market-oriented institutions, while firms with local governmental supports and industrial linkages are more likely to sustain. He et al. \cite{He2018} later analyzed firm-level data of manufacturing industries in China during 1998-2008. They found that regions tend to develop new industries that are technologically related to the existing portfolio. These results demonstrate that regional industrial development is a path-dependent process where industries related to pre-existing ones.

Recently, Gao \cite{Gao2017e} investigated how to maximize the learning from related industries (i.e., inter-industry learning) and neighboring regions (i.e., inter-regional learning) by leveraging the Brazilian labor data. He used a simple variant of the threshold model \cite{Watts2002} to simulate the diversification of industries on real networks. In the threshold model, a region or an industry will be activated if over half of its neighbors are already active \cite{Gao2015b}. For inter-regional learning, simulations are based on the Brazilian industry space \cite{Gao2017c}, and the set of initial industries are selected according to a turnable balancing index of core and periphery industries. Gao \cite{Gao2017e} found an optimal strategy that results in a good tradeoff between core and periphery industries in the initial activation. For inter-regional learning, simulations are based on the adjacent network of regions integrated with one spatial link being added between each pair of regions \cite{Gao2015b}, and the set of initial industries are randomly selected. The lengths of spatial links are determined by a turnable balancing index of nearby and distant regions. The result suggests an optimal strategy that makes a balance between nearby and distant regions in establishing new spatial connections. These findings demonstrate that there are optimal strategies for both channels that can maximize the learning effects in industrial diversification.

\begin{figure}[t]
  \centering
  \includegraphics[width=0.8\textwidth]{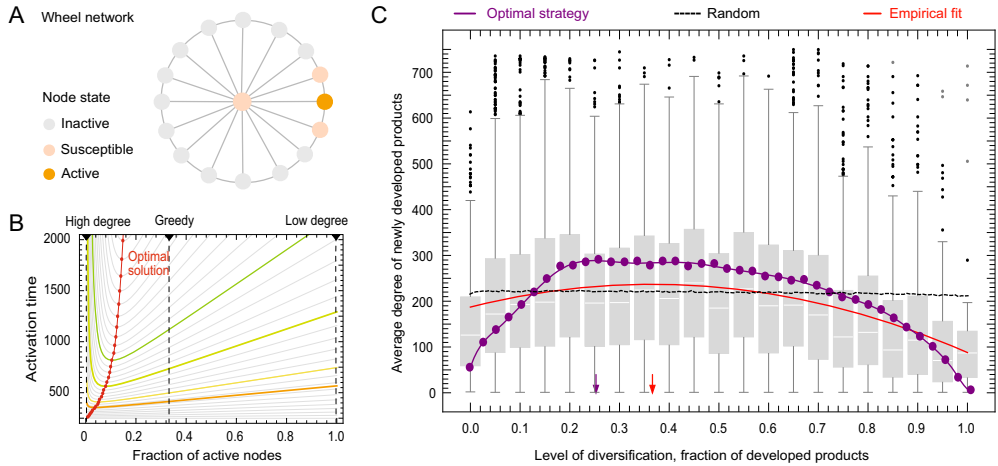}
  \caption{Strategic diffusion in networks, and the diversification of products. (A) A wheel network. (B) Time needed to activate all nodes in the wheel network as a function of the time when the hub is targeted (i.e., the fraction of active nodes). The red line indicates the optimal time based on the strategic diffusion. (C) Box-plot presenting the average degree of newly developed products in the product space as a function of the fraction of developed products. The red line shows the fit for the empirical values, and the purple line shows the optimal mix of greedy and high degree strategies obtained via numerical simulations. The black line shows the null model baseline which uses the random strategy. Figure after \cite{Alshamsi2018}.}
  \label{Fig_4_2_3}
\end{figure}

As suggested by many empirical studies, countries and regions are likely to develop economic activities that have close relatedness to what they have already developed, yielding the principle of relatedness \cite{Hidalgo2018}. In other words, the probability of developing a new industry in a region increases with the density of the region's developed industries that are related to the industry. As the produce space \cite{Hidalgo2007} and industrial space \cite{Gao2017,Gao2017e} have a core-periphery structure, the difficulty and opportunity of developing produces and industries at different locations of the space are different. Alshamsi et al. \cite{Alshamsi2018} explored the optimal diversification strategies in the produce space (see Figure~\ref{Fig_4_2_3}). They showed that the high-degree strategy, i.e., always targeting the potentially products with the highest degree (e.g., products in the core), will result in a long activation time, while the low-degree strategy, i.e., always targeting the potentially products with the lowest degree (e.g, products in the periphery), will miss the opportunity for a rapid development. In order to minimize the total time needed to develop all products, they proposed a method named strategic diffusion to identify products that are optimal to target at each time step. The optimal strategy targets core produces during a narrow and specific time window, which comes earlier than we previously thought (e.g., the time by the greedy strategy). They analyzed the international trade data and demonstrated that the countries' strategies to diversify their products are close to the optimal ones. The time that countries target core products, however, is later than the one suggested by the model, showing that countries can still save the total time of developing all products.

The path-dependent process of regional diversification suggests that regions have more opportunities to develop industries that have high relatedness to their pre-existing ones \cite{Boschma2017b}, while the strategic diffusion suggests that countries can optimize their development paths by targeting highly connected but somewhat unrelated activities at a certain time \cite{Alshamsi2018}. The development of unrelated economic activities is particularly significant for the catching-up growth in developing economies as it is usually hard for them to jump from periphery to core areas in the product and industry space, say the space conditions the development \cite{Hidalgo2007}. Regarding this point, Zhu et al. \cite{Zhu2017} explored the development paths of regions in the heterogeneous industry space built on the exporting data of Chinese firms during 2002-2011. They studied whether developing regions can catch up by breaking the path-dependent trajectories and jumping farther into core areas of the uneven industry space. They demonstrated that developing regions can make a farther jump to new industries in a path-breaking way, and the reliance of technological relatedness can be transcended by internal innovations and extra-regional linkages. These findings suggest that less developed economies should pay more attention to improving other factors (such as infrastructure and education, government supports and extra-regional linkages) to promote their jumping capability in the catching-up growth.

Processes of unrelated diversification are also important for economic development, and economies can benefit from entering unrelated activities. Boschma et al. \cite{Boschma2017c} argued that a theory of regional diversification should also accounts for the processes of unrelated diversification. They suggested to pay attention to the role of agency in institutional entrepreneurship and enabling factors at different spatial scales. In particular, they discussed four regional diversification trajectories including two related diversification (replication and exaptation) and two unrelated diversification (transplantation involves and saltation stands). Pinheiro et al. \cite{Pinheiro2018} identified the periods that countries entered unrelated products by analyzing the diversification paths of 93 countries in product exports during 1965-2014. They found that countries tend to enter unrelated products when they have high levels of human capital and during their intermediate level of economic development. Moreover, countries that entered more unrelated products experienced a significant increase in economic growth, showing the positive gain to target unrelated activities at a specific development stage. All the above results indeed ask for more intelligent strategies for economic diversification by balancing related and unrelated activities in development.

\subsection{Urban scalings and perception}

The availability of large-scale and quantitative data from socioeconomic systems and image database has enhanced our perception of urban landscape and surrounding environment. In this subsection, we will summarize empirical observations and theoretical explanations of scaling laws of urban population with urban metrics (e.g., crime rate, employment, innovation and economic activity). Then, we will review recent applications of novel data on inferring the function of urban areas. Next, we introduce crowdsourcing methods and computational vision techniques to measure livability, safety and inequality, to infer the status of urban life, and to quantify the changes of urban streetscapes. Finally, we will introduce recent progresses on urban computing for better development in urban areas.

\subsubsection{Scaling laws for cities}

Empirical observations in economics suggest the Zipf's law for cities in most countries. That is, the number of cities with populations greater than $N$ is proportional to $1/N$. Formally, $P(\text{size} > N) = Y_0 / N^{\beta} $, with $\beta \simeq 1$ and $Y_0$ being a constant. Gabaix \cite{Gabaix1999} provided a simple explanation for the emergence of such Zipf's law. They demonstrated that the power-law exponent $\beta = 1$ is necessarily led by the most natural conditions on the Markov chain. Using the maximum likelihood estimation (MLE) method, Clauset et al. \cite{Clauset2009} estimated the power-law exponent for the populations of US cities in 2000, finding a rank-size slope of $-0.73$ (i.e., $\beta = 0.73$). Later, Small et al. \cite{Small2011} tested the Zipf's law based on a unique proxy for anthropogenic development, specifically, the temporally stable nighttime lights (NTLs) from the DMSP-OLS. They found that the estimated $\beta$ ranges from $0.95$ to $1.11$, suggesting that Zipf's law holds for spatial extent of anthropogenic development at global scales.

\begin{figure}[t]
  \centering
  \includegraphics[width=0.45\textwidth]{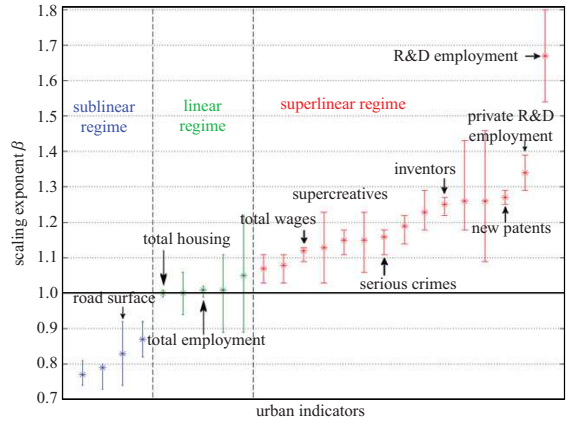}
  \caption{Scaling exponents for urban metrics versus city size. Scaling exponents $\beta$ found for China, Germany and USA are shown with 95\% confidence interval for different urban indicators. Scaling exponents are colour-coded according to their regime: sublinear in blue, linear in green, and superlinear in red. Figure from \cite{Arcaute2015} with data coming from \cite{Bettencourt2007}.}
  \label{Fig_4_3_1}
\end{figure}

Urban scaling laws provide a quantitative connection between urbanization and economic development, which is common to all cities around the world. By analyzing datasets from urban systems in the US, Germany and China, Bettencourt et al. \cite{Bettencourt2007} found that many diverse urban variables fit power-law functions of population size with scaling exponents $\beta$ falling into distinct universality classes. Using total population $N(t)$ to estimate the size of the city at time $t$, the power-law scaling takes the form
\begin{equation}
 Y(t) = Y_0 N(t)^{\beta},
\end{equation}
where $Y(t)$ denotes a certain metric on social activities or material resources at time $t$, and $Y_0$ is the normalization constant. They found a pervasive property of urban organization with exponents falling into three categories (see Figure~\ref{Fig_4_3_1} for the results summarized by Arcaute et al. \cite{Arcaute2015}): $\beta > 1$ (superlinear), $\beta = 1$ (linear), and $\beta < 1$ (sublinear). In particular, they showed that $\beta \approx 1$ is usually associated with individual human needs such as housing, employment and household electrical consumption, $\beta \approx 1.2 > 1$ is associated with social currencies such as information, innovation and wealth, and $\beta \approx 0.8 < 1$ is associated with infrastructure such as road surface, gasoline stations and length of electrical cables.

Urban scaling laws have been widely observed in emissions and infrastructures. Louf and Barth{\'e}lemy \cite{Louf2014b} found that the CO$_{\text{2}}$ emission scales superlinearly with $N$ in the US in 2012 ($\beta=1.26$) and the OECD countries in 2008 ($\beta=1.21$). Oliveira et al. \cite{Oliveira2014} analyzed data of CO$_{\text{2}}$ emissions in the US during 1999-2008 and found a superlinear scaling (with an average exponent $\langle \beta \rangle =1.46$) across all cities. Delong and Burger \cite{Delong2015} found that energy use scales superlinearly with $N$ in Sweden, England and Wales (E\&W), the US, and the world. Samaniego and Moses \cite{Samaniego2008} analyzed the structure of road networks in 425 US cities. They found that road network capacity per capita is independent of city size measured both by population and spatial extent of the urban area. Batty \cite{Batty2013} analyzed the road network of cities in E\&W and found a superlinear scaling ($\beta \approx 1.09$) of road accessibility with $N$. Louf et al. \cite{Louf2014c} analyzed the data of about 140 subways and over 50 railway networks across the world. They found that the length of subway networks scales superlinearly ($\beta \approx 1.13$) while the yearly ridership of railway networks scales linearly with the number of stations. For the UK and urban California, Masucci et al. \cite{Masucci2015} found that both the total length $L(N)$ and the area $A(N)$ of street networks scale almost linearly with $N$, and the urban scalings persist in space and time.

A body of literature have demonstrated the urban scaling of crime in cities. Alves et al. \cite{Alves2013} analyzed data of homicides in Brazilian cities and found that the number of homicides scales superlinearly ($\beta \approx 1.15$) with $N$. They further proposed an approach to unveil relations between crime and urban metrics using the distance between the actual homicide number and the expected number from the scaling law. Banerjee et al. \cite{Banerjee2015} analyzed the data of US cities and found that crime scales superlinearly ($\beta=1.26$) with $N$. They gave the explanation that the number of polices scale sublinearly while the number of generated crimes scales linearly. After analyzing monthly police crime reports in E\&W, Hanley et al. \cite{Hanley2016} found four types of scaling behaviors based on population density: non-urban scaling, accelerated scaling ($\beta_L < \beta_H$), inhibited scaling ($\beta_L > \beta_H >0$) and collapsed scaling ($\beta_L > \beta_H$, with $\beta_H<0$), where $\beta_L$ and $\beta_H$ are the scaling exponents for low and high population density, respectively. Oliveira et al. \cite{Oliveira2017} analyzed the disaggregated criminal data from the US and UK. They found that the crime concentration does not scale with the city size, and the crime distribution in a city follows a power-law distribution with exponent depending on the crime type.

In most of these aforementioned literature, the word ``city'' refers to a larger agglomeration around the central city, which is socioeconomic unit instead of administrative definition. In fact, there are alternative definitions of city boundaries. Arcaute et al. \cite{Arcaute2015} developed a framework to produce a system of cities by clustering small statistical units. They found that the scaling exponent $\beta$ gives mild deviations from linearity in E\&W, suggesting that economic intricacies are not fully grasped by the urban population $N$. Van Raan et al. \cite{VanRaan2016b} analyzed the urban scalings in the Netherlands and found a superlinear scaling ($\beta \approx 1.15$) of GDP with $N$ for major cities. After considering three separate modalities, they found that municipalities perform better than urban agglomerations and urban areas with the same population, showing that cities with a municipal reorganization are likely to perform better. Bettencourt and Jose \cite{Bettencourt2016} applied new harmonized definitions of functional urban areas to examine scalings, finding that pooling together cities from different urban systems can better identify scaling behaviors in European cities.

Social ties of cities also exhibit scaling behaviors. Pan et al. \cite{Pan2013} found that the density of social ties $T(\rho)$ scales superlinearly with urban population density $\rho$. The social-tie density is given by $T(\rho) = \rho \ln \rho + (C-1)\rho$, where $C=2\ln r_{\max} + \ln \pi + 1$ with a unique $r_{\max}$ for each city. In particular, $\beta$ fells into a narrow band $1.1 \leq \beta \leq 1.3$, where $\beta = 1.21$ for the AIDS/HIV prevalence in the US cities and $\beta=1.26$ for the total GDP per square kilometer in the European cities. Moreover, the superlinear scaling ($\beta > 1$) is led by the increase in $\rho$, and the diffusion rate along social ties can accurately reproduce urban scalings. After analyzing mobile phone data of 31 Spanish cities, Louail et al. \cite{Louail2014} found that the number of activity centers scales sublinearly ($\beta<1$) with $N$. Markus et al. \cite{Schlapfer2014} analyzed the nationwide communication records in Portugal and the UK. They found that the total number of contacts and communication activities scale superlinearly with $N$. Recently, Leit{\~a}o et al. \cite{Leitao2016} studied the existence of nonlinear scaling by developing a statistical framework to account fluctuations. They found that $\beta$ does not only depend on the fluctuations contained in the datasets but also on the assumptions of models and the heavy-tailed distribution of city sizes.

Several explanations have been proposed for the origin of urban scalings. Arbesman et al. \cite{Arbesman2009} explained the observed superlinear scaling in the relations between population size and innovation by a network model, where the number of long-distance ties associated with a city is proportional to its population and these ties provide the potential for innovation. The model yields a reasonable range of the scaling exponent, suggesting socially distant ties as a powerful force of the superlinear scaling. Later, Gomez-Lievano et al. \cite{Gomez2012} built a statistical framework to explore how urban scaling laws emerge and relate to Zipf's law. Using data of homicides in three cities, they derived the conditional probability density $P(Y|N)$ for the number of homicides $Y$ in a city with population $N$ by exploiting the Bayes' rule
\begin{equation}
 P(Y|N) = \frac{P(N|Y)P(Y)}{P(N)},
\end{equation}
where $P(Y)$ is the distribution of homicides in cities, and $P(N|Y)$ is the conditional probability for the populations of cities with a given number of homicides. After studying the statistical properties of $P(Y)$ and $P(N|Y)$, they found that scaling laws emerge as the expectation value of $Y$, which is a function of $N$. Moreover, the knowledge of the distribution $P(Y)$ can be used to predict the Zipf's exponent from the statistics of urban metrics.

To better understand the origin of urban scalings, Bettencourt \cite{Bettencourt2013} developed a framework to estimate scaling exponents without modeling infrastructure. In a city with land area $A$ and population $N$, the strength of local interactions between people in an area $a_0$ is denoted as $g$. The basis ideas behind their model are summarized as follows. First, the number of local interactions per person is given by $a_0 \ell \cdot N/A$, where $N/A$ is the population density, and $\ell$ is the length of travel. Then, a city's total social output $Y$ is given by
\begin{equation}
 Y = \bar{g} \cdot N \cdot a_0 \ell \cdot \frac{N}{A} = G \cdot \frac{N^2}{A},
\end{equation}
where $G \equiv \bar{g} a_0 \ell$, $\bar{g}$ is the average social output per interaction, and $N$ is the population size. Next, the total cost to mix the city is $T= \varepsilon L N = \varepsilon A^{1/2} N$, where $\varepsilon$ is a force per unit time, and $L = A^{1/2}$ is the cost per person. The cost should be covered by each individual, $y=Y/N$, requiring $y \simeq T/N$. This implies $A(N) = a N^{\alpha}$ with $\alpha = 2/3 <1$ (sublinear scaling) and $a = (G/\varepsilon)^{\alpha}$. Thus, they obtained $Y = Y_0 N^{\beta}$, where $\beta=2-\alpha = 4/3 >1$ (superlinear scaling) and $Y_0 = G^{1-\alpha} \varepsilon^{\alpha}$. Yakubo et al. \cite{Yakubo2014} provided explanations for both scalings based on a geographical scale-free network. The individual activity $y_{ij}$ of two connected nodes $i$ and $j$ depends on their Euclidean distance $l_{ij}$. The urban metric $Y(N)$ scales superlinearly when $y_{ij}$ increases with $l_{ij}$ (e.g., for creative productivities), $Y(N)$ scales sublinearly when $y_{ij}$ decreases with $l_{ij}$ (e.g., for infrastructures), and $Y(N)$ scales linearly when the geographical constraint is strong enough.

\begin{figure}[t]
  \centering
  \includegraphics[width=0.8\textwidth]{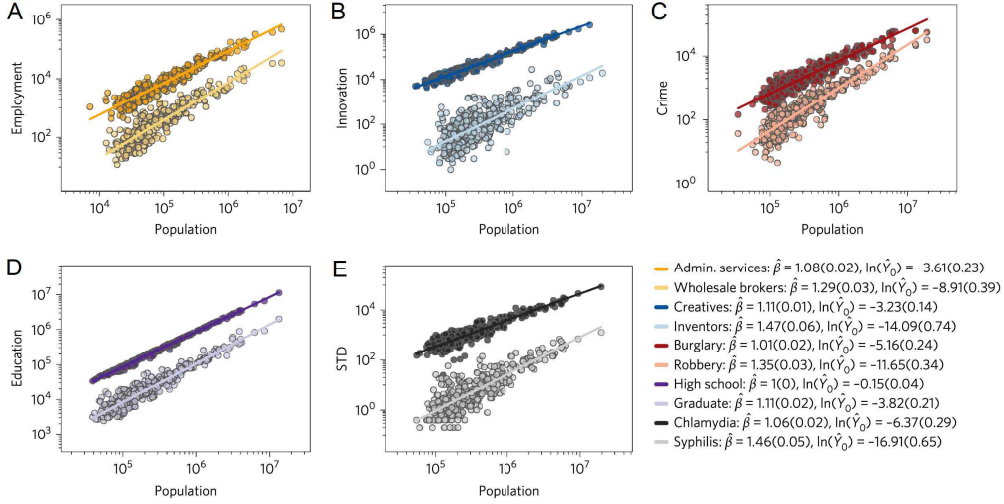}
  \caption{Explanations of ten different urban scaling phenomena. (A) Employment in two industries. (B) Two types of innovative activities. (C) Two types of violent crime. (D) People with a given educational level. (E) Two sexually transmitted diseases (STDs). Lines represent the best fit of the model $E\{Y | N \} = Y_0 N^{\beta}$. Hat ($\char`\^$) denotes a statistical estimate of a parameter. Figure from \cite{Gomez2016}.}
  \label{Fig_4_3_2}
\end{figure}

Recently, alternative explanations of urban scalings have been suggested. Gomez-Lievano et al. \cite{Gomez2016} showed that the number of people engaged in each phenomena scales as a power law with population size (see Figure~\ref{Fig_4_3_2}). Accordingly, they proposed a simple model to explain urban scalings, where each phenomenon depends on a number of factors $M$, an individual requires one factor with probability $q$, and a city provides one factor with probability $r$. The aggregated output of a given phenomenon is modeled by $Y = \sum_{j=1}^N X_j$, where $N$ is the population size, $X_j =1$ if individual $j$ gets all required $m$ factors, and $X_j =0$ otherwise. If $m$ factors have been required, an individual will not require the other $M-m$ factors. Given a city with $m$ factors, the probability of an individual involved in the activity is $P\{X_j = 1| M_{\text{city}} = m\} = (1-q)^{M-m}$, where $M_{\text{city}} = B(M,r)$ is a binomially distributed random variable. The expected value of $Y$ is given by
\begin{equation}
 E\{Y\} = N \sum_{m=0}^{M} P\{X_j = 1| M_{\text{city}} = m\}P\{M_{\text{city}} = m\} \approx NP,
\end{equation}
and the variance of $Y$ is given by
\begin{equation}
 \text{Var} \{Y\} \approx E\{Y\}^2 \big(  \frac{1}{E\{Y\}} - \frac{1}{N} + \frac{1}{p^q} -1 \big),
\end{equation}
where $P \equiv e^{-Mq(1-r)}$ (see Ref. \cite{Gomez2016} for details). By assuming that $r = a+b\ln (N)$, the scaling function $E\{Y\} = Y_0 N^{\beta}$ can be obtained. The model suggests that phenomena requiring more factors will scale more superlinearly (i.e., with larger $\beta >1$). Ribeiro et al. \cite{Ribeiro2017} proposed an explanation of urban scalings based on the interactions between individuals and the fractal dimension of cities. Their framework can reproduce the urban scaling for infrastructure (sublinear) and social indicators (superlinear). Using a spatial attraction and matching growth mechanism, Li et al. \cite{Li2017c} proposed a unified model that can reproduce the spatial scalings for population, total road length, and total number of socioeconomic interactions. Their model presents consistent results with empirical data and explains the origins of sublinear and superlinear scalings.

Urban scaling laws have been applied to differentiate urban economic productivity and quantify intrinsic diversity of urban economic activities. Based on the deviations from general scaling laws, Bettencourt et al. \cite{Bettencourt2010} proposed new metrics of a city's dynamics and urban performance (e.g., personal income, patents and violent crime). Formally, the deviation $\xi_i$ of a metric is quantified by the residuals \cite{Batty2008}
\begin{equation}
 \xi_i = \log \frac{Y_i}{Y(N_i)} = \log \frac{Y_i}{Y_0 N_{i}^\beta},
\end{equation}
where $Y_i$ is the observed value of metric $i$ for an arbitrary city, $Y(N_i)$ is the average value of urban metrics, and $N_i$ is the population size. The scale-adjusted metropolitan indicator (SAMI) $\xi$ is dimensionless and independent of $N$. Moreover, SAMI captures the specific dynamics of a city and represents its performance relative to other cities. This method provides a promising way to rank cities without the population size bias. Lobo et al. \cite{Lobo2013} derived a new expression for the total factor productivity (TFP) of urban areas. The scale-adjusted urban TFP is well-approximated by $\xi_{i}^A \approx (1-\alpha)(\xi_{i}^W - \xi_{i})$, where $\alpha$ is the production factors, $\xi_{i}^W$ is the SAMI for total labor income, and $\xi_{i}$ is the SAMI for total capital income. They found a systematic dependence of urban productivity on population size. Youn et al. \cite{Youn2016} analyzed records of establishments in the US urban areas and found that the total establishment number scales linearly with the city size. Further, they proposed a framework to measure the intrinsic diversity of economic activities, revealed the universal scaling distribution of business types, and presented a simple mathematical derivation of the universality.

\subsubsection{Unfolding urban functional areas}

In regional and urban economic development, a variety of data from remotely sensing (RS), mobile phones (MPs) and online social media (SM) have been used to map the function of regions and capture the urban structure. In the following, we will introduce the applications of nighttime lights (NTLs) data from the DMSP-OLS to measure the spatio-temporal urban dynamics. Then, we will review literature that leveraged novel data sources (e.g., MPs, Twitter, and check-ins) to uncover the inherent characteristics of functional regions and to predict regional economic status. At last, we will summarize recent progresses on the analysis and prediction of house price based on the data of satellite imagery and subway in addition to the traditional market data.

NTLs data have been used to monitor the dynamics of urban structure. Sutton \cite{Sutton2003} developed a measure of urban sprawl based on NTLs imagery. The urban sprawl is scale-adjusted to an urban area's total population, and the areal extent of metropolitan areas is measured based on the NTLs of the US. They found that inland and midwestern cities have more urban sprawl than west coast cities. This work sheds some insights to the spatial patterns of urban sprawl that is difficult to be precisely defined. Pandey et al. \cite{Pandey2013} extracted urban areas and monitored urbanization dynamics of India based on NTLs data from the DMSP-OLS and the SPOT vegetation \cite{Hagolle2005}. They employed the SVM-based classification algorithm to extract urban land extent from NTLs (see also Ref. \cite{Cao2009}) and verified the results by global urban extent map and Google Earth images. The state-wise increase in urban area is consistent with the change in urban population and GDP, showing the applicability of NTLs to quantify urban patterns.

Based on NTLs data, Zhang and Seto \cite{Zhang2011} monitored urban changes at the regional scale. They applied an iterative unsupervised classification method to analyze the NTLs data from the DMSP-OLS during 1992-2008 and mapped urbanization dynamics in China, Japan, India and the US. They found that India had higher growth rates than China between 1992 and 2000, while China experienced higher rates of urban growth than India between 2000 and 2008. Frolking et al. \cite{Frolking2013} analyzed the data from the SeaWinds microwave backscatter power return (PR) \cite{Long2013} and the DMSP-OLS. They found different evolution patterns of urban structure between India and China. Chinese cities rapidly expanded their built-up infrastructure in both height and extent with increased urban population density. By comparison, Indian cities were primarily built out with larger areas of lower population density. These results suggest two distinct trajectories of urban growth.

Recently, Li et al. \cite{Li2016} analyzed the spatio-temporal urban dynamics by applying a linear regression method to time-series from the DMSP-OLS for the southeast US. They found that the newly built urban areas can be effectively detected, while the urban expansion cannot be explained solely by population growth. Huang et al. \cite{Huang2016b} mapped sub-pixel urban expansion of Chinese cities based on the data from the DMSP-OLS and the Moderate Resolution Imaging Spectroradiometer (MODIS) \cite{Schneider2010}. They applied the random forest regression model to estimate sub-pixel urban percentage with the high quality calibration information derived from the Landsat data. The estimation of urban land area can be improved by including data from the MODIS and DMSP-OLS. Based on a set of landscape metrics, Liu et al. \cite{Liu2016b} explored the general spatio-temporal patterns of urbanization by examining 16 world-wide cities during 1800-2000. They uncovered several common urbanization patterns. For example, urban landscape becomes more fragmented, diverse and complex in the urbanization process.

MP data have been used to estimate the characteristics of functional regions. Based on a MP dataset consisting of 431 million calls and the involved locations of mobile base towers, Chi et al. \cite{Chi2016} constructed a cell-based spatially embedded interaction network of regions in Heilongjiang province, China. The cells (Voronoi polygons) are used to approximate the service area of a mobile base tower. They calculated the betweenness centrality \cite{Freeman1977} of a cell $k$ in the unweighted interaction network by
\begin{equation}
  C_B (k) = \sum_{i \neq j \neq k}^{N} \frac{\sigma_{ij} (k)}{\sigma_{ij}},
\end{equation}
where $\sigma_{ij}$ is the number of shortest paths from nodes $i$ to $j$, and $\sigma_{ij}(k)$ is the number of shortest paths passing through node $k$. They found that cells with high $C_B$ are distributed linearly across the province. After applying a community detection algorithm \cite{Girvan2002}, they found a two-level hierarchical organization embedded in the interaction network, where the bottom-level communities respect the county boundaries, and the top-level communities respect the prefecture-level unit boundaries. Moreover, almost every community has a cell with high $C_B$ at the commercial center or the government seat. Toole et al. \cite{Toole2012} studied dynamic land usages based on the temporal activity patterns of MP users. They demonstrated that supervised classification of labeled MP zoning data exhibits reasonable accuracy in identifying clusters of locations.

\begin{figure}[t]
  \centering
  \includegraphics[width=0.58\textwidth]{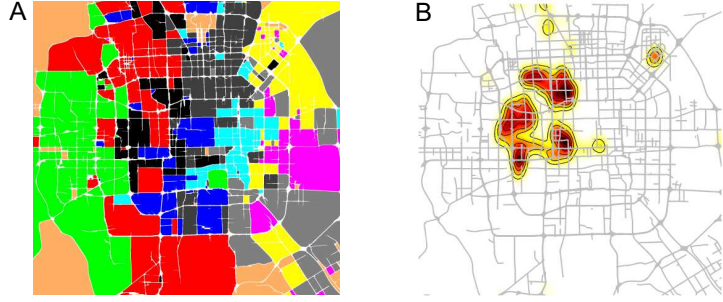}
  \caption{Identification of functional regions in a city. (A) Functional regions. Regions have similar functions are filled with the same color. (B) Intensity of a function. The functionality intensity of developed commercial/entertainment (a kind of function) areas in Beijing is illustrated. The darker part suggests a higher intensity. Figure from \cite{Yuan2012}.}
  \label{Fig_4_3_3}
\end{figure}

Different functional regions in urban areas can be identified by data about points of interests (POIs) and human mobility. Yuan et al. \cite{Yuan2012} proposed a framework named DRoF (Discovers Regions of different Functions) to discover regions of different functions in a city (see Figure~\ref{Fig_4_3_3}). They segmented a city into disjointed regions by major roads and employed a topic-based model to infer the functions of each region. The topic-based model treats a urban region as a document, regards mobility patterns as words, deems a function as a topic and uses POIs as metadata. In this way, they represented a region by a distribution of functions and denoted a function by a distribution of mobility patterns. Next, they clustered regions by the topic distribution and identified the intensity of each function. Experiments based on POI datasets and taxicab-based GPS trajectory datasets of Beijing demonstrated that their method outperforms baseline methods solely using POIs or human mobility. Further, Yuan et al. \cite{Yuan2015} extended the DRoF framework by introducing the concept of latent activity trajectory (LAT) to capture citizens' socioeconomic activities. Their new method employs a morphological approach to segment a city and applies a collaborative-filtering-based approach to learn the location semantics from POIs. Their method exhibits improved performance in identifying functional zones using location and mobility mined from semantics LAT.

Data from location-based social networks (LBSNs) have been used to identify land usages and derive commercial locations. Based on Twitter's spatial (geotagged) and temporal (time-stamped) data, Frias-Martinez et al. \cite{Frias2012c} presented methods to automatically determine land usages and locate urban POIs. Preliminary validation in Manhattan suggests geotagged tweets as a powerful data source for characterizing urban landscapes. Frias-Martinez et al. \cite{Frias2014} further used unsupervised learning to cluster regions with similar patterns of tweeting activities. They verified the new method in Manhattan, London and Madrid based on tweeting activities and ground truthing information for land usages. Recently, Lloyd and Cheshire \cite{Lloyd2017b} estimated locations of retail centers from geotagged tweets. They used an adaptive kernel density estimation to identify retail-related tweets and examined their spatial attributes. They showed that areas of elevated retail activities can be well located by retail-related tweets. Soliman et al. \cite{Soliman2017} analyzed 39 million geotagged tweets in Chicago and found that the majority's temporal patterns of tweeting at key locations are significantly associated with the types of land usage. They proposed a novel approach that can classify key locations into types of land usage with an overall accuracy of 0.78.

Check-in data have also been used to analyze regional structure and predict urban economic status. Shen and Karimi \cite{Shen2016} analyzed check-in data to explicitly portray urban structure. They proposed a novel framework to characterize urban streets with land-use connectivity indices and introduced a model to package three principal dimensions of an urban function network into one integrated index (see Tianjin in China as an example in Ref.~\cite{Shen2016}), which can explicitly describe the inherent function structure and the regions' typology across scales. Zhi et al. \cite{Zhi2016} proposed a model to infer functional regions from about 15 million check-in records during a year-long period in Shanghai, China. They used a novel low-rank approximation model to identify a series of latent spatio-temporal activity structures and obtained a series of underlying associations between the spatial and temporal activity patterns. The model is applicable to estimate functional regions including commercial dominant areas, developed residential areas, and developing residential areas.

\begin{figure}[t]
  \centering
  \includegraphics[width=0.55\textwidth]{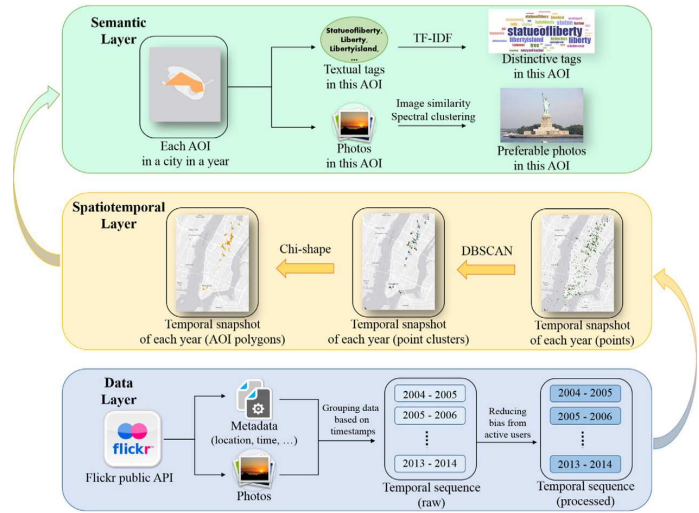}
  \caption{A three-layer coherent framework for extracting urban areas of interest (AOI) from geotagged Flickr data. The framework covers data pre-processing, point clustering, area construction, and semantics enrichment. Figure from \cite{Hu2015}.}
  \label{Fig_4_3_4}
\end{figure}

Urban areas of interest can also be extracted from online volunteered geographic information such as VGI, POIs, and geotagged photos. By utilizing VGI-based POI data obtained from Yahoo! in part of the Boston metropolitan area, Jiang et al. \cite{Jiang2015} applied a local maximum likelihood estimation to determine disaggregated land usage. They showed that employment estimations based on VGI-based POI data (Yahoo!) match estimations based on proprietary business establishment databases. Their method provides an alternative to estimate disaggregated land usages in a timely manner as POI data can be obtained at a high frequency with a low cost. Based on geotagged photos from Flickr, Hu et al. \cite{Hu2015} developed a coherent framework to extract and understand urban areas of interest (AOI). Their framework covers data pre-processing, point clustering, area construction, and semantics enrichment (see Figure~\ref{Fig_4_3_4}). They applied the method called density-based spatial clustering for applications with noise (DBSCAN) \cite{Ester1996} to identify AOI, employed the method named term frequency and inverse document frequency (TF-IDF) \cite{Salton1988} to extract distinctive textual tags for understanding AOI and designed a workflow to select photos capturing a preferable view of an AOI. Their framework provides a better identification of AOI and helps understand dynamics of AOI.

Affected by location and neighborhood in regional and urban areas, house price is a sensitive and important signal in real estate market and even regional economic development. Many methods have been proposed to analyze and predict house price, for example, the hedonic house price model \cite{Sirmans2005}. After analyzing purchase records from housing market in Chengdu, China, Xin and Zheng \cite{Xin2014} found that the spatial structure in housing data sets is important for house price estimation. They proposed a spatial hedonic model and successfully applied it to estimate house price index at the urban zone level. Zhang et al. \cite{Zhang2016c} studied how urban village removal affects nearby house price in Beijing, China. They found that an average urban village discounts house price by 2.5\%, while its removal increases house price by over 3.3\%. Zheng et al. \cite{Zheng2016b} studied the impact of subway transit on local house price by analyzing data of subway construction history, restaurant activities in station neighborhoods and rental housing transactions in Beijing, China. They found that changes in restaurant activities capture 20-40\% of house price appreciation in neighborhood of new subway stations.

Online searches and satellite imagery data have also been used to track urban housing market dynamics. Based on Google search activity data, Zheng et al. \cite{Zheng2016} constructed a real estate confidence index to measure the view on future housing market trend. They tested how the confidence index is associated with prices of newly built housing units in 35 Chinese cities during the past 10 years. The confidence index can predict the sale of new houses and the increase of local house price, while it has heterogeneous impacts on local real estate outcomes. Recently, Bency et al. \cite{Bency2017} proposed a deep learning framework to predict house price based on satellite imagery. They extracted features from satellite imagery by training deep convolutional neural networks (DCNNs) \cite{Lecun2015} to capture the neighborhood effects and then trained multiple models to regress house price using the extracted features. After validating results based on POI data, they demonstrated that leveraging neighborhood information embedded in satellite images can improve the accuracy of house price prediction.

\subsubsection{Perceiving urban environment}

Visual appearances of urban spaces are thought to have significant effects on psychological states of inhabitants, behaviors of citizens, and socioeconomic outcomes in neighborhoods. Recently, high spatial resolution data have been used to quantify the perception of urban environment and its relation to residents' health. The Google Street View \cite{Anguelov2010} provides street-level panoramic imagery captured in hundreds of cities on a global scale, allowing to audit neighborhood environments more easily. Rundle et al. \cite{Rundle2011} tested the feasibility of collecting data from Google Street View in 2008 to audit the environments of neighborhood in New York City (NYC). They found that measurements based on Google Street View exhibit higher levels of concordance in pedestrian safety, traffic and parking and infrastructure for active travel compared to field audit data in 2007. The result suggests the promising application of street views for auditing neighborhood environments in a more rapid way but with a lower cost.

Geotagged images collected from Google Street View and other websites have been leveraged to quantity urban perception using crowdsourcing methods, where human observers make a variety of perceptual inferences about images of places based on their prior knowledge and experiences. Salesses et al. \cite{Salesses2013} presented a method to measure the urban perception of safety, class and uniqueness in two US cities (Boston and NYC) and two Austrian cities (Linz and Salzburg) based on hundreds of geotagged images. They created a website to collect perception data by asking human observers to do pairwise comparison of two randomly selected images in response to questions, for example, on safety: ``Which place looks safer?'' (see Place Pulse, http://pulse.media.mit.edu). With the collected perception data, they calculated a $Q$-score for image $i$ on question $u$ by
\begin{equation}
  Q_{i,u} = \frac{10}{3} \left(  W_{i,u} + \frac{1}{n_{i}^{w}} \sum_{j_1 = 1}^{n_{i}^{w}} W_{j_1 u} -  \frac{1}{n_{i}^{l}} \sum_{j_2 = 1}^{n_{i}^{l}} W_{j_2 u} +1 \right),
\end{equation}
where $W_{i,u} = w_{i,u} / (w_{i,u} + l_{i,u} + t_{i,u})$ and $L_{i,u} = l_{i,u} / (w_{i,u} + l_{i,u} + t_{i,u})$ are respectively the win (W) and loss (L) ratios of image $i$ with $n_{i}^{w}$ and $n_{i}^{l}$ being respectively the total number of wins and losses. The $Q$-score takes the value in $[0, 10]$ with $Q=10$ meaning the highest level of safety. They found that the range of perceptions elicited by images from Boston and NYC is wider, suggesting the two US cities are perceptually more unequal than Linz and Salzburg. Moreover, the spatial variation of urban perception helps explain violent crimes in NYC zones at zip-code resolution. Later, Ordonez and Berg \cite{Ordonez2014} predicted human perceptions of safety, uniqueness and wealth in urban places by applying classification and regression models to the Place Pulse dataset and another crowd-sourced dataset of street view images. They found that perceptual predictions are highly correlated with official crime and wealth statistics.

\begin{figure}[t]
  \centering
  \includegraphics[width=0.6\textwidth]{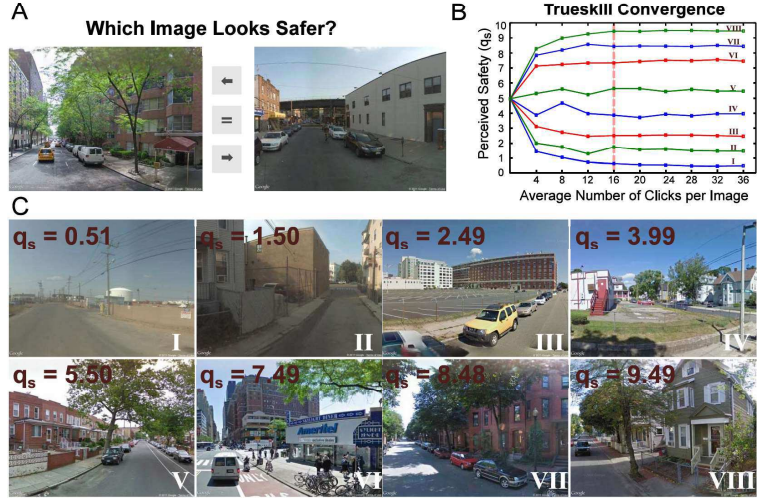}
  \caption{Perceived safety scores converted from pairwise image comparisons. (A) The pairwise image comparisons from a crowdsourced study, the Place Pulse \cite{Salesses2013}. (B) The Trueskill \cite{Herbrich2006} converges to a stable $Q$-score of perceived safety after about 16 clicks. (C) The images are ranked by their $Q$-scores of perceived safety, which is between 0 and 10. Figure from \cite{Naik2014}.}
  \label{Fig_4_3_5}
\end{figure}

Novel computational tools have been used to predict urban perception from street images. Naik et al. \cite{Naik2014} trained a scene understanding model (named \emph{Streetscore}) based on data from an online survey with 7000 participants to predict the perceived safety of a streetscape using generic image features (see Figure~\ref{Fig_4_3_5}). They used the Microsoft Trueskill algorithm \cite{Herbrich2006} to convert ratings from the Place Pulse dataset to a $Q$-score for each image and then trained the $v-$support vector regression ($\nu$-SVR) model \cite{Scholkopf2000} using input feature vectors $\mathrm{x}$ and their corresponding labels $y$ to predict the $Q$-score of each image. The goal of SVR \cite{Cortes1995,Smola2004} is to approximate $y$ by a regression function $f(\mathrm{x})$. Here, $f(\mathrm{x}) = (\mathrm{w} \cdot \mathrm{w}) + b$. The key idea of $\nu$-SVR is to guarantee that the number of predictions with an error over $\epsilon$ is less than $\nu$ through minimizing the following function
\begin{equation}
\begin{array}{ll}
    \min & \frac{1}{2}\parallel \mathrm{w}\parallel^2 + C \cdot \big( \nu \epsilon + \frac{1}{K} \sum_{i=1}^{K} (\xi_i + \xi_{i}^{*}) \big)\\
    \textrm{s.t.} & \left\{
        \begin{array}{l}
        \big( \mathrm{w} \cdot \mathrm{x} + b \big) - y_i \leq \epsilon + \xi_i \\
        y_i - \big( \mathrm{w} \cdot \mathrm{x} + b \big) \leq \epsilon + \xi_{i}^{*} \\
        \epsilon \geq 0, \xi_{i}^{*} \geq 0 \\
        \end{array}
    \right.
\end{array}
.
\end{equation}
The model including all features exhibits an accuracy of $R^2=0.5676$ in predicting the $Q$-score of safety. After scoring about 1 million street view images using the \emph{Streetscore}, they created high resolution perception maps for 21 US cities. However, Porzi et al. \cite{Porzi2015} argued that taking a ``ground-truth'' function in the regression model \cite{Salesses2013} or using the Trueskill scoring function \cite{Naik2014} to predict scores of images based on users' judgments will bring intrinsic bias. It is hard to explain the true distribution of users' pairwise judgments by using one of many possible scoring functions. On the other hand, training an algorithm for prediction using a scoring function will leave no independent data for error assessments as the scoring function is constructed based on all users' judgments. To address these issues, they developed a ranking framework that employs a convolutional neural network (CNN) to train algorithms directly on users' ratings. Their CNN architecture uses a novel pooling layer to automatically discover mid-level visual patterns that have the strongest correlations with urban perception. Evaluations on the Place Pulse dataset demonstrates the advantages of their method on predicting perceived safety of images.

Yet, only using human judgments of images, it is still difficult to define some concepts in urban perceptions (e.g., beauty, happiness and quietness). To this end, Quercia et al. \cite{Quercia2014} explored visual assessments to better understand people's visual perceptions and developed methods to automatically extract aesthetically informative features from city scenes. They built a crowdsourcing website (http://urbangems.org) to collect user ratings of London's streets and then translated the ratings into quantitative measures of urban perception on three qualities: beautiful, quiet and happy. Next, they employed image processing techniques to determine visual cues that may cause the perception of a street. After checking the association with the three qualities, they found that the most positive visual cue is the amount of greenery. Arietta et al. \cite{Arietta2014} developed a method to automatically identify relations between a city's visual appearance and its non-visual attributes. They spatially interpolated the data of (location, value of attribute) to obtain the attribute values and then detected discriminative visual elements of the attribute by building some SVMs \cite{Boser1992}. Next, they trained attribute predictors by employing a nonlinear SVR \cite{Cortes1995,Smola2004} to learn a set of weights over these elements. They found that visual elements are predictive to many city attributes including crime rates, population density and tree presence. The attribute predictors estimate the theft rate on average 33\% more accurate than humans.

Computer vision methods have been increasingly used to learn deep features for scene recognition at the presence of the availability of large datasets. The introduction of deep convolutional neural networks (DCNNs) \cite{Lecun2015} has dramatically improved the performance in computer vision tasks, for instance, extracting urban perceptions from images. Seresinhe et al. \cite{Seresinhe2017} quantified the beauty of outdoor places by analyzing online crowdsourced ratings of over 200,000 images of Great Britain and features extracted from a scene-centric image dataset \cite{Zhou2014}. They found that places with natural features and man-made structures are considered more scenic. Then, they trained a neural network (named Scenic CNN) to predict the beauty of scenes for new places. The Scenic CNN can automatically identify natural and built scenic places such as the Big Ben in London. Dubey et al. \cite{Dubey2016} created a new global crowdsourced dataset (Place Pulse 2.0) consisting of over 1 million pairwise comparisons for more than 110 thousand images from 56 cities in 28 countries. Images are scored along perceptual attributes: safe, lively, boring, wealthy, depressing, and beautiful. They designed the Streetscore-CNN (SS-CNN) to predict the winner in the task of a pairwise comparison and then designed the Ranking SS-CNN (RSS-CNN) to better understand the fine-grained differences between image pairs. Experimental results show that RSS-CNN outperforms SS-CNN in predicting perceptual attributes. Moreover, models trained to predict one visual attribute can be used to predict other visual attributes with a fair accuracy.

Albert et al. \cite{Albert2017} identified patterns in urban environments based on satellite imagery from Google Maps Static. They applied computer vision techniques based on DCNNs to classify images. Their results show good agreement with public benchmark data in green urban areas, water bodies, urban fabric, airports, etc. Moreover, they found that deep features of urban environments extracted from satellite imagery exhibit good performance on comparing neighborhoods across several cities, suggesting their method's ability of transfer learning. Tracewski et al. \cite{Tracewski2017} investigated whether a deep learning network trained on scene characteristics can be used to classify volunteered photos for land cover characterization. They applied a simple post hoc weighting approach and a complex decision tree approach to extract land cover information from the network. They found that a general neural network without specifical training can achieve modest levels of classification accuracy, suggesting the usage of well-validated methods without doing a long and costly training exercise. Lef{\'e}vre et al. \cite{Lefevre2017} explored the use of multiview imagery combining overhead and ground views on scene analysis. They suggested to integrate remote sensing, computer vision and machine learning for better urban observation.

Mobile phone (MP) data have also been used to visually profile cities and analyze urban lives. Based on GPS locations recorded by a smartphone app and self-rated happiness of over 20,000 subjects in the UK, Guillen et al. \cite{Guillen2013} explored the relationship between happiness and natural environments associated with GPS locations. They found that participants exposed to green environments or natural habitat types are significantly happier than those to urban environments. De Nadai et al. \cite{Nadai2016b} explored whether safer looking neighborhoods are more lively in two major Italian cities (Rome and Milan). They defined metrics for human activity or liveliness in an area based on MP billing and operation data. Then, they employed DCNNs trained on the Place Pulse dataset to predict the scores of perceived safety based on streetscape imagery. Finally, they explored the relation between safety perception and liveliness using the spatially corrected ordinary least squares regressions. They found that neighborhoods perceived safer are more lively, while population demographics affect the perception. Moreover, street facing windows and greenery also contribute to safety perception. Recently, Harvey and Aultmanhall \cite{Harvey2016} reviewed approaches on measuring urban streetscapes for livability such as the uses of Internet-enabled surveys, streetscape images and social media.

\begin{figure}[t]
  \centering
  \includegraphics[width=0.75\textwidth]{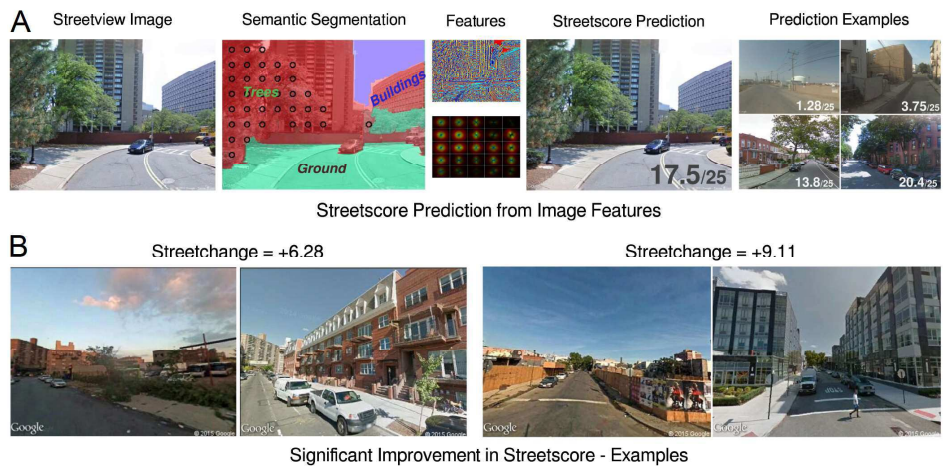}
  \caption{The computation of Streetchange. (A) Computing Streetscore using a regression model based on two image features (GIST and texton maps), which are extracted from pixels of four object categories (ground, buildings, trees, and sky) using semantic segmentation. (B) Computing Streetchange by the difference between the Streetscores of a pair of images captured in 2007 and 2014. Significantly positive Streetchange is usually associated with major construction. Figure after \cite{Naik2017}.}
  \label{Fig_4_3_6}
\end{figure}

Physical appearances of neighborhoods are not static, but changing over time. Naik et al. \cite{Naik2017} introduced a computer vision method to understand physical dynamics of cities based on street views at different times. They collected over one million image cutouts for street blocks in 2007 and 2014 for five large US cities and matched the 2007 panel with the 2014 panel according to geographical locations of image cutouts. Then, they predicted the perceived safety of street view images using a variant of the Streetscore algorithm \cite{Naik2014} and obtained Streetchange by calculating changes in streetscores of images from 2007 to 2014 (see Figure~\ref{Fig_4_3_6}). After exploring the connection between Streetchange and socioeconomic changes, they found supportive evidences to the three classical theories of urban change: (i) both education and population density predict physical improvements in neighborhoods, supporting theories of human capital agglomeration; (ii) physical proximity to city centers and attractive neighborhoods predicts neighborhood improvement, supporting the \emph{invasion theory} \cite{Burgess1925} that improvements in a neighborhood will spillover to adjacent areas; (iii) neighborhoods with better initial appearances have larger improvements, supporting the \emph{tipping theory} \cite{Schelling1969} that nicer neighborhoods will get better.

Recently, data from other platforms have also been used to analyze neighborhood change and to infer socioeconomic levels. Glaeser et al. \cite{Glaeser2018b} explored the potential use of Yelp data in tracking neighborhoods change and forecasting local economic activity. They found that the entry of Starbucks recorded by Yelp data is indicative of house price growth in the US, and gentrifying neighborhoods can attract more upscale establishments. Moreover, Streetchanges of perceived neighborhood quality are predictive of changes in local economy. Brelsford et al. \cite{Brelsford2018} explored the topology and spatial evolution of neighborhoods based on a diverse set of detailed urban maps. They found that neighborhoods in developed cities fall into the same topological class while urban slums display different topological characteristics. Moreover, it is possible to build a street network in existing slums that can guarantee universal connectivity at minimal disruption and construction costs. Indeed, urban forms are predictive to socioeconomic status. Venerandi et al. \cite{Venerandi2018} explored the relations between metrics of urban form and socioeconomic status extracted from five openly accessible datasets. They found that urban form can explain up to 70\% of the variance in the official Index of Multiple Deprivation (IMD) of six major UK cities. Moreover, they observed some patterns of more deprived UK neighbourhoods such as higher population density, more regular street patterns, and more dead-end roads.

\subsubsection{Urban computing for better lives}

In the emerging interdisciplinary field named \emph{urban computing} \cite{Zheng2014,Zheng2019}, the unobtrusive and continuous improvement of urban lives have been increasingly studied in recent years. By leveraging data generated in cities, urban computing connects computer sciences with conventional city-related fields through technologies of urban sensing, data analytics and visualization in a recurrent process, with applications to social science, economy, urban planning, transportation, and so on. Recently, Zheng et al. \cite{Zheng2014} reviewed the general framework and key challenges of urban computing from a computer science perspective. They classified the applications into seven categories (i.e., urban planning, transportation, environment, public safety and security, energy, social, and economy) and summarized the typical technologies into four folds (i.e., urban sensing, urban data management, knowledge fusion across heterogeneous data, and urban data visualization). Later, Calabrese et al. \cite{Calabrese2015} reviewed the techniques related to the use of mobile phone (MP) networks for urban sensing. They provided recommendations on datasets and techniques for specific applications. Regarding the improvement of urban lives, Glaeser et al. \cite{Glaeser2018} described a variety of new urban data sources and illustrated how these data can be used to improve the quality of urban services.

Understanding what creates a better urban life is critical if anyone would like to make some improvements. Jacobs \cite{Jacobs1961} suggested the urban physical environment as an essential factor for urban vitality. Specifically, the presence of pedestrians at all times of the day create life, while the elimination of pedestrian activity causes death. Therefore, the improvement of urban life in large cities is associated with the diversity of physical environments that require four essential conditions: mixed use, small blocks, buildings diverse, and building concentration. Recently, by collecting pedestrian activity through surveys and employing multilevel binomial models, Sung et al. \cite{Sung2015} empirically verified the role of Jacobs's four conditions in urban diversity in Seoul, South Korea. De Nadai et al. \cite{Nadai2016} later verified the necessity of Jacobs's four conditions for the promotion of urban lives in six Italian cities. They used a proxy for urban vitality extracted from MP records, land usage mapped from satellite images, and socio-demographic information collected from national census and the Open Street Map project. Their work provides a new way to test traditional urban theories in fine-grained details.

Recent literature have studied urban segregation of people with different socioeconomic status. Using geotagged tweets in Louisville, Shelton et al. \cite{Shelton2015} developed an approach to study intra-neighborhood segregation, mobility and inequality. After analyzing the everyday activity spaces of different groups, they proposed to understand Louisvillian neighborhoods by the fluid, porous and actively produced. After exploring the socio-spatial segregation in Beijing, Wang et al. \cite{Wang2012b} found significant differences between residents inside and outside the so-called privileged enclaves in the usages of time. They suggested scientists should pay more attention to how different social groups actual use urban spaces and spend their time in terms of everyday activities. Yip et al. \cite{Yip2016} analyzed the mobility patterns of people in Hong Kong that are tracked by a mobile phone app. They found that the interactions of people with other income groups are limited. Rich people tend to move to rich neighbourhoods, while poorer people move to poorer neighbourhoods. Louf and Barth{\'e}lemy \cite{Louf2016} provided a direct definition of residential segregation and showed that richer class in high density zones is over represented. In particular, they suggested density as a relevant factor for understanding urban income structure and explaining differences observed in cities.

\begin{figure}[t]
  \centering
  \includegraphics[width=0.6\textwidth]{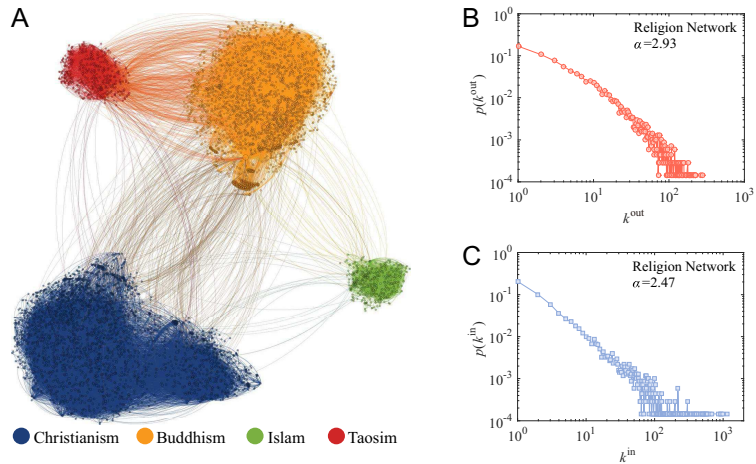}
  \caption{Structure of the religion network. (A) Structural layout of the network neglecting the directions of links, where blue, orange, green and red nodes denote Christians, Buddhists, Islamists and Taoists, respectively. (B) The out-degree distribution in a log-log plot. (C) The in-degree distribution in a log-log plot. Figure from \cite{Hu2019}.}
  \label{Fig_4_3_62}
\end{figure}

Data from social networks have been used to study religious segregation and urban indigenization. Hu et al. \cite{Hu2019} quantified religious segregation by analyzing religious social network based on Weibo. They found that the religious network is highly segregative (see Figure~\ref{Fig_4_3_62}), and the extent of religious segregation is higher than racial segregation. In addition, 46.7\% of cross-religion connections are probably related to charitable issues, suggesting the role of charitable activities in promoting cross-religion communications. Yang et al. \cite{Yang2017} identified the distinct mobility patterns of natives and non-natives in five large cities in China by analyzing about 1.37 million check-ins. They found that the distribution of location visiting frequencies is relatively homogeneous for natives as they usually check in repeatedly at locations of personal importance. By contrast, the distribution is more heterogeneous for non-natives as they tend to visit popular locations. With this insight, Yang et al. \cite{Yang2017} proposed a so-called indigenization coefficient to estimate the likelihood of an individual to be a native or to what extent an individual behaves like a native, which is based solely on check-in behaviors. Such method can be applied in estimating the time required for non-natives to behave the same as natives, as well as in enhancing the prediction accuracy of human mobility (i.e., the next check-in location).

The Schelling model \cite{Schelling1971} is widely used to explain the emergence of racial segregation. A small preference to alike neighbors at the individual level can lead to large-scale segregation at the collective level through neighborhood tipping. Sahasranaman et al. \cite{Sahasranaman2016} studied the dynamics of transformation from segregation to mixed wealth cities using a variation of the Schelling model, in which the movement of agents to neighborhoods should satisfy the threshold condition: the wealth of agents is not lesser than the wealth of the threshold proportion of their neighbors in the new neighborhoods. They found that wealth-based segregation occurs and persists, however, the dynamics can yield a persistent mixed wealth distribution in the tolerance condition that small proportions of disallowed moves (where threshold condition is not satisfied) are introduced. The results suggest to enable a small fraction of disallowed moves to drive the transformation to mixed wealth cities. Later, Sahasranaman et al. \cite{Sahasranaman2017} extended the above model and studied the long-term patterns of neighborhood economic status. They found that very poor and very rich neighborhoods tend to retain economic status than middle-wealth neighborhoods.

\begin{figure}[t]
  \centering
  \includegraphics[width=0.75\textwidth]{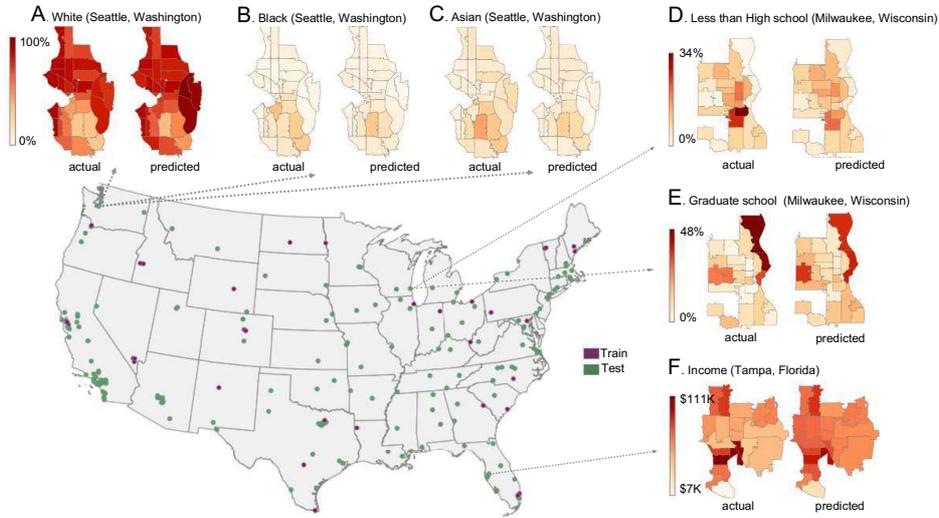}
  \caption{Estimations of socioeconomic status from car attributes. The model was trained in counties (shown in purple on the map) and used to estimate demographic variables at the zip code level for all cities (shown in green). Maps of actual versus predicted values are shown for the percentages of (A-C) Black, Asian, and White people, (D) people with less than a high school degree, (E) people with graduate degrees, as well as for (F) the median household income. Figure after \cite{Gebru2017}.}
  \label{Fig_4_3_7}
\end{figure}

Deep-learning-based computer vision techniques have been applied to analyze digital imagery, which provides a faster and cheaper alternative of community survey. Gebru et al. \cite{Gebru2017} proposed a method to estimate socioeconomic trends from 50 million street view images in 200 US cities (see Figure~\ref{Fig_4_3_7}). They automatically detected 22 million distinct vehicles from images using the object recognition algorithm \cite{Felzenszwalb2010} and then deployed CNNs \cite{Krizhevsky2017} to determine features of vehicles and classify each vehicle into one of the 2,657 fine-grained categories. Using the resulting data, they estimated race and education levels by training a logistic regression model and estimated income and voter preferences by employing a ridge regression model. Compared to the American Community Survey, their demographic estimates exhibit satisfied accuracy at the city level (e.g., $r=0.82$ for the estimates of median household income, $r=0.87$ for the percentage of Asians, and $r=0.70$ for the percentage of people with a graduate degree). The method can also provide a good accuracy at a more fine-grained zip code resolution, for example, the estimation of the percentage of Asians yields a correlation $r=0.77$ at zip code resolution for Seattle. In addition, the method can accurately estimate voter preference. For example, a city tend to vote for a Democrat (88\% chance) if encountering more sedans than pickup trucks during a drive through the city.

The demand for facilities (such as hospitals, airports and malls) increases as the development of cities. Locations of facilities are ideally determined by the necessities of people who live nearby. Theoretical derivations starting from assumptions of least cost \cite{Mycielski1963} and minimum time \cite{Stephan1984} have suggested that the optimal density of facilities $D(r)$ at certain position $r$ scales as a power law ($\alpha=2/3$) with the density of relevant population $\rho(r)$. After analyzing the distribution of over 400 nongovernmental and service establishments in the US, Stephan \cite{Stephan1988} found that $\alpha$ is more close to 1 than to the widely observed $2/3$ for governmental establishments. Gastner and Newman \cite{Gastner2006} found a slope of 0.66 in the log-log plot of $D(r)$ versus $\rho(r)$ in the US and presented an analytic solution based on density-dependent map projections. Recently, Um et al. \cite{Um2009} explored the scaling $D \sim \rho^{\alpha}$ and found that $\alpha$ depends on facility types. By proposing a microscopically mechanism model, they demonstrated that public facilities driven by social opportunity cost have an exponent $\alpha \approx 2/3$, whereas private facilities driven by profit have an exponent $\alpha \approx 1$. The distributions of the optimal positions of public or private facilities on real US map predicted by their model agree well with the empirical data.

Fine-grained data describing business activities have been used to identify the optimal locations for new retail stores and improve the layout of amenities (e.g., schools, restaurants, cafes, and libraries) in urban areas. Karamshuk et al. \cite{Karamshuk2013} predicted the popularity of retail stores in NYC using machine learning approaches based on check-in data collected from Foursquare. They tested the predictive power of various features extracted from the check-in data including geographic features and mobility features, finding that the strongest indicators of popularity are the presence of user attractors and retail stores of the same type. Recently, Hidalgo and Casta{\~n}er \cite{Hidalgo2015} studied the neighborhood-scale agglomerations of amenities by building a network of amenities, named amenity space, based on the precise locations of millions of amenities across 47 US cities. They introduced a clustering algorithm to identify neighborhood-scale agglomerations and mapped the amenity space by connecting pairs of amenities that are more likely to co-locate in the same neighborhood (see Ref. \cite{Hidalgo2007} for a similar method). Based on the amenity space, they further built a recommender system \cite{Lu2012b} to recommend missing amenities in the neighborhood, given its current pattern of specialization. Their method provides new avenues for the optimal layout of facilities within cities.

The analysis of bank card transactions provides a new way to estimate regional socioeconomic status and improve urban spatial equity. Louail et al. \cite{Louail2017} analyzed a database of card transactions in two Spanish cities (Madrid and Barcelona) and then proposed a bottom-up approach to redistribute money flows for equality situations through redirecting a limited fraction of individual shopping trips. They constructed the ``individual-business'' bipartite spatial network, where the edges correspond to card transactions. Then, they performed the rewiring of individual transactions by redirecting them to the same business category located in different neighborhoods, with the goal to re-balance the commercial income among neighborhoods and with the preservation of human mobility properties (see Ref. \cite{Louail2017} for details). They found that reassigning only 5\% of individual transactions can reduce more than 80\% spatial inequality between neighbourhoods and can even improve other sustainability indicators like total distance traveled and spatial mixing. Their work illustrates an excellent implementation of crowdsourcing the ``Robin Hood effect'' \cite{Poddar2012}, a process through which capital is redistributed to reduce inequality.

\section{Individual socioeconomic status and attributes}
\label{Sec4}

\subsection{Individual socioeconomic level}

Socioeconomic level is an indicator typically defined as a combination of income related variables to characterize an individual's social and economic status. As such, individual socioeconomic level serves as an indication of the purchasing power, and thus it is important to the design and evaluation of social policies. In this subsection, we will introduce literature on how to infer individual socioeconomic levels from the ownership and usage of mobile phones (MPs), social media (SM) data, bank transitions, human mobility patterns and individual behavioral traits.

\subsubsection{Mobile phone and credit card usage}

The computation of socioeconomic indices faces some challenges such as the large expenses in acquiring a whole country's data, the long-time delay of the census data, and the lack of high-quality data in developing economies. Thanks to the development of information and communication technology, MPs are now becoming ubiquitous and frequently used even in the world's poorest countries and regions \cite{Donner2006}. MPs serve as important sensors of human activities, which should be closely related to individual socioeconomic status. Using behavioral variables extracted from MP data, Soto et al. \cite{Soto2011} predicted individual socioeconomic level by employing models constructed with the support vector machines and random forests. Using only 38 features, their model can achieve a classification accuracy up to 80\% in determining socioeconomic levels of about 0.5 million citizens. With promising applications, their method can be a cost-effective complement to traditional socioeconomic estimation techniques.

The ownership and usage of MPs in the past decade are not uniform across populations and continents. Blumenstock and Eagle \cite{Blumenstock2010b} provided a quantitative perspective on the socioeconomic structure of individual MP usage in Rwanda between 2005 and 2009. After analyzing data collected through interviews and merged with MP call histories, they found that MP owners are considerably wealthier, better educated and more predominantly male. Most notably, they found some differences in MP adoption between the relatively poor and rich. For example, richer people have a larger number of calls, a greater total length of calls, more MP-using days, a larger number of contacts, and so on. Their results demonstrate that MPs are owned and used by the privileged strata of Rwanda society (see also Ref. \cite{Blumenstock2010c}). Wesolowski et al. \cite{Wesolowski2012b} analyzed a survey of MP ownership and usage across Kenya in 2009 to understand the social and geographical heterogeneity of MP usage patterns. They found distinct regional, gender-related and socioeconomic variations, with particularly low ownership among rural communities and poor people. In particular, there is a nonlinear relationship between MP ownership and sharing behavior across counties, where MP sharing practices are extremely common in rural areas.

Novel methods have been proposed to infer individual socioeconomic levels from MP data. Blumenstock \cite{Blumenstock2014} explored the extent to which MP data can be used to predict an individual's socioeconomic status by analyzing the combined data of MP records and phone-based interviews in Rwanda during 2009-2010. They found significant correlations between asset ownership and some MP-derived measures including phone usage, social network structure and geographic mobility. The first principal component of 97 metrics of MP usages can explain 34.63\% of the variance in the asset categories. Moreover, simple classification methods are able to predict the fixed household characteristics and whether the respondent owns assets. The prediction accuracy for television ownership is over 85\%. Agarwal et al. \cite{Agarwal2018} analyzed 350 million MP call logs and found that phone-based features have significant predictive power for an individual's financial risk with an accuracy of about 65\%. Recently, deep learning technologies are also used to classify individual socioeconomic status based on large-scale MP datasets. Sunds{\o}y et al. \cite{Sundsoy2016} explored how socioeconomic levels can be accurately classified by implementing a multi-layer feed-forward deep learning architecture \cite{Lecun2015} without any manual operations on feature selection. The new model using location traces as the sole input achieves an average AUC value about 0.77 in separating individuals into high and low socioeconomic status.

MPs can generate rich data about financial histories of mobile money. For example, airtime credit is the money on MP devices, which can be used for purchase (e.g., calls, texts and data) and be transferred to others. Based on the history of airtime credit purchases and MP communications in 2012, Gutierrez et al. \cite{Gutierrez2013} estimated the relative income of individuals and the diversity of income for fine-grained regions in C{\^o}te d'Ivoire. They quantified individual purchasing behavior by calculating the variation in the purchase amounts through the coefficient of variation (CV):
\begin{equation}
  CV = \sigma / \mu ,
\end{equation}
where $\sigma$ is the standard deviation, and $\mu$ is the mean value. They found that some people make few big purchases, while others make many small purchases. Thereby, they hypothesized that the frequency and size of purchases are correlated with individual income as the poorer may not buy lots at once due to the lack of enough airtime credit. After analyzing the social network built on the MP communications, they found a certain homophyly in terms of purchase averages, where people belonging to the same community tend to have the same amount of average purchase. Recently, behavioral patterns revealed by MP usages have been used to predict default among borrowers. For example, Bj{\"o}rkegren and Grissen \cite{Bjorkegren2017} proposed a method to predict the likelihood of repayment using behavioral features derived from MP transaction records in a Caribbean country. The method achieves an AUC of 0.76 and 0.77 respectively for banked and unbanked consumers without formal financial histories.

Purchasing behaviors documented by bank accounts and credit cards are also predictive to individual socioeconomic status. By analyzing bank administrative data and survey data, Prina \cite{Prina2015} found that the access to free savings accounts can generate welfare effects. After randomly giving free access of bank accounts to many female household heads in Nepal, they found that physical proximity and zero fees lead to high take-up and usage rates of savings accounts, resulting in the overall improvement of financial situation for low-income households. These results suggest that the access to formal financial services can lead to household improvements. Dong et al. \cite{Dong2016} analyzed millions of individual credit card transactions in two low-to-middle income countries. They found that patterns of purchase activity are strongly correlated with socioeconomic status. They defined two measures of purchase diversity at the district-level (the outgoing purchase diversity and the incoming purchase diversity) by averaging the purchase diversity defined by the Shannon entropy (see Ref. \cite{Eagle2010} for details) over corresponding individuals. They found a positive correlation between both purchase diversities and socioeconomic status, e.g., $r=0.77$ for the European country and $r=0.53$ for the Latin American country in case of outgoing purchase diversity.

\begin{figure}[t]
  \centering
  \includegraphics[width=0.8\textwidth]{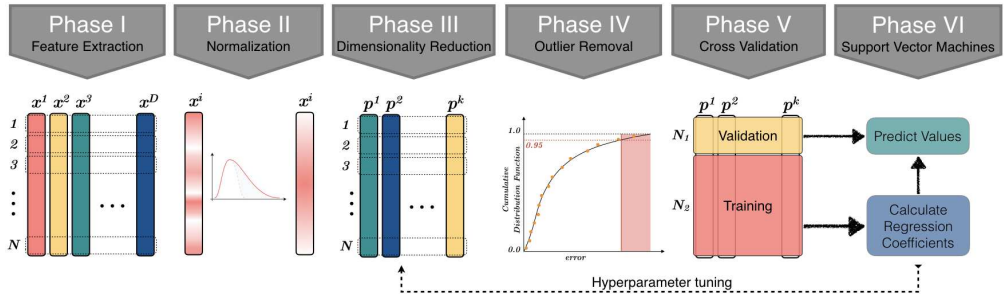}
  \caption{The modular machine learning workflow. The individual financial transactions are used in the predictive modeling of official socioeconomic indicators. The support vector machines (SVM) is employed based on the reduced features space using the principal component analysis (PCA). Figure from \cite{Hashemian2017}.}
  \label{Fig_5_1_2}
\end{figure}

Hashemian et al. \cite{Hashemian2017} connected individual financial transactions with the socioeconomic description of regions by analyzing a dataset of bank card transactions in Spain. They built a feature space after extracting 33 microeconomics indicators and predicted official socioeconomic indices by employing a modular machine learning workflow (see Figure~\ref{Fig_5_1_2}). Their model performs decently, e.g., in predicting regional GDP per capita with an accuracy of 0.729 using only seven principal components of the reduced features space. Sobolevsky et al. \cite{Sobolevsky2017} analyzed the same Spanish transaction data and found a clear correlation between individual spending behaviors and six official socioeconomic indicators. They created a feature space consisting of 35 microeconomics indicators derived from the bank card transactions. They further applied the principal component analysis (PCA) \cite{Jolliffe2002} to build a reduced feature space, based on which they used a generalized linear model (GLM) \cite{Mccullagh1989} to predict socioeconomic indices. Their model well predicts all the considered quantities. The average accuracy on the training set is about 0.70 for the province-level GDP prediction. By analyzing credit card purchasing data, Clemente et al. \cite{Clemente2018} found that male and young adults are of higher probability to use credit card than female and old adult. Moreover, the median credit card record expenditure is correlated with the average monthly wage at the district level.

\subsubsection{Social profile and network structure}

Individual socioeconomic status can be inferred from social media (SM) data and the network structure of online social networks. For example, there are wide discussions on the embeddedness of economic activities in social networks \cite{Granovetter2005}. On the other hand, users' behaviours on social network are affected by their socioeconomic status and health consciousness. For example, the non-adaptive coping responses to health-related messages in the mass media are negatively correlated with socioeconomic status \cite{Iversen2006}. Wangberg et al. \cite{Wangberg2008} explored the relations between individual Internet use and socioeconomic status based on two survey datasets. They found that Internet is a plausible mediator between subjective health and socioeconomic variables.

Messages posted to SM have been used to estimate users' social class and socioeconomic status. Filho et al. \cite{Filho2014} proposed a method to predict a user's social class based on interactions on Foursquare and messages on Twitter  with a hypothesize that richer users usually visit wealthier places. In their method, each neighbourhood is assigned with a social class according to the income. They mapped the coordinates of Foursquare interactions and tweets to neighbourhoods and assigned every visited place with a social class. A user's social class is estimated by the most frequently visited class. Further, they employed classification models to predict users' social classes using textual features extracted from tweets. The average F1 value of their models varies from 0.57 to 0.73, depending on the classification segments. Based on a large dataset of Twitter users annotated with income, Preo{\c{t}}iuc-Pietro et al. \cite{Preoctiuc2015} built a model to predict income using user profile, psycho-demographic, emotion and shallow textual features. They found that high earnings are indicated by intelligence, high education, male and old age. Low-income users express more disgust emotions and sadness, have high posting rate, use more swear words, and post more URLs. In contrast, high-income users express more fear and anger, have more retweets and talk more about justice and politics. They proposed nonlinear models that can predict user income with an accuracy of 0.633 using a combination of all features.

Based on UK Twitter users' profiles and tweets, Lampos et al. \cite{Lampos2016} proposed a method to classify users into the upper, middle or lower socioeconomic status. They mapped each user to a socioeconomic status by utilizing the standard occupation classification hierarchy \cite{Elias2010}. They extracted five categories of features from tweets: behavior, impact, profile, tweets, and topics. They performed the classification by employing a nonlinear learning approach, in which a composite Gaussian process (GP) is used. Given the input $\mathrm{x} \in \mathcal{R}^d$, the aim of the GP method is to learn a function $f: \mathcal{R}^d \rightarrow \mathcal{R}$:
\begin{equation}
f(\mathrm{x}) \sim \mathcal{GP} \big( m(\mathrm{x}), k(\mathrm{x}, \mathrm{x}') \big) ,
\end{equation}
where $m(\cdot)$ is the mean function. The covariance kernel $k(\cdot,\cdot)$ for feature category is given by
\begin{equation}
k(\mathrm{x}, \mathrm{x}') = \left( \sum_{n=1}^{C} k_{SE}(\mathrm{c}_n, {\mathrm{c}'}_n) \right) + k_{N}(\mathrm{x}, \mathrm{x}') ,
\end{equation}
where $\mathrm{x} = \{ \mathrm{c}_1, \ldots, \mathrm{c}_C \}$ is the input of $C$ feature categories, and $k_{N}(\mathrm{x}, {\mathrm{x}'}) = \theta_{N}^2 \times \delta(\mathrm{x}, {\mathrm{x}'})$ models noise with $\delta$ being the Kronecker delta function. The $k_{SE}(\mathrm{x}, {\mathrm{x}'})$ defines the squared exponential kernel, which is given by
\begin{equation}
k_{SE}(\mathrm{x}, {\mathrm{x}'}) = \theta^2 \exp \left(- \parallel \mathrm{x} - \mathrm{x}' \parallel _{2}^{2} / (2 \mathcal{L}^2) \right) .
\end{equation}
The GP model yields accuracies of 0.7509 and 0.8205 respectively for 3-way and binary classification scenarios.

The structure of mobile phones (MP) communication networks has been used to predict individual socioeconomic status. Leo et al. \cite{Leo2015} analyzed a coupled dataset of MP call detailed records (CDRs) and bank credit information of individuals living in Mexico. They found that the identified socioeconomic classes from the structure and evolution of the communication network are strongly correlated with typical consumption patterns. Moreover, they found positive correlations between people regarding their economic status and further confirmed the social stratification in social structure. Fixman et al. \cite{Fixman2016} analyzed CDRs and account balances for over 10 million clients of a Mexican bank and found a strong socioeconomic homophily in Mexico. Users linked in the MP communication network are more likely to have similar income. Further, they proposed a Bayesian approach to predict individual income, where individuals are distinguished into either low income group ($H_1$) or high-income group ($H_2$) according to their income ($g_s$). For user $q^j$, the amount of outgoing calls $a_{i}^{j}$ to the category $H_i$ is calculated. The Beta distribution $B^j$ for the probability of belonging to a given category is defined based on $a_{i}^{j}$:
\begin{equation}
  B^{j}(x;\alpha^j, \beta^j) = \frac{1}{B(\alpha^j, \beta^j)}x^{\alpha^j -1} \cdot (1-x)^{\beta^j -1},
\end{equation}
where $\alpha^j = \alpha_{1}^{j} + 1$ and $\beta^j = \alpha_{2}^{j} + 1$ are parameters of the Beta distribution
\begin{equation}
  B(\alpha, \beta) = \frac{\Gamma(\alpha)\cdot \Gamma(\beta)}{\Gamma(\alpha+\beta)} ,
\end{equation}
where $\Gamma (\alpha)$ is the gamma function. After obtaining the Beta distribution for the probability of belonging to the high-income category, the lowest five percentile $p_{\text{lower}}$ for this probability is found. The user's income category is set to $H_1$ if $p_{\text{lower}}$ is above a given threshold $\tau$ and $H_2$ if otherwise. Their method achieves an accuracy of AUC=0.71 with $\tau=0.4$ in classifying users into low- and high-income groups.

\begin{figure}[t]
  \centering
  \includegraphics[width=0.65\textwidth]{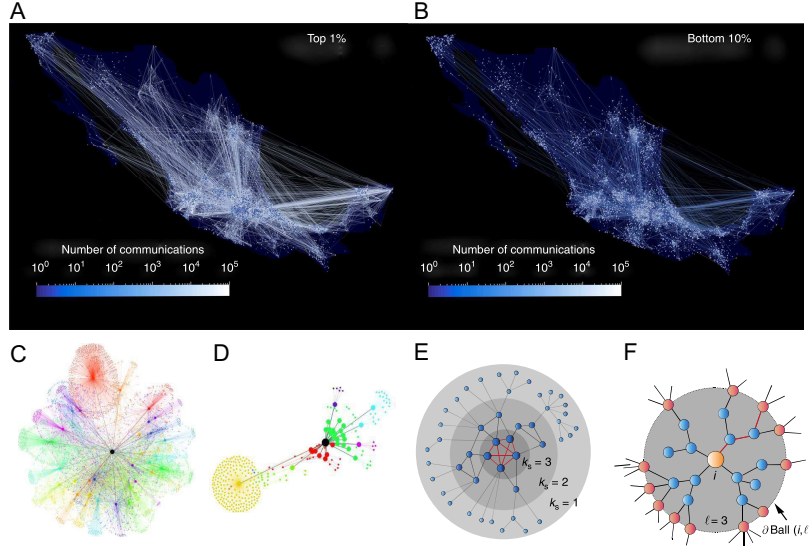}
  \caption{Communication networks of the population (A) in the top 1\% and (B) in the bottom 10\% of credit limit classes. Nodes are communities with the size denoting the number of bank clients inside each community. The colour and thickness of the edges reflects the number of communications. Examples of the ego-network for an individual (C) in the top 1\% wealthy class and (D) in the bottom 10\% class. (E) Illustration of the k-shell network decomposition. (F) Illustration of the calculation of the CI index. The CI Ball$(i,l)$ of radius $l=3$ around the central node $i$ (yellow) is the set of nodes inside the sphere, and $\partial$Ball is the set of nodes on the boundary (brown). The value of CI is the degree-minus-one of the central node times the sum of the degree-minus-one of the nodes at the boundary of the sphere of influence. Figure from \cite{Luo2017}.}
  \label{Fig_5_1_1}
\end{figure}

For a Latin American country's whole population, Luo et al. \cite{Luo2017} built a giant connected social network among 107 million users based on data about MPs and residential communications, and estimated individuals' financial status based on the combined credit limit of their credit cards. They found that people in the top economic class have higher diversity in communicating with equally affluent people and in connecting remote locations (see Figures~\ref{Fig_5_1_1}A-D), suggesting the significant different communication patterns across economic classes. They further measured an individual's centrality in the network using the so-called collective influence (CI) \cite{Morone2015}, in addition to degree, PageRank \cite{Brin1998} and k-shell index \cite{Kitsak2010}. The k-shell index $k_s$ of a node is the location of the shell obtained by iteratively pruning all nodes with degree $k < k_s$ (see Figure~\ref{Fig_5_1_1}E). The CI index of an arbitrary node $i$ can be obtained by the optimal percolation theory \cite{Morone2015}, say
\begin{equation}
 \text{CI} = (k_i -1) \sum_{j \in \partial \text{Ball}(i,l)} (k_j -1),
\end{equation}
where $k_i$ is the degree of the node $i$, Ball$(i,l)$ is the set of nodes within radius $l$ centred at node $i$, and $\partial \text{Ball}(i,l)$ is the set of nodes at the radius $l$ on the boundary (see Figure~\ref{Fig_5_1_1}F). They found that individual economic status is highly correlated with centralities in the social network. The correlation for k-shell is $R^2 = 0.96$ and for CI ($l=2$) is $R^2 = 0.93$. The correlation for CI can be increased by including the ages of individuals. Formally, they defined a new index
\begin{equation}
\text{ANC}= \alpha \text{Age} + (1-\alpha) \text{CI}.
\end{equation}
ANC exhibits a hight correlation $R^2 = 0.99$ when $\alpha=0.5$ and $l=2$. They further quantified the diversity of an individual's links by DR=$W_{\text{out}} / W_{\text{in}}$ \cite{Burt1995}, where $W_{\text{out}}$ and $W_{\text{in}}$ are the total communications with this individual outside and inside his/her own community, respectively. They found that an age-diversity composite
\begin{equation}
\text{ADC}= \alpha \text{Age} + (1-\alpha) \text{DR}
\end{equation}
correlates well with individual economic status ($R^2 = 0.96$) when $\alpha=0.5$. The ability of communicating with individuals outside one's local tightly-knit social community and positioning oneself at network locations of high CI is suggestive to a high socioeconomic level.

By analyzing CDRs and income group of surveyed individuals, Jahani et al. \cite{Jahani2017} found that the structural diversity of ego networks exhibits a relatively strong correlation with the income of individuals. They built a communication network based on the CDRs, where the weights of links are the total number of calls between two connected individuals. Then, they calculated the structural diversity from the view of ego networks \cite{Wang2016e}. In particular, they measured the diversity of the alters (i.e., neighbors of the ego) by defining the weighted structural novelty, as
\begin{equation}
 M_i = \frac{ \sum_{j\in{N(i)}} (1- \sum_{q\in N(i) \cap N(j)} p_{iq}p_{qj})} {|N(i)|},
\end{equation}
where $i \neq j \neq q$, $N(i)$ is the set of ego $i$'s neighbors, and $p_{ij}$ is the proportion of time that ego $i$ spent on its neighbor $j$. They found a positive correlation of the structural diversity on income after controlling education, occupation, age and gender.

Data from online gaming platforms have also been used to explore the relations between individuals' positions in social networks and their economic outputs. Xie et al. \cite{Xie2017} analyzed a database recorded by 124 servers of a popular online role-playing game in China. They found that the position diversity of individuals in the game is positively correlated with their economic output and social status. For a given friendship network, a dependence network is built by removing the insignificant edges. In the dependence network, they identified 13 directed triadic motifs \cite{Milo2002}, within which 30 distinct motif positions \cite{Stouffer2012} are located. Accordingly, they obtained the motif position ratio profile $p_i = (p_{i,1},\ldots , p_{i,30})$ for an arbitrary individual $i$. For individual $i$, the individual position diversity $d_i$ is defined by the Shannon entropy,
\begin{equation}
d_{i} = - \sum_{j=1}^{30} p_{i,j} \ln (p_{i,j}).
\end{equation}
They found that $d_i$ is highly correlated ($r=0.63$) with individual economic output. Further, they applied the $k$-means algorithm \cite{Macqueen1967} to cluster individuals based on their position ratio profiles $Z_i$, where $Z_i$ is the $z$-sore of $p_i$ (see Ref. \cite{Xie2017} for details). The cluster centroid locations $P_k = ( P_{k,1}, \ldots, P_{k,30} )$ can be obtained for an arbitrary class $C_k$ after measuring the closeness between position ratio profiles $Z_i$. The cluster position diversity of individual $i$ is defined as $D_i = - \sum_{j=1}^{30} P_{i,j} \ln (P_{i,j})$. They found that economic outputs of classes increase with $D_i$. This work demonstrates that the structure of social network is predictive to individual economic outputs even in virtual world.

\subsubsection{Human mobility pattern}

The relations between individual socioeconomic status and human mobility patterns have been explored in some scenarios based on a variety of new data resources \cite{Gonzalez2008,Barbosa2018}. The relations are in two aspects. Socioeconomic status and demographic information have effects on mobility patterns of individuals \cite{Fan2017}. For example, Carlsson-Kanyama and Linden \cite{Carlsson1999} analyzed national travel survey data in Sweden. They found that middle-aged rich persons travel much farther, while low-income persons in general do not travel extensively. Propper et al. \cite{Propper2007} analyzed hospital episode statistics in England and found that individuals in higher deprived wards travel less far for hospital admission. Fan et al. \cite{Fan2017} revealed the correlation between social proximity and mobility similarity based on an LBSN dataset. On the other hand, large-scale mobility datasets (generated by MPs, digital cards, online platforms, and so on) have been used to quantify human mobility patterns at fine spatio-temporal scales, based on which individual socioeconomic status can be predicted. Lotero et al. \cite{Lotero2016} explored the role of socioeconomic differences in urban mobility from a multiplex perspective based on the origin-destination surveys carried out in two Colombian cities. Each city is represented by six multiplex networks, where each one represents the trips of individuals with a specific socioeconomic status (SES), from SES1 (low) to SES6 (high). In each multiplex network, layers correspond to different transportation modes (e.g., pedestrian and public transport), and layers are merged by subsequently adding the one representing the mostly used transportation mode. This process produces a mobility multiplex network (MMN) for each SES. Some structural measures of multiplex networks \cite{Boccaletti2014} are used to quantify the mobility pattern of each SES, and two overlap measures are defined to quantify the tendency of individuals with the same SES to use different transportation. They found that individuals belonging to SES1 display the smallest clustering coefficient and tend to avoid overlapping. Moreover, the poorest (SES1 and SES2) covers larger urban areas in a sparse way by using few and cheap transportation modes, the meddle (SES3 and SES4) covers most urban zones, and the elite (SES5 and SES6) covers smaller urban areas in a dense way by selecting costly transportation modes.

\begin{figure}[t]
  \centering
  \includegraphics[width=0.55\textwidth]{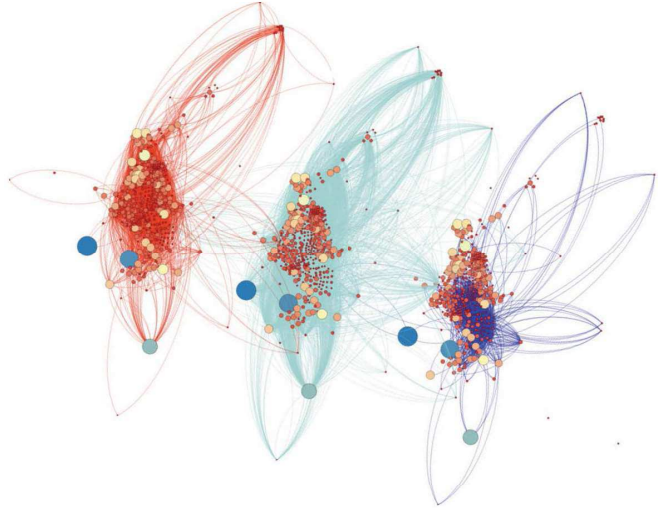}
  \caption{Three mobility networks of the Colombian city Medell{\'\i}n corresponding to three different socioeconomic status (SES). Networks from left to right correspond to SES1 (poorest class), SES3 (middle class) and SES6 (richest class), respectively. Nodes represent the origin and destination zones according to the origin-destination surveys. Sizes of nodes show the commuting strengths with colours representing their degrees. Figure from \cite{Lotero2016b}.}
  \label{Fig_5_1_3}
\end{figure}

Based on the same Colombian survey data, Lotero et al. \cite{Lotero2016b} further explored the temporal dependence of the trips performed by individuals with different SES. They found that the early-morning peak time delays (says rich do not rise early) and the midday peak becomes smoother as wealth increases. The strength of trips is more geographical dispersive for middle class but more localized for richest class (see Figure~\ref{Fig_5_1_3}). Moreover, the efficiency of urban mobility (defined by the ratio of the average distance traveled to the average time spent per trip) increases with the socioeconomic level. Carra et al. \cite{Carra2016} explored the relations between individual commuting distance and income based on datasets of national household surveys in Denmark, the UK and the US. Empirical results for Denmark and the UK confirm the prediction of basic equilibrium models \cite{Alonso1964} that individuals with a higher income have longer average commuting distances within a single city. The distribution of individual commuting distance across different countries is broad with a slow decaying tail that can be fitted by a power law with exponent $\gamma \approx 3$. Further, they proposed a new closest opportunity model for job searching process that can well predict the average commuting distance using individual income.

CDRs are increasingly used in human mobility analysis, which provides a chance to estimate individual socioeconomic status. Frias-Martinez et al. \cite{Frias2012b} analyzed CDRs in a Latin American country and found that population with higher socioeconomic levels have larger mobility ranges compared to population with lower socioeconomic levels. In particular, socioeconomic level is highly correlated with six human mobility variables including the traveled distance and radius of gyration \cite{Gonzalez2008}. The latter is defined as
\begin{equation}
r_{g}(t) = \sqrt{\frac{1}{n(t)} \sum_{i=1}^{n}(r_i - r_{\text{cm}})^2},
\end{equation}
where $r_i$ with $i= \{1,\ldots,n(t)\}$ is the vector of phone tower $i$'s coordinates, and $r_{\text{cm}} = \sum_{i=1}^{n} r_i / n(t)$ is the vector of the mobility trajectory's center of mass. There is a high correlation ($r=0.58$) between the number of used phone towers and the socioeconomic level. Further, they proposed a model that can estimate the tower-level socioeconomic status with an adjusted $R^2=0.72$. Later, Frias-Martinez et al. \cite{Frias2013} analyzed a large-scale dataset of CDRs and suggested to predict future values of socioeconomic indicators based on human behavioral patterns. Using the multivariate regression analysis, they found that mobility variables perform better than consumption variables on predicting future socioeconomic indicators. Moreover, multivariate time-series models can yield high accuracy, for example, the training $R^2$ value is about 0.68 in predicting the total assets.

\begin{figure}[t]
  \centering
  \includegraphics[width=0.9\textwidth]{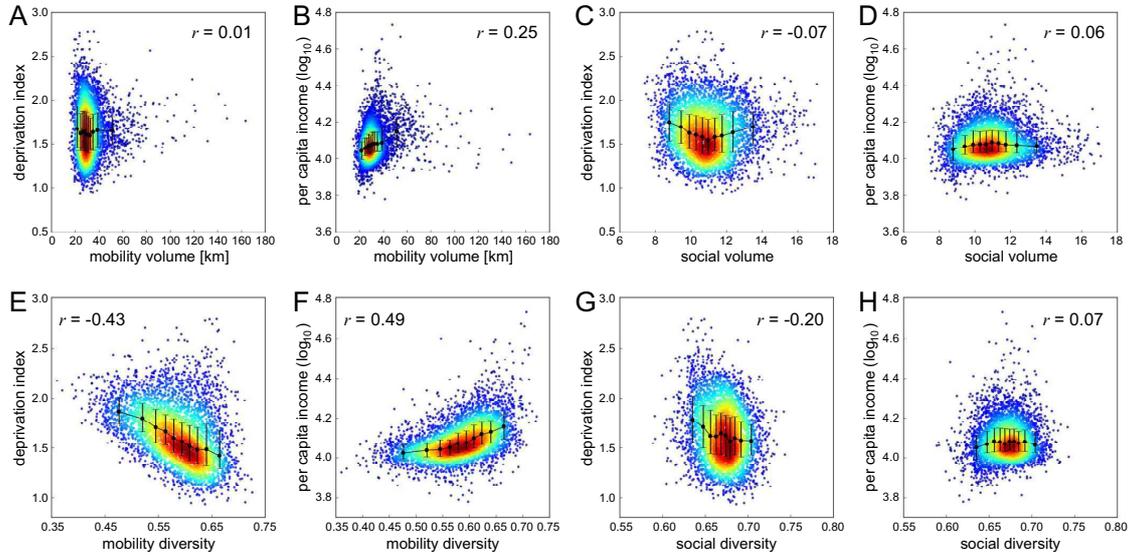}
  \caption{Relations between aggregated diversity measures and socioeconomic indicators. Mobility volume versus (A) deprivation index and (B) per capita income. Social volume versus (C) deprivation index and (D) per capita income. Mobility diversity versus (E) deprivation index and (F) per capita income. Social diversity versus (G) deprivation index and (H) per capita income. Black circles and error bars correspond to the mean and standard deviation for ten equal-sized municipalities. The $p$-value of the Pearson correlation coefficient $r$ is under 0.001 in all panels. Figure after \cite{Pappalardo2016}.}
  \label{Fig_5_1_4}
\end{figure}

Based on CDRs of about 20 million users in France, Pappalardo et al. \cite{Pappalardo2015} explored the relations between human mobility patterns and socioeconomic development. For each individual, they extracted a measure of mobility volume and a measure of mobility diversity from the CDRs. For individual $i$, the mobility volume (MV) is defined by the radius of gyration $r_{g}(i)$, and the mobility diversity (MD) \cite{Song2010} is measured by the Shannon entropy,
\begin{equation}
MD(i) = - \frac{\sum_{e\in E} p(e) \log p(e)}{\log N},
\end{equation}
where $e = (a,b)$ stands for a trip from the origin base station $a$ to the destination base station $b$, $E$ is the set of origin-destination pairs with size $N$, and $p(e)$ is the probability of a movement along $e$. After aggregating the measures at the municipality level, they found that MD exhibits the strongest correlation with socioeconomic indicators (Pearson coefficient $r=0.49$ for per capita income, $r=0.49$ for primary education rate, $r=-0.43$ for deprivation index, and $r=-0.17$ for unemployment rate), showing that individuals living in developed municipalities have a high mobility diversity. Moreover, there is a negative correlation ($r=-0.38$) between MD and MV, suggesting that people living in less developed municipalities have to travel in search of activities out of their municipalities. Pappalardo et al. \cite{Pappalardo2016} further added two measures into consideration, namely, the social volume (SV) \cite{Onnela2007} and the social diversity (SD) \cite{Eagle2010}. They found that MD exhibits a higher correlation with socioeconomic indicators than SV, SD and MV (see Figure~\ref{Fig_5_1_4}). Moreover, MD adds a predictive power to models that use social and mobility measures in the prediction of socioeconomic indicators.

Similarly, Florez et al. \cite{Florez2018} extracted urban mobility patterns from CDRs covering 1.5 million users and studied many characteristics of the commuting network. They found a significantly positive correlation ($r=0.47$) between the income rank of a group and its commuting trip diversity \cite{Pappalardo2016}. Moreover, poor people travel longer to work and spend more time in commuting. Yang et al. \cite{Yang2018} derived six mobility indicators from large-scale MP data and proposed a data fusion approach to approximate the aggregated socioeconomic status of MP users. They found that richer groups tend to travel longer in Boston but shorter in Singapore, however, different socioeconomic classes in both cities exhibit similar diversity of individual travel and activity patterns. Hong et al. \cite{Hong2016} introduced a LDA-based topic modeling framework \cite{Blei2003} to estimate socioeconomic levels based on MP data. They employed LDA to extract latent recurring patterns of population behaviors (topics) from individual behaviors (words) tracked by MPs. The spatio-temporal MP data are used to model individual mobility across regions, and the latent features are used to predict socioeconomic levels under a supervised approach (PMBSEL-sLDA) and an unsupervised approach (PMB-LDA). In predicting regional socioeconomic labels, both approaches exhibit good accuracy ($R^2=0.7802$ for PMBSEL-sLDA and $R^2=0.7188$ for PMB-LDA) and outperform traditional approaches using pre-determined features by about 9\% in the best case.

Data recorded by financial systems and smart cards have been used to estimate individual financial status. Singh et al. \cite{Singh2015} analyzed a dataset of economic transactions and found an intricate connection between individual financial outcomes and their spatio-temporal traits such as exploration and engagement. Models including such features improve the comparable demographic models by 30\% to 49\% in predicting future financial difficulties. Lenormand et al. \cite{Lenormand2015} analyzed geotagged credit-card transactions of individuals living in Barcelona and Madrid. They found that younger and older people exhibit differences in traveled distance and purpose of travel, showing the effects of demographic characteristics on mobility patterns. Recently, Zhu et al. \cite{Zhu2018} inferred economic attributes of urban rail transit passengers from their mobility patterns and personal attributes. They proposed a mobility-to-attribute framework that integrates smart card data (for extracting individual mobility patterns), house prices and shop consumer prices (for estimating economic status). They found that passengers' income is negatively correlated with their commuting distance and transit frequency, suggesting that the daily trip of low-income passengers depends strongly on the metro. In a word, it is a promising method to estimate individual socioeconomic status from financial transactions and transportation modes.

\subsection{Employment and performance}

Employment and performance are of high relevance to national prosperity, and resignation may result in great losses for companies. Previous studies based on survey data have found that employees' job satisfaction and organizational commitment are predictive to their turnover \cite{Shahnawaz2009}, and job performance is curvilinearly related to turnover \cite{Sturman2012}. A variety of linear and nonlinear models have been employed to predict unemployment rate based on statistics and jobless claims. Recently, new methods have been proposed to analyze unemployment by utilizing novel data such as Internet search queries, mobile phone (MP) records, and social media (SM) posts. In addition to unemployment, individual and group performances are also of particular interest. As the availability of new data and methods, our understanding of performance has been remarkably improved \cite{Barabasi2018}. In this subsection, we will briefly introduce recent progresses on unemployment prediction and performance analysis.

\subsubsection{Search queries indicate unemployment}

The Internet provides rich information about people's wants, needs and concerns on a continual basis, and thus it is possible to mine employment-related information and estimate unemployment rate from Internet search queries. Literature have found that some unemployed individuals use the Internet to seek jobs, and human behaviors reflected by search queries are predictive to unemployment data. Ettredge et al. \cite{Ettredge2005} analyzed employment-related Internet search queries recorded by the WordTracker metasearch engines and unemployment data collected from the US Bureau of Labor Statistics. They found that job search queries are positively correlated with the official unemployment data, and the explanatory power of the used regression model decreases with the increase of the lead time. However, the correlation between search queries and unemployment rate holds only for males of age 20 and over. Their work demonstrates the limitation of search queries in predicting unemployment rate.

Based on data from the Google Insights (GI), Askitas and Zimmermann \cite{Askitas2009} explored the utilization of search queries to predict unemployment rate in Germany. They developed a time-series causality approach to regress monthly unemployment rate against search activity. They found that search queries exhibit strong correlations with unemployment rate, showing that search queries are helpful for economic behavior prediction even under complicated and varying situations. Similarly, Choi and Varian \cite{Choi2009} explored how Google search queries of unemployment-related topics, classified by GI and GT \cite{Choi2012}, are related with the initial jobless claims, a widely accepted indicator of unemployment rate in the US. They found that a GT-based long-term model can help predict initial claims seven days ahead of the official release. D'Amuri \cite{DAmuri2009} leveraged job search queries to predict the quarterly unemployment rate in Italy. The unemployment-related queries are based on GI, and the official unemployment rates are collected from the Italian Labor Force Survey. They found that 1\% increase in GI is associated with 0.44\% increase in the unemployment rate, and GI-included models perform fairly well.

To comprehensively test GI's predictive power for the US unemployment rate, D'Amuri and Marcucci \cite{DAmuri2017} analyzed over five hundred time-series forecasting models on an out-of-sample forecasting task. They found that models using GI as a leading indicator significantly outperform traditional ones. In particular, GI-augmented models exhibit the mean squared error (MSE) 29\% lower when forecasting at one step ahead, and the MSE decreases by 40\% when forecasting at three steps ahead. Using unemployment-related Google search queries, Xu et al. \cite{Xu2012} proposed a neural-network-based forecasting method for unemployment rate with features mined by several feature selection algorithms. Their method exhibits a higher accuracy than other benchmark methods. Xu et al. \cite{Xu2013} further developed a framework to forecast unemployment trend using data mining tools. They extracted employment-related activities from search queries and reduced the data dimension by employing feature selection models. Then, they used neural networks and SVRs \cite{Cortes1995,Smola2004} to model the relations between search activities and unemployment rate and selected the predictive data mining method with the best feature subset. Their method exhibits an outstanding performance in predicting unemployment rate.

\begin{figure}[t]
  \centering
  \includegraphics[width=0.55\textwidth]{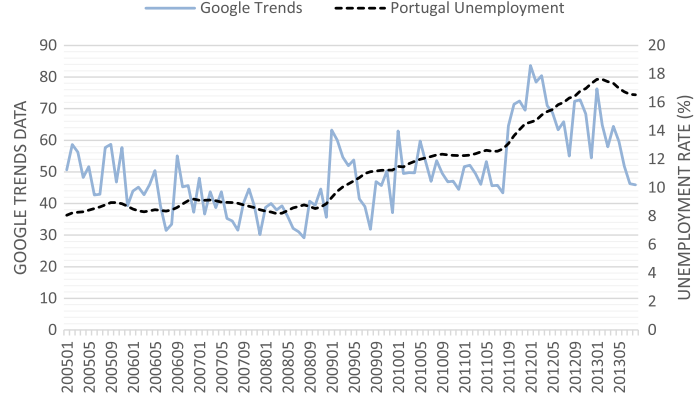}
  \caption{Official unemployment rate versus Google Trends data in Portugal. Figure from \cite{Barreira2014}.}
  \label{Fig_5_2_1}
\end{figure}

Barreira et al. \cite{Barreira2014} explored the improvement of the unemployment nowcasting ability based on search queries from four countries in the south-western Europe. They found that the predictive ability of search queries differ by country and language. Data of GT can improve the nowcasting performance in three out of four considered countries (see Figure~\ref{Fig_5_2_1} for results of Portugal), but the predictive ability is different after considering different out-of-sample periods. In other words, the out-of-sample predictive ability can be improved by GT variables when they have significant in-sample differences. Li et al. \cite{Li2014b} proposed an ontology-based Web mining method to better predict unemployment rate based on Google search queries. The domain ontology captures unemployment-related concepts and their semantic relationships and thus contributes to the extraction of useful features. Their method outperforms some baseline methods such as the ARIMA model \cite{Montgomery1998}. Using a time-series approach, Vicente et al. \cite{Vicente2015} found that Google search queries remain useful in predicting unemployment rate when job destruction is skyrocketing such as the sharp increases in Spanish unemployment due to the economic crisis.

Job-related Google search queries are also predictive to unemployment rate in small and emerging economies. For the Visegrad Group countries, Pavlicek et al. \cite{Pavlicek2015} explored the relationship between the intensity of job-related Google search queries and the unemployment rate. Specifically, they used the linear regression model given by
\begin{equation}
\Delta \text{UR}_t = \alpha_0 + \alpha_1 \Delta \log(\text{GI})_t + \varepsilon_t ,
\end{equation}
where $\Delta \text{UR}_t$ is the first difference of the unemployment rate, $\Delta \log(\text{GI})_t$ is the first logarithmic difference of the search queries, $\alpha_i$ (i=0,1) is the regression coefficient, and $\varepsilon_t$ is the error term at time $t$. They found that job-related search queries can track changes of unemployment rate. Further, they proposed a nowcasting model for unemployment rate based on the job-related search queries, as
\begin{equation}
\Delta \text{UR}_t = \beta_0 + \sum_{i=1}^{L}\beta_i \Delta \log(\text{UR})_{t-i} + \sum_{j=0}^{L}\gamma_1 \Delta \log(\text{GI})_{t-j} + \varepsilon_t ,
\end{equation}
where a three-month lag ($L=3$) in the unemployment rate is assumed available (see Ref. \cite{Pavlicek2015} for details). Similar methods have been used to predict unemployment rates in developed and emerging economies such as Italy \cite{Falorsi2015} and Turkey \cite{Chadwick2015}. The best model augmented with Google search queries performs more accurate (38.4\% out-of-sample and 47.7\% in-sample) than the benchmark autoregressive model in predicting monthly unemployment rates.

Web search queries have also been used to forecast youth unemployment rates. Fondeur and Karam\'e \cite{Fondeur2013} analyzed unemployment-related Google search queries and official claimant count data in France. They proposed a statistical model that considers the non-stationarity and multiple frequencies in the data and estimated it with the diffuse Kalman filter \cite{Durbin2001}. They found that search queries can improve unemployment predictions for the 15- to 24-year old French unemployed population. By integrating search queries into the ARIMA model, Kwon and Jung \cite{Kwon2016} developed a model to predict the Korean youth unemployment rate. Their linear regression model is given by
\begin{equation}
\log(X_t) = \alpha_0 + \alpha_1 \log(X_{t-1}) + \alpha_2 \log(X_{t-12}) + \sum_{k=1}^n \beta_k \log(q_{t}^{k}) + \varepsilon_t ,
\end{equation}
where $\log(X_t)$ is the logarithm of the youth employment rate, $q_{t}^{k}$ is the value of the search volume index, $\alpha_i$ (i=0,1,2) is the regression coefficient, $\beta_k$ is the coefficient of the query value, and $\varepsilon_t$ is the error term at time $t$. The model exhibits an explanatory power of about 80\% for the Korean youth unemployment rate. Naccarato et al. \cite{Naccarato2018} predicted the Italian youth unemployment rate based on Google search queries and official labor data. They employed two time-series models, namely, the ARIMA model using the labor data and the vector autoregressive (VAR) model \cite{Lutkepohl2006} using both data. They found that VAR outperforms ARIMA in the forecasting error.

Recently, some novel employment-related indicators are proposed for labor markets. Baker and Fradkin \cite{Baker2017} developed the Google Job Search Index (GJSI), which is a job search activity index based on Google search queries. GJSI is a useful complement to existing measures as it is available in real time, it has a higher geotemporal resolution, and it suffers less from sampling bias. In addition, GJSI correlates with Google search queries for ``jobs'', displays the holiday effects and reveals the search intensities of different groups. Together with the introduction of search queries, large-scale administrative records can also provide new opportunities to understand unemployment. For example, Guerrero and Lopez \cite{Guerrero2017} developed a data-driven model of unemployment dynamics by taking the advantages of fine-grained details of administrative data.

\subsubsection{Other sources relevant to unemployment}

The availability of large-scale social media (SM) data enables us to better explore unemployment. Based on a large Twitter dataset, Antenucci et al. \cite{Antenucci2014} created a SM signal of job loss, which tracks the initial unemployment insurance claims at medium and high frequencies. They constructed a real-time SM job loss index from the principal components of the collected unemployment-related phrases like ``lost my job'' in tweets. Results demonstrate the usefulness of SM in constructing indicators of economic activity, more specifically, unemployment status. Proserpio et al. \cite{Proserpio2016} analyzed over 1.2 billion Twitter posts collected from US users during 2010 and 2015. To identify users that gained a new job or lost a job, they searched for tweets containing relevant text strings (e.g., ``started my new job'' and ``I lost my job''). In particular, they explored the relations between psychological well-being and the macroeconomic shock of employment instability. They found that SM is able to capture and track changes in psychological variables over time. They proposed a behavioral model that leverages these changes to predict the US unemployment rate. Their model improves the prediction accuracy by about 25\% and 49\% compared to the baseline autoregressive model for employed and unemployed samples, respectively.

\begin{figure}[t]
  \centering
  \includegraphics[width=0.7\textwidth]{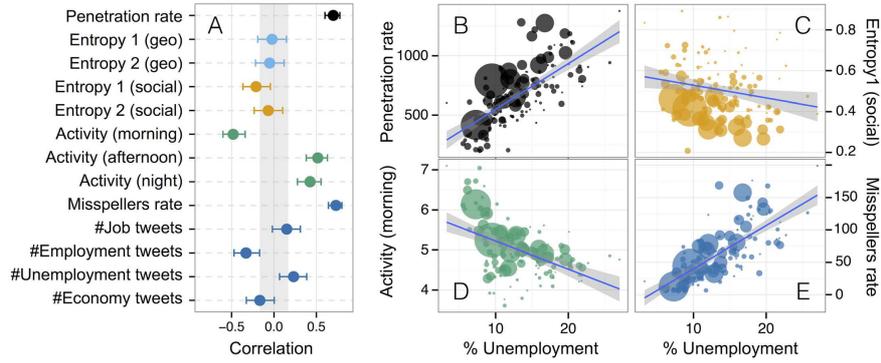}
  \caption{The extracted Twitter metrics and the unemployment rate. (A) Correlation coefficients between behavioral patterns and unemployment rates. Error bars correspond to 95\% confidence intervals. (B-E) Relations between the values of 4 selected variables and the unemployment rate in each geographical communities. Size of the points is proportional to the population in each geographical community. Solid lines correspond to linear fits to the data. Figure from \cite{Llorente2015}.}
  \label{Fig_5_2_2}
\end{figure}

Based on geotagged tweets collected from Spain, Llorente et al. \cite{Llorente2015} investigated whether the unemployment incidence information can be revealed from behavioral patterns underlying human mobility, activity and communication. They constructed individual behavioral features by defining four sets of metrics based on tweets (see Figure~\ref{Fig_5_2_2}). For SM technology adoption, they found a strong and positive correlation between unemployment and the Twitter penetration rate defined by the fraction of Twitter users in the national census. For the SM activity, the correlation between unemployment and the percentage of tweets is strongly negative in the morning while positive in the afternoon. For the SM content, there is a positive and strong correlation between unemployment and the fraction of misspellers. For the diversity of SM interactions defined by entropy \cite{Eagle2010}, areas with less diverse communication patterns have larger unemployment rates. Further, they employed a simple linear regression model to predict regional unemployment rates using these variables. The model exhibits a strong predictive power $R^2=0.62$ for ages below 25 and $R^2=0.52$ for ages between 25 and 44.

Digital exhaust of human activities left on SM platforms can provide important insights to models of employment-related indicators. Bok{\'a}nyi et al. \cite{Bokanyi2017} studied how unemployment and employment statistics of counties in the US are encoded in the daily rhythm of people. They collected 63 million geotagged tweets posted between January and October 2014 from the contiguous US and aggregated them to form a workday tweeting activity pattern with hourly resolution for each county. They found that hourly activities during the daytime (6 am - 8 pm) correlate negatively with unemployment and correlate positively with employment. These results are in accordance with previous findings for Spain that higher morning tweeting activities indicate lower unemployment rates \cite{Llorente2015}. Further, they decomposed the tweeting pattern of a county into a linear combination of two universal patterns: one group with regular working hours, and the other group who wake up later and stay up until late in the evening. Formally, the predicted activity $x_{i}^{(k)}$ of county $k$ in hour $i$ is given by
\begin{equation}
x_{i}^{(k)} = \alpha^{(k)} A_i + \left( 1-\alpha^{(k)} \right) B_i,
\end{equation}
where $\alpha^{(k)}$ and $1-\alpha^{(k)}$ are the mixing proportions for the two universal patterns $A$ and $B$, respectively. Bokanyi et al. \cite{Bokanyi2017} found that the mixing ratio defines a country-specific measure that correlates significantly with unemployment ($r = -0.34\pm0.02$) and employment ($r = 0.46\pm0.02$). The result demonstrates that daily rhythms of tweets exhibit predictive power for whether individuals have regular working lifestyles.

Turnover of employees may have network effects on the attitudes of stayers. Krackhardt and Porter \cite{Krackhardt1986} analyzed a communication network questionnaire and found the snowball effect that turnover occurs in clusters. On the other hand, the structural information of social networks is predictive to individual employment status. Feeley and Barnett \cite{Feeley1997} studied three social network models of employee turnover and found evidences supporting the Erosion model \cite{Feeley1997} that employees located on the periphery of a social network are more likely to leave their position. Mossholder et al. \cite{Mossholder2005} analyzed a sample of health care employees and found that network centrality is predictive to turnover. Based on survey data, Feeley \cite{Feeley2000} found that the employees with large degrees and betweennesses are less likely to leave their jobs. Feeley et al. \cite{Feeley2010} proposed an improved Erosion model to explain the observed negative correlation between the network centrality and the probability of employee turnover.

\begin{figure}[t]
  \centering
  \includegraphics[width=0.6\textwidth]{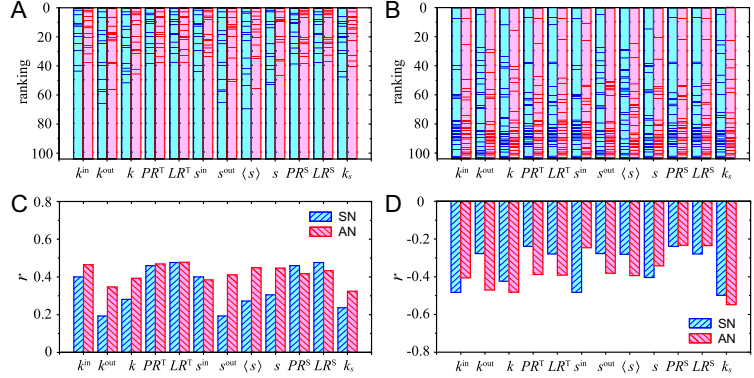}
  \caption{The relations between network centrality measures and employees' career development. (A) and (B) utilize horizontal lines to show the ranking of promoted and resigned employees by centrality measures, respectively. (C) and (D) present the Pearson correlations $r$ between centrality measures and the probability of promotion and resignation, respectively. The centrality of an individual is measured by in-degree ($k^{\text{in}}$), out-degree ($k^{\text{out}}$), degree ($k$), PageRank index ($PR^T$) and LeaderRank index ($LR^T$) in unweighted networks and the metrics in-strength ($s^{\text{in}}$), out-strength ($s^{\text{out}}$), strength ($s$), average strength ($\langle s \rangle $), PageRank index ($PR^S$) and LeaderRank index ($LR^S$) in weighted networks. The k-core index ($k_s$) \cite{Kitsak2010} is applied in unweighted networks. Figure from \cite{Gao2014}.}
  \label{Fig_5_2_3}
\end{figure}

Recent availability of large-scale and reliable network data has made it possible to better infer individuals' employment intentions from their centralities in social networks. Based on the social network data collected from a platform used by a Chinese company, Gao et al. \cite{Gao2014} built two directed networks: social network (SN) and action network (AN), where links indicate social connections and work-related interactions among employees, respectively. After linking network features to human resource data, they found that the most negatively correlated features with turnover are degree ($k$), out-degree ($k^{\text{out}}$) and k-core value ($k_s$) \cite{Dorogovtsev2006} in AN as well as in-degree ($k^{\text{in}}$), in-strength ($s^{\text{in}}$) and k-core value ($k_s$) in SN (see Figure~\ref{Fig_5_2_3}). Moreover, the employees with high in-degrees ($k^{\text{in}}$) are likely to be promoted, and the strongly and positively correlated features with promotion are PageRank index ($PR$) \cite{Brin1998} and LeaderRank index ($LR$) \cite{Lu2011}. Based on the same data, Yuan et al. \cite{Yuan2016} further studied the predictability of employee career development. They employed a binary logistic regression model to predict the promotion or resignation of employees based on the network structural features. In the model, the conditional probability of promotion or resignation is given by
\begin{equation}
  P(1|\vec{x})=\frac{1}{1+e^{-(b_0+\sum_{i}^{m}b_ix_i)}} ,
\end{equation}
where $\vec{x}=(x_1, \ldots ,x_m)$ is the vector of structural features, and $\{b_0,b_1, \ldots ,b_m\}$ are the coefficients estimated based on the data. After predicting employee career development using the model, they found that features of AN have stronger predictive power for both turnover and promotion than features of SN.

Mobile phone (MP) data have been used to analyze and predict unemployment. Based on call detail records (CDRs) in two European countries, Toole et al. \cite{Toole2015} developed a methodology to track employment shocks in nearly real time. They proposed a structural break model to detect mass layoffs based on the drop of calling activity near the plant. The laid-off users are identified by calculating the Bayesian probability weights based on the observed changes of calling patterns. Formally, user $i$'s probability of laid off is given by
\begin{equation}
  P(\text{laid off})_i=\frac{\gamma P(\Delta \hat{q}| \Delta q=d)}{\gamma P(\Delta \hat{q}| \Delta q=d) + (1-\gamma)\gamma P(\Delta \hat{q}| \Delta q=0)} ,
\end{equation}
where $\Delta q = q_{\text{pre}} - q_{\text{post}}$ is the difference in the fraction of days on which a user made a call near the plant in 50 days prior to the layoff, $\gamma$ is the prior that an individual is a non-resident worker at the plant, and $d$ is the threshold used for the alternative hypothesis. The method correctly identifies the portion of laid-off users. The social interactions of laid-off individuals are less stable and experience significant decline. For example, the total number of calls drops 51\%, and the number of outgoing calls drops 54\%. The mobility of laid-off individuals generally declines. For example, the number of uniquely visited towers decreases by 17\%, and the radius of gyration \cite{Frias2012b} decreases by 20\% relative to the random sample. Further, Toole et al. \cite{Toole2015} employed regression models to predict the province-level unemployment rate. Their predictions exhibit high correlations with present unemployment rate ($r=0.95$) and unemployment rate ($r=0.85$) one-quarter in the future.

The effectiveness of MP data in predicting individual employment status can be confirmed by external validations. Sunds{\o}y et al. \cite{Sundsoy2017} derived a set of features from MP logs in a South-Asian developing country, which reflects users' social, financial and mobility patterns. They employed several machine learning algorithms to predict individual employment status of different profession groups using these features. They found that individual employment status can be predicted with an average accuracy of 0.675 for all profession groups, but the accuracy varies for different groups. The model can also predict whether phone users are unemployed with an accuracy up to 0.735, which is over 30 times better than the random guess. Moreover, they showed that individual employment can be aggregated and mapped geographically to the cell tower level. Almaatouq et al. \cite{Almaatouq2016} studied whether district-level unemployment rate can be predicted by behavioral features extracted from CDRs in Riyadh. They found that district-level unemployment rate exhibits strong correlation with behavioral features, specifically, $r = 0.53$ for the number of MP records, $r = 0.49$ for the percentage of night calls and $r = -0.40$ for the social diversity \cite{Eagle2010}. These results indicate that the unemployment rate can be well estimated based on massive MP data in a cost-effective way.

\subsubsection{Individual and group performance}

Performance of individuals and groups is one of the central concerns of organizations in human resource management. Individual performance is loosely related to the frequently used tokens of success as it is a measure capturing a performer's actions \cite{Yucesoy2016}. The quantification of performance remains challenging as it is usually hard to track and analyze individual actions due to the lack of high-quality data. On the other hand, the structure of social and collaboration networks can affect an individual's performance. Through a field study, Sparrowe et al. \cite{Sparrowe2001} found that centrality in advice networks is positively correlated with individual performance, while the density of hindrance network is negatively correlated with group performance. Ahuja et al. \cite{Ahuja2003} examined the determinants of individual performance in virtual R\&D groups. They found that network centrality has a stronger predictive power for individual performance than individual characteristics. After analyzing data of surveys representing virtual 35 teams, Kirkman et al. \cite{Kirkman2004} found a positive correlation between team empowerment and performance, while face-to-face interaction can moderate their relationship.

The network effects on performance have been revealed by traditional survey and small-scale data in different contexts. Cross and Cummings \cite{Cross2004} built information and awareness networks based on two e-mail surveys. They found that network structures are associated with individual job performance in knowledge-intensive work. In particular, betweenness centrality in both networks is predictive of individual performance. Duch et al. \cite{Duch2010} analyzed the performance of soccer players and found that flow centrality is a powerful quantification of individual and team performance. They developed a method of social network analysis to quantify individual performance in the context of soccer. In two studies involving 699 people, Woolley et al. \cite{Woolley2010b} showed that a group's performance can be explained by the collective intelligence factor, which is not determined by individual intelligence but correlated with the proportion of females in the group ($r=0.23$), the equality in distribution of conversational turn-taking ($r=-0.41$) and the average social sensitivity of group members ($r=0.26$). Bear and Woolley \cite{Bear2011} explored the role of gender diversity in group performance and found that the presence of female can improve group collaboration and practical consequences.

After analyzing social network data collected through questionnaire in China, Cai et al. \cite{Cai2014} found that the structure of employees' informal network other than formal network has a significant impact on their performance. In particular, a brokerage's performance is greater affected by direct contacts than indirect contacts. Taking into account the multiplex structure of employee social networks, Cai et al. \cite{Cai2018} showed that a nuanced multiplex network model can provide a richer explanation of employee performance than a single-layer model. They built a superimposed multiplex network (SMN) and an unfolded multiplex network (UMN) based on five different categories of employee relationships. They found that different types of social relations have different effects, where employees with high degrees and large eigenvector centralities in the weighted UMN are more likely to perform well. Not only network structure but also team size is predictive to performance. Through an online experiment, Mao et al. \cite{Mao2016} found that larger teams have better performance than an equivalent number of independent workers in completing complex tasks, while team members exert lower overall efforts than independent workers.

Recent works have focused on the interactions between group members using novel data collected by digital devices. Pentland \cite{Pentland2012} analyzed data of many project/industry teams and found that a team's communication pattern is the most important predictor of its performance. They suggested energy, engagement and exploration as three key communication dynamics that affect team performance after analyzing data collected by wearable electronic sensors (see Ref. \cite{Olguin2009} for details). Moreover, there exists an ideal pattern for each team, and thus a team's performance can be improved by adjusting its communication behavior towards the ideal. Watanabe et al. \cite{Watanabe2012} analyzed data collected by sociometric badges in a call center environment and found that team performance is correlated with the activity level while resting other than the activity level while working. The result suggests a way to improve team performance by enhancing members' face-to-face communications. By analyzing similar data of call centers in China, Tjosvold et al. \cite{Tjosvold2014} found that individuals in cooperative teams have higher productivity. The number of phones answered by members of cooperative teams increases about 40\% compared to control ones.

\begin{figure}[t]
  \centering
  \includegraphics[width=0.45\textwidth]{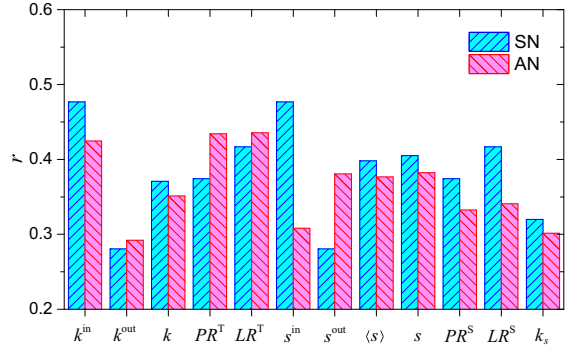}
  \caption{The Pearson correlations between employees' performance and their centralities in the social network (SN) and action network (AN). The centrality of an individual is measured by in-degree ($k^{\text{in}}$), out-degree ($k^{\text{out}}$), degree ($k$), PageRank index ($PR^T$) and LeaderRank index ($LR^T$) in unweighted networks and the metrics in-strength ($s^{\text{in}}$), out-strength ($s^{\text{out}}$), strength ($s$), average strength ($\langle s \rangle $), PageRank index ($PR^S$) and LeaderRank index ($LR^S$) in weighted networks. The k-core index ($k_s$) \cite{Kitsak2010} is applied in unweighted networks. Figure from \cite{Gao2014}.}
  \label{Fig_5_2_4}
\end{figure}

Social network data have been used to examine the relations between employees' performance and their social network structure. Gao et al. \cite{Gao2014} built two networks, namely, social network (SN) and action network (AN), based on data from a social network platform used by a Chinese company (see also Ref. \cite{Yuan2016}). They found that centralities of employees in SN, on average, have stronger correlations with performance than those in AN (see Figure~\ref{Fig_5_2_4}). The most correlated metrics to employees' performance in SN are in-degree and weighted in-degree with the Pearson correlation $r \approx 0.48$. By comparison, the most correlated metrics in AN are in-degree, PageRank and LeaderRank with $r \approx 0.42$. De Montjoye et al. \cite{Montjoye2014} explored how network structure affects teams' problem solving abilities in real working environment. They found that only the strongest ties in within-team networks and the members' extended information networks affect the performance of teams.

The analysis and evaluation of researchers' performance is an important part of scientometrics. Recently, an interdisciplinary field named ``science of science'' (SciSci) \cite{Zeng2017,Fortunato2018} emerges, aiming to quantify, understand and predict scientific discoveries and the resulting outcomes of individuals and groups. After analyzing millions of papers and patents, Wuchty et al. \cite{Wuchty2007} found the increasing dominance of teams in the production of knowledge. In particular, teams produce more frequently cited and high-impact outcomes than individuals do. Jones et al. \cite{Jones2008b} further found that cross-university collaborations overall improve paper quality. Collaborations involving top-tier universities produce the highest-quality papers, while weak-weak combinations produce even worse papers than independent researches. De Stefano et al. \cite{Stefano2013} examined three co-authorship networks of Italian academic statisticians. They found that centrality measures are positively correlated with scientific performance, while local clustering coefficient has a negative influence. Lungeanu et al. \cite{Lungeanu2014} analyzed grant proposals and found that successful teams tend to have members with longer tenure, lower institutional tier, more female gender, less prior citation relationships, and so on.

Toward an objective measure of individual scientific performance, many metrics have been considered such as total paper count, citations per paper, impact factors of published journals, and so on. One metric that enjoys a spectacularly quick success is h-index \cite{Hirsch2005}, which is defined as the largest integer $h$ such that there are at least $h$ papers with $\ge h$ citations each. Hirsch \cite{Hirsch2007} showed that h-index outperforms other indicators in predicting future scientific achievement of individuals. Later, Radicchi et al. \cite{Radicchi2008} found the universality of citation distributions across disciplines indicated by a universal curve of the relative indicator $c_f=c/c_0$, where $c$ is the citation counts, and $c_0$ is the average citation per article for discipline $f$. As $c_f$ is an unbiased indicator for citation performance, they introduced a generalized h-index named $h_f$ index (see Ref. \cite{Radicchi2008} for details) that is suitable for comparing scientists across disciplines. Abbasi et al. \cite{Abbasi2010} proposed the researcher collaboration index (RC-Index) and the community collaboration index (CC-Index) to identify researchers who may be suitable to lead research projects. Till far, there are many variants of h-index to evaluate the performance of scientists (see review article \cite{Bornmann2011}).

Wang et al. \cite{Wang2013} developed a mechanistic model to predict a paper's ultimate impact by a single parameter inferred from its early citation history. The unfolded universal temporal pattern in such citation dynamics of papers can be used to evaluate scientific impacts of individuals. Sinatra et al. \cite{Sinatra2016} quantified the evolution of individual scientific impact based on publication records and career profiles. They developed a quantitative model of scientific impact and proposed a factor $Q$ to capture a scientist's sustained ability to publish high-impact papers. Interestingly, the factor $Q$ is a fingerprint for scientists and independent to their career stages. The $Q$-model can predict future time evolution of individual scientific impact. After analyzing scientists' career trajectories, Deville et al. \cite{Deville2014b} found that individuals moved from elite to lower-rank institutions tend to experience modest decrease in scientific performance, while movements towards elite institutions do not bring subsequent performance gain. Shen and Barab\'asi \cite{Shen2014} proposed a credit allocation algorithm that can accurately measure the relative credits for different coauthors. Jia et al. \cite{Jia2017} analyzed publication records and found an exponential distribution of changes in research interests. They further developed a random-walk-based model to accurately reproduce the empirical observations.

\begin{figure}[t]
  \centering
  \includegraphics[width=0.75\textwidth]{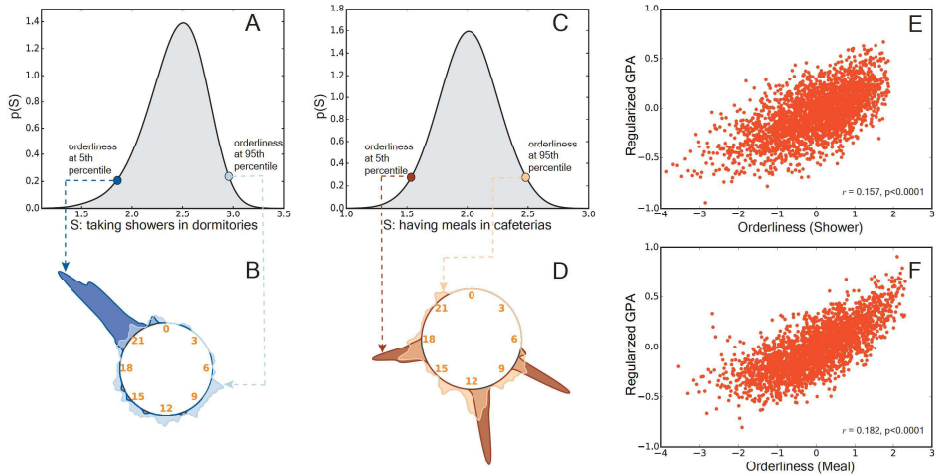}
  \caption{Distributions of actual entropies of students in (A) taking showers and (C) having meals. The behavioral clocks of two students at the 5th percentile and the 95th percentile are shown for (B) taking showers and (D) having meals. Correlations between regularized GPA and (E) regularized orderliness (Shower) as well as (F) regularized orderliness (Meal). The corresponding Spearman's rank correlation coefficients $r$ and the level of statistical significance $p$ are also shown in the plots. Figure after \cite{Cao2018}.}
  \label{Fig_5_2_5}
\end{figure}

Researchers have also tried to connect students' educational achievements with their behavioral patterns. Based on passive sensing data from smart phones, Wang et al. \cite{Wang2015b} explored individual behavioral differences among a group of students with different performance. They found a number of important behavioral factors that are significantly correlated with term and cumulative GPA, such as conversational interaction, class attendance, and studying hours. Using these behavioral features, a linear regression model can predict cumulative GPA with $r=0.81$. Using behavioral records collected by students' smart cards, Cao et al. \cite{Cao2018,Cao2019} quantified the relations between behavioral patterns and academic performance. They introduce orderliness, a novel metric based on the actual entropy \cite{Kontoyiannis1998,Xu2019}, to measure the regularity of each student's campus lifestyle based on the temporal records of having meals and taking showers (see Figure~\ref{Fig_5_2_5}). They found that orderliness exhibits a significantly positive correlation with GPA, and it can remarkably improve the prediction accuracy of students' academic performance at the presence of other behavioral indicators. Similarly, Yao et al. \cite{Yao2019} collected longitudinal behavioral data of 6,597 students through smart cards and proposed three behavioral features, namely, orderliness, diligence, and sleep patterns. They further built a multi-task predictive framework that can well predict student's academic performance by utilizing proposed behavioral features.

\subsection{Demographics and personal variables}

Demographic attributes of individuals have remarkable effects on their socioeconomic status, while traditional methods of individual profiling based on surveys and censuses are costly and follow a long-time delay. Recently, data from novel sources such as social media (SM) and mobile phones (MPs) have been used alternatively to predict individual demographic attributes. Moreover, individual behaviors on social networking platforms have been used to estimate individual mental states such as emotion, depression and suicidal intent. Meanwhile, these novel data sources have also been leveraged to predict personality and evaluate reputation of individuals.

\subsubsection{Demographic inference}

Understanding demographics of individuals has important applications in estimating socioeconomic outcomes. Beyond traditionally used census, data from MPs and Twitter have been used to infer demographic attributes such as gender and age. Using a rich set of features extracted from tweets, Rao et al. \cite{Rao2010} developed a stacked-SVM-based classification algorithm to classify latent user attributes, giving an accuracy 0.723 on gender classification. The algorithm outperforms the baseline ngram-only model with accuracy 0.687 and the SVM-based binary classifier with accuracy 0.718 \cite{Garera2009}. Based on a Twitter dataset labeled with gender, Burger et al. \cite{Burger2011} applied some statistical models to predict gender using features of both word- and character-level ngrams. They found that the most informative feature is a user's full name, which provides an accuracy 0.891 in gender classification. Moreover, tweets contribute more than user description in predicting a user's gender. Test classifier with all features extracted from user profile and texts exhibits an accuracy about 0.92.

By calculating gender-name association scores, Liu and Ruths \cite{Liu2013} explored the link between gender and first name in English tweets. They found that including first name can improve the accuracy of gender inference by about 20\%. They further developed a method to identify gender-labels without analyzing user profile or textual content. Ciot et al. \cite{Ciot2013} assessed gender inference methods based on non-English tweets. They found that existing machinery can address the gender inference problem, and including language-specific features can make accuracy gains. Volkova et al. \cite{Volkova2015} applied machine learning and natural language processing techniques to predict personal attributes. They trained log-linear models using lexical features extracted from 200 tweets per user profile and identified male gender with an accuracy 0.8. Culotta et al. \cite{Culotta2015} predicted the demographics of Twitter users based on whom they follow. Specifically, they labeled demographics of visitors to over 1,500 websites. Then, they predicted demographics of Twitter users by a regression model using the information of users following the websites' accounts on Twitter. Montasser and Kifer \cite{Montasser2017} developed a method to predict a region's demographics based on the characteristics of geotagged tweets in that region. Using lexical features extracted from tweets, their method predicts census-based race data at the block level with an average accuracy 0.692.

Content and network features of SM users are predictive of their occupations. Huang et al. \cite{Huang2015c} developed an integration framework to infer users' occupations from their social activities on Weibo. They proposed a content model to identify beneficial content features and used a network model to identify user latent affiliations. Their model that integrates both network and content exhibits an accuracy 0.80 on occupation inference. Preotiuc-Pietro et al. \cite{Preotiuc2015} employed linear and nonlinear models to predict a user's occupational class based on latent features extracted from tweets. They found that text features can improve the performance, and the best model can give an accuracy of over 0.50 for a 9-way occupation classification. Sloan et al. \cite{Sloan2015} developed two methods to extract social class of occupation from the profiles of UK Twitter users. Their methods can identify certain occupational groups such as professionals.

MP data have been used to infer demographic information. Frias-Martinez et al. \cite{Frias2010b} analyzed call detail records (CDRs) and found that male and female users are significantly different in behavioral and social variables such as duration of calls and degree in social networks. They proposed a semi-supervised classification algorithm that can identify gender with an accuracy up to 0.80. From CDRs, Herrera-Yag\"ue and Zufiria \cite{Herrera2012} extracted 22 features that are most relevant to gender. They found that females tend to have a higher average call length, a larger median of call duration and more messages sent per relationship. They tested several machine learning schemes and found that SVM using call features performs the best with an accuracy over 0.60 for gender prediction. After analyzing daily communication patterns extracted from one billion CDRs, Dong et al. \cite{Dong2014} found that female users pay more attention to cross-generation interactions. They proposed a factor graph model name WhoAmI that accounts for the interrelations to infer gender and age. The WhoAmI outperforms benchmark methods by about 10\% in terms of F1 measure. Dong et al. \cite{Dong2017} generalized WhoAmI to infer any number of interrelated attributes and proposed a coupled WhoAmI method to predict demographics across two mobile operators. Their new methods exhibit accuracies up to 0.80 and 0.73 in predicting gender and age, respectively.

\begin{figure}[t]
  \centering
  \includegraphics[width=0.6\textwidth]{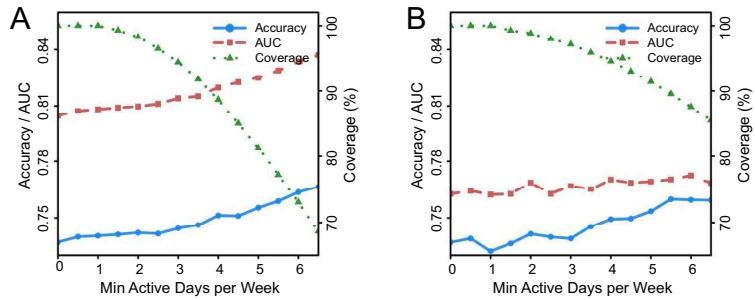}
  \caption{Accuracy and AUC of gender prediction in (A) the European and (B) the south Asian country as a function of the minimum required active days per week. As the minimum required threshold increases, the coverage of the CDR data decreases while the algorithm performance measured by the accuracy and AUC improves slightly. Figure from \cite{Jahani2017b}.}
  \label{Fig_5_3_1}
\end{figure}

After extracting user behavioral and social variables from Mexican CDRs, Sarraute et al. \cite{Sarraute2014} evaluated several machine learning algorithms for gender prediction. They found that the logistic regression and linear SVM perform the best with an accuracy 0.814. Including individual calling patterns and communication network structures, machine learning algorithms can also predict other demographic variables such as age. Jahani et al. \cite{Jahani2017b} developed a framework to predict individual characteristics based on over 1400 features extracted from CDRs. They found that machine learning algorithms trained with only 10,000 users are sufficient to predict gender with an accuracy up to 0.884 in a south Asian developing country and with an accuracy up to 0.797 in an European developed country. The performance can be slightly improved by increasing the minimum required days that users are active per week (see Figure~\ref{Fig_5_3_1}). Moreover, their method can be used to predict other demographic variables such as age with $R^2 = 0.47$ in a multi-classification task.

Based on GPS data collected from MPs, Akter and Holder \cite{Akter2017} proposed a graphical-feature-based framework to improve demographic prediction. They constructed user-wise networks, in which nodes are location categories and edges represent users' movements between location categories. Then, they extracted relevant graphical features from the graph representation and trained a SVM to identify gender. Their method with the optimal set of features exhibits an accuracy 0.8599 in gender classification. Wang et al. \cite{Wang2017b} analyzed data of recorded MP connections to access points (APs) in two campuses and found that users' demographic attributes can be solely inferred from their spatiotemporal AP-trajectories. They developed a method named Sinfer to infer the social network of users based on the co-occurrence events of AP-trajectories. Further, they proposed a tensor-factorization-based method named Dinfer to predict users' demographic attributes using the social network learned by Sinfer. Their method gives an F1 measure about 0.7 in gender prediction. Felbo et al. \cite{Felbo2017} developed the convolutional network (ConvNet) architecture to transform MP data into high-level features for each week and then aggregated patterns across weeks by reusing the same convolutional filters. They designed a 2-step model (ConvNet-SVM) using an SVM with a radial basis function kernel, which slightly outperforms the state-of-the-art method, with accuracies 0.797 and 0.631 in predicting gender and age, respectively.

Digital traces on other platforms have also been used to infer demographic information. Kosinski et al. \cite{Kosinski2013} analyzed a dataset of Facebook Likes and found that users' online behaviors are predictive of their personal attributes such as ethnicity, age and gender. Zhong et al. \cite{Zhong2015} analyzed profiles of 159,530 SM users and found that a variety of demographic attributes can be inferred from check-ins, including education background, marital status, gender, and age. They proposed a location-to-profile framework that outperforms benchmark methods in inferring user profile. Using data of Facebook and Pokec social networks, Lin et al. \cite{Lin2016} proposed an algorithm that can predict user profile with a high accuracy. Recently, Ren et al. \cite{Ren2018} studied demographic prediction using different types of data including cyber, physical and social behaviors. They found that including cyber-physical-social behaviors into an SVM model can significantly increase the accuracy for gender prediction. Readers are encouraged to read a recent review about demographic attribute prediction based on online digital traces \cite{Hinds2018}.

\subsubsection{Personality analysis}

Personality is usually quantified by the five-factor model (FFM) \cite{Mccrae1992,Digman2003}, which suggests five broad dimensions to characterize human personality: conscientiousness, agreeableness, neuroticism, openness, and extraversion. Yet, FFM-based analyses are mainly driven by personality survey data. Ross et al. \cite{Ross2009} explored the association between personality and Facebook usage patterns. They found that neuroticisms prefer to post photos on their profiles, and extraversions tend to report membership in more groups. Correa et al. \cite{Correa2010} investigated the relations between social media (SM) activity and three personality factors (extraversion, neuroticism, and openness). They found that extraversion and openness are positively correlated with SM activity, while neuroticism is negatively correlated with SM activity. Yet, these results are affected by age and gender. For example, extraversion is strongly correlated with SM activity among the young adult cohort.

Golbeck et al. \cite{Golbeck2011} demonstrated that personality of users can be inferred from their Facebook profiles. They found a positive correlation ($r = 0.264$) between conscientiousness and words surrounding social processes. They employed two machine learning algorithms to predict personality factors using 74 features and found that each factor can be predicted within on average 11\% of its actual value. Bachrach et al. \cite{Bachrach2012} examined the relations between personality of users and their Facebook profiles based on data of 180,000 users. They found that openness and neuroticism are positively correlated with the number of likes and group associations of users, while the correlation is negative for conscientiousness. They developed a multivariate linear regression to predict each factor based on multiple profile features. Their model exhibits reasonable prediction accuracy for some factors such as extraversion ($R^2 = 0.33$) and neuroticism ($R^2 = 0.26$). They further applied several more sophisticated machine learning methods but found that the improvement of predication accuracy is very limited.

\begin{figure}[t]
  \centering
  \includegraphics[width=0.6\textwidth]{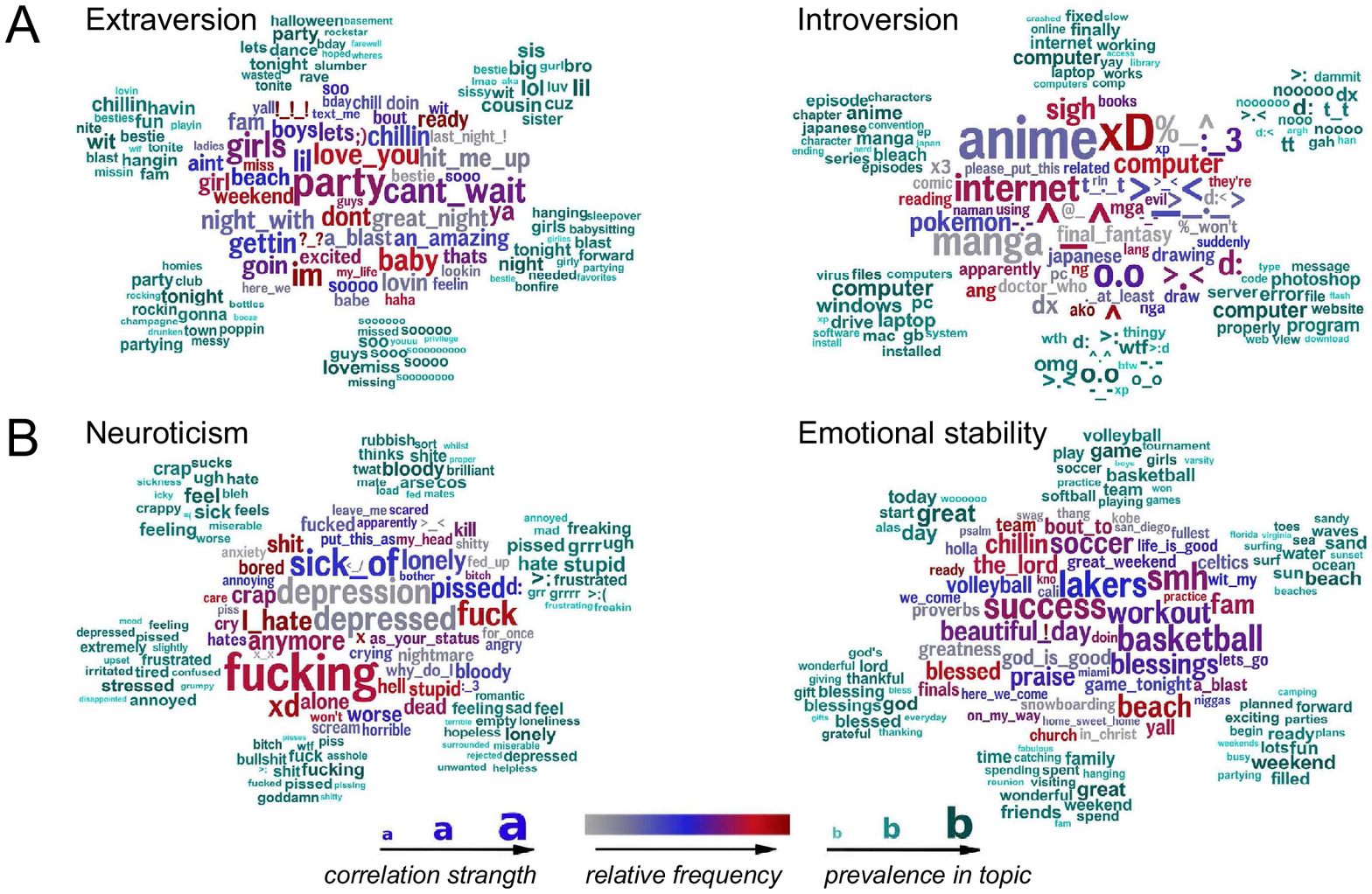}
  \caption{Word clouds for extraversion and neuroticism. (A) Language distinguishing extraversion (left, e.g., ``party'') from introversion (right, e.g., ``computer''). (B) Language distinguishing neuroticism (left, e.g. ``hate'') from emotional stability (right, e.g., ``blessed''). Figure after \cite{Schwartz2013}.}
  \label{Fig_5_3_2}
\end{figure}

After surveying the personality and the Facebook behaviors of 184 undergraduates, Seidman \cite{Seidman2013} found that belongingness-related behaviors can be best predicted by high agreeableness and neuroticism, and more frequent use of Facebook is highly correlated with extraversion. Self-presentational behaviors and motivations can be predicted by low conscientiousness and high neuroticism, while the motivation to express self-aspects can mediate the relationship between neuroticism and self-disclosure. Schwartz et al. \cite{Schwartz2013} analyzed millions of Facebook messages and found that language features can distinguish demographic and psychological attributes (see Figure~\ref{Fig_5_3_2}). They extracted 700 million instances by developing a method of open-vocabulary analysis and then linked them with personality. They found that neurotic people disproportionately use the phrase ``sick of'', while extraverts tend to use social words such as ``party''. Further, they employed the ridge regression \cite{Hoerl2000} to predict each factor using the open-vocabulary features. Their method exhibits fine performance, for example, the accuracy of predicting openness is $R^2 \approx 0.18$.

Data from Twitter have also been used to study individual personality. Preo{\c t}iuc-Pietro et al. \cite{Preotiuc2016} developed a multi-modal approach to quantify observable user behaviors on Twitter using natural language processing and image analysis. After extracting online behavioral features capturing text use, profile picture posting and general profile information, they explored how these features are related to dark traits of human personality \cite{Paulhus2002} including narcissism, machiavellianism and psychopathy. They found that trust and positive sentiments are correlated with narcissism, a wide range of negativity and morbidity are correlated with psychopathy, and allowing advertisements in personal communications is correlated with machiavellianism. Further, they built a linear regression model to predict the dark triad traits with a robust performance of Pearson coefficient $r \approx 0.25$ for all three traits. Similarly, semantic content of Facebook updates has also been used to predict the dark traits \cite{Garcia2014}. Guntuku et al. \cite{Guntuku2017} explored the relations between personality and online image posting based on about 1.5 million images posted on Twitter. They found that image content of posts and likes can predict the big five personality traits. For example, the performance measured by the Pearson correlations are 0.530 and 0.566 in predicting neuroticism and conscientiousness, respectively.

Based on tweets and profile images of over 66,000 Twitter users, Liu et al. \cite{Liu2016d} estimated the big five personality. They used the state-of-the-art text prediction methods developed by Schwartz et al. \cite{Schwartz2013} to predict personality based on tweets and employed deep learning methods (see Ref. \cite{Zhou2013}) to extract facial features from profile pictures. After linking the text-based personality to the facial features, they found that users using appealing profile images are likely to have high openness, profile images with faces are good indicators of higher conscientiousness, and neurotic people have a strong tendency not to present faces. Moreover, these interpretable visual features can be harnessed to predict personality traits with robust accuracy, for example, the Pearson correlation is 0.189 in predicting conscientiousness. Segalin et al. \cite{Segalin2017} analyzed the effectiveness of visual features on the prediction of personality traits based on Facebook profile pictures. They found that agreeable individuals and extroverts tend to use warm colored pictures, while neurotic individuals tend to post pictures of indoor places. Based on these visual features, they developed a classification method to predict personality traits, exhibiting a mean classification accuracy of about 0.60 for each of the five traits.

\subsubsection{Online reputation evaluation}

Reputation has received considerable attention recently in a variety of disciplines \cite{Masum2004}. Mui et al. \cite{Mui2002} provided an overview of reputation studies across disciplines by summarizing existing notions of reputation and discussing their advantages. Among existing notions, individual reputation, referring to the judgment of an individual's impression by others, is a valuable asset in online social lives. For example, Zacharia et al. \cite{Zacharia2000} studies reputation in an on-line community and found that reputation is related to ratings received by an individual from others. So far, there have been many computational reputation models \cite{Sabater2005} and many reputation systems for online services \cite{Resnick2000,Josang2007}. In the following, we will introduce some basic notations for online rating systems and review recent literature that evaluate user reputation.

An online rating system consisting of $m$ users and $n$ objects can be described by a weighed bipartite network $G=\{U, O, E\}$, where $U=\{U_{1}, U_{2}, \ldots, U_{m}\}$, $O=\{O_{1}, O_{2}, \ldots, O_{n}\}$ and $E=\{E_{1}, E_{2}, \ldots, E_{l}\}$ are sets of users, objects and ratings, respectively. Naturally, the bipartite network can be represented by a rating matrix $A$, whose element $A_{i\alpha}$ is the rating given by user $i$ to object $\alpha$. Here, Greek and Latin letters are used for object-related and user-related indices, respectively. Reputation evaluation model aims to assign user $i$ with reputation $R_{i}$ by analyzing the bipartite network $G$. Ranking-based reputation evaluation methods can be roughly classified into two categories, namely, quality-based ranking methods and group-based ranking methods.

The quality-based ranking methods assume that a most objective rating can best reflect true quality $Q_{\alpha}$ of an object $\alpha$. Due to the lack of true quality information, the weighted average rating is used as a proxy. Formally, the estimated object quality $\hat{Q}_{\alpha}$ of object $\alpha$ is given by
\begin{equation}
\label{eq:QIR}
\hat{Q}_{\alpha}= \frac{\sum_{i\in{U_{\alpha}}}{R_{i}A_{i\alpha}}}{\sum_{i\in{U_{\alpha}}}{R_i}},
\end{equation}
where $U_{\alpha}$ is the set of users who have rated object $\alpha$, and $R_{i}$ is the reputation of user $i$. The most straightforward method is the iterative refinement (IR) \cite{Laureti2006}, which calculates user reputation and object quality in an iterative way. The reputation of a user is inversely proportional to the difference between the vectors of the user's ratings and the estimated object qualities, as
\begin{equation}
\label{eq:dIR}
  IR_i = \left(\frac{1}{k_{i}} \sum_{\alpha\in {O_{i}}} (A_{i\alpha} - \hat{Q}_{\alpha})^2 + \varepsilon \right)^{-\beta},
\end{equation}
where $O_{i}$ is the set of objects rated by user $i$, $k_i$ is the degree of user $i$, $\beta$ is a tunable parameter, and $\varepsilon$ is a small quantify to avoid zero value of the summation. User reputation is initialized as $IR_{i}=1/n$ and iteratively updated by Eq.~(\ref{eq:QIR}) and Eq.~(\ref{eq:dIR}) (setting $R_i \leftarrow IR_i$ in Eq.~(\ref{eq:QIR})) until user reputation $IR_i$ and object quality $\hat{Q}_{\alpha}$ converge. Note that, $IR_i$ should be normalized after every iteration.

Zhou et al. \cite{Zhou2011} proposed a correlation-based ranking (CR) method, which is more robust under spamming attacks. In the CR method, reputation of a user is iteratively determined by the correlation between the vectors of the ratings $A$ and the estimated object qualities $\hat{Q}$. For user $i$, a so-called temporal reputation $TR_i$ is calculated by the Pearson correlation, as
\begin{equation}
\label{eq:TRI}
TR_i=\frac{1}{k_{i}}\sum_{\alpha\in{O_i}}{\left(\frac{A_{i\alpha}-\mu(A_i)}{\sigma(A_i)}\right)}{\left(\frac{\hat{Q}_{\alpha}-\mu(\hat{Q}_i)}{\sigma(\hat{Q}_i)}\right)},
\end{equation}
where $\mu(A'_{i})=\sum_{\alpha} A'_{i\alpha} / k_{i}$ and $\sigma(A'_{i})=\sqrt{\sum_{\alpha}(A'_{i\alpha}-\mu(A'_{i}))^2 / k_{i}}$ are the mean value and standard deviation of $A'_{i}$, respectively. The reputation is set as $CR_i = 0$ if $TR_i$ is smaller than 0, and $CR_i = TR_i$ otherwise. In this process, user reputation $CR_i$ and object quality $\hat{Q}_{\alpha}$ are iteratively updated according to Eq.~(\ref{eq:QIR}) and Eq.~(\ref{eq:TRI}) (setting $R_i \leftarrow CR_i$ in Eq.~(\ref{eq:QIR})). To start the iteration, user reputation is initialized as $CR_{i} = k_{i}/n$, where $k_{i}$ is the degree of user $i$.

Under the CR framework, Liao et al. \cite{Liao2014} proposed an iterative algorithm with reputation redistribution (IARR) by enhancing the influences of users with high reputation. The user reputation in IARR is calculated by nonlinearly redistributing the reputation given by CR, as
\begin{equation}
\label{eq:RRI}
IARR_i=CR_i^{\theta} \cdot \frac{\sum_j CR_j}{\sum_j CR_j^{\theta}},
\end{equation}
where $\theta$ is a tunable parameter. Note that, IARR degenerates to CR when $\theta=1$. Liao et al. \cite{Liao2014} further proposed an enhanced iterative algorithm named IARR2 by introducing two penalty factors. Specifically, they modified the calculation of object quality $\hat{Q}$ in Eq.~(\ref{eq:QIR}) by
\begin{equation}
\label{eq:QIR2}
\hat{Q'}_{\alpha}= \max \limits_{i\in{U_{\alpha}}} \{R_{i}\} \cdot \hat{Q}_{\alpha},
\end{equation}
and they modified the calculation of temporal reputation $TR$ in Eq.~(\ref{eq:TRI}) by
\begin{equation}
\label{eq:TRI2}
{TR'}_i = \frac{\log k_i}{\max \limits_{j} \{ \log k_j \}} \cdot TR_i.
\end{equation}
In a word, IARR eliminates the noisy information by reducing the influence of users with low reputation, while IARR2 gradually filters out the influence of less reliable users in the iterations.

Liu et al. \cite{Liu2015b} proposed an improved iterative algorithm (IRUA) to rank user reputation by taking into account the role of users' activity patterns. Specifically, IRUA considers the maximum degree of the users who have rated an object and estimates the quality of object $\alpha$ by
\begin{equation}
\label{eq:QIRUA}
\hat{Q}^{\prime\prime}_{\alpha} = \max \limits_{i\in{U_{\alpha}}} \left \{ \frac{k_i}{n} \right \} \cdot \hat{Q}_{\alpha},
\end{equation}
where $k_i$ is the degree of user $i$, and $n$ is the total number of objects. The reputation of user $i$ is updated by considering both the temporal reputation $TR_i$ calculated by Eq.~(\ref{eq:TRI}) and the user degree $k_i$. The reputation of user $i$ is given by
\begin{equation}
IRUA_i=\left\{
\begin{array}{ll}
 \left( \frac{k_i}{k_{\max}} \right) ^{\theta} \cdot TR_i , & TR_i \ge 0 \\
 0, & TR_i < 0
\end{array}
\right.
\end{equation}
Here, $k_{\max}$ is the maximum degree of all users, and $\theta$ is a tunable parameter that enhances the reputation of large-degree users when $\theta > 0$.

Recently, Liu et al. \cite{Liu2017} proposed a parameter-free reputation ranking method based on the beta probability distribution (RBPD). Firstly, a rating $A_{i\alpha}$ is transformed to the extent of fanciness $A'_{i\alpha}$ by a normalization method: $A'_{i\alpha} = 2(A_{i\alpha}-A_{i}^{\min})/(A_{i}^{\max}-A_{i}^{\min})$ if $A_{i}^{\max} \neq A_{i}^{\min}$, and $A'_{i\alpha}=0$ otherwise, where $A_{i}^{\max}$ and $A_{i}^{\min}$ are respectively the maximum and minimum ratings given by user $i$. Then, the Bayesian analysis is used to determine the number of fair ratings $s$ and unfair ratings $f$ given by user $i$ based on $A'_{i\alpha}$. A rating is fair if it is consistent with the majority of other users' opinions on the corresponding object (see Ref. \cite{Liu2017} for details). Finally, the RBPD method defines the user reputation as the probability of giving fair ratings to objects. Formally, the reputation of user $i$ is given by
\begin{equation}
\label{eq:RBPD}
RBPD_i = \frac{s_i + 1}{k_i + 2},
\end{equation}
where $s_i$ is the number of fair ratings given by user $i$. The RBPD method updates $RBPD_i$ by Eq.~(\ref{eq:RBPD}) and $\hat{Q}_{\alpha}$ by Eq.~(\ref{eq:QIR2}) in an iterative manner until object qualities become stable.

The group-based ranking methods define user reputation by the group sizes after grouping all users by their rating similarities. Instead of following the traditional assumption that each object has only one rating that best reflects its quality, the group-based ranking methods underlie that one object should accept multiple reasonable ratings since the users' ratings are subjective and can be affected by many factors \cite{Tian2012}. For a discrete rating system with $\Omega =\{\omega_{1}, \omega_{2}, \ldots, \omega_{z}\}$, Gao et al. \cite{Gao2015} proposed a group-based ranking (GR) method that involves the following steps: (i) group users according to their ratings; (ii) calculate the group size matrix $\Lambda$, where $\Lambda_{s\alpha}$ is the number of users who rated object $\alpha$ with rating $\omega_{s}$; (iii) build the rating-rewarding matrix $\Lambda^{\ast}_{s\alpha}=\Lambda_{s\alpha}/k_{\alpha}$, where $k_{\alpha}$ is the degree of object $\alpha$; (iv) map the original rating matrix $A$ to the rewarding matrix $A'$, where $A'_{i\alpha} = \Lambda^{\ast}_{s\alpha}$ if $A_{i\alpha}=\omega_{s}$. If $A_{i\alpha}=0$ (i.e., the user $i$ didn't rate the object $\alpha$), the value of $A'_{i\alpha}$ is null and should be ignored in the following calculation; (v) calculate the reputation $GR_i$ of user $i$ by dividing the mean value of $A'_{i}$ by its standard deviation. Formally, $GR_i$ is defined as
\begin{equation}
\label{eq:GR}
    GR_{i}=\frac{\mu(A'_{i})}{\sigma(A'_{i})},
\end{equation}
where $\mu(\cdot)$ and $\sigma(\cdot)$ are mean value and standard deviation, respectively. As presented, GR assigns users with high reputation if they always fall into large rating groups.

\begin{figure}[t]
  \centering
  \includegraphics[width=0.55\textwidth]{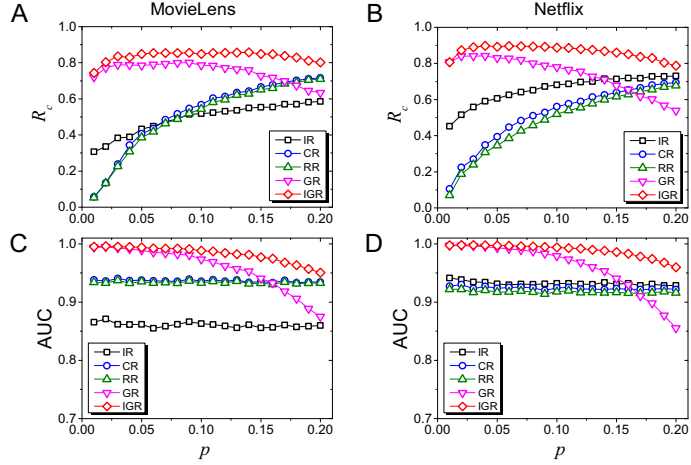}
  \caption{Performance of different reputation ranking methods. All methods are evaluated on the MovieLens (A and C) and Netflix (B and D) datasets, in which $p$ ratio of ground truthing low-reputation users are assigned with the minimum and maximum allowable ratings. Two evaluation metrics are used, namely, recall ($R_c$) and AUC. Figure from \cite{Gao2017d}.}
  \label{Fig_5_3_3}
\end{figure}

Later, Gao and Zhou \cite{Gao2017d} proposed an iterative group-based ranking (IGR) method by introducing an iterative reputation-allocation into the GR method. In IGR, the sizes of user rating groups are weighted by the reputation of users, where ratings from users with higher reputation have larger influences. Formally, the weighted sizes of user rating groups are calculated by
\begin{equation}
\label{eq:Lam}
\Lambda_{s\alpha}=\sum^{m}_{i=1}IGR_{i} \cdot B^{(i)}_{s\alpha},
\end{equation}
where $IGR_{i}$ is the reputation of user $i$ calculated in the previous step, $B^{(i)}$ is the rating-object matrix of user $i$, and $m$ is the total number of users. Here, the rating-object matrix $B^{(i)}_{s\alpha} = 1$ if $A_{i\alpha}=\omega_{s}$, and $B^{(i)}_{s\alpha}=0$ otherwise (see Ref. \cite{Gao2017d} for details). Following GR, the rating-rewarding matrix is calculated by $\Lambda^{\ast}_{s\alpha}=\Lambda_{s\alpha}/k_{\alpha}$, and the rewarding matrix is mapped by $A'_{i\alpha} = \Lambda^{\ast}_{s\alpha}$. The user reputation $IGR_i$ is re-allocated via $IGR_{i}=\mu(A'_{i}) / \sigma(A'_{i})$. IGR iteratively updates the weighted group sizes by Eq.~(\ref{eq:Lam}) and the user reputation by Eq.~(\ref{eq:GR}) until convergence. The reputation of every user $i$ is initialized with the same value ($IGR_i=1$). Experimental results on the MovieLens and Netflix datasets demonstrate the improved accuracy and robustness of IGR in ranking low reputation users (see Figure~\ref{Fig_5_3_3}).

Recently, Dai et al. \cite{Dai2018} proposed a group-based ranking method based on the user preference, named PGR method, which is a variant of the original GR method. PGR is based on the idea that the preferences of online users are diverse when they give ratings to objects. For example, large-degree users tend to give low ratings \cite{Zhang2015}. Different from GR where users are grouped by their ratings, PGR divides users into groups by their rating preferences. First, a user rating $A_{i\alpha}$ is transformed to a mapped rating $A'_{i\alpha}$ by a normalization method: $A'_{i\alpha} = (A_{i\alpha}-\mu(A_{i}))/(A_{i}^{\max}-A_{i}^{\min})$ if $A_{i}^{\max} \neq A_{i}^{\min}$, and $A'_{i\alpha}=0$ otherwise, where $\mu(A_{i})$ is the average rating given by user $i$. The matrix $A'$ measures user preference of giving high or low ratings. Then, a new rating matrix $A^{\prime\prime}$ is constructed by transforming $A'_{i\alpha}$ to new ratings ${A^{\prime\prime}}_{i\alpha}$, where a linear mapping is used to ensure that the sets of rating values for $A^{\prime\prime}$ and $A$ are the same (see Ref. \cite{Dai2018} for details). Finally, the user reputation $PGR$ is calculated under the framework of GR (replacing $A$ by $A^{\prime\prime}$).

Besides these aforementioned quality- and group-based methods, scientists have proposed some other user reputation evaluation methods for online rating systems. For example, Fouss et al. \cite{Fouss2010} proposed a probabilistic reputation model based on transaction ratings. Liao et al. \cite{Liao2012} developed a general ranking method to evaluate user reputation in online communities. Li et al. \cite{Li2015} proposed a topic-biased user reputation model for online rating systems. Li et al. \cite{Li2012} reviewed some reputation ranking methods and further proposed six new reputation-based algorithms which exhibit better effectiveness, efficiency and robustness. Indeed, many factors can affect the performance of reputation ranking methods such as the resolution of ratings. After analyzing the effects of discrete and continuous ratings, Medo and Wakeling \cite{Medo2010} found that the overall performance of reputation ranking can be improved by increasing noise in ratings when the rating resolution is low. Recently, Liao et al. \cite{Liao2017} reviewed both static and time-aware ranking algorithms and emphasized the benefits by including the temporal dimension.

\subsubsection{Emotion and health analysis}

People are accustomed to express feelings through social media (SM). Texts posted on SM have been used to track the emotion intensity of individuals. Thelwall et al. \cite{Thelwall2010} examined the extent to which emotion is present in online comments. They classified positive and negative emotions of an initial set of 2,600 human-classified comments from US users in MySpace. They found that two thirds of the comments express positive emotions and a minority contain negative emotions, showing that MySpace is an emotion-rich environment. Thelwall et al. \cite{Thelwall2010b} proposed a new method named SentiStrength, which applies machine learning approaches to extract positive and negative sentiment strength from informal text. Applied to MySpace comments, SentiStrength can predict positive and negative emotions with accuracies 0.606 and 0.728, respectively. The result demonstrates the feasibility of using SM to predict emotions of online users.

Twitter provides a rich data source for individual emotional inference. Pak and Paroubek \cite{Pak2010} performed a linguistic analysis of tweets and trained a sentiment classifier to determine positive, neutral and negative sentiments for a document. They found that SM users describe different emotions using different syntactic structures. Their methods using N-gram and POS-tags (part-of-speech) as features are efficient in identifying emotional sentiments. Mislove et al. \cite{Mislove2010} developed color-coded cartograms to track the mood of each state in the US based on over 300 million tweets. They found that the highest level of happiness occurs in the early morning and late evening, weekends are happier than weekdays, and the west coast is happier than the east coast. Bollen et al. \cite{Bollen2011} extracted six mood states (tension, depression, anger, vigor, fatigue, and confusion) from tweets. They found that popular events can affect many dimensions of public moods. These results suggest that SM can be used to track real-time emotive landscape and trend.

Wang et al. \cite{Wang2012c} created an emotion-labeled dataset of about 2.5 million tweets and identified emotions by training two machine learning algorithms, namely, LibLinear \cite{Fan2008} and multinomial naive Bayes \cite{Witten2011}. They found that the most effective features are unigrams, big rams, sentiment/emotion-bearing words, and POS information. With a training data containing about 2 million tweets, the algorithms can achieve the highest accuracy 0.6557 on emotion identification. Larsen et al. \cite{Larsen2015} constructed a so-called ``We Feel'' system to analyze variations in emotional expression on Twitter. The system collected 2.73 billion emotional tweets over a 12-week period and classified emotional words into six primary emotion categories with 25 subgroups of secondary emotions. The system can detect emotional responses to significant events and identify depression burdens from emotions expressed on Twitter. Jones et al. \cite{Jones2016} found an increase in post-event negative emotion expression on Twitter after mass violence, suggesting the effects of traumatic events on user emotion.

Recently, Mohammad and Bravomarquez \cite{Mohammad2017} annotated a dataset of tweets with emotion intensities, where the best-worst scaling technique is used to improve the annotation consistency \cite{Kiritchenko2017}. They found that emotion-word hashtags often impact emotion intensity. Further, they developed a regression model to explore the usefulness of features in emotion intensity detection and found that word embedding and lexicon features are the best indicators ($r=0.66$). Madisetty and Desarkar \cite{Madisetty2017} employed an ensemble of three machine learning methods to determine the emotion intensity of tweets. Specifically, SVR \cite{Cortes1995,Smola2004} uses lexicon and word embedding features, CNN \cite{Kim2014} uses word embedding features, and XGBoost \cite{Chen2016} uses word n-gram and char n-gram features. The ensemble method outperforms some baseline methods by giving the Spearman rank correlation 0.725.

\begin{figure}[t]
  \centering
  \includegraphics[width=0.75\textwidth]{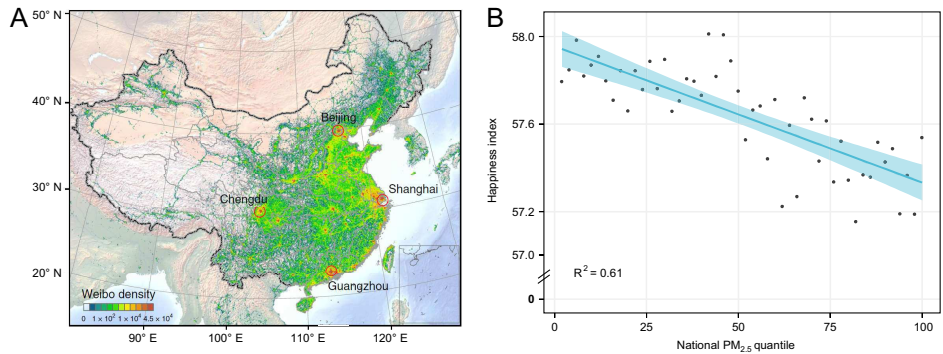}
  \caption{The geography of Weibo tweets, and the relationship between PM2.5 concentration and happiness index. (A) The spatial distribution of geotagged Weibo tweets in China. (B) The relationship between PM2.5 concentration and the happiness index. The happiness index ranges from 0 to 100. The PM2.5 concentration is divided into 50 groups, and the median happiness index value for each group is represented as a dot. The dots are fitted by the downwards sloping line with the blue shaded area represents the 95\% confidence interval. Figure from \cite{Zheng2019b}.}
  \label{Fig_5_3_4}
\end{figure}

Data from other online SM platforms, such as Facebook and Weibo, can also be used to analyze emotion. Settanni and Marengo \cite{Settanni2015} collected self-report measures of stress, anxiety and depression of over 200 adult Facebook users from North Italy, and explored the relationship between users' posts on Facebook and their emotional well-being. Through correlation analyses, they found that individuals who have higher levels of anxiety and depression will express negative emotions on Facebook more frequently. Moreover, the use of positive emotions has a negative correlation with the level of stress. By sentimentally analyzing 210 million geotagged tweets collected from Weibo, Zheng et al. \cite{Zheng2019b} constructed a daily city-level happiness index and further explored its relation to daily local PM2.5 concentrations. After exporting the data for 144 Chinese cities in 2014, they found that the happiness index is negatively correlated with PM2.5 concentrations (see Figure~\ref{Fig_5_3_4}). They work demonstrates the possibility of capturing users' emotional expressions and providing real-time feedback about life concerns using SM data.

Rich information provided by SM has been used to study mental health. Coppersmith et al. \cite{Coppersmith2014} explored the linguistic signals relevant to specific disorders and mental health. They gathered mental-illnesses-related data and replicated previous findings based on the linguistic inquiry word count (LIWC) \cite{Pennebaker2007}, suggesting the relevance of tweets to mental health. Coppersmith et al. \cite{Coppersmith2015} further built machine learning classifiers using self-reported statements of diagnosis to identify Twitter users with ten serious mental conditions such as post-traumatic stress disorder (PTSD), generalized anxiety disorder (Anxiety), and eating disorders (Eating). Their classifiers exhibit reasonable performance for most mental conditions, for example, the highest precision is 0.85 for Anxiety and 0.75 for Eating, respectively. Based on the posts in mental health forums on Reddit, Balani and de Choudhury \cite{Balani2015} proposed an algorithm that can detect the levels of self-disclosure with an accuracy 0.78 using content features.

Literature have also leveraged SM data to detect depression. Moreno et al. \cite{Moreno2011} modeled the association between demographics and depression disclosures on Facebook. They found that 25\% of Facebook user profiles display depressive symptoms, and 2.5\% of them meet the criteria for a major depressive episode. Park et al. \cite{Park2013} identified features related to depression by analyzing data of 55 Facebook student users in Korea. They found that the response of users to tips has a positive correlation ($r = 0.278$) with the CES-D scale (the ground-truth depressive symptomatology \cite{Radloff1977}), while the correlation ($r = -0.237$) is negative for the number of friends. De Choudhury et al. \cite{DeChoudhury2014} developed models to predict a mother's onset of post-partum depression (PPD) based on the Facebook data shared by 165 new mothers. They applied stepwise logistic regressions to predict the likelihood of PPD during the postpartum period using features extracted from the prenatal period alone. The model utilizing all features provides the most explanatory power, and the prediction accuracy can be improved by including the information in the early postnatal phase. The model using information of both the prenatal period and the early postnatal phase can explain about 48\% of the variance in the data.

After compiling a set of 476 depressed users on Twitter, de Choudhury et al. \cite{DeChoudhury2013} extracted behavioral features from their tweets over a year before the onset of depression. They found that some useful signals such as the decrease in social activity can characterize the onset of depression for individuals. Further, they built an SVM classifier with a RBF kernel \cite{Duda2000} that can predict depression with an average accuracy about 0.70. Based on Weibo data, Wang et al. \cite{Wang2013b} developed a depression detection model by considering features of users only. They calculated the depression inclination of each post using a sentiment analysis method and constructed the detection model using features of depressed users. Their classifiers can detect depression users with an accuracy about 0.80. Tsugawa et al. \cite{Tsugawa2015} explored the effectiveness of using Twitter activity to characterize the depression level. They found that depressed users can be predicted with 0.69 accuracy using features extracted from their activity histories on Twitter. Moreover, two months is the optimal length of observation data from Twitter to identify depression.

\begin{figure}[t]
  \centering
  \includegraphics[width=0.95\textwidth]{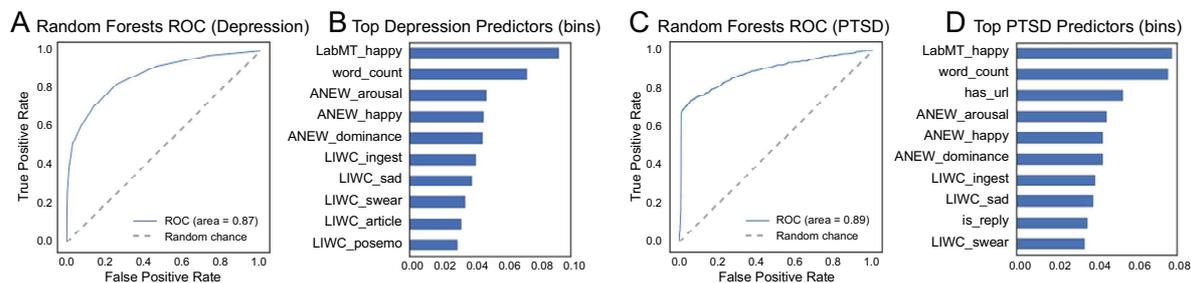}
  \caption{ROC curve and top predictors from the random forests algorithm, for depression (A and B) and PTSD (C and D) samples. Predictors with names ending with ``\_happy'' are happiness measures. LIWC predictors refer to the occurrence of semantic categories. Figure after \cite{Reece2017b}.}
  \label{Fig_5_3_5}
\end{figure}

Reece and Danforth \cite{Reece2017} identified markers of depression from photographs posted to Instagram by 166 individuals. They extracted different features from the photographs and determined the strength of each feature using the Bayesian logistic regression. They found that depressed individuals have a higher posting frequency, post more comments, prefer to post photos with less face count, and are more likely to use filters. Further, they developed a suite of supervised machine learning algorithms that can achieve an F1 measure 0.647 in identifying depressed individuals. Reece et al. \cite{Reece2017b} predicted depression and PTSD users (see Figure~\ref{Fig_5_3_5}) based on depression histories and Twitter behaviors of 204 individuals. After extracting several features from tweets, they trained supervised machine learning classifiers to identify depressed individuals. They found that a 1200-tree random forests classifier performs the best, where about 88.2\% of PTSD predictions are correct. Further, they trained a hidden Markov model that can infer the onset of depression from tweets several months prior to clinical diagnosis.

Suicidality is a serious mental illness, and suicide attempt has extremely negative socioeconomic consequences. SM can be used as a surveillance tool to track suicidal behaviours. After analyzing Korean Naver blog data, Won et al. \cite{Won2013} found that dysphoria-related and suicide-related posts are predictive of suicide frequency under a multivariate model. In particular, dysphoria variables display a low variability and long secular trend, while suicide variables show a high variability and reaction to celebrity suicide events. Sueki \cite{Sueki2015} explored the relations between suicidal behaviours and suicide-related tweets using logistic regressions. They found that tweeting ``want to commit suicide'' is strongly related to suicide attempts. Abboute et al. \cite{Abboute2014} identified suicidal risky behaviors from tweets using language processing and classification methods. After collecting suspect tweets according to nine topics that suicidal people usually talk about \cite{Gunn2012}, they classified into risky and non-risky tweets using six classifiers. They found that the naive Bayes performs the best in classifying risky tweets.

Tweets have been increasingly used to identify risky behaviors and track factors of suicide. Jashinsky et al. \cite{Jashinsky2014} found that southern and eastern states of the US have lower proportion of suicide-related tweets. The relevant tweets have a strong correlation (Spearman's rank correlation $rho =0.53$) with actual age-adjusted suicide data. Burnap et al. \cite{Burnap2015} evaluated a number of classifiers in classifying suicidal content and topics on Twitter. After extracting structural, lexical, emotional and sentiment features from tweets, they built an ensemble classifier based on the outcome of baseline classifiers. Their ensemble classifier can achieve an F1 measure 0.69 in classifying suicidal ideation. Benton et al. \cite{Benton2017} proposed a multitask learning approach using a neural architecture to predict mental conditions based on Twitter data. Their approach improves all baselines models by giving an AUC about 0.84 at the best case.

Internet search queries have also been used to estimate suicide rates. Kristoufek et al. \cite{Kristoufek2016} explored the effectiveness of using Google Trends (GTs) to estimate suicide statistics in England. They found that larger search volumes of the term ``suicide'' indicate more suicides, while more searches for the term ``depression'' indicate fewer suicides. Moreover, suicide estimates based on GTs are much better than predictions based on previous suicide data. Tran et al. \cite{Tran2017} validated GTs on predicting suicide rates in the US, Germany, Austria and Switzerland. They found that the associations between search volumes of suicide-related queries and suicide rates are weak. In particular, search volumes of the query ``suicide'' fail to show associations with suicide rates in the US. Thereby, they argued that the queries should be specific rather than broad in order to improve the performance of GTs-based suicide estimation.

Recently, Robinson et al. \cite{Robinson2016} reviewed literature that used SM for suicide identification and prevention. They summarized thirty articles published between 1991 and 2014 and concluded that to accurately assess suicide risk are still challenging. Mohr et al. \cite{Mohr2017} provided a review of mental health studies based on personal sensing, focusing on data from SM, smartphones, wearable devices, and so on. They provided a model for translating raw sensor data into behavioral markers that are related to mental health. Melia et al. \cite{Melia2018} evaluated the effectiveness of interventions for suicide prevention based on mobile technology. They claimed that data-driven mental health analyses have technological advances but ethical challenges.

\section{Situational awareness and disaster management}
\label{Sec5}

\subsection{Public health and epidemic surveillance}

Public health surveillance is critical to social and economic systems since disease outbreaks may bring a huge burden on economics and a great damage on individual well-being. Scientists have shown that the origins of infectious diseases are significantly correlated with socioeconomic, environmental and ecological factors \cite{Jones2008}, and the human behavioural responses play an important role in epidemic spreading \cite{Brockmann2013,Zhang2013}. A quantitative understanding of epidemic spreading is necessary for improving public health surveillance. To this point, a variety of models have been proposed to describe epidemic spreading on different types of networks \cite{Pastor2015,Wang2017,Wang2018c,Wang2019} and to understand the interplay between human mobility patterns and epidemic dynamics \cite{Funk2010,Belik2011}. Further studies have shown the possibility of locating the sources of diffusions with high credibility \cite{Pinto2012}, for example, by designing a time-reversal backward spreading algorithm \cite{Shen2016b}. Recently, novel large-scale data and mathematical models have deepen our understanding of epidemic dynamics, helped to map epidemic activity \cite{Shaman2012} and suggested better immunization strategies to control epidemic spreading \cite{Zhang2014,Wang2016c,Chen2018}. Indeed, the increasingly available data streams have been integrated into public health surveillance \cite{Althouse2015} and have contributed to the optimization of epidemic surveillance at multiple resolutions \cite{Lee2018}. In this section, we will briefly introduce recent works that leveraged Internet search queries, social media (SM) data and mobile phone (MP) data to advance nearly real-time epidemic monitoring and surveillance.

\subsubsection{Search queries for epidemic surveillance}

Real-time forecasts of influenza-like illness (ILI) outbreaks are usually hindered by the difficulties in collecting and analyzing a large volume of digital data in a timely manner. Traditional surveillance, relaying on collections of clinicians' records and medical claims, is limited by data coverage, spatial resolution and long-time delay in delivering analysis results \cite{Lee2018,Santillana2014}. Recent advances in information technology have made it possible to collect large-scale data of search queries that are related to public health. Useful health statistics regarding infectious disease activity can be yielded from health-related web searches. For example, Eysenbach \cite{Eysenbach2006} developed a method to analyze data collected from Google during the 2004/2005 flu season in Canada. They found that the number of clicks on links that triggered by entering ``flu'' or ``flu symptoms'' in Google is well correlated with the traditional disease surveillance data ($r=0.91$). Similarly, Polgreen et al. \cite{Polgreen2008} found a correlation between the frequency of influenza-related searches in Yahoo! and the influenza activity. They built up linear models that can predict influenza increases three weeks in advance in the US. Pelat et al. \cite{Pelat2009} compared search trends of Google queries related to three infectious diseases in a French network. They found that search query data can also be utilized for infectious disease surveillance in a non-English-speaking country.

Google launched the Google Flu Trends (GFT) in 2008 as an Internet-based influenza surveillance tool, which uses aggregated Google search data to estimate ILI activity in real time. Specifically, Ginsberg et al. \cite{Ginsberg2009} analyzed Google search queries with influenza-like symptoms and proposed a method that can estimate the current level of weekly influenza activity in each US region with a reporting lag of about one day. Later, Cook et al. \cite{Cook2011} evaluated the accuracy of the GFT model by comparing weekly estimates of ILI activity with the official data reported by the US ILINet. They found a high correlation ($r \approx 0.94$) between the models' estimates (both the original and the updated GFT models) and the ILINet data, before and during the surveillance period. Yet, scientists have also pointed out some limitations of the original GFT model \cite{Goel2010}, for example, it predicted over double the proportion of doctor visits for ILI than the Centers for Disease Control and Prevention (CDC) did in 2013. Olson et al. \cite{Olson2013} studied the reliability of GFT from 2003 to 2013 and argued that GFT may not provide reliable surveillance for seasonal or pandemic influenza. Changes in Internet search behavior and differences between the periods of GFT model-fitting and prospective usage diminish the performance of the original GFT model. Meanwhile, Lazer \cite{Lazer2014} suggested two issues that may contribute to GFT's drawbacks. One is the big data hubris as most data are not the output of instruments designed to ensure data's validity and reliability for scientific analysis. The other is the instability of the algorithm as changes made by engineers and consumers will affect the tracking of GFT.

The original GFT model has been well improved in recent years, and Google search queries have been widely used to nowcast influenza outbreaks. Based on real-time data with external information from GFT, Dugas et al. \cite{Dugas2013} developed a practical influenza forecast model that can provide accurate prediction of influenza cases. They developed the model by the generalized linear autoregressive moving average (GARMA) methods \cite{Benjamin2003} with the negative binomial distribution. Formally, the GARMA$(p,q)$ model is given by
\begin{equation}
\log(\mu_t) = X'_{t-1}\beta + \sum_{i=1}^p \phi_i \left[\log(y_{t-1}) - X'_{t-1-i}\beta \right] + \sum_{j=1}^q \theta_j \left[ \log\left(\frac{y_{t-j}}{\mu_{t-j}}\right) \right],
\end{equation}
where $\mu$ is the expected value of response $(y)$, and $X$ is the vector of external variables with primes standing for transpose. The parameters $\beta$, $\phi$ and $\theta$ of the model can be estimated from the training data. The model GARMA(3,0) can predict weekly influenza cases for 83\% of the estimates with the out-of-sample verification, showing the capability of GFT in influenza forecasting. Using hospital influenza test results, Araz et al. \cite{Araz2014} validated the usefulness of GFT in forecasting ILI-related emergency department visits. After testing five forecasting models, they found that linear regression models perform significantly better when including GFT data during 2008-2012.

Using Google searches between Jan 2010 and Sep 2013, Preis and Moat \cite{Preis2014} built dynamic models to estimate the current level of influenza outbreak before the release of official data. They added GFT time series to an autoregressive integrated moving average (ARIMA) model \cite{Makridakis1998} as an external regressor. The ARIMA model contains three autoregressive (AR) terms and two moving average (MA) terms (i.e., ARIMA(3,0,2)). Generally, the ARIMA($p,d,q$) model is given by
\begin{equation}
y_{t+h} = \theta_0 + \sum_{i=1}^{p+d} \phi_i y_{t+h-i}  + \sum_{j=1}^q \theta_j \varepsilon_{t+h-j} + \varepsilon_{t+h},
\end{equation}
where $p$ is the number of AR terms, $q$ is the order of the non-seasonal MA lags, $d$ is the number of non-seasonal differences, $h$ is the period in the future, $y_t$ is the ILI level at week $t$, $\epsilon_t$ is the white noise random error, and $\phi_i$ and $\theta_j$ are parameters to be estimated from data \cite{Box2015,Xu2017}. They found that the in-sample forecasting mean absolute error (MAE) of the model is about 14\% smaller than the baseline model without the GFT data. Teng et al. \cite{Teng2017} developed a dynamic forecasting model to predict Zika Virus based on web searches from the Google Trends (GTs). They found a strong correlation between Zika-related GTs and the cumulative numbers of reported cases. Further, they constructed an ARIMA (0,1,3) model using online search data as the external regressor that can improve the predication accuracy. Xu et al. \cite{Xu2017} explored the predictive utility of Google search data in forecasting new ILI cases in Hong Kong by testing some individual models including generalized linear model (GLM) \cite{Mccullagh1989}, ARIMA \cite{Makridakis1998} and deep learning (DL) with feedforward neural networks (FNN) \cite{Lecun2015}. They found that DL with FNN are the best-performed algorithms in predicting the influenza peaks.

\begin{figure}[t]
  \centering
  \includegraphics[width=0.8\textwidth]{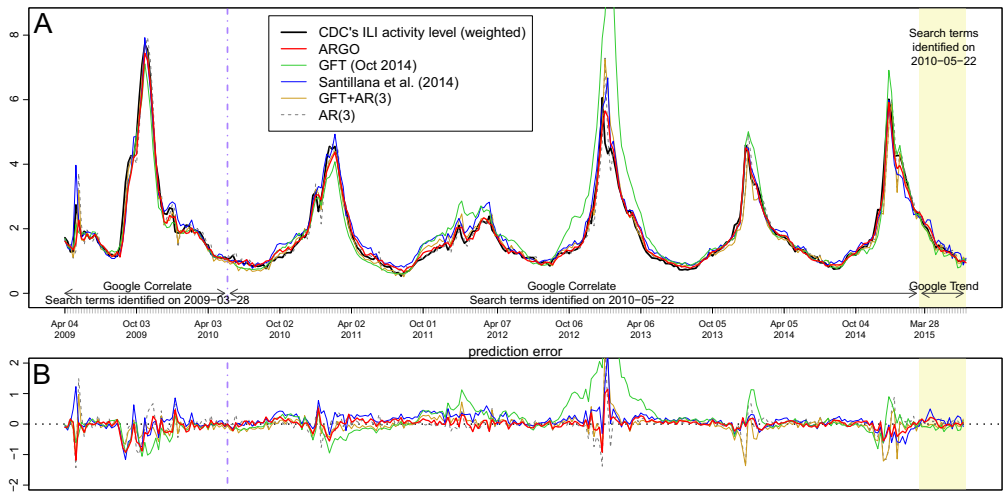}
  \caption{Comparison between estimations of ILI activity. (A) The estimated ILI activity level from ARGO (thick red), in comparison with the CDC's ILI activity level (black). The estimates from GFT (green), method of Santillana et al. \cite{Santillana2014} (blue), GFT plus AR(3) model (yellow), and AR(3) model (dashed gray) are also presented. The dash-dotted purple vertical line separates Google Correlate data into two periods. (B) The error of ILI activity estimation, defined as the estimated value minus the CDC's ILI activity level. Figure from \cite{Yang2015}.}
  \label{Fig_6_1_1}
\end{figure}

Data of Google search queries in other countries and from other platforms have also been used to improve infectious disease surveillance. Based on data of ILI-related Google search queries and historical CDC's ILI activity reports, Yang et al. \cite{Yang2015} developed a methodological framework to produce retrospective estimates of ILI levels. Their multivariate linear regression modeling framework named ARGO (AutoRegression with GOogle search data) is expressed as
\begin{equation}
y_t = \mu_y + \sum_{j\in J}\alpha_j y_{t-j} + \sum_{k\in K} \beta_k X_{k,t} + \epsilon_t ,
\end{equation}
where $y_t = \log(c_t + 1)$ is the dengue case counts $c_t$ at time $t$, $X_{k,t}$ is the Google search frequency of query term $k$ at time $t$, $J$ is the set of auto-regressive lags, $k$ is the set of Google query terms, and $\epsilon_t \overset{\text{iid}}{\sim} \mathcal{N}(0,\sigma^2)$. The ARGO model performs better in tracking ILI activity than some benchmark methods such as the GFT launched in 2014 (see Figure~\ref{Fig_6_1_1}). Later, Yang et al. \cite{Yang2017b} extended ARGO to predict dengue cases in multiple countries/states, showing its nearly real-time ability to estimate dengue activity in data-poor environments. Yuan et al. \cite{Yuan2013} monitored influenza epidemics in China using Baidu search queries. They found a significant correlation between the selected composite keyword index and the case survey of Chinese influenza. Further, they fitted a linear model that can predict influenza cases one-month ahead for the first eight months of 2012 with the mean absolute error less than 11\%. Li et al. \cite{Li2017b} developed a dengue Baidu search index (DBSI), based on which they proposed a predictive model of dengue fever in Guangzhou, China. Their dengue early warning system combining DBSI with traditional surveillance and meteorological data can improve the capability of dengue case prediction.

\subsubsection{Online posts for disease surveillance}

Messages posted on Twitter are a new data source for disease surveillance. Chew and Eysenbach \cite{Chew2010} analyzed over 2 million Twitter posts with keywords related to pandemic during the 2009 H1N1 outbreak. They found that Tweets containing ``H1N1'' increased from 8.8\% to 40.5\%, showing the potential of social media (SM) data on conducting infodemiology studies for public health. Culotta \cite{Culotta2010} identified influenza-related messages from over 500,000 messages spanning 10 weeks on Twitter. Then, they applied several regression models to examine the relations between the identified tweets and the CDC reported influenza statistics. They found that the best method exhibits a high correlation ($r=0.78$) with the CDC statistics. However, not all tweets containing influenza-related terms are suggestive to influenza outbreaks. For example, Aramaki et al. \cite{Aramaki2011} found that 42\% of tweets containing the word ``influenza'' are unrelated influenza tweets which do not refer to actual influenza outbreaks. Further, they developed an SVM-based classifier \cite{Cristianini2000} to extract the mention of actual influenza patients from influenza-related tweets. Their method exhibits a very high correlation ($r=0.89$) in detecting actual influenza outbreaks.

Signorini et al. \cite{Signorini2011} collected a large sample of public tweets between April 29 and June 1, 2009 that contains pre-specified terms and the ground truth ILI data reported by the CDC. They trained a support vector regression (SVR) model \cite{Cortes1995,Smola2004} on weekly term-frequency statistics. Their method produces estimates of national ILI values with an average error of 0.28\%. Using 318,379 influenza-related tweets generated by 101,853 users, Salath{\'e} and Khandelwal \cite{Salathe2011} trained a machine learning algorithm to automatically judge sentiments of tweets. They found that sentiments expressed in tweets are positively correlated ($r=0.78$) with the official CDC-estimated vaccination rates at the regional level. Later, Lamb et al. \cite{Lamb2013} discriminated flu tweets that report actual infection from those that express concerned awareness of flu. Similarly, Broniatowski et al. \cite{Broniatowski2013} developed an influenza infection detection algorithm that can automatically identify relevant tweets. The estimates based on identified relevant tweets exhibits a strong correlation ($r=0.93$) with the CDC data, and their method can detect weekly deceleration of influenza prevalence with 85\% accuracy.

Information embedded in Twitter stream has been shown very helpful in detecting rapidly evolving public concern with respect to the ILI emergence and transmission. Using over 287 million Korean tweets posted from October 2011 to September 2012, Kim et al. \cite{Kim2013} developed regression models to track actual ILI epidemics and predict their activity levels. They maximized the correlation with the official reported ILI data by choosing a subset of markers and their weights using the LASSO regression method \cite{Tibshirani1996}. Their model has a significant improvement in prediction performance at the initial phase of ILI peak. Paul et al. \cite{Paul2014} found that models incorporating influenza-related tweets can reduce the forecasting error by 17-30\% compared with a baseline model using historical ILI data only. Moreover, models relying on tweets perform better in estimating influenza prevalence than models using data from GFTs. Aslam et al. \cite{Aslam2014} analyzed 159,802 tweets collected from 11 US cities and found that tweets can serve as a supplementary surveillance tool for influenza with increased accuracy.

Topic analysis and topic models have been used to infer health concerns from Twitter data. Culotta \cite{Culotta2014} collected 4.31 million tweets of users in the US top-100 most populous counties and performed a linguistic analysis of geographical Twitter activity. They found a significant correlation on testing data for 6 of 27 health statistics, and they proposed models using Twitter-derived information, which can improve prediction accuracy for 20 of 27 health statistics. Chen et al. \cite{Chen2016b} proposed a temporal topic model to infer hidden biological states of users from their tweets. They developed an EM-based learning algorithm to model users' hidden epidemiological states. Their algorithm utilizing tweets from 15 countries in South America can learn meaningful word distributions and state transitions. Moreover, their algorithm can give better predictions of flu trends and flu peaks by aggregating states of users. Kagashe et al. \cite{Kagashe2017} performed a topic analysis of widely used medicinal drugs during the 2012-2013 influenza season. They constructed an SVM classifier using dependency words as features to extract tweets that are suggestive to consumed drugs. Their model significantly outperforms some well-known benchmarks such as the lexicon-based model. Further, they extracted trending topics from drug-mentioning tweets using the LDA model \cite{Blei2003} and found that the topic information of widely-consumed drugs can enhance seasonal influenza surveillance.

Researchers have developed several disease surveillance systems relying on data from Twitter and other SM platforms. Lee et al. \cite{Lee2013} described a surveillance system to automatically predict seasonal disease outbreaks and monitor cancer activity levels based on tweets in the US. The resulted disease surveillance maps clearly show the distributions and timelines of disease types, symptoms and treatments. Meanwhile, Dredze et al. \cite{Dredze2014} presented a platform (HealthTweets.org) to share the latest surveillance results based on Twitter with public health officials. This platform provides three main visualizations including temporal health trends, specific locations and maps with a geographical view of health trends in the world. Indeed, health informatics datasets gathered from SM platforms have been increasingly applied to improve health care (see recent reviews \cite{Grajales2014,Santillana2015} and the references therein). To improve disease surveillance, Santillana et al. \cite{Santillana2015} suggested to leverage data from multiple sources such as online search, SM and traditional data. By combining ILI activity predictors of each data source, they developed an ensemble learning approach that outperforms GFT and autoregressive models by producing earlier estimates with a comparable accuracy.

\begin{figure}[t]
  \centering
  \includegraphics[width=0.5\textwidth]{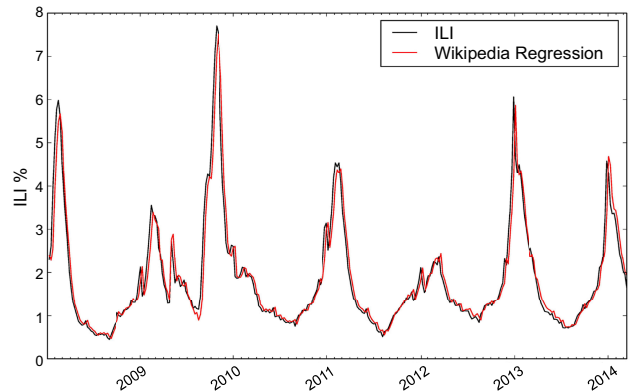}
  \caption{Regression of Wikipedia access logs to the officially released ILI data. The figure presents the linear regression of one week prior ILI observation, a constant term, and the access logs of five Wikipedia articles related to influenza to the current ILI observation. The regression of Wikipedia access logs is highly correlated to the true ILI outcome. Figure from \cite{Hickmann2015}.}
  \label{Fig_6_1_2}
\end{figure}

Pageviews of disease-related Wikipedia articles have also been used to forecast seasonal influenza. By monitoring the rate of views on some specific Wikipedia articles, Mciver and Brownstein \cite{Mciver2014} developed a Poisson model that can accurately estimate ILI levels in the US in a timely manner. They collected Wikipedia article view data from December 2007 to August 2013 (294 weeks) and developed a generalized linear model to estimate ILI activity levels in the American population. Their model including 35 variables selected by the LASSO regression method can forecast the peaking weeks of ILI activity within a season more accurate than GFT. Later, Generous et al. \cite{Generous2014} extended the above work for health purposes. In the same manner, they collected the Wikipedia article access logs from March 2010 to February 2014 and then applied a Poisson model fitted by the LASSO regression method to estimate ILI levels in the US. The estimates by their model exhibits up to $R^2 = 0.92$ in predicting the official data with about one month in advance and with high feasibility across locations. Similarly, Hickmann et al. \cite{Hickmann2015} leveraged Wikipedia access logs to create a weekly forecast for ILI activity levels during the 2013-2014 influenza season. Their linear regression model includes the weekly request data for influenza Wikipedia article and the previous week's ILI data as independent variables. They found that Wikipedia article access logs have a high correlation with historical ILI records (see Figure~\ref{Fig_6_1_2}), and their method can accurately predict official reported ILI levels several weeks before the release.

The entire Wikipedia editing histories and article content are available \cite{Zha2016}, which provides a very rich data source for disease surveillance. Fairchild et al. \cite{Fairchild2015} showed that Wikipedia content serves a centralized open-source monitoring and repository system, where disease-related data can be collected in real time. They used standard natural language processing techniques to identify key phrases in the content of disease-related Wikipedia articles such as death and hospitalization counts. Their method achieves an F1 of 0.753 in identifying relevant entities. Further, they analyzed articles of the 2014 West African Ebola virus disease epidemic and found that Wikipedia can provide detailed time series data, which are closely aligned with the data reported by WHO. Recently, Priedhorsky et al. \cite{Priedhorsky2017} evaluated the use of Wikipedia access logs and category links for measuring global disease. They compared thousands of individual models for testing the effects of semantic article selection, forecast horizon, amount of training data and model staleness based on Wikipedia data across six diseases and four countries. They found that the accuracy and robustness of disease estimation can be effectively improved by using minimal-configuration, language-agnostic article selection process based on semantic relatedness. Similarly, Sharpe et al. \cite{Sharpe2016} provided a comparative analysis of Google, Twitter and Wikipedia data for influenza surveillance during the 2012-2015 influenza seasons. They detected seasonal change points by performing the Bayesian change point analysis \cite{Barry1993} and calculated the sensitivity and positive predictive values (PPV) for each data source. They found that Wikipedia data have fewer change points in common with the CDC's ILI data, while they present the lowest sensitivity (33\%) and PPV (40\%) compared to Google and Twitter data.

\subsubsection{Mobile phone records for epidemic prediction}

Mobile phone (MP) data are a valuable source for studying the spreading dynamics of infectious diseases such as malaria, rubella, dengue, cholera and Chagas. The transmission of infectious diseases is significantly affected by human movements as infected individuals that make long-distance travels may transmit disease to healthy population in other regions. Therefore, quantifying human movements is critical to perceive and predict epidemic diffusion. Large-scale MP data have been leveraged to study human movements, which improves our ability of modeling and predicting the spreading of infectious diseases. Based on call detail records (CDRs) of 770,369 users for three months, Tatem et al. \cite{Tatem2009} estimated human travelling patterns and the imported malaria risk in Zanzibar. They extracted information on users' travelling and staying among Tanzania regions from CDRs, showing that imported malaria risks are heterogeneously distributed, where a very few people account for most of the malaria risk. Moreover, the likely sources and rates of malaria importation can be predicted by MP-derived human movement patterns in combination with the malaria endemicity data.

Based on data of 1.5 million Kenyan MP users, Wesolowski et al. \cite{Wesolowski2012} explored the dynamics of human carriers that drive malaria parasite spreading in Kenya. They mapped MP users to cell towers in settlements and assigned each settlement a malaria endemicity class according to a high resolution map of malaria prevalence. Then, they built travel networks of people and parasites between settlements and regions, focusing on the parasite importation by returning residents and by visitors from risky regions. Next, they examined directional and net movements of people and parasites between settlements by analyzing asymmetries between ``source'' and ``sink'' settlements. They found that the capital city Nairobi and its surroundings are a major destination for both humans and parasites. Moreover, returning residents play an important role in importing parasites to major parasite sinks, and some local transmission may occur in residential and less developed areas on the periphery of the city. Wesolowski et al. \cite{Wesolowski2015} further explored the CDRs data and found that human mobility measures extracted from MP data can predict which region is lacking preventive care. Tatem et al. \cite{Tatem2014} predicted malaria risk by analyzing CDRs of about 1.5 million users and case-data of risk maps in Namibia. They found that most of the northern Namibian areas are major sources and sinks of parasites, and the heterogeneity in human movement patterns results in the variation in the risk connectivity of sources and sinks. Moreover, targeting control on malaria risk connectivity in certain areas with larger exporters (sources) of infections can largely affect their surrounding areas. These works highlight the value of MP and complimentary data on perceiving and controlling infectious disease.

\begin{figure}[t]
  \centering
  \includegraphics[width=0.55\textwidth]{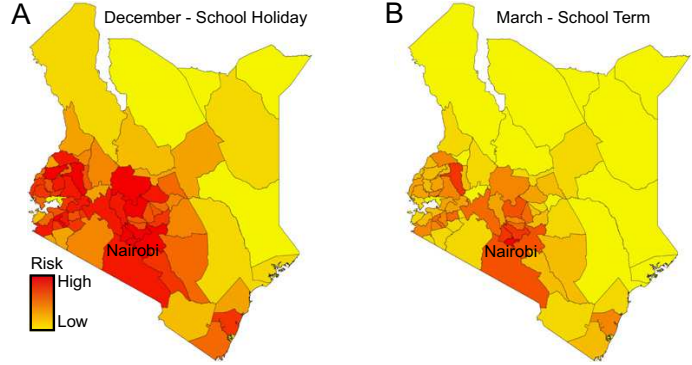}
  \caption{The seasonal variability in the risk of rubella importation in Kenya. There are large amounts of population flux and consequently the risk of rubella importation per district (A) during the major holiday and a school term break (December), while the rubella importation risk decreases (B) during the school term. The western Kenyan districts are with higher risks during the year, and the capital city Nairobi consistently remains at a high risk of importation. Figure from \cite{Wesolowski2015b}.}
  \label{Fig_6_1_3}
\end{figure}

Multiple data sources have been intergraded to better understand disease spreading and implement targeted surveillance. Wesolowski et al. \cite{Wesolowski2014} collected travel data of 2,650 individuals in two Kenyan districts through a malariometric survey and compared it with the data of about 1.5 million MP users \cite{Wesolowski2012}. They found that both survey and MP data can predict the amount of imported malaria and identify the major routes, however, the two datasets have a wide divergence in terms of the travel magnitude. For example, travel surveys can provide information on demographics, travel motivations and destinations \cite{Yan2013c}, but they tend to under-estimate the volume and range of human mobility. Although the distributions of human movements estimated based on surveys and MPs are highly correlated with each other, Tizzoni et al. \cite{Tizzoni2014} observed a systematic overestimation of commuting traffic in MP data. That is because MP-sampled population can not perfectly represent the general population. In addition, human mobility proxies perform differently in approximating commuting patterns for disease spread at different resolutions, suggesting that the chosen of mobility proxies should account for the epidemic situation under study. Regarding this point, Wesolowski et al. \cite{Wesolowski2015b} showed that MP-derived seasonal and spatial travel patterns are predictive for disease transmission. They produced dynamic importation risk maps for rubella in Kenya (Figure~\ref{Fig_6_1_3}) using MP data. Their work demonstrates the effectiveness of MP data on identifying critical drivers of epidemics on relevant spatial and temporal scales.

In-depth analyses of human mobility patterns help improve the prediction of geographic spread and timing of epidemics. Based on data of about 40 million MP users, Wesolowski et al. \cite{Wesolowski2015c} quantified human travel patterns underlying the spread of dengue virus in Pakistan. They estimated the travels of users based on the MP data and then fitted an ento-epidemiological model \cite{Lourenco2014} to the reported dengue cases in southern Pakistan. They estimated the timing of the first imported case in northern Pakistan by employing an epidemiological approach, and they found that MP-based mobility estimates can predict the geographic spread and timing of epidemics. Further, they generated fine-scale dynamic risk maps by combining estimates of seasonal dengue importation and transmission suitability maps. Formally, the epidemic risk for a location $x$ is given by
\begin{equation}
\text{risk}(x) = \sum_{t=1}^{N} Z_{x}(T_t) Y_{x,t},
\end{equation}
where $Z_{x}(T_t)$ is the environmental suitability for dengue, and $Y_{x,t}$ is the importation of infected travelers on day $t$. Based on MP data of about 2.9 million users in Haiti, Bengtsson et al. \cite{Bengtsson2015} explored the influence of population mobility on the spatial evolution of a large-scale cholera outbreak. They predicted the risk of a new outbreak in an area by calculating the infectious pressure based on the human mobility. They found that the infectious pressure at outbreak onset is linearly correlated ($r \approx 0.3$) with the average daily number of reported cases within seven days of the new outbreak. Finger et al. \cite{Finger2016} collected MP data of about 150,000 users to estimate human mobility fluxes during the 2005 Senegal cholera outbreak period. They developed a mechanistic model that takes into account human mobility and other drivers of cholera disease transmission such as rainfall. They found that mass gatherings during the initial phase of the outbreak can significantly affect the spreading of waterborne diseases cholera.

Human mobility patterns derived from MP data have been introduced into the mathematical modeling of infectious disease transmission. By analyzing the MP data \cite{Montjoye2014b} released for the Data for Development (D4D) project, Tompkins and Mccreesh \cite{Tompkins2016} identified the characteristics of human movements involving overnight stays, which are relevant for malaria transmission. They found that about 60\% of people have regular destinations that they visit repeatedly, and the number of overnight journeys peaks at a distance of 50 km. Further, they proposed an agent-based migration model by adapting a gravity model to describe overnight journeys. Their model can well reproduce general population mobility patterns driving malaria transmission. Based on CDRs during the chagas spreading in Argentina and Mexico, de Monasterio et al. \cite{Monasterio2016} analyzed mobility patterns of users and predicted the movements among different regions. They detected possible risk zones of chagas disease and produced the risk maps for two Latin American countries. Recently, Wesolowski et al. \cite{Wesolowski2017} explored how seasonal variation in human movement affects infectious disease dynamics based on MP data collected from Kenya, Namibia and Pakistan. They found that major national holidays will lead to seasonal fluctuations in human mobility, which further result in seasonal fluctuations of the country-scale connectivity. Using a spatial diffusion model (see Ref. \cite{Wesolowski2017}), they evaluated the consequences of directional asymmetries and seasonal variation in travels as well as pathogen characteristics on epidemic spreading, and found that the spreading speed depends not only on the pathogen's characteristics but also on the month that the pathogen is imported.

Panigutti et al. \cite{Panigutti2017} built two human mobility networks in France, namely, the MP commuting network and the census commuting network. Then, they compared 658,000 simulated epidemic outbreaks generated using a reaction-diffusion (RD) metapopulation model based on the two mobility networks. The RD dynamics are separated into two components, home time and work time. The number of susceptible, infected and recovered individuals \cite{Hethcote2000} who live in district $i$ and work in district $j$ are respectively defined as $S_{ij}$, $I_{ij}$ and $R_{ij}$. The spreading
rate is $\beta$. The force of infection during home time $\lambda_{i}^\text{home}$ is defined as
\begin{equation}
\lambda_{i}^\text{home} = \beta \frac{ I_{ii} + \sum_{j\in \nu_i} I_{ij} }{ N_{ii} + \sum_{j\in \nu_i} N_{ij} } ,
\end{equation}
where the sums run over the neighbourhood of district $i$: $j\in \nu_i$. The force of infection during work time $\lambda_{i}^\text{work}$ is defined as
\begin{equation}
\lambda_{i}^\text{work} = \beta \frac{ I_{ii} + \sum_{j\in \nu_i} I_{ji} }{ N_{ii} + \sum_{j\in \nu_i} N_{ji} } ,
\end{equation}
where $N_{ij} = S_{ij} + I_{ij} + R_{ij}$ is the total number of commuters living in district $i$ and working in district $j$, and $N_{ii}$ is the number of residents in district $i$ who also work in district $i$. The numbers of susceptible individuals in district $i$ during home and work time are respectively given by $S_{i}^\text{home} = S_{ii} + \sum_{j\in \nu_i} S_{ij}$ and $S_{i}^\text{work} = S_{ii} + \sum_{j\in \nu_i} S_{ij}$ \cite{Panigutti2017}. Simulation results show that MP data are more reliable in describing human movements in central regions. Moreover, it is essentially important to obtain an accurate estimation of epidemiologically relevant mobility patterns in the seed area in order to capture future spreading patterns of the outbreak. Mari et al. \cite{Mari2017} explored the drivers of endemic schistosomiasis by parameterizing a spatially explicit network model based on a large dataset of MP traces. They found that the epidemic prevalence may be reduced by moderate mobility while increased by either high or low mobility. Moreover, environmental and socioeconomic heterogeneities play a crucial role in capturing the spatial epidemic prevalence, and the inclusion of human mobility patterns can significantly improve the predictive power for the infectious disease spreading.

There are opportunities yet challenges of leveraging MP data to link human mobility patterns with infectious disease dynamics. Wesolowski et al. \cite{Wesolowski2016} pointed out some limits such as the availability of MP data, the biases on MP ownership, and the lack of large-scale demographic or social identifiers. Meanwhile, they suggested some opportunities including the fine-scale individual movements across large numbers of individuals, the new data of GPS for understanding social connections, and the data access for public health interventions. In a broader perspective, Jones et al. \cite{Jones2018} reviewed some challenges and potential opportunities of using CDRs for public health research. They pointed out that the majority of previous studies paid attentions to below middle-income countries, CDRs were mainly used in aggregated form to estimate population movement, while public views on using CDRs for public health research were lacking. To address these issues, they suggested to develop an ethically founded framework to gain public views and to better integrate routine health records with validated CDRs in future public health research.

\subsection{Emergency and disaster monitoring}

Along with increased urbanization and changing climate, many areas are now facing an unprecedented number of emergent events and natural disasters, which pose numerous threats to human life and economic development. It urges rapid situational awareness and efficient management strategies to reduce human suffering and economic losses \cite{Huang2015e,Yu2018}. In rural areas, assessments of natural hazards usually follow a delay, resulting in difficulties of disaster response and relief. In urban areas, detections of emergent events (such as terrorist attacks, riots and large-scale demonstrations) and natural disasters (such as earthquakes, floods and hurricanes) are critical not only for governments' rapid disaster response \cite{Imran2015} but also for in-depth understanding of human behaviors in extreme situations that will further help in better designing strategies in disaster relief \cite{Bellomo2016}. In addition to theoretical methods \cite{Zhou2017,Han2018}, novel data sources have been leveraged to improve emergency awareness and disaster management such as remote sensing (RS) \cite{Plank2014}, mobile phone (MP) \cite{Calabrese2011}, and social media (SM) \cite{Castillo2016}, with remarkable advantages of low acquisition cost, real-time updates and high spatio-temporal resolutions.

\subsubsection{Remote sensing for disaster assessment}

Mapping natural hazard and disasters is critical for emergence response, disaster relief and crisis-management support \cite{Voigt2007}. However, disaster management relying on site surveys and field observations usually requires many resources, follows a long-time delay and is constrained by time and space. Fortunately, these problems can be tackled by using the increasingly available RS data as they update timely with low cost, have high spatial and temporal resolutions and capture a wide field of view \cite{Plank2014}. In recent years, a number of pioneering works have illustrated the utilization of RS data in combination with image processing techniques for a rapid damage assessment of natural disasters such as earthquakes, flooding, wildfire and landslides \cite{Joyce2009}. The RS-based disaster management is necessary for supporting, for example, damage assessment and relief priority map. In the following, we will introduce some applications of RS data for earthquake damage assessment and flood monitoring.

High-resolution satellite images can be used to detect changes of ground surface and buildings before and after earthquake. This opens new opportunities for earthquake damage assessment at the level of settlements and buildings. Using synthetic aperture radar (SAR) interferometry images obtained by the ERS-1 satellite, Massonnet et al. \cite{Massonnet1993} captured the ground surface movements caused by the 1992 Landers earthquake in California, which agree well with surveying measured displacements. Miura and Midorikawa \cite{Miura2006} detected locations of newly constructed buildings from high-resolution satellite images and updated GIS building inventory data. They conducted the building damage assessment for a scenario earthquake in Metro Manila, Philippines. Marin et al. \cite{Marin2015} developed an approach to detect building changes before and after earthquakes from very high resolution (VHR) SAR images. They extracted information on changes by analyzing the exploitation of expected backscattering properties of buildings. Validated using spotlight images of two Italian cities, their approach shows a high reliability in identifying demolished buildings.

Satellite images from RS have been used to map earthquake exposure and conduct damage assessment after the 2010 Haiti Earthquake. Based on pre- and post-event VHR satellite imagery, Corbane et al. \cite{Corbane2011} produced a building damage assessment map for the 2010 Haiti Earthquake. They compared the reliability of this area-based map to the detailed damage assessment derived from the post-event aerial imagery. Result suggests that satellite-based damage assessment maps are able to capture the damage pattern especially in heavily damaged areas, however, they cannot provide sufficient information to quantify damage intensity. Uprety and Yamazaki \cite{Uprety2012} detected buildings that were damaged during the 2010 Haiti Earthquake by calculating the backscattering difference (as well as correlation) between two SAR images taken before and after the earthquake. Later, Ehrlich et al. \cite{Ehrlich2013} showed that building damages can be detected from pre- and post-disaster VHR imagery, and the measured damages can provide vulnerability information related to the structural fragility of building stocks. Tian et al. \cite{Tian2015} developed a novel method to monitor after-disaster building damages. Two post-event satellite stereo imageries were combined with digital surface models, panchromatic images were segmented, and a rule-based classification was used to identify collapsed buildings.

Satellite imagery has also been used to locate affected areas after large earthquakes in China. One of the most severe natural disasters in China is the 12 May 2008 Wenchuan Earthquake, which changed the entire landscape of the affected area. Based on VHR satellite images, Liou et al. \cite{Liou2010} investigated landslides and their consequences following the Wenchuan earthquake. They identified structural deformation of land areas and rupture of dams and suggested some precautionary measures to avoid further destruction. Tong et al. \cite{Tong2012} proposed an approach to detect collapsed buildings based on satellite stereo image pairs taken before and after the Wenchuan earthquake. They detected an individual collapsed building by the height differences and identified the region of collapsed buildings by differentiating the digital elevation models generated from the image pairs. Their method can accurately estimate the status of collapsed storeys and determine the collapsed region with an overall accuracy of over 90\% assessed using an error matrix \cite{Congalton1991}. The 2010 Yushu Earthquake is another large earthquake causing many buildings collapsed in China. For the Yushu earthquake, Jin et al. \cite{Jin2011} derived a building damage assessment from VHR SAR images. After analyzing the features of collapsed, partial collapsed and non-collapsed buildings in the SAR images, they counted the number of buildings with different damage levels and built the damage index for each block. Shi et al. \cite{Shi2015} extracted multiconfiguration features to distinguish the fallen and intact structures. They employed a random-forest framework to quantify the importance of each feature to improve the accuracy.

\begin{figure}[t]
  \centering
  \includegraphics[width=0.6\textwidth]{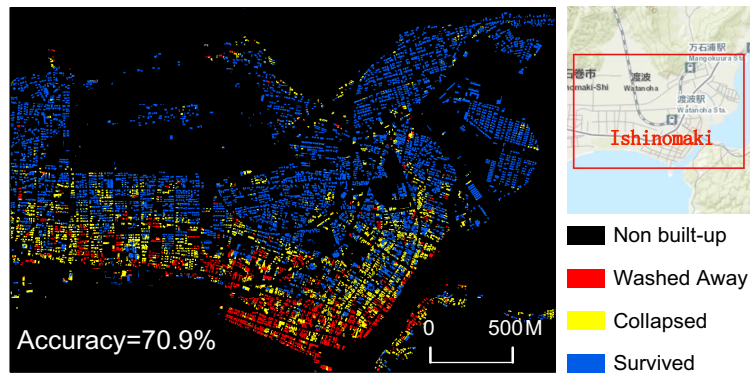}
  \caption{Results of the damage map using the U-net convolutional network. The south-eastern part of Ishinomaki city is used as the validation area during the 2011 Tohoku Earthquake-Tsunami. The figure presents the survived areas (blue), contrasting with collapsed areas (yellow), washed away areas (red), and not built-up areas (black), identified by the U-net convolutional network. The overall accuracy of damage classification is 0.709. Figure from \cite{Bai2018}.}
  \label{Fig_6_2_1}
\end{figure}

Recently, deep learning algorithms have been introduced to analyze RS data for rapid earthquake damage mapping \cite{Lecun2015,Schmidhuber2015}. For the 2010 Haiti Earthquake, Cooner et al. \cite{Cooner2016} evaluated the effectiveness of several deep learning algorithms in detecting earthquake damage. They found that spatial texture and structure features extracted from satellite images are more important than spectral information in classification. Multilayer feedforward neural network can detect damaged buildings with an error rate below 40\%. By combining convolution neural networks (CNNs) and multiscale segmentation, Sun et al. \cite{Sun2017} proposed a method to map earthquake damage from VHR images. They firstly trained CNNs to get initial classification about original images, and then combine the initial classification with the results of multiscale segmentation, to obtain class-based segmented images with different scales. Their approach performs very well in mapping the Wenchuan earthquake damage. Fujita et al. \cite{Fujita2017} applied CNNs to detect damage buildings from pairs of satellite image patches taken before and after the 2011 Tohoku Earthquake-Tsunami. Their CNN-based detection system can classify washed-away buildings with accuracy in the range 94-96\%. Recently, Bai et al. \cite{Bai2018} developed a deep learning algorithm to map damage, where the U-net convolutional network is employed to semantically segment VHR satellite images. Their algorithm can classify damage with an overall accuracy 0.709 based on pre- and post-disaster images of the Tohoku tsunami (see Figure~\ref{Fig_6_2_1}). In addition, the damage map can be updated in every 2-15 minutes when images are available.

RS data play an indispensable role in operational practice of earthquake response practice. Through analyzing high-resolution satellite imagery, not only public health needs can be rapidly assessed, but also relief priority map can be suggested. For some isolated earthquake areas with uneven damage distribution, RS-based damage assessment can promote the effectiveness of rescue efforts. By leveraging GIS technology and high-resolution satellite images, Zhao et al. \cite{Zhao2018} developed an assessment framework to rapidly evaluate health loss. Their method can estimate casualties and injuries with an accuracy 0.77 within a few hours for the 2008 Wenchuan earthquake. Moreover, they can identify damaged medical institutions, mark high-risk areas of schistosomiasis and map temporary settlements for victims. Hamid et al. \cite{Hamid2018} determined the degree of building damage based on the texture features extracted from pre- and post-event VHR satellite imagery. They proposed an algorithm that can produce relief priority maps after earthquakes. For the Varzaghan earthquake in Iran, their method achieves a general accuracy 0.88 in classifying damaged buildings into three classes of negligible damage, substantial damage and heavy damage with relief priorities from low to high. Their relief priority maps can guide rescue teams under limited resources after earthquakes. Recently, Quinn et al. \cite{Quinn2018} reviewed applications of machine learning approaches for refugee settlement mapping based on RS data. The combination of machine learning algorithms and RS data provides a way to better coordinate humanitarian relief.

Flood is one of the most serious disasters which can destroy homes, cause mudslides and take human lives. Flood management requires timely awareness of flood situation, locating flooded areas and implementing damage relief \cite{Klemas2015}. Satellite- and aircraft-based RS can provide the required information with high spatial and temporal resolutions. In recent years, RS has become a useful tool in flood management \cite{Sanyal2004}. By integrating satellite images with ancillary information from GIS, Brivio et al. \cite{Brivio2002} proposed a procedure to estimate flooded areas at the peak time. For the 1994 flood in northern Italy, their method can identify 96.7\% of flooded areas compared with the official reference map. Groeve \cite{Groeve2010} described a method to early detect floods from satellite imagery. For the 2009 Southern Africa flood and the 2010 Namibia flood, their method can provide early flood warning up to 30 days by monitoring upstream areas and detect floods two hours after their occurrences. Skakun et al. \cite{Skakun2014} assessed flood risk based on time series of satellite images. They calculated the relative frequency of inundation (RFI) based on the flooded areas extracted from satellite images and the maximum flood extent images produced for previous flood events. The RFI map serves as a hazard map for flood risk assessment. Their method can identify cities and villages with the highest flood risk for the Namibia flood.

RS technology can produce high-resolution flood hazard maps for regions lacking of ground based system to monitor rainfall and river discharge. Giustarini et al. \cite{Giustarini2015} combined multiple annual satellite observations and continuous spatially-distributed hydrodynamic model to map flood hazard with a high spatial resolution. According to the study on the UK Severn River flood, their method exhibits advantages in high-resolution flood mapping against the reference map computed by the hydraulic modeling approach. Moreover, their method has merit on the flexibility, where any type of RS images can be included as the inputs. Kwak \cite{Kwak2017} derived the so-called synchronized floodwater index (SfWi) from annual time-series optical satellite data to detect the maximum extent of a nationwide flood. For the 2015 monsoon season, they revealed the propensity of flood risk in three major rivers by analyzing the spatio-temporal dynamics of the maximum flood extent. Flood areas suggested by SfWi are small but accurate. Rahman and Di \cite{Rahman2016} reviewed the applications of the state-of-the-art RS techniques for flood management. On flood risk assessment, RS can be used to assess flood risk, exposure and vulnerability. On flood emergency planning, RS has contributed to flood warning system, rescue and relief operation, post flood damage assessment, and policy making.

The production of high-precision flood maps requires the integration and classification of information coming from different RS data sources. D'Addabbo et al. \cite{DAddabbo2018} applied Bayesian networks to monitor flood based on multi-temporal and multi-sensor RS data. They used Bayesian networks to perform a data fusion procedure of different types of satellite imagery and ancillary data. Other open-access data can also supplement commercial satellite imagery and ground-based data in developing regions where relevant data are usually sparse. Ekeu-wei and Blackburn \cite{Ekeu2018} introduced applications of open-access RS data (such as altimetry, DEMs, optical and radar images) on mapping flood. Using Nigeria as a case study, they evaluated the significance of open-access datasets in flood risk assessment and found that open-source data provided by the private sector play a key role in assessing flood risk especially for data sparse regions. Readers are encouraged to read a recent book by Refice et al. \cite{Refice2018}, who reviewed the state-of-the-art RS techniques and useful tools for flood hazard monitoring.

\subsubsection{Mobile phones for emergency management}

Mobile phone (MP) is a useful tool not only to facilitate the daily communications between users but also to record their geographic positions with high spatial and temporal resolutions. As shown in the previous sections, MP data is powerful in inferring individual socioeconomic status, analyzing offline human mobility and exploring online activity patterns. However, these quantitative features of human activities are usually under normal and stationary circumstances. Yet, human behavioral patterns are intuitively different under unfamiliar situations, especially during emergent events. Therefore, human activity patterns revealed by MP data have promising applications on detecting, monitoring and managing emergent events \cite{Gething2011}. On the one hand, analyzing information flow recorded by MPs can help detect emergent events in real time. People under extreme circumstances will change their calling and behavioral patterns (see Ref. \cite{Jiang2013} for calling and behavioral patterns under normal circumstances). Dramatic behavioral changes can be treated as signals of emergent events, with high-resolution time and locations. On the other hand, MP data can help provide real-time emergency monitoring by analyzing mobility patterns. In a word, MP data can be used to quantify human emergency behavior and tack population evacuation.

Large-scale MP datasets have potential applicability in real-time situation awareness and emergency detection. Bagrow et al. \cite{Bagrow2011} explored human societal response to external perturbations based on a country-wide dataset of MP communications covering about ten million users. They compared the real-time changes in mobile communication patterns between eight emergencies and eight non-emergencies, and found that calling activities under emergencies spike rapidly and decay immediately after the event, and the call volume decays exponentially with the spatial distance. Moreover, people affected by emergencies propagate the emergency information rapidly and globally through social networks. Moumni et al. \cite{Moumni2013} explored social response to the 2012 Oaxaca earthquake in Mexico based on call detail records (CDRs) for two weeks. They analyzed four different variables including call volume, call duration, social activity and mobility, showing three stages of the social response: a spike of many short calls in five minutes after the earthquake, a reduced activity of short calls lasting one to two hours and a moderate increase in call and duration volumes lasting about 5 hours. Moreover, users tend to moderately increase mobility during the earthquake.

MP data can be used to explore collective call behaviors and information flow following emergencies. Based on MP activities of about 10 million users in an European country, Gao et al. \cite{Gao2014b} studied the information spreading and the changes of users' communication patterns during emergencies. In consistent with the previous findings \cite{Bagrow2011,Moumni2013}, they observed a sharp increase in call volume during an emergent event. To explain the volume spikes, they analyzed reciprocal communications by decomposing them into call-forward and call-back. They found that call-back response and dissemination of emergency information have an effect on the magnitude of volume spikes but the former is the dominant component. Moreover, the observed reciprocal communications, in particular, the call-back response during emergent events, can not be explained by the inherent reciprocity in social networks under normal circumstance. Taking three days' CDRs during the 2012 Xinjiang earthquake in China, Yu et al. \cite{Yu2015} analyzed the collective call patterns of people who experienced the earthquake. They found that earthquake significantly increased many indices of call patterns such as call volume and call duration. In particular, call volume gets increased more significantly. From a spatial perspective, people made more local calls on the earthquake day. From a temporal perspective, local call volume raised rapidly within two hours after the earthquake, and large volume of distant calls last a whole day.

Calling behavior changes during flooding events can be tracked by MPs. By analyzing CDR data and RS data of the 2009 Tabasco floods in Mexico, Pastor-Escuredo et al. \cite{Pastor2014} revealed abnormal human communication patterns during and after the events. They identified the floods from satellite images and reconstructed a flood impact map. To detect abnormalities in the communication activity from the CDR data, they calculated the variation metric at the cell tower by comparing the MP activity $x(t)$ during the disaster against their characteristic variation obtained during the baseline (BL) period. Formally, the variation metric $x_{\text{norm}}$ is defined by
\begin{equation}
x_{\text{norm}}=\frac{x(t)-\mu_{\text{BL}}}{\sigma_{\text{BL}}},
\end{equation}
where $\mu_{\text{BL}}$ and $\sigma_{\text{BL}}$ are respectively the mean value and standard deviation of the activity during the baseline period. They found that the variation metric spikes in most flooded areas and shows consistence to the flood impact map, suggesting that the CDR-based variation metric can be used for situation awareness. Based on CDRs of Senegal, Hong et al. \cite{Hong2018} explored how different levels of floods affect MP call volumes and communication network features. They found that people increase their calling activities during floods, and the call volumes are positively correlated with the flooding intensity. Moreover, large cities with more recurring floods have more introversion communications within their neighborhoods, suggesting that people living in larger cities rely more on local communities.

Rapid emergency detection based on MP data can facilitate humanitarian response and reduce the toll of extreme events. Based on the combined data of MP activities and official event records in Rwanda, Dobra et al. \cite{Dobra2015} proposed an efficient system that can detect days with anomalous behavioral patterns under many emergent and non-emergent events. They found that days with increased anomalous behaviors suggest joyous events, while days with decreased anomalous behaviors suggest emergent events. Thus, the type of events can be identified by examining the increases and decreases in anomalous behaviors. Moreover, they confirmed that the behavioral responses have significant spatial and temporal variances. Recently, Gundogdu et al. \cite{Gundogdu2016} proposed an approach based on Markov modulated Poisson process to detect behavioral anomalies from CDRs of 5 million users. They found that different types of events have spatial and temporal differences in call volume changes, and emergent events can be distinguished from non-emergent events by comparing their associated calling patterns. Moreover, communication activity is more significant than mass movement in event detection.

\begin{figure}[t]
  \centering
  \includegraphics[width=0.6\textwidth]{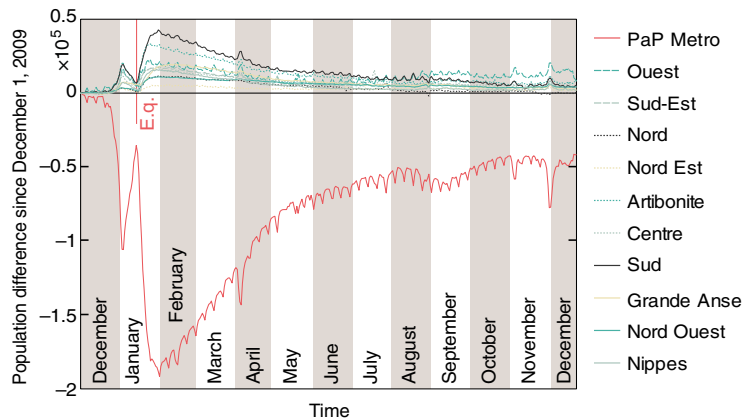}
  \caption{Population movements following the 2010 Haiti earthquake. The figure presents the changes in the number of individuals in the various provinces (color coded curves) before, during and after the earthquake. The red vertical line marks the date of the earthquake. The red curve highlights the capital PaP Metro. Figure from \cite{Lu2012}.}
  \label{Fig_6_2_2}
\end{figure}

In addition to analyzing communication behaviors, MP data have been increasingly used to model human mobility perturbations, assess population displacements and improve emergency responses during large-scale disasters. The subscriber identify module (SIM) cards are useful in tracking human movements in real time. Using records of 1.9 million SIM cards before and after the 2010 Haiti earthquake, Bengtsson et al. \cite{Bengtsson2011} estimated the trends and magnitudes of population movements after the January earthquake and during the October cholera outbreak. They found that SIM cards could provide valid estimates of the distribution, magnitude and trends in population displacements. In particular, the geographic distribution of the estimated population moving out of the capital city Port-au-Prince (PaP) is highly heterogeneous and consistent to the retrospective survey data, while the destination of population movements out of the cholera outbreak areas is heavily concentrated. Based on the same dataset, Lu et al. \cite{Lu2012} explored the predictability of population displacements after the Haiti earthquake. They found that the population in PaP decreases by 23\% in the three-month period after the earthquake due to population movements (see Figure~\ref{Fig_6_2_2}). Although people' mobility patterns sharply changed with a growth in travel distances and trajectory sizes, the predictability of population movements remains high. In particular, they found a high correlation between the destinations of people who left PaP during the first three weeks and their mobility patterns during normal times. People tend to move to locations where they have significant social bonds. The results suggest a high-level predictability of human spatial regularities even under extreme events \cite{Kenett2012}, which will help predict disaster responses and manage population movements.

Accurate prediction of human emergency behaviors is critical to disaster management and societal reconstruction. Song et al. \cite{Song2013} constructed an enormous set of GPS mobile sensor data recording about 1.6 million people's activities during the 2011 Japan Earthquake and the Fukushima nuclear accident. They revealed the short-term and long-term evacuation behaviors after the disasters. For example, the population in affected areas substantially decreased by more than 50\% during the first 12 days and stabilized 81 days after the earthquake. Further, they trained a general probabilistic model to automatically predict population evacuations in affected cities. Their method can find new mobility features and predict large population movements with an accuracy about 0.80. Song et al. \cite{Song2014} further found that human emergency mobility after disasters is correlated with their mobility patterns during normal times and highly affected by many factors including the intensity of disaster and social relationships. Taking into account the above factors, they developed a model that performs better than some well-know baseline models on predicting human emergency mobilities. The model outperforms their previous method \cite{Song2013} by up to 11.08\% in human mobility prediction. These results suggest that human mobility under extreme events is more predictable than our intuition.

MP data have also been combined with other data sources to analyze human behavioral changes and assess population displacement during emergencies. Based on MP records and satellite images, Bharti et al. \cite{Bharti2015} analyzed population movements during the 2010 C\^ote d'Ivoire internal political conflict. The two datasets strongly agree with each other on estimating average population sizes pre- and post-conflict, and they complement each other on estimating long- and short-term population dynamics throughout the crisis. Based on CDRs of 12 million MP users, Wilson et al. \cite{Wilson2016} explored the rapid assessments of the national-level population displacements after the 2015 Nepal Earthquake. They uncovered the patterns of return to earthquake affected areas. An estimated number of 390,000 people in earthquake affected areas evacuated immediately to surrounding areas, in particular to those with high populations. Most people will gradually return to their hometown after the earthquake, while less than 15\% people were still away from their home three months after the earthquake. Ghurye et al. \cite{Ghurye2016} analyzed the changes of mobility patterns and communication behaviors based on CDRs during the 2012 Rwanda flood season and found that disasters disrupted mobility patterns. During the first three weeks, the number of victims who left their hometown reaches its peak at about 10 millions, and the recovery to normal patterns takes over two months.

\subsubsection{Social media for situational awareness}

When people facing extraordinary events, their collective attentions and behaviors will emerge on social media (SM). By tracking Twitter users' hashtags, He and Lin \cite{He2017b} quantified the shift of human collective attentions under exogenous shocks. They found that the co-occurrence network of users' hashtags exhibits a strong community structure before the event, while a few hashtags will suddenly appear in many tweets and thus become hubs after the event. Sano et al. \cite{Sano2013} analyzed the keyword appearance rate based on more than 1.8 billion Japanese blog entries. They found that the functional forms of decay and growth of keyword appearance that peaked on a certain day exhibit power laws with the various exponents values between $-0.1$ and $-2.5$. In particular, the absolute exponent value is less than one for some keywords of news during extraordinary events.

SM is a valuable source of information for gaining situational awareness, detecting and locating emergent events, improving disaster response and enhancing relief efforts. Yet, the extremely high volume of messages generated during crises urges for automatic methods that can extract relevant and valuable disaster-related information from SM posts \cite{Wang2018b}. For example, Kireyev et al. \cite{Kireyev2009} collected two datasets from Twitter during earthquakes and applied topic models to analyze earthquake-related tweets. Imran et al. \cite{Imran2013} utilized the state-of-the-art machine learning techniques to extract disaster-related information from tweets generated during the 2011 Joplin tornado. Based on a large set of tweets, Olteanu et al. \cite{Olteanu2014} created a lexicon consisting of crisis-related terms that frequently appear in messages posted during various crisis situations. Such a crisis lexicon can be used to improve the recall but maintain the precision in the sampling of crisis-related tweets. Further, they showed how to automatically identify the terms describing a given crisis based on the crisis lexicon. For disaster response and relief, Ashktorab et al. \cite{Ashktorab2014} proposed a Twitter-mining tool named Tweedr that can rapidly extract relevant information from tweets posted during disasters. The Tweedr pipeline has three phases, where disaster-related tweets are identified in the classification phase, similar tweets are merged in the clustering phase, and tokens and phrases of damage information are extracted in the extraction phase. The Tweedr can identify 12 crises events occurred in the US since 2006.

The utilization of SM data has transformed the methodology of earthquake detection and early warning \cite{Allen2012}, where the distribution of shakings can be mapped in minutes from earthquake-related posts. Acar et al. \cite{Acar2011} studied earthquake information sharing on Twitter by analyzing the tweets posted near two disaster-struck areas during the 2011 Tohoku Earthquake. They found that people in directly affected areas tweeted to announce their uncertain and unsafe situation, while people in remote areas tweeted to inform followers that they are safe. Toriumi et al. \cite{Toriumi2013} analyzed 360 million pre- and post-disaster tweets for the 2011 Tohoku Earthquake. They found that users changed their main purpose of using Twitter from communication to information sharing after the disaster. In particular, critical information was widely retweeted while non-emergency tweets were decreased. For the 2012 Indonesia Earthquake, Chatfield and Brajawidagda \cite{Chatfield2012} identified 6,383 earthquake-related tweets and performed a social network analysis of the information flows on Twitter. They showed that Twitter can be utilized as an early warning network, where the followers of governmental agencies will retweet warnings immediately. Dong et al. \cite{Dong2018} analyzed information diffusion on Weibo after two earthquakes in China. They found that strangers play an important role in spreading earthquake-related news, and verified users dominantly influence information diffusion on Weibo.

People post many earthquake-related messages on SM soon after they feel shaking. Social media users indeed serve as social sensors with their posts being the sensory information. Sakaki et al. \cite{Sakaki2010} proposed a method to detect target events by leveraging the real-time and geographical nature of Twitter. They employed a support vector machine (SVM) to classify tweets related to target events and proposed a probabilistic spatio-temporal model for each event. Then, they estimated the centers of events by applying location estimation methods such as particle filtering \cite{Hightower2004}. Their method can promptly detect 93\% of earthquakes in Japan. Later, Sakaki et al. \cite{Sakaki2013} developed a system based on the above method, which can send alarm e-mails about an earthquake to relevant people much faster than the official agency. Similarly, Earle et al. \cite{Earle2012} developed an earthquake detection algorithm relying solely on tweets. They found that the peak of tweets containing ``earthquake'' is correlated with the event time. A short-term-average over long-term-average algorithm can effectively detect 48 globally-distributed earthquakes based on five months' tweets. Their algorithm runs very fast, with 75\% of detections accomplished within two minutes of the event time. For Australia and New Zealand, Robinson et al. \cite{Robinson2013} built a sensitive earthquake detector by monitoring earthquake-related tweets. They located earthquakes by examining tweets that contributed to an alert. Their detector can identify earthquake with an accuracy of about 0.81 in terms of F1 score at the best case. SM has multi-level functionalities during earthquakes such as interpersonal communications and information sharing \cite{Jung2014}, and thus it can be utilized to design effective monitoring and warning systems.

\begin{figure}[t]
  \centering
  \includegraphics[width=0.8\textwidth]{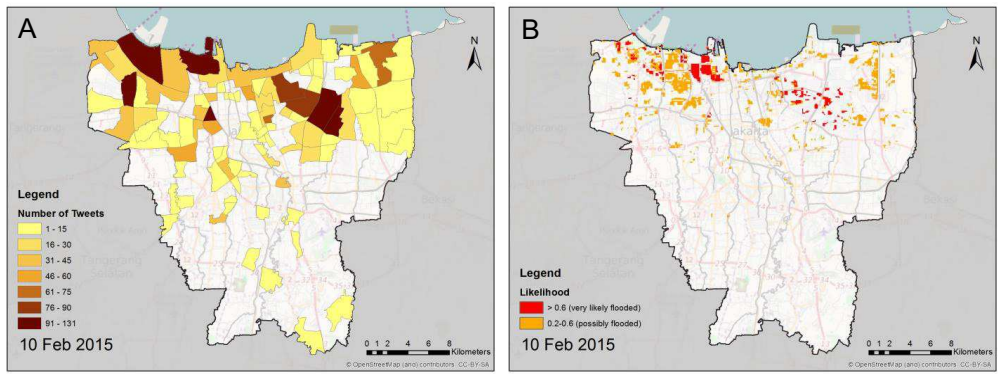}
  \caption{Mapping floods based on Twitter data. (A) Total number of flood-related tweets that contain a water depth. (B) Derived flood likelihood map. Figure from \cite{Eilander2016}.}
  \label{Fig_6_2_3}
\end{figure}

SM data have been increasingly used in monitoring and mapping floods in a timely manner. Vieweg et al. \cite{Vieweg2010} analyzed tweets containing case-insensitive terms of the 2009 Red River Floods. They identified some high-level situational features of information generated during emergencies. Similarly, Cheong and Cheong \cite{Cheong2011} analyzed tweets generated during the Australian 2010-2011 floods and revealed interesting features of interactions between Twitter users during the crisis. They found that local authorities, officials and volunteers are influential players of the online communities. Later, de Albuquerque et al. \cite{DeAlbuquerque2015} analyzed tweets generated during the 2013 River Elbe Flood. They found that the appearances of flood-related tweets show a general spatial pattern that flood-related tweets are strongly correlated with distance to flood events and relative water level. Many independent observations reporting the same flood on Twitter are more reliable. Following this principle, Eilander et al. \cite{Eilander2016} explored physical characteristics of floods and applied filtering and geo-statistical methods to assess the reliability of tweets over all flooded areas based on multiple observations. They developed an approach to construct a flood probability map. When tested in Jakarta, their approach can detect 93\% of flood locations in the regions where people tweeted about water depth (see Figure~\ref{Fig_6_2_3}). Recently, Arthur et al. \cite{Arthur2018} leveraged tweets to detect and locate flood events in the UK. They collected tweets containing flood-related terms and located flood events by analyzing many indicators such as mentioned place names and GPS coordinates. They produced high-quality flood event maps based on the relevant geotagged tweets and validated the flood maps by official data. Similarly, Li et al. \cite{Li2018} identified the spatio-temporal patterns of flood-related tweets for the 2015 South Carolina floods. They further proposed a kernel-based model to map the possibility of floods based on the water height mentioned in tweets.

In addition to tweets, data from other SM platforms can also used to map and predict floods. Tkachenko et al. \cite{Tkachenko2017} analyzed tags of Flickr images uploaded during floods. They found that flood-related tags are correlated with hydrologically themed tags and then revealed connections between risk-signalling and generic environmental semantics. Further, they showed that environmental semantics derived from volunteered geographic data are helpful for improving flood warning. For the 2014 UK flood, Rosser et al. \cite{Rosser2017} collected geotagged photographs from Flickr and satellite imagery. They developed a Bayesian statistical model to estimate the probability of flood inundation by combining SM and remote sensing (RS) data. Their method can effectively predict spatial flood inundation with an AUC value over 0.93. Wang et al. \cite{Wang2018} collected flood-related tweets from Twitter and flooding photos from the crowdsourcing platform MyCoast. They applied natural language processing (NLP) methods \cite{Manning1999} to extract location names and flood depth from tweets and then employed convolutional neural networks (CNN) \cite{Lecun2015} to automatically classify flooding photos. They found that the combination of tweets and crowdsourcing photos can provide high-resolution flood monitoring. Recently, observations on many SM platforms have been used to map and model floods \cite{Assumpcao2018}, showing the valuable means of rapidly estimated flood maps in improving situational awareness after floods.

\begin{figure}[t]
  \centering
  \includegraphics[width=0.8\textwidth]{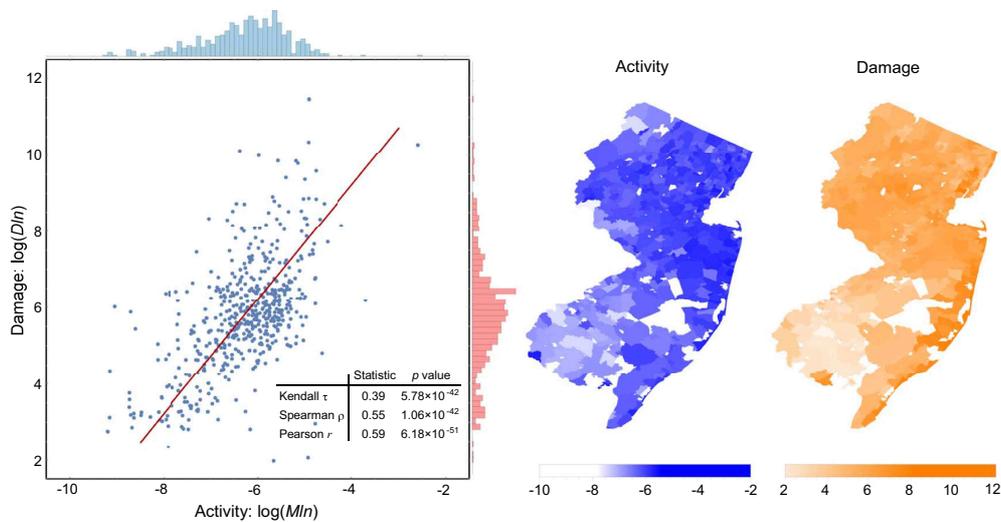}
  \caption{Correlations between Twitter activity and Hurricane Sandy damage and their spatial distributions. The per-capita Twitter activity and damage both follow a quasi log-normal distribution. The correlations between activity and damage is shown for New Jersey. Figure from \cite{Kryvasheyeu2016}.}
  \label{Fig_6_2_4}
\end{figure}

The 2012 Hurricane Sandy disaster is one of the costliest disasters in the US history. Preis et al. \cite{Preis2013} studied users' attention to Hurricane Sandy by examining photos on Flickr with related tags, titles or descriptions. They found that the moving average of the normalized number of photos related to Hurricane Sandy (under one day window) has a striking correlation (Kendall's $\tau = -0.37$) \cite{Kendall1938} with the atmospheric pressure in New Jersey during the hurricane period, suggesting Flickr as a real-time sensor to track collective attention during emergencies. Based on 9.7 million geotagged tweets containing hurricane-related keywords, Kryvasheyeu et al. \cite{Kryvasheyeu2016} performed a multiscale analysis of Twitter activity before, during and after Hurricane Sandy. They found that the number of tweets containing keywords is strongly correlated with the peak of wind power on the day of hurricane landfall, and the magnitude of tweet activity increases with the proximity to the hurricane path. Moreover, there is a moderate correlation between the per-capita Twitter activity and the economic damage inflicted by Hurricane Sandy. The Kendall and Spearman rank correlations respectively reach 0.39 and 0.55 at the level of zip code tabulation areas (ZCTA) for New Jersey (see Figure~\ref{Fig_6_2_4}). Moreover, they found that sentiment in Twitter is a weak predictor of economic damage. Gruebner et al. \cite{Gruebner2017} sentimentally analyzed 344,957 geotagged tweets from the greater New York City (NYC) area and extracted basic emotions during Hurricane Sandy. They found that sadness is the most pronounced emotion during the whole hurricane period, while surprise and fear peak on the day of hurricane landfall. Further, they detected space-time clusters of excess risk of multiple basic emotions. Their work has taken a significant step towards improving mental health surveillance after disasters.

SM data have been applied to quantify human mobility perturbation and to improve resilience analysis of hurricane disasters. Wang and Taylor \cite{Wang2014c} studied the perturbation and resilience of human movement after Hurricane Sandy landed in NYC by analyzing tweets. They found that long-distance movements are significantly decreased during the first day. Moreover, human displacements follow a truncated power-law distribution with power-law exponent being 1.73 in the first day and being around 1.19 in the following 11 days. These results suggest that human mobility patterns are significantly perturbed by Hurricane Sandy. After further calculating the radius of gyration and the center of mass of each individual's movement trajectory \cite{Gonzalez2008}, they found a high correlation between values of these human mobility parameters under perturbation and those in steady states. These findings imply an inherent resilience of human mobility patterns during hurricanes and may be helpful for developing strategies to enable more effective evacuations. Based on content of geotagged tweets, Middleton et al. \cite{Middleton2014} developed a real-time crisis-mapping platform at the building- and street-level resolutions. After applied to Hurricane Sandy, their platform can produce the crisis mapping with an F1 value 0.53 compared to the official post-event impact assessment. Recently, Zou et al. \cite{Zou2018} analyzed Twitter activities during Hurricane Sandy and showed an improved estimation on post-hurricane damage.

Mining real-time and large-scale SM data can also provide rapid situational awareness with precise locations for many other disasters, such as drought, wildfires and snowstorms. Tang et al. \cite{Tang2015} found that governmental agencies used SM platforms as communication channels for information sharing during the 2014 California drought. In particular, Twitter plays an important role in expediting drought risk information dissemination. Boulton et al. \cite{Boulton2016} tested the feasibility of mapping wildfires from SM data. After analyzing wildfire-related posts collected from Twitter and Instagram, they found a positive correlation between SM activity and officially reported fire occurrences. Specifically, the correlations are $r=0.529$ for Twitter and $r=0.716$ for Instagram. Further, they analyzed the spatial and temporal features of the wildfire-related tweets and found that hotspots of wildfire-related Twitter activity are very likely to be the wildfire locations. Hong et al. \cite{Hong2017} analyzed geotagged tweets generated during 18 snowstorms on the US east coast. They found that local governments used Twitter mainly for preparedness and response, while citizens used Twitter mainly for sharing their opinions about snowstorms. As a summary, SM is a rich data source that can be utilized to monitor disaster events, fasten damage assessment, improve disaster management and reduce economic losses.

\section{Discussions}
\label{Sec6}

Ranging from international trade networks and global inequality of household income and individual wealth, from perception and prediction of socioeconomic status to prevention and protection of public health and natural disaster, through rich examples, this review shows how to dig out novel insights about socioeconomic development based on large-scale real data and advanced analysis tools like data mining and machine learning. These insights are usually not easy to be obtained by traditional methods.

At the same time, as an emerging branch of science, the studies of computational socioeconomics still confront some shortcomings and challenges. Methodologically speaking, the problems are twofold.

Firstly, the data quality, especially the authenticity of the data, cannot be fully guaranteed. In most world-known social media platforms, a considerable fraction of users are not human beings but artificial bots (named as \emph{social bots} in literature) \cite{Ferrara2014}. These bots do not just simply bring some noises, but they may be controlled by some commercial or academic institutions, and be assigned with some certain tasks. Therefore, these bots can largely affect the tendency of public sentiment and accelerate the spreading of rumors \cite{Vosoughi2018}. Accordingly, the results obtained from data sets polluted by social bots are probably far different from the reality. In addition, when a certain project has successfully drawn world-wide attention to its results (e.g., how to detect influenza epidemics using Google search query data \cite{Ginsberg2009}), some players in academic community may intentionally generate artificial data to disturb the reported models and algorithms (maybe not for manipulation, but for fun). Such smart jokes will also reduce the data authenticity and result in incorrect conclusions \cite{Lazer2014}. In a word, data are not collected from the pure land, and thus we'd better implement effective pretreatment to improve the quality of data before analysis and take into account the possible risks caused by noisy and unreal data before reporting any results to the public.

Secondly, the applicability and relevance of results are limited. First of all, through some data resources can cover a huge number of samples (even scaling with the whole population, such as mobile phones, Facebook, Twitter, WeChat, Weibo, etc.), these samples are not drawn randomly and thus cannot represent the whole population. For example, children, old and poor people are less engaged in the Internet and mobile Internet, resulting in the less chance to cover them by the above-mentioned data resources. Therefore, we have to carefully clarify the range of validity of the findings \cite{Ruths2014}. Secondly, socioeconomic problems are highly affected by local landscapes of religion, culture and politics, and thus a certain conclusion validated in one region may be not applicable in some other regions. For example, by analyzing the religion network consisted of believers in Weibo, Hu et al. \cite{Hu2019} showed that religions in China are highly segregated, while about a half cross-religion links in Weibo are related to charitable issues. However, since the organization of religions in China is different from many other countries, whether the findings reported in Ref. \cite{Hu2019} are suitable for other countries asks for further validations. Lastly, even for the same dataset, the statistical regularities observed at the group level cannot be indiscriminately imitated at the individual level, or vice versa \cite{Yang2017}. This is very similar to the Simpson's paradox \cite{Good1987} in statistics, where a trend appears in several different groups of data will disappear or reverse when these groups are combined.

In addition to the aforementioned shortcomings, we think the following five open issues are worth for future studies.

Firstly, we should try to design novel indices with a high ability in explanation and prediction. Currently, a considerable fraction of studies in computational socioeconomics applied new data resources and advanced analysis tools to estimate and predict routine socioeconomic indicators, such as GDP \cite{Tacchella2018}. Such works are very valuable and easy to be accepted by the traditional socioeconomic community, while an obvious drawback is that any new methods cannot outperform the original statistical methods corresponding to the target indices. For example, even a complicated and advanced algorithm based on large-scale multiple-resource data can achieve an accuracy of 0.99 in estimating GDP, it is still worse than the original statistical method for GDP, whose accuracy is 1. Regarding this point, the designs of Economic Complexity Index \cite{Hidalgo2009} and Fitness \cite{Tacchella2012} are successful examples. Very recently, satellite-based remote sensing data are also used to measure the level of manufacturing (see, for example, the China Satellite Manufacturing Index produced by SpaceKnow \cite{Reinstein2019}). In a word, in addition to using novel data sources to estimate known data, we'd better design some new indices that make use of the novel datasets and well reflect valuable and important socioeconomic status behind the data. This issue still needs more attention and effort.

Secondly, it is valuable to re-evaluate the correctness and validation of traditional socioeconomic theories by some data-driven methods. Many traditional socioeconomic theories originate from simplified models of socioeconomic insights or extensions of known theories, which have not been validated by real data or have just been validated by small datasets in a highly limited range. It lacks comprehensive evaluation based on large-scale data that support in-depth analysis across different social formations, political systems, ideologies, culture traditions, economic levels, and so on. Using novel methods to critically evaluate traditional theories can also help computational socioeconomics to attract sufficient attention from traditional socioeconomics and thus to accelerate the integration of socioeconomics and computational socioeconomics.

Thirdly, we should reveal the underlying causal relationships and provide theoretical insights. The majority of findings reported in this review are only about the correlations between data features and targeted metrics. No matter how strong the correlation is and whether the data features are successfully utilized to predict the tendency of some targeted metrics, if one cannot find solid and robust causal relationships according to the inspiration from correlations, the theoretical value is limited \cite{Duncan2012}. Only if some future studies in computational socioeconomics could find out significant causal relationships via in-depth analysis of large-scale data, and then build up theoretical models that can withstand strict evaluations, the foundation of computational economics is considered to be solid.

Fourthly, we need to verify newly reported theories by controlled experiments. Controlled experiments play a central role in the methodology of socioeconomics and sociology. Although this review emphasizes on the studies based on unobtrusive data, it does not suggest the abandonment of controlled experiments. In fact, to design and implement controlled experiments is a very effective and sometimes necessary way towards uncovering solid causal relationships. Different from routine experiments, we can launch experiments covering huge population through the Internet and/or mobile Internet (e.g., Bond et al. \cite{Bond2012} reported a 61-million-person experiment about social influence and political mobilization via Facebook). We can analyze large-scale data to find out potentially key factors, which can be utilized to better design and guide the following experiments.

Fifthly, we should try our best to find applications of theoretical and empirical analyses in practice. In despite of some effective methods, a couple of significant metrics and a few interesting conclusions developed by computational socioeconomics, overall speaking, the majority of known works have not found enough real applications or just generated some armchair strategies. However, the long-term value and vitality of computational socioeconomics depend on whether it can at least lead to some successes in practice. We suggest building up a system that can provide high-resolution and real-time monitoring about socioeconomic status and thus can detect possible risks at an early stage, so that governmental administrators, enterprisers, investors and other related people can make better decisions accordingly. This monitoring system should make use of multiple data sources, including the novel sources like mobile phones, satellites and social media platforms and the traditional sources like the data from economic censuses, questionnaire surveys and statistical yearbooks. Of particular importance, we suggest putting forward a series of data-driven policy suggestions (for example, the work by Alshamsi et al. \cite{Alshamsi2018} is a representative attempt but still far from practical policies) and implementing these suggested policies in some qualified regions (even though the most critical issue is whether the heads of the corresponding governments support such blaze new trails), which is the way having the chance to make really big contributions. The classical economics failed to provide effective prescriptions to the world-wide poverty (for example, looking at the results of the huge efforts by the World Bank in fighting with the poverty of these so-called developing countries after the World War II \cite{Danaher1994,Pincus2002}), while if computational socioeconomic studies can find a new path to help impoverished people, its contribution to the world is tremendous.

We believe this review offers many valuable enlightenments to researchers in socioeconomics and other branches of social science. However, besides the booming development of computational socioeconomics, one should calmly notice that many important methods and conclusions introduced here still have not been accepted as a part of socioeconomics. Indeed, researchers doing works related to computational socioeconomics, currently being scattered in many disciplines, have not yet been seriously treated as challengers to the traditional socioeconomic methodology. In a word, we are still at the very early stage of the paradigm shifting of social science driven by big data and artificial intelligence. The way towards a quantitative version of social science is long and rugged, but undoubtedly, changes are happening and becoming increasingly fierce.

\section*{Acknowledgements}
\addcontentsline{toc}{section}{Acknowledgements}

This work was partially supported by the National Natural Science Foundation of China (Grant Nos. 61433014, 61603074, 61673086 and 61703074) and the Science Promotion Programme of UESTC (No. Y03111023901014006). The authors acknowledge the support of the Swiss National Science Foundation (Grant No. 182498) during this collaboration. J.G. acknowledges the China Scholarship Council for a scholarship (No. 201606070051) and the Collective Learning Group at the MIT Media Lab for hosting.

\addcontentsline{toc}{section}{References}

\bibliographystyle{elsarticle-num}

\bibliography{ref}

\end{document}